\documentclass[12pt]{article}
\usepackage{jheppub}
\usepackage{ytableau}
\usepackage{comment}
\usepackage{array}
\usepackage{longtable}
\usepackage[vcentermath]{youngtab}
\usepackage{tikz}
\usetikzlibrary{calc,arrows.meta,decorations.pathreplacing}

\tikzset{
  gauge/.style={circle,draw,fill=yellow!20,minimum size=8.6mm,inner sep=0.5pt,font=\small},
  flavor/.style={rectangle,draw,fill=cyan!10,minimum size=7mm,inner sep=1.5pt,font=\small},
  qdots/.style={inner sep=0.5pt,font=\small},
  mirror/.style={{Latex[length=2.2mm,width=3.6mm]}-{Latex[length=2.2mm,width=3.6mm]},line width=1.8pt,black},
  quiver/.style={baseline=(current bounding box.center)}
}
\newcommand{\qcell}[1]{\raisebox{0pt}[\dimexpr\height+5pt\relax][\dimexpr\depth+4pt\relax]{\scalebox{0.64}{#1}}}
\newcommand{\qSQCD}[2]{\begin{tikzpicture}[quiver]
\node[gauge](a) at (0,0){$#1$}; \node[flavor](b) at (1.35,0){$#2$}; \draw (a)--(b);
\end{tikzpicture}}
\newcommand{\qSQCDmirror}[1]{\begin{tikzpicture}[quiver]
\node[gauge](a) at (0,0){$1$}; \node[qdots](b) at (1.15,0){$\cdots$}; \node[gauge](c) at (2.3,0){$N$};
\node[qdots](d) at (3.45,0){$\cdots$}; \node[gauge](e) at (4.6,0){$1$}; \node[flavor](f) at (2.3,-1.25){$#1$};
\draw (a)--(b)--(c)--(d)--(e); \draw (c)--(f);
\end{tikzpicture}}
\newcommand{\qSQCDmirrorNB}{\begin{tikzpicture}[quiver]
\node[gauge](a) at (0,0){$1$}; \node[qdots](b) at (1.15,0){$\cdots$}; \node[gauge](c) at (2.3,0){$N$};
\node[qdots](d) at (3.45,0){$\cdots$}; \node[gauge](e) at (4.6,0){$N$}; \node[qdots](g) at (5.75,0){$\cdots$};
\node[gauge](h) at (6.9,0){$1$}; \node[flavor](f1) at (2.3,-1.25){$1$}; \node[flavor](f2) at (4.6,-1.25){$1$};
\draw (a)--(b)--(c)--(d)--(e)--(g)--(h); \draw (c)--(f1); \draw (e)--(f2);
\end{tikzpicture}}
\newcommand{\qTSUN}{\begin{tikzpicture}[quiver]
\node[gauge](a) at (0,0){$1$}; \node[qdots](b) at (1.15,0){$\cdots$}; \node[gauge](c) at (2.3,0){$N\!-\!1$};
\node[flavor](d) at (3.65,0){$N$}; \draw (a)--(b)--(c)--(d);
\end{tikzpicture}}
\newcommand{\qADHM}[1]{\begin{tikzpicture}[quiver]
\node[gauge](a) at (0,0){$N$}; \node[flavor](b) at (1.4,0){$#1$}; \draw (a)--(b);
\draw (a.110) .. controls +(-0.45,0.9) and +(0.45,0.9) .. (a.70);
\end{tikzpicture}}
\newcommand{\qNecklace}{\begin{tikzpicture}[quiver]
\foreach \i/\a in {1/90,2/162,3/234,4/306,5/18} \node[gauge](n\i) at (\a:1.35){$N$};
\node[flavor](f) at (0,2.6){$1$};
\draw (n1)--(n2)--(n3); \draw (n4)--(n5)--(n1); \draw[dotted,thick] (n3)--(n4); \draw (n1)--(f);
\end{tikzpicture}}
\newcommand{\qUSpADHM}{\begin{tikzpicture}[quiver]
\node[gauge,font=\scriptsize,inner sep=0.3pt](a) at (0,0){$USp(2N)$}; \node[flavor,font=\scriptsize](b) at (1.9,0){$SO(2l)$}; \draw (a)--(b);
\draw (a.115) .. controls +(-0.5,1.0) and +(0.5,1.0) .. (a.65);
\end{tikzpicture}}
\newcommand{\qDquiver}{\begin{tikzpicture}[quiver]
\node[gauge](n0) at (-1.05,0.9){$N$}; \node[gauge](n1) at (-1.05,-0.9){$N$};
\node[gauge](n2) at (0,0){$2N$}; \node[qdots](d) at (1.15,0){$\cdots$}; \node[gauge](n3) at (2.3,0){$2N$};
\node[gauge](n4) at (3.35,0.9){$N$}; \node[gauge](n5) at (3.35,-0.9){$N$};
\node[flavor](f) at (-2.4,0.9){$1$};
\draw (f)--(n0)--(n2)--(d)--(n3)--(n4); \draw (n1)--(n2); \draw (n3)--(n5);
\end{tikzpicture}}

\makeatletter
\newcommand{\wtilde}[1]{\mathpalette\wtilde@{#1}}
\newcommand{\wtilde@}[2]{%
  \ifx#1\displaystyle\def\wt@s{1}\else\ifx#1\textstyle\def\wt@s{1}\else\def\wt@s{0.7}\fi\fi
  \tikz[baseline=(wt.base)]{%
    \node[inner sep=0pt,outer sep=0pt](wt){$\m@th#1#2$};%
    \draw[line width=\wt@s*0.45pt,line cap=round]
      let \p1=(wt.north west), \p2=(wt.north east), \n1={\x2-\x1},
          \n2={\wt@s*min(1.2pt+0.012*\n1,3.2pt)}, \n3={\wt@s*2pt} in
      (\x1,\y1+\n3) .. controls (\x1+0.33*\n1,\y1+\n3+2.2*\n2) and (\x1+0.67*\n1,\y1+\n3-1.2*\n2) .. (\x2,\y1+\n3+\n2);%
  }}
\makeatother

\newcommand\be{\begin{equation}}
\newcommand\ee{\end{equation}}

\preprint{
}

\title{Mirror symmetry of $\mathcal{N}=(0,4)$ boundary conditions with Nahm poles}
\abstract{
We propose mirror symmetry between $\mathcal{N}=(0,4)$ boundary conditions with Nahm poles and Neumann boundary conditions with boundary Fermi multiplets
in 3d $\mathcal{N}=4$ non-Abelian gauge theories, including $U(N)$ SQCD, $T[SU(N)]$ and the ADHM theories.
We obtain the half-indices of the Nahm pole boundary conditions from those of the Dirichlet boundary conditions by Higgsing,
where the boundary fugacities are specialized along the principal embedding of $\mathfrak{su}(2)$.
We check that they agree with the half-indices of the dual Neumann boundary conditions.
In the twist limits they give closed-form formulas for the vacuum characters of the boundary vertex operator algebras of Costello and Gaiotto.
We find that the unflavored vacuum characters are eta-quotients times quasimodular forms, which we write explicitly in terms of Eisenstein series for all our examples.
}

\author{Tadashi Okazaki}

\emailAdd{tokazaki@seu.edu.cn}
\affiliation{
School of Physics and Shing-Tung Yau Center, Southeast University,\\
Yifu Architecture Building, No.2 Sipailou, Xuanwu district, Nanjing, Jiangsu, 210096, China}
\allowdisplaybreaks[1]
\begin{document}
\maketitle

\section{Introduction and summary}
\label{sec_intro}
Supersymmetric boundary conditions in 3d $\mathcal{N}=4$ gauge theories
which preserve 2d $\mathcal{N}=(0,4)$ supersymmetry have a remarkably rich structure.
In Type IIB string theory, 3d $\mathcal{N}=4$ gauge theories are realized by D3-branes
which are suspended between NS5-branes and intersect with D5-branes \cite{Hanany:1996ie}.
The $\mathcal{N}=(0,4)$ boundary conditions are realized by the D3-branes ending on
additional NS5- or D5-branes extending along different directions,
which we call NS5$'$- and D5$'$-branes \cite{Chung:2016pgt,Hanany:2018hlz,Gaiotto:2019jvo,Okazaki:2019bok}.
Since S-duality exchanges the NS5- and D5-branes as well as the NS5$'$- and D5$'$-branes,
it predicts dualities between the $\mathcal{N}=(0,4)$ boundary conditions of mirror pairs of 3d $\mathcal{N}=4$ theories
\cite{Intriligator:1996ex,deBoer:1996mp,Hanany:1996ie}.
Moreover, these $\mathcal{N}=(0,4)$ boundary conditions are expected to admit deformations compatible with one of
the two distinct topological twists of the bulk theories, i.e. the H- and C-twists \cite{Costello:2018fnz}.
With these deformations the boundary local operators form vertex operator algebras (VOAs) \cite{Costello:2018fnz}.

The half-indices of 3d supersymmetric field theories \cite{Gadde:2013wq,Yoshida:2014ssa,Dimofte:2017tpi} count the boundary local operators.
They provide stringent tests of the proposed dualities,
and their twist limits are related to the vacuum characters of the boundary VOAs \cite{Costello:2018fnz}.
In \cite{Okazaki:2019bok} the dualities of $\mathcal{N}=(0,4)$ boundary conditions were proposed for 3d $\mathcal{N}=4$ Abelian gauge theories and tested by computing the half-indices.
For SQED$_{N_f}$ and its mirror quiver gauge theory,
the Neumann boundary condition with boundary Fermi multiplets is dual to the Dirichlet boundary condition of the mirror theory.
Conversely, the Dirichlet boundary condition is dual to the Neumann boundary condition with boundary Fermi multiplets of the mirror theory.
When several D3-branes end on a single D5$'$-brane,
the 3d vector multiplet obeys the Nahm pole boundary condition \cite{Chung:2016pgt}.
Like its counterpart in 4d $\mathcal{N}=4$ super Yang-Mills (SYM) theories \cite{Gaiotto:2008sa},
it is a boundary condition for non-Abelian gauge theories
which allows the scalar fields in the vector multiplet to have a singular profile near the boundary.
Hence the dualities for non-Abelian gauge theories require the half-indices of the Nahm pole boundary conditions.
It was also argued in \cite{Costello:2018fnz} that the Dirichlet boundary conditions can be modified to the Nahm pole boundary conditions,
which should descend to the ``oper-like'' boundary conditions for the twisted theory.

In this paper, we propose a formula for the half-index of the $\mathcal{N}=(0,4)$ Nahm pole boundary condition
for 3d $\mathcal{N}=4$ gauge theories.
It is based on a three-dimensional analogue of the Higgsing procedure employed for the Nahm pole boundary conditions in 4d $\mathcal{N}=4$ SYM theory \cite{Gaiotto:2019jvo}
(see also \cite{Hatsuda:2024lcc,Hatsuda:2025yzp,Hatsuda:2026ysv}).
The procedure originates from the construction of the superconformal indices with surface defects in 4d $\mathcal{N}=2$ theories \cite{Gaiotto:2012xa}.

The Nahm pole boundary condition is realized as the IR endpoint of the RG flow triggered by the nilpotent vev of the boundary moment map
of the Dirichlet boundary condition.
As the starting point we use the half-index of the $\mathcal{N}=(0,4)$ Dirichlet boundary condition for non-Abelian gauge theories,
in which each boundary monopole sector is weighted by the bare monopole operators,
whose spin and charges, including the topological charges, are induced by the boundary anomaly \cite{Dimofte:2017tpi}
with the contribution of the vector multiplet (see section \ref{sec_Dirichlet}).
Since the vev breaks the boundary global symmetry, the half-index can only keep the fugacities of the symmetries preserved by the vev.
Accordingly, the boundary fugacities of the Dirichlet boundary condition are specialized along the principal embedding,
so that the components of the moment map with the vev carry no charges under these symmetries, including the spin and the R-symmetry.
These components give bosonic zero modes, which make the Dirichlet half-index divergent
in the boundary monopole sectors labeled by the dominant coweights.
Only these sectors contribute to the coefficient of the divergence, i.e. the residue at the pole,
which gives the Nahm pole half-index after the decoupled degrees of freedom are stripped off.
These degrees of freedom consist of the 2d $\mathcal{N}=(0,4)$ twisted hypermultiplets as the Goldstone modes of the broken boundary symmetry
and the 2d Fermi multiplets arising from the splitting of the bulk hypermultiplets into representations of the principal $\mathfrak{su}(2)$.
Thus we propose that the Nahm pole boundary conditions are obtained by Higgsing the boundary global symmetry of the Dirichlet boundary condition,
just as the Neumann boundary conditions with boundary degrees of freedom are obtained by gauging it \cite{Okazaki:2019bok}.
We test this proposal by comparing the resulting half-indices with those of the dual Neumann boundary conditions, which are computed independently.

Most of the contributions of the vector multiplet and the hypermultiplets cancel against
those of the Goldstone modes and the split Fermi multiplets respectively,
so that the half-index takes a simple form, which is written directly in terms of the gauge theory data (see section \ref{sec_UN_simple}).
For unitary gauge groups it is a multiple bilateral basic hypergeometric series,
in which each boundary monopole sector is weighted by the squared principal specialization of the Macdonald polynomial of each $U(N)$ gauge group divided by its norm,
with the parameters used for the half-indices of 4d $\mathcal{N}=4$ $U(N)$ SYM theory \cite{Hatsuda:2025yzp}.
The weights of the boundary monopole sectors are those of the Dirichlet boundary condition at the specialized boundary fugacities.
Only the sector of zero flux is different,
and it is given by the top components of the hypermultiplets and the Slodowy slice \cite{Gaiotto:2008sa} of the vector multiplet,
i.e. its fluctuations transverse to the nilpotent orbit (see section \ref{sec_Nahm_def}).

We propose and test the dualities of the $\mathcal{N}=(0,4)$ boundary conditions for the following four families of 3d $\mathcal{N}=4$ gauge theories.

\begin{enumerate}

\item \textbf{$U(N)$ SQCD with $N_f$ flavors and its mirror}

We consider $U(N)$ SQCD with $N_f\ge 2N$ flavors, which is a good theory in the sense of \cite{Gaiotto:2008ak}.
Its mirror is the linear quiver gauge theory with $N_f-1$ gauge nodes \cite{Intriligator:1996ex,Hanany:1996ie},\footnote{
In the quiver diagrams the circles and squares are the gauge and flavor nodes,
and the lines and the loops are the (bi)fundamental and adjoint hypermultiplets (see section \ref{sec_conv}).}
\begin{align*}
\qSQCD{N}{N_f}
\qquad\quad
\begin{tikzpicture}[quiver]
\node[gauge](a1) at (0,0){$1$}; \node[gauge](a2) at (1.3,0){$2$}; \node[qdots](a3) at (2.45,0){$\cdots$};
\node[gauge](a4) at (3.6,0){$N$}; \node[qdots](a5) at (4.75,0){$\cdots$}; \node[gauge](a6) at (5.9,0){$N$};
\node[qdots](a7) at (7.05,0){$\cdots$}; \node[gauge](a8) at (8.2,0){$2$}; \node[gauge](a9) at (9.5,0){$1$};
\node[flavor](f1) at (3.6,-1.3){$1$}; \node[flavor](f2) at (5.9,-1.3){$1$};
\draw (a1)--(a2)--(a3)--(a4)--(a5)--(a6)--(a7)--(a8)--(a9); \draw (a4)--(f1); \draw (a6)--(f2);
\draw[decorate,decoration={brace,amplitude=5pt}] (3.15,0.62) -- (6.35,0.62) node[midway,above=5pt,font=\small]{$N_f-2N+1$};
\end{tikzpicture}
\end{align*}
The $N_f-2N+1$ nodes $U(N)$ form a plateau, and one flavor is attached to each of its two ends.
For $N_f=2N$ the plateau consists of the single node $U(N)$ with two flavors.
We find two types of dualities of boundary conditions.
The Neumann boundary condition with boundary Fermi multiplets for SQCD is dual to
the boundary condition of the mirror quiver consisting of the Nahm pole boundary conditions for the non-Abelian nodes
and the Dirichlet boundary conditions for the Abelian end nodes.
Conversely, the Nahm pole boundary condition for SQCD is dual to
the Neumann boundary condition with boundary Fermi multiplets of the mirror quiver.
For $N_f>2N$ the node of SQCD is unbalanced,
and the Neumann boundary condition for SQCD requires additional boundary Fermi multiplets in the fundamental representation.
For $N=1$ the dualities reduce to those for SQED$_{N_f}$ and its mirror in \cite{Okazaki:2019bok}.

\item \textbf{$T[SU(N)]$}

The theory $T[SU(N)]$ is the linear quiver gauge theory
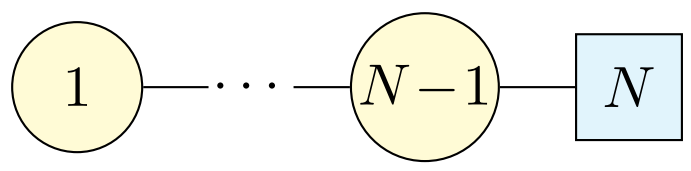
\begin{align*}
\qTSUN
\end{align*}
and it is self-mirror \cite{Gaiotto:2008ak}.
The Neumann boundary condition with boundary Fermi multiplets for $T[SU(N)]$ is dual to
the boundary condition for the mirror $T[SU(N)]$ consisting of the Dirichlet boundary condition for the $U(1)$ node
and the Nahm pole boundary conditions for the other nodes.
For $N=2$ it reduces to the duality in \cite{Okazaki:2019bok}.

\item \textbf{$U(N)$ ADHM theory with $l$ flavors}

The $U(N)$ ADHM theory has an adjoint hypermultiplet and $l$ fundamental hypermultiplets.
Its mirror is the necklace quiver gauge theory with $l$ nodes $U(N)$ \cite{deBoer:1996mp,Porrati:1996xi,deBoer:1996ck},
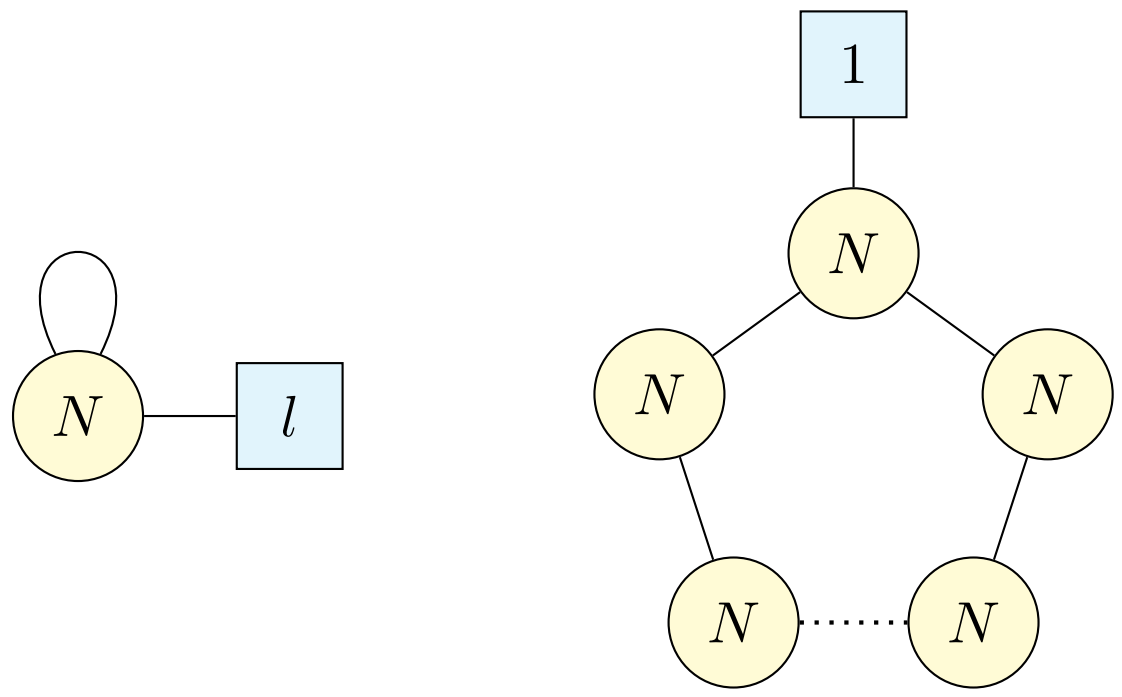
\begin{align*}
\qADHM{l}
\qquad\qquad
\qNecklace
\end{align*}
The Neumann boundary condition with boundary Fermi multiplets for the ADHM theory
is dual to the Nahm pole boundary condition for all the nodes of the mirror necklace quiver.
The boundary flavor symmetries of the Nahm pole boundary condition are identified with the flavor symmetries of the boundary Fermi multiplets
mixed with the flavor symmetry of the adjoint hypermultiplet.
Conversely, the Nahm pole boundary condition for the ADHM theory is dual to the Neumann boundary condition of the necklace quiver
with cross-determinant and fundamental Fermi multiplets \cite{Hanany:2018hlz}.
Their flavor symmetry is the boundary flavor symmetry of the Nahm pole mixed with the topological symmetry.
For $N=1$ the dualities reduce to those for the Abelian ADHM theory and its mirror in \cite{Okazaki:2025rjl}.

\item \textbf{$USp(2N)$ ADHM theory with $l$ flavors}

The $USp(2N)$ ADHM theory has an antisymmetric hypermultiplet and $l$ fundamental hypermultiplets.
Its mirror is the quiver gauge theory of the affine $D_l$ Dynkin diagram \cite{deBoer:1996mp,Porrati:1996xi,Kapustin:1998fa},
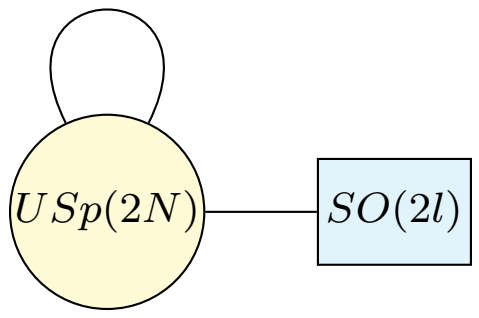
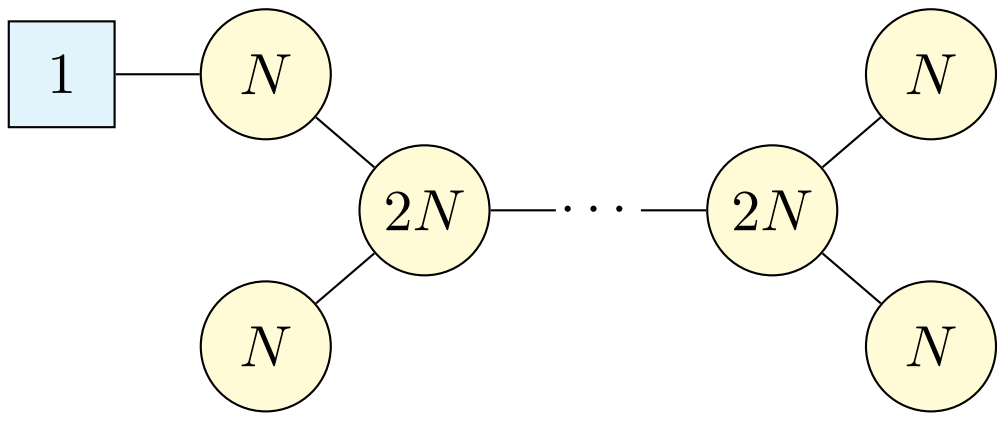
\begin{align*}
\qUSpADHM
\qquad\qquad
\qDquiver
\end{align*}
For $l\ge4$ the Neumann boundary condition with boundary Fermi multiplets for the $USp(2N)$ ADHM theory
is dual to the Nahm pole boundary condition for all the nodes of the mirror quiver.
For $N\ge2$ the R-charges of the Fermi multiplets for the end nodes mix with the flavor symmetry rotating them along the principal embedding.
Conversely, the Nahm pole boundary condition for the $USp(2N)$ ADHM theory is dual to the Neumann boundary condition of the mirror quiver
with a fundamental Fermi multiplet and $l$ Fermi multiplets whose gauge charges are the components of the simple roots of affine $D_l$.
The brane configuration with ON$^0$-planes suggests them, except the two Fermi multiplets at the ON$^0$-planes, which are fixed by the anomaly cancellation.

\end{enumerate}

The dual Neumann boundary conditions require specific boundary Fermi multiplets.
They are constrained by the cancellation of the boundary gauge anomaly and the matching of the boundary 't Hooft anomalies with the Nahm pole side,
and the half-indices confirm them.
In particular, the anomaly matching requires gauge-neutral Fermi multiplets and constrains their flavor charges.
For $U(N)$ SQCD, $T[SU(N)]$ and the $U(N)$ ADHM theory the Fermi multiplets agree with those read off from the brane configuration,
in which the gauge-neutral Fermi multiplets arise from the D3-D5 strings at the intersections of the D5-branes with the NS5$'$-brane \cite{Hanany:2018hlz}.
For the $USp(2N)$ ADHM theory with $N\ge2$ the anomalies do not determine uniquely the charges of the gauge-neutral Fermi multiplets of the end nodes.
They require the R-charges to mix with the flavor symmetry, but constrain only the sums of the squares of the charges.
With the mixing along the principal embedding described above,
the half-index of the Neumann boundary condition agrees with the proposed Nahm pole half-index.
The dual pairs are summarized in Table \ref{Table_pairs} (see section \ref{sec_conv} for the notation).

\begin{table}[t]
\centering
\footnotesize
\medmuskip=2mu
\setlength{\tabcolsep}{2pt}
\renewcommand{\arraystretch}{1.2}
\begin{tabular}{c|c|c|c}
theory & mirror & dual boundary conditions & section \\ \hline
\qcell{\qSQCD{N}{2N}} & \qcell{\qSQCDmirror{2}} &
\begin{tabular}{@{}c@{}}$(\mathcal{N}',N'+\Gamma+\eta)\leftrightarrow(\mathcal{D},\mathrm{Nahm}^{2N-3},\mathcal{D})$\\ $\mathrm{Nahm}'\leftrightarrow(\mathcal{N},N+\tilde{\eta})$\end{tabular} &
\begin{tabular}{@{}c@{}}\ref{sec_SQCD_fw}\\ \ref{sec_SQCD_rv}\end{tabular} \\ \hline
\begin{tabular}{@{}c@{}}\qcell{\qSQCD{N}{N_f}}\\ $N_f>2N$\end{tabular} & \qcell{\qSQCDmirrorNB} &
\begin{tabular}{@{}c@{}}$(\mathcal{N}',N'+\Gamma+\eta)\leftrightarrow(\mathcal{D},\mathrm{Nahm}^{N_f-3},\mathcal{D})$\\ $\mathrm{Nahm}'\leftrightarrow(\mathcal{N},N+\tilde{\eta})$\end{tabular} &
\begin{tabular}{@{}c@{}}\ref{sec_SQCD_nonbal_fw}\\ \ref{sec_SQCD_nonbal_rv}\end{tabular} \\ \hline
\qcell{\qTSUN} & \qcell{\qTSUN} & $(\mathcal{N}',N'+\Gamma+\eta)\leftrightarrow(\mathcal{D},\mathrm{Nahm}^{N-2})$ & \ref{sec_TSUN} \\ \hline
\qcell{\qADHM{l}} & \qcell{\qNecklace} &
\begin{tabular}{@{}c@{}}$(\mathcal{N}',N',N'+\Gamma+\eta)\leftrightarrow\mathrm{Nahm}^l$\\ $\mathrm{Nahm}'\leftrightarrow(\mathcal{N},N,N+\tilde{\Gamma}+\tilde{\eta})$\end{tabular} &
\begin{tabular}{@{}c@{}}\ref{sec_ADHM_l1},\,\ref{sec_ADHM_fw}\\ \ref{sec_ADHM_l1},\,\ref{sec_ADHM_rv}\end{tabular} \\ \hline
\qcell{\qUSpADHM} & \qcell{\qDquiver} &
\begin{tabular}{@{}c@{}}$(\mathcal{N}',N',N'+\Gamma+\eta)\leftrightarrow\mathrm{Nahm}^{l+1}$\\ $\mathrm{Nahm}'\leftrightarrow(\mathcal{N},N,N+\tilde{\Gamma}+\tilde{\eta})$\end{tabular} &
\begin{tabular}{@{}c@{}}\ref{sec_USp_fw}\\ \ref{sec_USp_rv}\end{tabular} \\
\end{tabular}
\caption{
The dual pairs of the Nahm pole and Neumann boundary conditions
for $U(N)$ SQCD with $2N$ and $N_f>2N$ flavors, $T[SU(N)]$ and the $U(N)$ and $USp(2N)$ ADHM theories with $l$ flavors, and their mirrors.
For the $USp(2N)$ ADHM theory the first pair requires $l\ge4$.
}
\label{Table_pairs}
\end{table}

In the twist limits the half-indices simplify (see section \ref{sec_VOA}).
The Neumann half-indices stay finite in the H-twist limit and the Nahm pole half-indices in the C-twist limit.
In these limits the formula for the Nahm pole half-indices gives closed forms as Jackson integrals, i.e. $q$-analogues of integrals (see section \ref{sec_VOA_closed}).
For $U(N)$ SQCD with $2N$ flavors the Jackson integral becomes a finite sum of products of theta functions over the choices of $N$ flavors.

The Neumann half-indices also simplify in the twist limit where they stay finite.
They are formally the contour integrals for the Schur indices \cite{Gadde:2011ik,Gadde:2011uv} of the 4d $\mathcal{N}=2$ gauge theories with the same gauge groups and hypermultiplets,
with the contributions of the boundary Fermi multiplets inserted in the integrands.
Through the dualities, the closed forms for the ADHM theories and their mirrors give free Fermi gas formulas
for the Schur indices of 4d $\mathcal{N}=4$ $U(N)$ SYM theory \cite{Bourdier:2015wda,Hatsuda:2022xdv}
and of 4d $\mathcal{N}=2$ necklace \cite{Bourdier:2015sga} and affine $D_l$ quiver gauge theories with their $U(1)$ flavor symmetries gauged by the boundary Fermi multiplets,
and a formula for the Schur index of the rank-$N$ $D_4$ theory as a Jackson integral (see section \ref{sec_VOA_closed_examples}).
To our knowledge, the formulas for the affine $D_l$ quivers and for the rank-$N$ $D_4$ theory with $N\ge2$ are new.

The closed forms also exhibit the modular properties of the unflavored twist limits (see section \ref{sec_VOA_qmod}).
In all our examples the sums reduce to polynomials in single sums, which are given by Eisenstein series and their derivatives.
We find that the unflavored characters are eta-quotients times quasimodular forms of mixed weight,
and that the weights of the products are at most the quaternionic dimensions of the Higgs branches.
For the affine $D_4$ quiver with $N=1$ the unflavored twist limit gives the unflavored Schur index of 4d $\mathcal{N}=2$ $SU(2)$ gauge theory with four flavors.
For $N\ge2$ it vanishes because of fermion zero modes, and the remaining factor is the unflavored Schur index of the rank-$N$ $D_4$ theory.
For $N=2$ and $3$ the latter has been given as a $q$-series \cite{Beem:2019tfp}, and we write it as an eta-quotient times a quasimodular form for $\Gamma^0(2)$.
We obtain these expressions from the modular differential equations there rather than from the half-indices, assuming that the Schur indices satisfy these equations.

These exact formulas lead us to propose the boundary VOAs.
Assuming that the relevant cohomologies are concentrated in degree zero, the twist limits are the vacuum characters of the boundary VOAs,
and the dualities give identities between them, which are consistent with \cite{Costello:2018fnz}.
For the Neumann boundary conditions the twist limits of the half-indices are the characters of the BRST reductions of symplectic bosons and complex fermions with respect to the gauge group.
The nilpotency of the BRST charge is equivalent to the cancellation of the boundary gauge anomaly.
Suppose that the symplectic bosons give the currents of the semisimple part of the gauge group at the level $-2h^{\vee}$,
which is the condition for the 4d $\mathcal{N}=2$ Lagrangian gauge theory with the same matter to be conformal,
and that the boundary Fermi multiplets are neutral under this part.
For unitary gauge groups the first condition means that all the nodes are balanced.
Then the reduction for the semisimple part gives the chiral algebra of this 4d theory \cite{Beem:2013sza},
and the remaining Abelian part, if any, is reduced together with the cross-determinant Fermi multiplets.

For the Nahm pole boundary conditions, in the twist limit the specialization of the boundary fugacities becomes the principal specialization in powers of $q$,
which is familiar from the characters of the quantum Hamiltonian reductions \cite{MR2013802}.
The decoupled Goldstone modes become the ghosts of a reduction of Drinfeld-Sokolov type \cite{Drinfeld:1984qv,MR1071340}
for the boundary VOA of the Dirichlet boundary condition, which is based on a Lie superalgebra formed by the boundary values of the gauge field, of the complex scalar field in the vector multiplet and of the hypermultiplets.
Its even part is the Takiff algebra of the Lie algebra of the gauge group \cite{Takiff:1971,Babichenko:2012uq}.
We conjecture that this reduction gives the Nahm pole boundary VOAs.
In the reduction the vev of the moment map pairs the fermionic currents into free fermions,
which correspond to the split Fermi multiplets and are expected to decouple.

The Neumann boundary conditions for SQCD and $T[SU(N)]$ are of the type considered in \cite{Costello:2018fnz}.
For $U(2)$ SQCD with four flavors our boundary VOA is, up to the decoupled gauge-neutral Fermi multiplets,
precisely one of the VOAs constructed in \cite{Costello:2018fnz}.
Its low-lying spectrum reproduces the currents and the spin-$\frac32$ fields predicted there.
The gauge-charged boundary Fermi multiplets give the current algebras for $U(N)$ SQCD and $T[SU(2)]$ in \cite{Costello:2018fnz}
and lattice VOAs for $T[SU(N)]$ with $N\ge3$ and for the mirror of $U(N)$ SQCD.
For the Nahm pole boundary condition for $U(N)$ SQCD the half-indices provide the vacuum characters of the ``oper-like'' boundary VOAs in \cite{Costello:2018fnz},
and the duality conjecturally identifies these VOAs with the BRST reductions of the boundary fields of the mirror quiver.
For the ADHM theory with one fundamental hypermultiplet the boundary VOA in the H-twist limit has the same supercharacter as
the vertex operator superalgebra associated with the Hilbert scheme of points on the plane \cite{Arakawa:2023cki}.
With $l$ fundamental hypermultiplets the boundary VOA in the C-twist limit is expected to be its generalization to the cyclic quiver variety,
in which the cross-determinant Fermi multiplets provide additional odd coordinates required for the BRST reduction.
For the $USp(2N)$ ADHM theory with four flavors the H-twist limit conjecturally identifies the Nahm pole boundary VOA of the affine $D_4$ quiver
with the chiral algebra of the rank-$N$ $D_4$ theory times free fields.

\subsection{Structure}
The paper is organized as follows.
In section \ref{sec_hind} we review the half-indices of the $\mathcal{N}=(0,4)$ boundary conditions following \cite{Okazaki:2019bok}.
We give the half-index of the Dirichlet boundary condition for non-Abelian unitary gauge theories with the boundary anomaly including the contribution of the vector multiplet.
For Abelian gauge theories we derive the minimal mirror identity.
In section \ref{sec_Nahm} we explain the Nahm pole boundary conditions in 3d $\mathcal{N}=4$ gauge theories, their boundary symmetries
and the splitting of the hypermultiplets, together with their brane construction and the rules for the dual Neumann boundary conditions.
We then formulate the Nahm pole half-index for $U(N)$ gauge theories by Higgsing and write it in a simple form.
We generalize the formula to quiver gauge theories with unitary gauge groups and to general gauge groups and hypermultiplets with principal Nahm poles, and discuss its properties and the boundary anomalies.
In sections \ref{sec_SQCD}, \ref{sec_TSUN} and \ref{sec_ADHM} we study the dualities for
$U(N)$ SQCD, $T[SU(N)]$ and the $U(N)$ ADHM theory respectively,
and write the Nahm pole half-indices explicitly by the proposed formula for all the examples.
In section \ref{sec_USpADHM} we apply the formula to the $USp(2N)$ ADHM theory
and propose its dualities with the Neumann and Nahm pole boundary conditions of the affine $D_l$ quiver.
In section \ref{sec_VOA} we derive the closed forms of the Nahm pole half-indices in the twist limits and study their quasimodularity.
We describe the boundary VOAs as the BRST reductions for the Neumann boundary conditions and as the reductions of Drinfeld-Sokolov type for the Nahm pole boundary conditions,
and compare them with \cite{Costello:2018fnz}.
We have checked the dualities by expanding both sides in $q^{\frac14}$ with generic fugacities in exact arithmetic
and by evaluating them numerically at fixed $q$ (see Appendix \ref{app_numerics} and Table \ref{Table_checks}).
In Appendix \ref{app_series} we exhibit the $q$-series expansions of the half-indices and of the unflavored characters.

\subsection{Future work}
\label{sec_future}

\begin{itemize}

\item It would be interesting to extend the analysis to other 3d $\mathcal{N}=4$ gauge theories with orthogonal and symplectic gauge groups,
such as those realized by the brane configurations with O3-planes \cite{Feng:2000eq,Gaiotto:2008ak}.
The formula passes the test for $USp(2N)$ in section \ref{sec_USpADHM},
and we expect that the Higgsing applies with the principal embedding into the Lie algebras of types $B$ and $D$ as well.
The spins in the Slodowy slices are then $k=1,3,\cdots,2N-1$ for $SO(2N+1)$ and $k=1,3,\cdots,2N-3$ and $N-1$ for $SO(2N)$.
As for $USp(2N)$, the center gives only a discrete boundary symmetry and the matter can contain half-hypermultiplets,
and the vector representation of $SO(2N)$ is reducible under the principal embedding.
The brane configurations with orientifold planes should also explain the boundary conditions of the $USp(2N)$ ADHM theory.
For the duality (\ref{USp_fw_dual}) it would be interesting to derive the gauge-neutral Fermi multiplets $\Gamma_{I,J}$
and the mixing of their R-charges with the flavor symmetry from the brane configuration with the O5$^-$-planes.

\item It would be interesting to decorate the Nahm pole boundary conditions and their dual Neumann boundary conditions with the BPS line operators
and to find the dualities of the line defect half-indices, which generalize those for the Abelian boundary conditions in \cite{Okazaki:2019bok} to the non-Abelian boundary conditions
(see \cite{Assel:2015oxa,Dimofte:2019zzj,Dey:2021jbf} for the mirror symmetry of the BPS line operators in the bulk and \cite{Okazaki:2019ony,Hayashi:2025guk} for its tests by the line defect indices).

\item In this paper we focus on the Nahm pole boundary conditions associated with the principal embedding.
For a general embedding of $\mathfrak{su}(2)$ labeled by a partition,
the vev also breaks the boundary symmetry along the off-diagonal blocks between the Jordan blocks,
so that the corresponding Goldstone modes have to be stripped off as well.
The formula in section \ref{sec_UN_simple} suggests that the sector of zero flux contains one component of each $\mathfrak{su}(2)$ multiplet in the adjoint representation,
as for the Slodowy slice.
For a general embedding the D3-branes end on several D5$'$-branes.
In the S-dual configuration they end on several NS5$'$-branes,
and the segments of the D3-branes bounded by the NS5- and NS5$'$-branes form brane boxes,
which support 2d $\mathcal{N}=(0,4)$ gauge theories at the boundary \cite{Hanany:2018hlz}.
Hence the dual Neumann boundary conditions are coupled to these 2d gauge theories.
It would be interesting to check the half-indices for general embeddings against the dual Neumann boundary conditions
and to test the consistency under the successive Higgsing along the partial order of the embeddings.

\item It would be interesting to prove the conjecture (\ref{qmod_weight}) on the quasimodularity of the unflavored twist limits
and to extend it to the orthogonal gauge groups and to other representations of the hypermultiplets.
For linear quivers whose adjacent ranks differ by more than one, the sums are not of the form in (\ref{tail_identity}).
For the Nahm pole boundary condition of the affine $D_l$ quiver with $N\ge2$ the unflavored characters vanish,
and it would be interesting to derive the closed forms (\ref{D4_Schur2}) and (\ref{D4_Schur3}) of the remaining factor, the unflavored Schur index of the rank-$N$ $D_4$ theory, from the half-indices and to extend them to $N\ge4$.

\item It would be interesting to study the twist limits of the Dirichlet half-indices of non-Abelian gauge theories and their mirror dual boundary conditions.
Since the unflavored limits do not exist, the boundary fugacities have to be kept.
We find that the twist limits are then theta functions times elliptic functions of the boundary fugacities,
which are polynomials in quasi-Jacobi forms \cite{MR2796409,oberdieck2012serre,MR3956895} and can be obtained by summing over the fluxes node by node.
We will report these results elsewhere.

\item It would be interesting to construct the boundary VOAs for the Nahm pole boundary conditions explicitly
as the reductions of Drinfeld-Sokolov type in section \ref{sec_VOA_DS}, including the boundary monopole sectors,
and to show that the cohomologies of the BRST and Drinfeld-Sokolov reductions vanish at non-zero ghost number.
It would also be interesting to examine the conjecture in \cite{Costello:2018fnz} that the algebra of bulk local operators is recovered from the self-Ext's of the vacuum module.
For the necklace quiver it remains to construct the chiral quantization of the equivariant Hilbert scheme with the odd line bundles from the cross-determinant Fermi multiplets
and to prove the factorization of the boundary VOA proposed in section \ref{sec_VOA_examples}.
This would generalize \cite{Arakawa:2023cki} for the single node and \cite{Arakawa:2026nut} for the Abelian case.

\end{itemize}

\section{Half-indices}
\label{sec_hind}

\subsection{Definition}
\label{sec_conv}
We follow the conventions in \cite{Okazaki:2019bok} (see also \cite{Gaiotto:2019jvo}).
The half-index is defined by\footnote{
The half-indices of $\mathcal{N}=(0,2)$ half-BPS boundary conditions in 3d $\mathcal{N}=2$ gauge theories were introduced in \cite{Gadde:2013wq,Yoshida:2014ssa} for the Neumann boundary conditions and in \cite{Dimofte:2017tpi} for the Dirichlet boundary conditions.
}
\begin{align}
\label{DEF_hind}
\mathbb{II}^{\mathcal{T}}_{\mathcal{B}}(t,x;q)&=\mathrm{Tr}\,(-1)^F q^{J+\frac{H+C}{4}}t^{H-C}x^f,
\end{align}
where the trace is taken over the local operators on the 2d boundary for the 3d $\mathcal{N}=4$ theory $\mathcal{T}$ with the boundary condition $\mathcal{B}$.
They are built not only from the 3d bulk fields but also from the 2d supermultiplets supported on the boundary, such as the boundary Fermi multiplets.
Here $J$ is the generator of the rotation in the 2d boundary,
$H$ and $C$ are the Cartan generators of the $SU(2)_H$ and $SU(2)_C$ R-symmetry groups,
and $f$ collectively stands for the Cartan generators of the global symmetries.
Throughout this paper the superscript and the subscript of $\mathbb{II}$ indicate the theory $\mathcal{T}$ and the boundary condition $\mathcal{B}$ respectively,
e.g. $\mathbb{II}^{U(N)+2N}_{\mathrm{Nahm}'}$ is the half-index of $U(N)$ SQCD with $2N$ flavors with the Nahm pole boundary condition.
We use the $q$-shifted factorials
\begin{align}
\label{qpoch_def}
(a;q)_n&=\prod_{k=0}^{n-1}(1-aq^k), &
(a;q)_{\infty}&=\prod_{k=0}^{\infty}(1-aq^k), &
(q)_n&=(q;q)_n,
\end{align}
and extend $(a;q)_n$ to all $n\in\mathbb{Z}$ by $(a;q)_n=(a;q)_{\infty}/(aq^n;q)_{\infty}$.
We write $(a^{\pm};q)_{n}=(a;q)_{n}(a^{-1};q)_{n}$
and $(q^{a\pm b}t^{c\pm d}x^{\pm};q)_{n}=(q^{a+b}t^{c+d}x;q)_{n}(q^{a-b}t^{c-d}x^{-1};q)_{n}$, including $n=\infty$.

The boundary conditions are realized by the D3-branes along $x^{0126}$,
the NS5- and D5-branes along $x^{012345}$ and $x^{012789}$,
and the NS5$'$- and D5$'$-branes along $x^{016789}$ and $x^{013456}$ \cite{Hanany:1996ie,Chung:2016pgt,Hanany:2018hlz,Okazaki:2019bok}.
The D3-branes ending on the NS5$'$-brane lead to the Neumann boundary conditions $\mathcal{N}'$ for the vector multiplets and $N'$ for the hypermultiplets.
The D3-branes ending on the D5$'$-branes lead to the Dirichlet boundary condition $D'$ for the hypermultiplets
and the Dirichlet boundary condition $\mathcal{D}'$ or the Nahm pole boundary condition $\mathrm{Nahm}'$ for the vector multiplets.
For the twisted multiplets arising in the S-dual configurations we drop the primes and write $\mathcal{N}$, $\mathcal{D}$, $\mathrm{Nahm}$ and $N$, $D$.
Thus the primes distinguish the boundary conditions for the vector multiplets and hypermultiplets from those for the twisted multiplets.
In the text we simply call them the Neumann, Dirichlet and Nahm pole boundary conditions, and the primes appear only in the notation.
As in \cite{Okazaki:2019bok}, we denote the boundary conditions by the pairs of those for the vector multiplets and hypermultiplets, e.g. $(\mathcal{N}',N'+\Gamma+\eta)$,
where $\Gamma$ and $\eta$ indicate the two types of boundary Fermi multiplets in section \ref{sec_dualrules}.
For the twisted multiplets we write these Fermi multiplets as $\tilde{\Gamma}$ and $\tilde{\eta}$.
For the quiver gauge theories whose nodes obey different boundary conditions,
we list the boundary conditions for the vector multiplets of the nodes, e.g. $(\mathcal{D},\mathrm{Nahm},\mathcal{D})$,
and omit those for the (twisted) hypermultiplets.
We write $\mathrm{Nahm}^n$ for $n$ consecutive nodes with the Nahm pole boundary conditions, e.g. $(\mathcal{D},\mathrm{Nahm}^{2N-3},\mathcal{D})$.
When all the nodes obey the Neumann boundary condition, we use the pair notation, e.g. $(\mathcal{N},N+\tilde{\eta})$.
For the $U(N)$ and $USp(2N)$ ADHM theories and their mirrors, we list the boundary conditions for the vector multiplets,
the adjoint, antisymmetric or bifundamental hypermultiplets and the fundamental hypermultiplets, e.g. $(\mathcal{N}',N',N'+\Gamma+\eta)$.

We denote the quiver gauge theories as in \cite{Okazaki:2019bok}.
$(N)$ and $[N_f]$ stand for the $U(N)$ gauge node and the $U(N_f)$ flavor symmetry,
and the lines for the (bi)fundamental hypermultiplets, e.g. $(N)-[2N]$ for $U(N)$ SQCD with $2N$ flavors.
In the quiver diagrams we draw the gauge nodes as circles and the flavor nodes as squares as in \cite{Okazaki:2019ony},
and a loop on a gauge node denotes the adjoint hypermultiplet (see Table \ref{Table_pairs}).
For the mirror description, in which the vector multiplets and the hypermultiplets are replaced by
the twisted vector multiplets and the twisted hypermultiplets,
we put a tilde over the theory in the superscripts of the half-indices, e.g. $\mathbb{II}^{\wtilde{(1)-(2)-(1)}}$ and $\mathbb{II}^{\wtilde{T[SU(N)]}}$.
Accordingly, the Higgs and Coulomb branches are exchanged and the half-indices are obtained by $t\rightarrow t^{-1}$.
The fugacities for the topological symmetries and for the flavor symmetries rotating the twisted hypermultiplets also exchange their roles,
so that the two sides of mirror symmetry share the same fugacities.
For example, $x_{\alpha}$ are the flavor fugacities of $U(N)$ SQCD and the topological fugacities of its mirror,
while $z_a$ are the topological fugacities of the former and the flavor fugacities of the latter.

\subsection{Building blocks}
\label{sec_multiplets}
The half-indices are built from the contributions of the 3d supermultiplets obeying the boundary conditions
and of the 2d supermultiplets supported on the boundary.
For the 3d supermultiplets we write the boundary conditions for the bosonic fields,
and those for the fermions follow from the $\mathcal{N}=(0,4)$ supersymmetry \cite{Chung:2016pgt,Okazaki:2019bok}.

\subsubsection{3d hypermultiplets}
The $\mathcal{N}=(0,4)$ Neumann and Dirichlet boundary conditions for the hypermultiplet \cite{Chung:2016pgt} are
\begin{align}
\label{hyper_bc}
N':&\quad D_2q^A|_{\partial}=0, & D':&\quad q^A|_{\partial}=0, \qquad A=1,2,
\end{align}
where the complex scalar fields $q^A$ form the $SU(2)_H$ doublet.
In terms of the two 3d $\mathcal{N}=2$ chiral multiplets $X$ and $Y$ forming the hypermultiplet, we have $q^A=(X,Y^{\dagger})$,
so that $N'$ and $D'$ impose the Neumann and Dirichlet boundary conditions on both of them.
The half-indices for the hypermultiplet of flavor fugacity $x$ are \cite{Okazaki:2019bok}
\begin{align}
\label{hyper_half}
\mathbb{II}^{\textrm{HM}}_{N'}(t,x;q)&=\frac{1}{(q^{\frac14}tx^{\pm};q)_{\infty}}, &
\mathbb{II}^{\textrm{HM}}_{D'}(t,x;q)&=(q^{\frac34}t^{-1}x^{\pm};q)_{\infty}.
\end{align}

\subsubsection{3d vector multiplets}
The vector multiplet contains the real scalar field $\sigma$ and the complex scalar field $\varphi$ in the adjoint representation,
which form the $SU(2)_C$ triplet $\vec{\phi}=(\sigma,\mathrm{Re}\,\varphi,\mathrm{Im}\,\varphi)$.
The $\mathcal{N}=(0,4)$ Neumann boundary condition $\mathcal{N}'$ for the vector multiplet \cite{Chung:2016pgt} is
\begin{align}
\label{vec_bc}
F_{2\mu}|_{\partial}&=0, & \vec{\phi}|_{\partial}&=0, \qquad \mu=0,1.
\end{align}
The Dirichlet and Nahm pole boundary conditions are given in section \ref{sec_Nahm_def}.
The half-index of $\mathcal{N}'$ for the $U(N)$ vector multiplet is \cite{Okazaki:2019bok}
\begin{align}
\label{vec_N}
\mathbb{II}^{U(N)}_{\mathcal{N}'}(t;q)&=
\frac{(q)_{\infty}^N(q^{\frac12}t^2;q)_{\infty}^N}{N!}\oint \prod_{i=1}^N \frac{ds_i}{2\pi is_i}
\prod_{i\neq j}\left(\frac{s_i}{s_j};q\right)_{\infty}\left(q^{\frac12}t^2\frac{s_i}{s_j};q\right)_{\infty},
\end{align}
where the integration contour is the unit torus when the boundary has no 2d bosonic matter fields.
The half-index of the Dirichlet boundary condition $\mathcal{D}'$ is given in section \ref{sec_Dirichlet}.

\subsubsection{2d supermultiplets}
The index of the 2d Fermi multiplet of fugacity $x$ is $F(x;q)=(x;q)_{\infty}(qx^{-1};q)_{\infty}$ \cite{Okazaki:2019bok}.
The indices of the 2d $\mathcal{N}=(0,4)$ hypermultiplet and twisted hypermultiplet are \cite{Okazaki:2019bok}
\begin{align}
\label{2dHM}
I^{\textrm{2d HM}}(t,x;q)&=\frac{1}{(q^{\frac14}tx^{\pm};q)_{\infty}(q^{\frac34}t^{-1}x^{\pm};q)_{\infty}}, \\*
\label{2dtHM}
I^{\textrm{2d tHM}}(t,x;q)&=I^{\textrm{2d HM}}(t^{-1},x;q).
\end{align}

\subsection{Boundary anomalies}
\label{sec_anomaly}
The boundary anomaly polynomials are evaluated as in \cite{Dimofte:2017tpi,Okazaki:2019bok}.
We write the 4-form anomaly polynomial as a quadratic polynomial in the field strengths.
The field strength of each symmetry is denoted by the bold letter for its fugacity in the half-index.
For the gauge group $U(N)$ with the fugacities $s_i$, the field strength $\mathbf{s}$ has the components $\mathbf{s}_i$ in the Cartan subalgebra.
For the quiver gauge theories we write $\mathbf{s}^{(I)}_i$ for the node $U(N_I)$.
We first omit the terms with the field strengths of the R-symmetries, which we discuss at the end of this section.
As for the half-index, the superscript and the subscript of $\mathcal{A}$ indicate the theory $\mathcal{T}$ and the boundary condition $\mathcal{B}$,
e.g. $\mathcal{A}^{U(N)+2N}_{\mathrm{Nahm}'}$ is the boundary anomaly polynomial of $U(N)$ SQCD with $2N$ flavors with the Nahm pole boundary condition.

A 2d left-moving fermion contributes $+\mathbf{f}^2$ and a right-moving fermion contributes $-\mathbf{f}^2$,
where $\mathbf{f}$ is the sum of the field strengths weighted by the charges of the fermion.
Accordingly, the hypermultiplet contributes $-\mathbf{f}^2$ for $N'$ and $+\mathbf{f}^2$ for $D'$,
the boundary Fermi multiplet contributes $+\mathbf{f}^2$,
and the vector multiplet contributes $+\sum_{\alpha\in\Delta}(\alpha\cdot \mathbf{s})^2$ for $\mathcal{N}'$
and $-\sum_{\alpha\in\Delta}(\alpha\cdot \mathbf{u})^2$ for $\mathcal{D}'$, where $\Delta$ is the set of roots of the Lie algebra $\mathfrak{g}$ of the gauge group $G$.
For $G=U(N)$ they give $\alpha\cdot\mathbf{s}=\mathbf{s}_i-\mathbf{s}_j$ with $i\neq j$.
The FI coupling of the node $U(N_I)$ to the topological symmetry of field strength $\mathbf{z}_I$ contributes $-2\sum_i\mathbf{s}^{(I)}_i\,\mathbf{z}_I$.

Since the anomaly polynomial is invariant under the Weyl group, it is determined by its restriction to the Cartan subalgebra.
We write it in terms of the components $\mathbf{s}_i$, which enter as Abelian charges through the weights.
For $U(N)$ the quadratic invariants are $\mathrm{Tr}(\mathbf{s}^2)=\sum_i\mathbf{s}_i^2$ and $(\mathrm{Tr}\,\mathbf{s})^2=(\sum_i\mathbf{s}_i)^2$,
which give the trace notation in \cite{Dimofte:2017tpi,Okazaki:2025rjl}.
For example, $\sum_{\alpha\in\Delta}(\alpha\cdot\mathbf{s})^2=2N\,\mathrm{Tr}(\mathbf{s}^2)-2(\mathrm{Tr}\,\mathbf{s})^2$.

For the Dirichlet boundary condition $(\mathcal{D}',D')$ the gauge group $G$ gives rise to the boundary global symmetry $G_{\partial}$.
Its field strength $\mathbf{u}$ has the components $\mathbf{u}_i$ for the fugacities $u_i$ in the Dirichlet half-index.
Let the hypermultiplets carry the weights $\lambda=(w_{\lambda},f_{\lambda})$,
where $w_{\lambda}$ is the gauge weight and $f_{\lambda}$ is the flavor weight,
and let $\mathbf{x}$ be the field strength of the flavor symmetry.
The anomaly polynomial is
\begin{align}
\label{anom_DD}
\mathcal{A}^{\mathcal{T}}_{\mathcal{D}',D'}&=
\underbrace{\sum_{\lambda}(w_{\lambda}\cdot \mathbf{u}+f_{\lambda}\cdot \mathbf{x})^2}_{\textrm{hypermultiplets ($D'$)}}
\underbrace{-\sum_{\alpha\in\Delta}(\alpha\cdot \mathbf{u})^2}_{\textrm{vector multiplet ($\mathcal{D}'$)}}
\underbrace{-2\sum_{I,i}\mathbf{u}^{(I)}_i\,\mathbf{z}_I}_{\textrm{FI couplings}}.
\end{align}
The quadratic part in $\mathbf{u}$ is encoded by the symmetric matrix
\begin{align}
\label{keff}
k_{\rm eff}&=\sum_{\lambda}w_{\lambda}w_{\lambda}^T-\sum_{\alpha\in\Delta}\alpha\alpha^T,
\end{align}
which extends the gauge part of the matrix $k_{\rm eff}$ in \cite{Okazaki:2019bok} by the contribution of the vector multiplet.
For $U(N)$ we have $\sum_{\alpha\in\Delta}\alpha\alpha^T=2N\mathbf{1}_N-2\mathbb{J}_N$, where $\mathbb{J}_N=ee^T$ with $e=(1,\cdots,1)^T$ is the $N\times N$ matrix whose entries are all $1$.
Hence $k_{\rm eff}$ receives $-\sum_{\alpha\in\Delta}\alpha\alpha^T e=0$ on the vector $e$ of the $U(1)$ part
and $-\sum_{\alpha\in\Delta}\alpha\alpha^T v=-2Nv$ on the traceless vectors $v$ of the $SU(N)$ part.
A fundamental hypermultiplet contributes $\mathbf{1}_N$, i.e. $1$ on both parts.
Thus the vector multiplet shifts the level of $SU(N)$ by $-2N=-2h^{\vee}$ in the normalization in which a fundamental hypermultiplet contributes $1$,
and it does not shift that of $U(1)$.
This agrees with \cite{Costello:2018fnz},
in which the level of the C-twisted theory receives the contribution $-2h^{\vee}$ from the fermions in the vector multiplet.

The R-charges of the fermions give the terms with the field strengths $\mathbf{h}$ and $\mathbf{c}$ of the Cartan subgroups $U(1)_H\subset SU(2)_H$ and $U(1)_C\subset SU(2)_C$,
as for the R-symmetry in \cite{Dimofte:2017tpi}.
A fermion with the R-charges $H$ and $C$ contributes $\pm(\mathbf{f}+H\mathbf{h}+C\mathbf{c})^2$,
and a fermion of a bulk multiplet contributes half of it at the boundary.
The fermion of a boundary Fermi multiplet has $H=C=0$, except for the Fermi multiplets in section \ref{sec_USp_fw_N}.
The fermions of the hypermultiplet form a doublet of $SU(2)_C$ with $H=0$,
and those of the vector multiplet transform as $(\mathbf{2},\mathbf{2})$ under $SU(2)_H\times SU(2)_C$,
while $H$ and $C$ are exchanged for the twisted multiplets.
Hence the hypermultiplet contributes $-\mathbf{f}^2-\mathbf{c}^2$ for $N'$ and $+\mathbf{f}^2+\mathbf{c}^2$ for $D'$,
and the vector multiplet contributes $\sum_{\alpha\in\Delta}(\alpha\cdot\mathbf{s})^2+\dim G\,(\mathbf{h}^2+\mathbf{c}^2)$ for $\mathcal{N}'$
and $-\sum_{\alpha\in\Delta}(\alpha\cdot\mathbf{u})^2-\dim G\,(\mathbf{h}^2+\mathbf{c}^2)$ for $\mathcal{D}'$.
The terms for the Nahm pole boundary conditions are given in section \ref{sec_Nahm_anom}.

\subsection{Dirichlet half-indices}
\label{sec_Dirichlet}
The half-index of the $\mathcal{N}=(0,4)$ Dirichlet boundary condition receives the contributions from the boundary monopole operators \cite{Bullimore:2016nji,Dimofte:2017tpi}.
For the gauge group $G=\prod_I U(N_I)$ with the hypermultiplets of weights $\lambda=(w_{\lambda},f_{\lambda})$, we take
\begin{align}
\label{Dhalf}
\mathbb{II}^{\mathcal{T}}_{\mathcal{D}',D'}(u)
&=\sum_{m}\underbrace{(-q^{\frac12})^{m^Tk_{\rm eff}m}u^{k_{\rm eff}m}\prod_{\lambda}x^{w_{\lambda}(m)f_{\lambda}}\prod_{I}z_I^{-\sum_im^{(I)}_i}}_{\textrm{bare monopole operators}}
\nonumber\\*
&\times \underbrace{\frac{1}{[(q)_{\infty}(q^{\frac12}t^{-2};q)_{\infty}]^{\mathrm{rk}\,G}}\prod_{\alpha\in\Delta}\frac{1}{(q^{1+\alpha(m)}u^{\alpha};q)_{\infty}(q^{\frac12+\alpha(m)}t^{-2}u^{\alpha};q)_{\infty}}}_{\textrm{vector multiplets ($\mathcal{D}'$)}}
\nonumber\\*
&\times \underbrace{\prod_{\lambda}(q^{\frac34\pm w_{\lambda}(m)}t^{-1}(u^{w_{\lambda}}x^{f_{\lambda}})^{\pm};q)_{\infty}}_{\textrm{hypermultiplets ($D'$)}},
\end{align}
where the sum runs over the cocharacters $m=(m^{(I)}_i)$ of $G$ and $u=(u^{(I)}_i)$ is the fugacity for $G_{\partial}$.
As in (\ref{DEF_hind}), $x$ is the fugacity for the flavor symmetry,
and $z_I$ are the fugacities for the topological symmetries.
The contributions from the vector multiplet and the shift $u\rightarrow q^mu$ in the matter contributions are those in \cite{Okazaki:2019bok}.

The matrix (\ref{keff}) enters (\ref{Dhalf}) through the contribution of the bare monopole operators.
As in \cite{Dimofte:2017tpi}, their spin and charges are induced by the boundary anomaly (\ref{anom_DD}).
The effective Chern-Simons level $k_{\rm eff}$, including the shift from the vector multiplet, gives the spin and the electric charges $k_{\rm eff}m$.
The mixed terms give the flavor charges and, through the FI couplings, the topological charges.
For $U(N)$ SQCD with $N_f$ flavors we have, for the flux $m=(m_i)$,
\begin{align}
\label{keff_SQCD}
(k_{\rm eff}m)_i&=(N_f-2N)m_i+2\sum_{j=1}^N m_j.
\end{align}
For a linear or circular quiver gauge theory, the diagonal block of $k_{\rm eff}$ for a balanced node $U(N_I)$, i.e. a node with $2N_I$ flavors including the bifundamental hypermultiplets, is $2\mathbb{J}_{N_I}$,
and each bifundamental hypermultiplet contributes the matrix whose entries are all $-1$ to the off-diagonal blocks.
Hence $(k_{\rm eff}m)^{(I)}_i$ for a balanced node does not depend on $i$,
\begin{align}
\label{keff_balanced}
(k_{\rm eff}m)^{(I)}_i&=2\sum_jm^{(I)}_j-\sum_jm^{(I-1)}_j-\sum_jm^{(I+1)}_j,
\end{align}
where $m^{(0)}=m^{(n+1)}=0$ for a linear quiver with $n$ nodes and the labels are periodic for a circular quiver.
If all the nodes are balanced, then
\begin{align*}
\frac12m^Tk_{\rm eff}m&=\sum_I\Bigl(\sum_im^{(I)}_i\Bigr)^2-\sum_I\sum_{i,j}m^{(I)}_im^{(I+1)}_j
\end{align*}
is an integer, so that $m^Tk_{\rm eff}m$ is even.
In general, since $n^2\equiv n$ mod $2$ and the roots come in pairs $\pm\alpha$, we have $(-1)^{m^Tk_{\rm eff}m}=(-1)^{\sum_{\lambda}w_{\lambda}(m)}$,
so that the sign in (\ref{Dhalf}) amounts to a choice of the signs of the topological fugacities $z_I$.
Within each balanced node the quadratic form $m^Tk_{\rm eff}m$ is degenerate,
e.g. $k_{\rm eff}=2\mathbb{J}_N$ for $U(N)$ SQCD with $2N$ flavors.
Nevertheless, the sum over the fluxes converges as a power series in $q^{\frac14}$.
As seen in (\ref{D_simple}) below, the Gaussian factor is absorbed into the flux-shifted contributions of the multiplets,
and the summand behaves as $(q^{\frac12}t^{-2})^{\Delta(m)}$ with the dimension
\begin{align}
\label{monopole_dim}
\Delta(m)&=\frac12\sum_{\lambda}|w_{\lambda}(m)|-\sum_{\alpha\in\Delta^+}|\alpha(m)|
\end{align}
of the bare monopole operator of flux $m$ \cite{Borokhov:2002cg,Gaiotto:2008ak}, where $\Delta^+$ is the set of the positive roots of $G$.

The Dirichlet half-index (\ref{Dhalf}) takes a simpler form, in which the sign and the Gaussian factor are absorbed into finite products.
We find
\begin{align}
\label{D_simple}
&\mathbb{II}^{\mathcal{T}}_{\mathcal{D}',D'}(u)
\nonumber\\*
&=\underbrace{\prod_{\lambda}(q^{\frac34}t^{-1}(u^{w_{\lambda}}x^{f_{\lambda}})^{\pm};q)_{\infty}}_{\textrm{hypermultiplets ($D'$) at }m=0}\;
\underbrace{\frac{1}{[(q)_{\infty}(q^{\frac12}t^{-2};q)_{\infty}]^{\mathrm{rk}\,G}\prod_I\prod_{i\neq j}(q\frac{u^{(I)}_i}{u^{(I)}_j};q)_{\infty}(q^{\frac12}t^{-2}\frac{u^{(I)}_i}{u^{(I)}_j};q)_{\infty}}}_{\textrm{vector multiplets ($\mathcal{D}'$) at }m=0}
\nonumber\\
&\times\sum_{m}\underbrace{\prod_Iz_I^{-\sum_im^{(I)}_i}}_{\textrm{topological charges}}
\underbrace{\prod_I\prod_{1\le i<j\le N_I}(q^{\frac12}t^{-2})^{m^{(I)}_i-m^{(I)}_j}
\frac{(q\frac{u^{(I)}_j}{u^{(I)}_i};q)_{m^{(I)}_j-m^{(I)}_i}(q^{\frac12}t^{-2}\frac{u^{(I)}_j}{u^{(I)}_i};q)_{m^{(I)}_j-m^{(I)}_i}}
{(\frac{u^{(I)}_j}{u^{(I)}_i};q)_{m^{(I)}_j-m^{(I)}_i}(q^{\frac12}t^{2}\frac{u^{(I)}_j}{u^{(I)}_i};q)_{m^{(I)}_j-m^{(I)}_i}}}_{\textrm{vector multiplets ($\mathcal{D}'$)}}
\nonumber\\*
&\times\underbrace{\prod_{\lambda}(q^{\frac14}t^{-1})^{w_{\lambda}(m)}
\frac{(q^{\frac14}tu^{w_{\lambda}}x^{f_{\lambda}};q)_{w_{\lambda}(m)}}{(q^{\frac34}t^{-1}u^{w_{\lambda}}x^{f_{\lambda}};q)_{w_{\lambda}(m)}}}_{\textrm{hypermultiplets ($D'$)}},
\end{align}
where the sum runs over all the cocharacters $m$.
The prefactor is the contribution of the sector $m=0$, for which the summand is $1$.

To derive (\ref{D_simple}), consider the sector $m$ in (\ref{Dhalf}).
For $i<j$ in the node $U(N_I)$ let $a=u^{(I)}_j/u^{(I)}_i$ and $\nu=m^{(I)}_j-m^{(I)}_i$.
Let $e_i$ be the weight of the $i$-th component of the fundamental representation.
The contributions of the roots $\pm(e_j-e_i)$ to the vector multiplet are
\begin{align}
\label{D_vec}
\frac{1}{(q^{1\pm\nu}a^{\pm};q)_{\infty}(q^{\frac12\pm\nu}t^{-2}a^{\pm};q)_{\infty}}
&=\frac{q^{\nu^2}a^{2\nu}}{(qa^{\pm};q)_{\infty}(q^{\frac12}t^{-2}a^{\pm};q)_{\infty}}
(q^{\frac12}t^{-2})^{-\nu}\frac{(qa;q)_{\nu}(q^{\frac12}t^{-2}a;q)_{\nu}}{(a;q)_{\nu}(q^{\frac12}t^2a;q)_{\nu}}.
\end{align}
For a weight $\lambda$ let $b=u^{w_{\lambda}}x^{f_{\lambda}}$ and $n=w_{\lambda}(m)$.
The hypermultiplet gives
\begin{align}
\label{D_hyp}
(q^{\frac34\pm n}t^{-1}b^{\pm};q)_{\infty}
&=(-1)^nq^{-\frac{n^2}{2}}b^{-n}(q^{\frac34}t^{-1}b^{\pm};q)_{\infty}(q^{\frac14}t^{-1})^n\frac{(q^{\frac14}tb;q)_n}{(q^{\frac34}t^{-1}b;q)_n}.
\end{align}
Both identities hold for all integers $\nu$ and $n$.
Since $k_{\rm eff}=\sum_{\lambda}w_{\lambda}w_{\lambda}^T-\sum_{\alpha\in\Delta}\alpha\alpha^T$,
the contribution of the bare monopole operators splits into the parts induced by the vector multiplet and by the hypermultiplets.
The part induced by the vector multiplet is $q^{-\nu^2}a^{-2\nu}$ for each pair of roots $\pm(e_j-e_i)$,
which is cancelled by the factor $q^{\nu^2}a^{2\nu}$ in (\ref{D_vec}).
The part induced by the hypermultiplet of weight $\lambda$ is $(-1)^nq^{\frac{n^2}{2}}b^n$,
which is cancelled by the factor $(-1)^nq^{-\frac{n^2}{2}}b^{-n}$ in (\ref{D_hyp}).
The remaining factors give (\ref{D_simple}).
Hence the spin and the electric and flavor charges of the bare monopole operators are absorbed into the finite products of the multiplets which induce them,
and only their topological charges remain as monomials in the summand.

\subsection{Minimal mirror identity}
\label{sec_abelian}
\label{sec_minimal}
The simplest Abelian duality in \cite{Okazaki:2019bok} relates the Dirichlet boundary condition $(\mathcal{D}',D')$ for SQED$_1$
to the Neumann boundary condition for the free twisted hypermultiplet with a boundary Fermi multiplet.
The half-indices read
\begin{align}
\label{SQED1_DD}
\mathbb{II}^{\textrm{SQED}_1}_{\mathcal{D}',D'}(t,u,z;q)
&=\underbrace{\frac{1}{(q)_{\infty}(q^{\frac12}t^{-2};q)_{\infty}}}_{\textrm{vector multiplet ($\mathcal{D}'$)}}
\sum_{m\in\mathbb{Z}}\underbrace{(-q^{\frac12})^{m^2}u^mz^{-m}}_{\textrm{bare monopole operators}}
\nonumber\\*
&\times \underbrace{(q^{\frac34+m}t^{-1}u;q)_{\infty}(q^{\frac34-m}t^{-1}u^{-1};q)_{\infty}}_{\textrm{hypermultiplet ($D'$)}}, \\*
\label{tHM_NF}
\mathbb{II}^{\textrm{tHM}}_{N+\tilde{\eta}}(t,u,z;q)
&=\underbrace{\frac{1}{(q^{\frac14}t^{-1}z^{\pm};q)_{\infty}}}_{\textrm{twisted hypermultiplet ($N$)}}\;\underbrace{F(q^{\frac12}zu^{-1};q)}_{\tilde{\eta}}.
\end{align}
The equality of (\ref{SQED1_DD}) and (\ref{tHM_NF}) was called the minimal mirror identity in \cite{Okazaki:2019bok}.
It underlies the dualities for SQED$_{N_f}$.
It follows from Ramanujan's ${}_1\psi_1$ summation formula \cite{MR2128719} (see also \cite{Gaiotto:2019jvo})
\begin{align}
\label{1psi1_Ramanujan}
\sum_{p\in\mathbb{Z}}\frac{(a;q)_p}{(b;q)_p}w^p
&=\frac{(q;q)_{\infty}(\frac{b}{a};q)_{\infty}(aw;q)_{\infty}(\frac{q}{aw};q)_{\infty}}{(b;q)_{\infty}(\frac{q}{a};q)_{\infty}(w;q)_{\infty}(\frac{b}{aw};q)_{\infty}}.
\end{align}
By (\ref{D_simple}) the Dirichlet half-index (\ref{SQED1_DD}) is the ${}_1\psi_1$ series itself,
\begin{align}
\label{SQED1_simple}
\mathbb{II}^{\textrm{SQED}_1}_{\mathcal{D}',D'}(t,u,z;q)
&=\underbrace{(q^{\frac34}t^{-1}u^{\pm};q)_{\infty}}_{\textrm{hypermultiplet ($D'$) at }m=0}\;
\underbrace{\frac{1}{(q)_{\infty}(q^{\frac12}t^{-2};q)_{\infty}}}_{\textrm{vector multiplet ($\mathcal{D}'$) at }m=0}
\nonumber\\*
&\times \sum_{m\in\mathbb{Z}}\underbrace{z^{-m}}_{\textrm{topological charge}}\;
\underbrace{(q^{\frac14}t^{-1})^m\frac{(q^{\frac14}tu;q)_m}{(q^{\frac34}t^{-1}u;q)_m}}_{\textrm{hypermultiplet ($D'$)}}.
\end{align}
Hence (\ref{1psi1_Ramanujan}) with $a=q^{\frac14}tu$, $b=q^{\frac34}t^{-1}u$ and $w=q^{\frac14}t^{-1}z^{-1}$ gives the minimal mirror identity
\begin{align}
\label{minimal_mirror}
\mathbb{II}^{\textrm{SQED}_1}_{\mathcal{D}',D'}(t,u,z;q)&=\mathbb{II}^{\textrm{tHM}}_{N+\tilde{\eta}}(t,u,z;q).
\end{align}
Note that the sign $(-q^{\frac12})^{m^2}$ in (\ref{SQED1_DD}) is essential for the cancellation leading to (\ref{SQED1_simple}).
Since the identities of the half-indices for the dualities of SQED$_{N_f}$ in \cite{Okazaki:2019bok} follow from the minimal mirror identity, they are also proven.

\section{Nahm pole half-indices}
\label{sec_Nahm}

\subsection{Nahm pole boundary conditions}
\label{sec_Nahm_bc}

\subsubsection{Definition}
\label{sec_Nahm_def}
The half-BPS boundary conditions in 3d $\mathcal{N}=4$ gauge theories preserve
either 2d $\mathcal{N}=(2,2)$ or 2d $\mathcal{N}=(0,4)$ supersymmetry,
and both types are realized by brane configurations \cite{Chung:2016pgt}.
We consider the $\mathcal{N}=(0,4)$ boundary conditions.
They preserve the whole R-symmetry $SU(2)_H\times SU(2)_C$ as the R-symmetry of the 2d $\mathcal{N}=(0,4)$ supersymmetry,
whereas the $\mathcal{N}=(2,2)$ boundary conditions preserve only $U(1)_H\times U(1)_C$ \cite{Bullimore:2016nji}.
For the $\mathcal{N}=(0,4)$ boundary conditions the vector multiplet admits the Neumann boundary condition $\mathcal{N}'$
and the Nahm pole boundary condition $\mathrm{Nahm}'$, which contains the Dirichlet boundary condition $\mathcal{D}'$ as the trivial case
\cite{Chung:2016pgt,Okazaki:2019bok}.
They arise from the D3-branes ending on the NS5$'$-brane and on the D5$'$-branes respectively.
Consider the Nahm pole boundary condition for the vector multiplet of the node $U(N_I)$.
We put the boundary at $x^2=0$ and the bulk at $x^2>0$.
The $SU(2)_C$ triplet $\vec{\phi}$ of the vector multiplet describes the positions of the D3-branes along $x^{345}$ inside the D5$'$-brane.
At the boundary it obeys the $SU(2)_C$ covariant Nahm equation
\begin{align}
\label{Nahm_eq}
F_{\mu\nu}|_{\partial}&=0, & D_2\vec{\phi}+\vec{\phi}\times\vec{\phi}|_{\partial}&=0, \qquad \mu,\nu=0,1,
\end{align}
where $(\vec{\phi}\times\vec{\phi})_1=[\phi_2,\phi_3]$ and similarly for the other components by cyclic permutations.
As in 4d $\mathcal{N}=4$ SYM theory \cite{Gaiotto:2008sa}, the Nahm equation has the singular solution
\begin{align}
\label{Nahm_profile}
\vec{\phi}&=\frac{\vec{t}}{x^2}, & [t_1,t_2]&=t_3, & [t_2,t_3]&=t_1, & [t_3,t_1]&=t_2.
\end{align}
The Nahm pole boundary condition requires $\vec{\phi}$ to approach (\ref{Nahm_profile}) near the boundary up to a gauge transformation.
Since the matrices $t_a$ obey the commutation relations of $\mathfrak{su}(2)$,
they define a homomorphism $\rho:\mathfrak{su}(2)\rightarrow\mathfrak{u}(N_I)$, which labels the Nahm pole.
Up to conjugation, $\rho$ is fixed by the decomposition of the fundamental representation $\mathbb{C}^{N_I}$ of $U(N_I)$
into irreducible representations of $\mathfrak{su}(2)$.
We extend $\rho$ complex linearly to $\mathfrak{sl}(2,\mathbb{C})$ with the basis $\sigma_3$ and $\sigma^{\pm}=\frac12(\sigma_1\pm i\sigma_2)$,
where $\sigma_a$ are the Pauli matrices.
Up to normalization, the complex scalar field then behaves as $\varphi\sim\rho(\sigma^+)/x^2$
and the real scalar field as $\rho(\sigma_3)/x^2$.
The matrix $\rho(\sigma^+)$ is nilpotent, and its Jordan blocks have the sizes of the irreducible representations in $\mathbb{C}^{N_I}$.

The homomorphism is determined by how the D3-branes end.
When the $N_I$ D3-branes end on a single D5$'$-brane,
$\mathbb{C}^{N_I}$ is the irreducible representation of spin $\frac{N_I-1}{2}$ and $\rho$ is the principal embedding.
Then $\frac12\rho(\sigma_3)$ has the eigenvalues $\frac{N_I-1}{2},\frac{N_I-3}{2},\cdots,-\frac{N_I-1}{2}$,
and $\rho(\sigma^+)$ consists of a single Jordan block, i.e. it is a regular nilpotent element.
We call such a Nahm pole regular.
When the D3-branes end on several D5$'$-branes,
the irreducible representations have the dimensions given by the numbers of the D3-branes ending on each D5$'$-brane.
When each D3-brane ends on its own D5$'$-brane, $\rho$ is trivial and we get the Dirichlet boundary condition $\mathcal{D}'$.
In this paper we consider the principal embedding and leave general embeddings for future work.

At the boundary the Nahm pole breaks the gauge group $U(N_I)$ to the subgroup commuting with $\rho$ \cite{Gaiotto:2008sa},
which is its center $U(1)$ for the principal embedding.
Since $\mathfrak{gl}(N_I)=\mathrm{im}\,\mathrm{ad}\,\rho(\sigma^+)\oplus\ker\mathrm{ad}\,\rho(\sigma^-)$,
the fluctuations of $\varphi$ around $\rho(\sigma^+)$ split into two parts.
Those along the nilpotent orbit of $\rho(\sigma^+)$ are generated by the broken generators of the gauge group.
They are the Goldstone modes of the broken boundary symmetry, which decouple from the boundary condition (see section \ref{sec_Higgsing}).
The others are transverse to the orbit and lie in the Slodowy slice \cite{Gaiotto:2008sa}
\begin{align}
\label{Slodowy}
\mathcal{S}_{\rho}&=\rho(\sigma^+)+\ker\mathrm{ad}\,\rho(\sigma^-).
\end{align}
Under $\rho$ the adjoint representation decomposes into $\mathfrak{su}(2)$ multiplets,
and $\ker\mathrm{ad}\,\rho(\sigma^-)$ consists of their lowest weight components.
For the principal embedding the spins are $k=0,1,\cdots,N_I-1$, and $\mathcal{S}_{\rho}$ has dimension $N_I$.
Hence only the fluctuations in the Slodowy slice, i.e. the lowest weight component of each $\mathfrak{su}(2)$ multiplet, remain as the boundary degrees of freedom of the vector multiplet.
In the half-index the corresponding components of the boundary currents and the moment maps give the factor which we call the Slodowy slice (see section \ref{sec_UN_simple}).

We consider the Nahm pole boundary condition for the vector multiplets
together with the Dirichlet boundary condition $D'$ in (\ref{hyper_bc}) for the hypermultiplets,
which is realized by the brane configurations (see section \ref{sec_dualrules}).
We denote it by $(\mathrm{Nahm}',D')$, or simply by $\mathrm{Nahm}'$.
It arises from the Dirichlet boundary condition $(\mathcal{D}',D')$ by Higgsing the boundary symmetry (see section \ref{sec_Higgsing}),
which does not change the boundary condition for the hypermultiplets.
We therefore keep the notation $D'$ for the hypermultiplets.
The Higgsing nevertheless changes which components of the hypermultiplets give rise to boundary local operators,
as we explain in section \ref{sec_Nahm_hyper}.

\subsubsection{Boundary symmetries}
\label{sec_B}
For the Dirichlet boundary condition $\mathcal{D}'$ the gauge field obeys $A_{\mu}|_{\partial}=0$ with $\mu=0,1$,
so that the gauge transformations are fixed at the boundary and $U(N_I)$ acts as the boundary global symmetry $U(N_I)_{\partial}$.
The triplet $\vec{\phi}$ obeys the Nahm equation (\ref{Nahm_eq}) without a pole, which does not fix its boundary value.
Hence the boundary values of $\vec{\phi}$ give rise to boundary local operators in the adjoint representation of $U(N_I)_{\partial}$, which form an $SU(2)_C$ triplet.
As in 4d $\mathcal{N}=4$ SYM theory \cite{Gaiotto:2008sa}, they play the role of the moment maps for $U(N_I)_{\partial}$,
and the boundary value of the complex scalar field $\varphi$ is the complex moment map.
The Nahm pole boundary condition corresponds to the nilpotent vev $\langle\varphi\rangle=\rho(\sigma^+)$ of the moment map,
which reflects the singular profile of $\varphi$ in (\ref{Nahm_profile}).
Since $\varphi$ carries the charge $C=2$ and is neutral under $U(1)_H$ and the boundary rotation,
the vev breaks $U(N_I)_{\partial}\times U(1)_C$ down to the commutant of $\rho$
times the combination of $U(1)_C$ and $\rho(\sigma_3)$ which annihilates $\rho(\sigma^+)$.
We call the latter the twisted $U(1)_C$ and denote its generator by
\begin{align}
\label{C_rho}
C_{\rho}&=C-\rho(\sigma_3).
\end{align}

For the principal embedding the commutant is the center of $U(N_I)_{\partial}$ generated by $\mathbf{1}_{N_I}$.
We call it $U(1)_{u^{(I)}}$ and denote its fugacity by $u^{(I)}$.
It is the boundary flavor symmetry which survives at the Nahm pole.
The component of the hypermultiplet with gauge weight $w_{\lambda}$ carries the $U(1)_{u^{(I)}}$ charge
given by the sum of the components of $w_{\lambda}$ for the node $I$.
Hence its fugacity under $\prod_IU(1)_{u^{(I)}}$ is $\bar{u}^{w_{\lambda}}=\prod_I(u^{(I)})^{\sum_iw^{(I)}_{\lambda,i}}$ with $\bar{u}=(u^{(I)})$.
The boundary monopole operator of flux $m$ carries the charge $\sum_i(k_{\rm eff}m)^{(I)}_i$ due to the boundary anomaly (\ref{anom_DD}).
For a general embedding the commutant is a product of unitary groups, one for each size of the Jordan blocks of $\rho_I(\sigma^+)$,
and the boundary fugacities are given by its Cartan elements.
In contrast, the regular Nahm pole in 4d $\mathcal{N}=4$ $U(N)$ SYM theory leaves no boundary flavor symmetry \cite{Gaiotto:2019jvo},
since the center of $U(N)_{\partial}$ acts trivially on the adjoint fields and there are no boundary monopole operators.

\subsubsection{Hypermultiplets}
\label{sec_Nahm_hyper}
Let the hypermultiplets transform in a representation $R$ of the gauge group $G$.
In terms of 3d $\mathcal{N}=2$ supersymmetry a hypermultiplet consists of two chiral multiplets $X$ and $Y$, which transform in $R$ and its conjugate,
and the 3d $\mathcal{N}=4$ vector multiplet consists of a 3d $\mathcal{N}=2$ vector multiplet and an adjoint chiral multiplet with the complex scalar field $\varphi$.
They couple through the superpotential
$\mathcal{W}=Y\varphi X$.
At the Nahm pole $\varphi$ behaves as $\rho(\sigma^+)/x^2$ (see section \ref{sec_Nahm_def}),
so that $\mathcal{W}$ gives the hypermultiplet the nilpotent complex mass $\rho_R(\sigma^+)$ up to the factor $1/x^2$,
where $\rho_R$ is the action of $\rho(\mathfrak{su}(2))$ on $R$ through the gauge group, on both factors for a bifundamental hypermultiplet.
To see which components become massive, we decompose $R$ into irreducible representations of $\mathfrak{su}(2)$ under $\rho_R$.
We call the part of the hypermultiplet in an irreducible representation of spin $J$ a \emph{spin-$J$ multiplet}.
For example, a fundamental hypermultiplet of $U(N_I)$ is a single spin-$\frac{N_I-1}{2}$ multiplet.
Each of the two 3d $\mathcal{N}=2$ chiral multiplets forming it has $2J+1$ components.
We denote them by $X_{\mu}$ and $Y_{\mu}$ with the weights $\mu=-J,\cdots,J$ under $\frac12\rho(\sigma_3)$.
Since $\rho_R(\sigma^+)$ raises the weight by one, the mass term in $\mathcal{W}$ takes the form
\begin{align}
\label{spinJ_mass}
\mathcal{W}&\supset\sum_{\mu=-J}^{J-1}c_{\mu}\,Y_{-\mu-1}X_{\mu}
\end{align}
with non-zero coefficients $c_{\mu}$.
Thus the $2J$ pairs $(X_{\mu},Y_{-\mu-1})$ become massive,
and only $X_J$ and $Y_J$, which are annihilated by $\rho(\sigma^+)$, remain unpaired.
We call them the \emph{top components} of the spin-$J$ multiplet.
The pole of the real scalar field gives them a real mass proportional to $J/x^2$, which is reflected in their charges under the twisted $U(1)_C$ below.
\begin{quote}
\textit{At the Nahm pole each spin-$J$ multiplet of the hypermultiplets keeps only its two top components unpaired,
and its other $4J$ components pair up and become massive.}
\end{quote}

A massive pair with the Dirichlet boundary condition does not disappear completely.
It leaves a 2d Fermi multiplet on the boundary \cite{Dimofte:2017tpi}, which we call a \emph{split Fermi multiplet}.
The split Fermi multiplets decouple from the other boundary degrees of freedom.
Apart from them, the boundary local operators from the hypermultiplet are built from the top components.

Since $\rho(\mathfrak{su}(2))$ commutes with $U(1)_{u^{(I)}}$ and the other boundary and flavor symmetries,
all the components of a spin-$J$ multiplet carry the same charges under them.
Their fugacity under these symmetries is $\bar{u}^{w_{\lambda}}x^{f_{\lambda}}$ for any weight $\lambda=(w_{\lambda},f_{\lambda})$ of the multiplet,
and we label the multiplet by $(J,\lambda)$.
In contrast, the components carry different charges $C_{\rho}$ in (\ref{C_rho}) under the twisted $U(1)_C$.
Since the top component has the highest weight $J$ under $\frac12\rho(\sigma_3)$, its charge $C_{\rho}$ differs from $C$ by $2J$ in absolute value,
e.g. its fermions with $C=\pm1$ have $C_{\rho}=\pm(1+2J)$ (see section \ref{sec_Nahm_anom}).
The same holds for the twisted hypermultiplets with $U(1)_C$ replaced by $U(1)_H$.
We see these features explicitly in the half-index in section \ref{sec_Higgsing}.

\subsubsection{Brane construction}
\label{sec_dualrules}
In the brane construction the D3-branes of the node $I$ end on the D5$'$-brane.
The hypermultiplets arising from the strings attached to these D3-branes then obey the Dirichlet boundary condition $D'$
\cite{Chung:2016pgt,Hanany:2018hlz} (see section \ref{sec_conv}).
The D5$'$-brane forms a 5-brane web with the NS5-branes and is divided into segments by them.
The world-volume $U(1)$ symmetry of the segment on which the node $I$ ends acts as $U(1)_{u^{(I)}}$ at the boundary,
while the overall $U(1)$ common to all the segments acts trivially.
For 4d $\mathcal{N}=4$ SYM theory, in contrast, the D5$'$-brane is not divided by NS5-branes and no such symmetry arises.

Under S-duality the D5$'$-brane maps to the NS5$'$-brane,
so that the Nahm pole boundary condition is dual to the Neumann boundary condition of the mirror theory with boundary Fermi multiplets.
The segments of the D5$'$-brane map to those of the NS5$'$-brane divided by the D5-branes.
Accordingly, in the dual Neumann boundary condition $U(1)_{u^{(I)}}$ is realized,
up to mixing with the bulk flavor and topological symmetries,
as the flavor symmetry of the boundary Fermi multiplets supported on the NS5$'$-brane.
The boundary Fermi multiplets are in one-to-one correspondence with the top components
of the (twisted) hypermultiplets which survive at the Nahm poles (see section \ref{sec_Nahm_hyper}).
The top component of a spin-$J$ multiplet $(J,\lambda)$ corresponds to a Fermi multiplet with fugacity $q^{\frac12}(\bar{u}^{w_{\lambda}}x^{f_{\lambda}})^{\pm1}$,
dressed by the charges under the gauge and flavor symmetries of the Neumann side.
For the $USp(2N)$ ADHM theory with $N\ge2$ the fugacities of some of them are further shifted along the principal embedding (see section \ref{sec_USp_fw_N}).
This is a non-Abelian generalization of the minimal mirror identity in section \ref{sec_minimal},
in which the hypermultiplet of SQED$_1$ with the Dirichlet boundary condition corresponds to the boundary Fermi multiplet.

\begin{figure}[t]
\centering
\scalebox{0.85}{
\begin{tikzpicture}
\fill[yellow!25] (-2.2,0) rectangle (0,2.4);
\fill[yellow!50] (0,0) rectangle (2.2,2.4);
\draw[thick,solid,black] (-2.8,0) -- (2.8,0) node[right] {\textrm{NS5}$'$};
\draw[thick,solid,black] (0,-0.6) -- (0,2.7) node[above] {\textrm{NS5} $(z)$};
\node at (-1.1,1.2) {$n_L$ \textrm{D3}};
\node at (1.1,1.2) {$n_R$ \textrm{D3}};
\node at (-1.1,-0.3) {$u$};
\node at (1.1,-0.3) {$u$};
\fill[black] (0,0) circle (2pt) node[below right] {$\eta$};
\node at (0,-1.2) {(a) junction of NS5- and NS5$'$-branes};
\end{tikzpicture}
}
\hspace{0.8cm}
\scalebox{0.85}{
\begin{tikzpicture}
\fill[yellow!25] (-2.2,0) rectangle (0,2.4);
\fill[yellow!50] (0,0) rectangle (2.2,2.4);
\draw[thick,solid,black] (-2.8,0) -- (2.8,0) node[right] {\textrm{NS5}$'$};
\draw[thick,dashed,black] (0,-0.6) -- (0,2.7) node[above] {\textrm{D5} $(x)$};
\node at (-1.1,1.2) {$n_L$ \textrm{D3}};
\node at (1.1,1.2) {$n_R$ \textrm{D3}};
\node at (-1.1,-0.3) {$u_L$};
\node at (1.1,-0.3) {$u_R$};
\fill[black] (0,0) circle (2pt) node[below right] {$\Gamma$};
\node at (0,-1.2) {(b) intersection of D5- and NS5$'$-branes};
\end{tikzpicture}
}
\caption{\textrm{
The local brane configurations supporting the boundary Fermi multiplets of the Neumann boundary conditions.
The horizontal and vertical directions are $x^6$ and $x^2$.
The solid and dashed vertical lines are NS5- and D5-branes, and the shaded regions are D3-branes.
The fugacities for the $U(1)$ symmetries are shown in the parentheses and below the NS5$'$-brane.
The dots denote the boundary Fermi multiplets.
}}
\label{fig_local}
\end{figure}

The correspondence is manifest in the brane configuration.
We take the configuration of the Neumann side which is S-dual to that of the Nahm pole side,
in which the D3-branes of each node end on a single D5$'$-brane.
Then the boundary Fermi multiplets are supported at the two types of local configurations in Figure \ref{fig_local} \cite{Hanany:2018hlz,Okazaki:2019bok}.
\begin{itemize}

\item At the junction of an NS5-brane and the NS5$'$-brane (Figure \ref{fig_local} (a)), a Fermi multiplet $\eta$ is localized \cite{Hanany:2018hlz}.
It couples to the D3-branes on the two sides of the NS5-brane only through the determinants.
It transforms in the determinant representation of the gauge group on one side and in the inverse determinant representation of that on the other side,
and we call it the cross-determinant Fermi multiplet.
It carries the $U(1)$ charges of the NS5-brane and of the NS5$'$-brane, i.e. $U(1)_z$ and $U(1)_u$ in Figure \ref{fig_local} (a).
Here $U(1)_u$ acts on the segment of the NS5$'$-brane between the adjacent D5-branes, which is not divided by the NS5-brane.
Since the NS5-brane is S-dual to a flavor D5-brane on the Nahm pole side,
this Fermi multiplet corresponds to the top component of the fundamental hypermultiplet from that D5-brane.

\item At the intersection of a D5-brane and the NS5$'$-brane (Figure \ref{fig_local} (b)),
each of the $\min(n_L,n_R)$ D3-branes passing through the D5-brane gives a Fermi multiplet $\Gamma$ from the D3-D5 strings.
It carries the $U(1)$ charge of the D5-brane and the opposite charges of the segments of the NS5$'$-brane on the two sides of the D5-brane,
i.e. $U(1)_x$ and $U(1)_{u_L}\times U(1)_{u_R}$ in Figure \ref{fig_local} (b).
On the Nahm pole side the D5-brane becomes the NS5-brane separating the nodes $U(n_L)$ and $U(n_R)$.
The bifundamental hypermultiplet between them decomposes into the spins from $\frac{|n_L-n_R|}{2}$ to $\frac{n_L+n_R}{2}-1$
and leaves $\min(n_L,n_R)$ top components, which correspond to these Fermi multiplets.

\end{itemize}
The two types are thus distinguished by the 5-brane which meets the NS5$'$-brane, an NS5-brane for the Fermi multiplet $\eta$ and a D5-brane for the Fermi multiplet $\Gamma$.
Equivalently, on the Nahm pole side the Fermi multiplet $\eta$ corresponds to a fundamental hypermultiplet and the Fermi multiplet $\Gamma$ to a bifundamental (or adjoint) hypermultiplet.
We use the symbols $\eta$ and $\Gamma$ for the Fermi multiplets in this sense throughout the paper, with subscripts distinguishing them.
A D3-brane suspended between two NS5-branes contributes to the gauge group,
while a D3-brane ending on a D5-brane carries the $U(1)$ symmetry of that D5-brane.
Hence the Fermi multiplets become gauge-neutral when the D3-branes involved end on D5-branes,
and they carry the flavor charges of these D5-branes.
This is the origin of the gauge-neutral Fermi multiplets for $U(N)$ SQCD and $T[SU(N)]$ (see Figures \ref{fig_SQCD} (a) and \ref{fig_TSUN} (a)).
For the circular quivers the D3-branes on the two sides of an NS5-brane are related by the flavor fugacity of the bifundamental (or adjoint) hypermultiplet,
which dresses the cross-determinant Fermi multiplet.
When the two sides belong to the same node, as for the ADHM theory, the cross-determinant Fermi multiplet is gauge-neutral (see Figure \ref{fig_ADHM} (a)).

These Fermi multiplets cancel the boundary gauge anomaly and match the boundary 't Hooft anomalies with the Nahm pole side,
up to the mixed terms between the flavor and topological symmetries which can be absorbed into background Chern-Simons terms.
For the balanced linear and circular quivers the non-Abelian parts of the boundary gauge anomaly cancel.
The remaining part depends only on the Abelian parts $\sum_i\mathbf{s}^{(I)}_i$ of the field strengths and is
$-\sum_I(\sum_i\mathbf{s}^{(I)}_i-\sum_i\mathbf{s}^{(I-1)}_i)^2$.
Here $I$ labels the NS5-branes, the $I$-th of which separates the nodes $U(N_{I-1})$ and $U(N_I)$,
with $N_0=N_{n+1}=0$ for the linear quivers of $n$ nodes and $I$ understood modulo $n$ for the circular quivers.
Each term is cancelled by the cross-determinant Fermi multiplet at the corresponding NS5-brane \cite{Hanany:2018hlz},
which carries the charges $+1$ and $-1$ under the determinants of $U(N_I)$ and $U(N_{I-1})$.
For $U(N)$ SQCD with $2N$ flavors they are the two cross-determinant Fermi multiplets, in agreement with \cite{Costello:2018fnz}.
The gauge-neutral Fermi multiplets are invisible from the gauge anomaly.
They are required by the 't Hooft anomalies for $U(1)_{u^{(I)}}$,
e.g. the terms $I(\mathbf{u}^{(I)}-\mathbf{u}^{(I+1)})^2$ with $I<N$ of the Nahm pole side for the mirror of $U(N)$ SQCD.
Their flavor charges are fixed by the terms without the boundary symmetries,
which must match up to the mixed terms between the flavor and topological symmetries.
For the mirror of $U(N)$ SQCD the Nahm pole side has no terms quadratic in $\mathbf{x}_{\alpha}$,
since $x_{\alpha}$ enter only through the FI couplings.
The gauge-neutral Fermi multiplets cancel such terms from the hypermultiplets and the cross-determinant Fermi multiplets.

The half-indices confirm the assignments fixed by the anomalies.
In all the cases we have tested, other assignments of the charges lead to different half-indices (see Appendix \ref{app_checks}).
For $U(N)$ SQCD, $T[SU(N)]$ and the ADHM theory the assignment coincides with that read off from the brane configuration.

\subsection{Nahm pole half-index}
\label{sec_UN}
We first consider a single node $G=U(N)$ with hypermultiplets in a representation $R$.
Under $\rho(\mathfrak{su}(2))$ the representation $R$ decomposes into the spin-$J$ multiplets $(J,\lambda)$ as in section \ref{sec_Nahm_hyper}.
We extend the formula to quivers and write it in terms of the root system in section \ref{sec_UN_simple}.

\subsubsection{Higgsing}
\label{sec_Higgsing}
Rather than defining the half-index directly from the singular boundary condition,
we regard the Nahm pole boundary condition as the endpoint of the RG flow from the Dirichlet boundary condition,
as in 4d $\mathcal{N}=4$ $U(N)$ SYM theory \cite{Gaiotto:2019jvo} (see also \cite{Hatsuda:2024uwt,Hatsuda:2024lcc}).
There the Dirichlet boundary condition has the boundary global symmetry $U(N)_{\partial}$,
and the Nahm pole boundary condition arises from the nilpotent vev of its moment map.
A vev breaks every symmetry under which it is charged, and the half-index can keep the fugacities only for the unbroken symmetries.
Hence the boundary fugacities are specialized so that the components with the vev are neutral,
i.e. they carry zero charge under all the symmetries counted in the trace, and their fugacities become $1$.
The residue of the Dirichlet half-index at the resulting pole is the Nahm pole half-index
times the contributions of the Goldstone modes of the broken $U(N)_{\partial}$, which are 3d hypermultiplets on the boundary.
We follow the same steps in 3d.
For the Dirichlet boundary condition $\mathcal{D}'$ with the boundary fugacities $u_i$,
the components $\varphi_{ij}$ of the moment map contribute to the half-index as $q^{\frac12}t^{-2}u_i/u_j$
through the factors $1/(q^{\frac12+m_i-m_j}t^{-2}u_i/u_j;q)_{\infty}$ in (\ref{Dhalf}).
The nilpotent vev $\langle \varphi\rangle=\rho(\sigma^+)$ triggers the RG flow to the Nahm pole boundary condition labeled by $\rho$.
In the basis where $\rho(\sigma^+)$ is upper triangular, the vev lies in the components $\varphi_{i,i+1}$,
which must be neutral, i.e. $q^{\frac12}t^{-2}u_i/u_{i+1}=1$.
Their $U(1)_C$ charge $C=2$ is then cancelled by their $U(N)_{\partial}$ charges,
so that the unbroken symmetry is the combination of $U(1)_C$ and $U(N)_{\partial}$ in section \ref{sec_B}.
For the principal embedding this leads to the specialization
\begin{align}
\label{specialization}
u_i&=uq^{\frac{\mu_i}{2}}t^{-2\mu_i}, & \mu_i&=i-\frac{N+1}{2}, & i&=1,\cdots,N,
\end{align}
which we abbreviate as $(u_i)=u(q^{\frac12}t^{-2})^{\rho}$,
where $(q^{\frac12}t^{-2})^{\rho}$ denotes the collection of the powers $(q^{\frac12}t^{-2})^{\mu_i}$.
Here $u$ is the fugacity for the center $U(1)_u$ of $U(N)_{\partial}$, which survives at the Nahm pole (see section \ref{sec_B}).
In this basis $\frac12\rho(\sigma_3)$ is the diagonal matrix with the entries $-\mu_i$.
Hence the powers $(q^{\frac12}t^{-2})^{\mu_i}$ implement the twist of $U(1)_C$ by $\rho(\sigma_3)$.
Namely, a state of gauge weight $w=(w_i)$ has the fugacity $u^{w}=\prod_iu_i^{w_i}$, which becomes
\begin{align*}
u^{w}&=u^{|w|}(q^{\frac12}t^{-2})^{w(\mu)}, & |w|&=\sum_iw_i, & w(\mu)&=\sum_iw_i\mu_i.
\end{align*}
Since the trace (\ref{DEF_hind}) contains $q^{\frac{C}{4}}t^{-C}$ with the Cartan generator $C$ of $SU(2)_C$,
the factor $(q^{\frac12}t^{-2})^{w(\mu)}$ amounts to the shift $C\rightarrow C+2w(\mu)$,
where $-w(\mu)$ is the eigenvalue of $\frac12\rho(\sigma_3)$ on the state.
Namely, $C$ is replaced by $C_{\rho}$ in (\ref{C_rho}), under which $\varphi_{i,i+1}$ are neutral.
For example, the $i$-th component of the fundamental representation has $|w|=1$ and $w(\mu)=\mu_i$,
while the $(i,j)$ component of the adjoint representation has $|w|=0$ and $w(\mu)=\mu_i-\mu_j=i-j$.
Their fugacities are $u_i=u(q^{\frac12}t^{-2})^{\mu_i}$ and $u_i/u_j=(q^{\frac12}t^{-2})^{i-j}$ respectively.
Thus the fundamental representation carries the charge $1$ under $U(1)_u$ and has $w(\mu)=-\frac{N-1}{2},\cdots,\frac{N-1}{2}$,
while the adjoint representation is neutral under $U(1)_u$ and has $w(\mu)=i-j$ for $i,j=1,\cdots,N$.
Hence they decompose into the spin-$J$ multiplets with
\begin{align}
\label{spins_UN}
\textrm{fundamental}:&\quad J=\frac{N-1}{2}, & \textrm{adjoint}:&\quad J=0,1,\cdots,N-1.
\end{align}
Together with the bifundamental representation in section \ref{sec_UN_simple},
these are all the representations of the hypermultiplets in the examples with unitary gauge groups.

More generally, consider a spin-$J$ multiplet $(J,\lambda)$ of the hypermultiplets (see section \ref{sec_Nahm_hyper}).
For $U(N)$ we write $(w_{\lambda,i})$ for the gauge weight $w_{\lambda}$ of a weight $\lambda=(w_{\lambda},f_{\lambda})$,
together with $w_{\lambda}(\mu)=\sum_iw_{\lambda,i}\mu_i$ and $w_{\lambda}(m)=\sum_iw_{\lambda,i}m_i$.
Then the weight $\lambda$ has the fugacity $\bar{u}^{w_{\lambda}}x^{f_{\lambda}}(q^{\frac12}t^{-2})^{w_{\lambda}(\mu)}$, where $\bar{u}^{w_{\lambda}}=u^{\sum_iw_{\lambda,i}}$ as in section \ref{sec_B}.
The factor $\bar{u}^{w_{\lambda}}x^{f_{\lambda}}$ is common to all the components of the multiplet,
while $w_{\lambda}(\mu)$ takes the values $-J,\cdots,J$ within the multiplet.
Hence one of the two chiral multiplets in $(J,\lambda)$ has the $2J+1$ components with fugacities $\bar{u}^{w_{\lambda}}x^{f_{\lambda}}(q^{\frac12}t^{-2})^{\mu}$, $\mu=-J,\cdots,J$,
and the other has the conjugate components.
They are distributed symmetrically around $\bar{u}^{w_{\lambda}}x^{f_{\lambda}}$ and its inverse respectively.

The requirement that the components with the vev are neutral only fixes the ratios $u_{i+1}/u_i=q^{\frac12}t^{-2}$.
The general solution is $u_i=u(q^{\frac12}t^{-2})^{\mu_i+c}$,
and a different choice of $c$ corresponds to the mixing of $U(1)_u$ with the twisted $U(1)_C$.
The symmetric choice $\sum_i\mu_i=0$ reflects $\mathrm{Tr}\,\rho(\sigma_3)=0$.
It is the canonical choice for which $U(1)_u$ has no mixed anomaly with the twisted $U(1)_C$.
Indeed, the mixed anomaly is proportional to $\sum_i(k_{\rm eff}e)_i\mu_i$ with the generator $e$ of $U(1)_u$.
Since $(k_{\rm eff}e)_i$ does not depend on $i$ by the Weyl invariance, it vanishes for $\sum_i\mu_i=0$.
In the dual description this corresponds to the canonical R-charges of the boundary Fermi multiplets,
and other choices of $c$ lead to the powers $(q^{\frac12}t^{-2})^{\pm c}$ in their fugacities.

The half-index of the Dirichlet boundary condition has a pole at the specialization (\ref{specialization}).
Unlike in 4d, it is a sum over the boundary monopole sectors,
and only the sectors with the non-decreasing fluxes $m_1\le\cdots\le m_N$ have the pole (see the example below).
Its residue factorizes into the half-index of the IR boundary condition and the contributions from the degrees of freedom decoupled along the RG flow
\cite{Gaiotto:2012xa,Gaiotto:2019jvo}.
There are two types of such degrees of freedom.
\begin{itemize}
\item \textbf{Goldstone modes}

The broken generators of $U(N)_{\partial}$ lead to the 2d $\mathcal{N}=(0,4)$ twisted hypermultiplets
with fugacities $q^{-\frac14}tu_j/u_i$, one for each pair $i<j$.
\item \textbf{Split Fermi multiplets}

At $(u_i)=u(q^{\frac12}t^{-2})^{\rho}$ the Dirichlet half-index of each spin-$J$ multiplet $(J,\lambda)$ in the sector of zero flux
contains the $2J$ split Fermi multiplets from the massive pairs (see section \ref{sec_Nahm_hyper}).
Dividing by them leaves only the top component
\begin{align}
\label{top_component}
\frac{\prod_{\mu=-J}^{J}(q^{\frac34\pm\frac{\mu}{2}}t^{-1\mp2\mu}(\bar{u}^{w_{\lambda}}x^{f_{\lambda}})^{\pm};q)_{\infty}}
{\prod_{\kappa=-(J-\frac12)}^{J-\frac12}F(q^{\frac{1+\kappa}{2}}t^{-2\kappa}\bar{u}^{w_{\lambda}}x^{f_{\lambda}};q)}
&=(q^{\frac34+\frac{J}{2}}t^{-1-2J}(\bar{u}^{w_{\lambda}}x^{f_{\lambda}})^{\pm};q)_{\infty}.
\end{align}
\end{itemize}

For example, consider $U(2)$ with a fundamental hypermultiplet of flavor fugacity $x$,
for which $u_1=uq^{-\frac14}t$ and $u_2=uq^{\frac14}t^{-1}$.
The vev $\langle\varphi\rangle=\sigma^+$ breaks $U(2)_{\partial}$ down to its center $U(1)_{u}$ with $u_1u_2=u^2$.
Under $U(1)_u$ the fundamental hypermultiplet carries charge $1$
and the boundary monopole operator of fluxes $(m_1,m_2)$ carries charge $(k_{\rm eff}m)_1+(k_{\rm eff}m)_2$.
Hence $U(1)_u$ survives as a boundary flavor symmetry (see section \ref{sec_B}).
In the sector with fluxes $(m_1,m_2)$ the factor $1/(q^{\frac12+m_1-m_2}t^{-2}\frac{u_1}{u_2};q)_{\infty}$ in (\ref{Dhalf})
becomes $1/(q^{m_1-m_2};q)_{\infty}$, which has the pole only for $m_1\le m_2$.
The same pole appears in the Goldstone mode $I^{\textrm{2d tHM}}(t,q^{\frac14}t^{-1};q)$, which contains $1/(1;q)_{\infty}$.
The Dirichlet half-index of the hypermultiplet becomes
\begin{align*}
(q^{\frac34\pm m_1}t^{-1}(u_1x)^{\pm};q)_{\infty}(q^{\frac34\pm m_2}t^{-1}(u_2x)^{\pm};q)_{\infty}
&=(q^{\frac12+m_1}ux;q)_{\infty}(q^{1-m_1}t^{-2}(ux)^{-1};q)_{\infty}\\*
&\times(q^{1+m_2}t^{-2}ux;q)_{\infty}(q^{\frac12-m_2}(ux)^{-1};q)_{\infty}.
\end{align*}
For $m_1=m_2=0$ the first and the last factors form the split Fermi multiplet $F(q^{\frac12}ux;q)$
and the other two give the top component $(qt^{-2}(ux)^{\pm};q)_{\infty}$ of the doublet.

\subsubsection{Prescription}
\label{sec_proposal}
Based on the Higgsing in section \ref{sec_Higgsing},
we propose that the half-index of the boundary condition $(\mathrm{Nahm}',D')$ is obtained by the following prescription.
\begin{enumerate}
\item Write down the Dirichlet half-index (\ref{Dhalf}).
\item Specialize $u_i=uq^{\frac{\mu_i}{2}}t^{-2\mu_i}$ as in (\ref{specialization}).
Then $u_j/u_i=(q^{\frac12}t^{-2})^{j-i}$, and the components $\varphi_{i,i+1}$ of the moment map, which contain the vev, become neutral.
\item Keep only the dominant sectors $m_1\le \cdots\le m_N$.
For $i=1,\cdots,N-1$, the factor $1/(q^{\frac12+m_i-m_{i+1}}t^{-2}u_i/u_{i+1};q)_{\infty}$ of the vector multiplet in (\ref{Dhalf})
then becomes $1/(q^{-n_i};q)_{\infty}$ with $n_i=m_{i+1}-m_i\ge 0$.
It is the contribution of the components $\varphi_{i,i+1}$ in step 2,
and it diverges because of their zero mode $1/(1-q^0)$.
Remove the zero mode by replacing $(q^{-n_i};q)_{\infty}$ with
$(q^{-n_i};q)'_{\infty}=(-1)^{n_i}q^{-\frac{n_i(n_i+1)}{2}}(q)_{n_i}(q)_{\infty}$, in which the factor $1-q^0$ is omitted.
\item Strip off the Goldstone modes, i.e. the 2d twisted hypermultiplets with fugacities $q^{-\frac14}tu_j/u_i$ for $1\le i<j\le N$,
\begin{align}
\label{2dtHM_UN}
&\prod_{1\le i<j\le N}I^{\textrm{2d tHM}}(t,q^{\frac{2(j-i)-1}{4}}t^{1-2(j-i)};q)
\nonumber\\*
&=\prod_{k=1}^{N-1}\frac{1}{[(q^{\frac{k}{2}}t^{-2k};q)_{\infty}(q^{1-\frac{k}{2}}t^{2k};q)_{\infty}(q^{\frac{1-k}{2}}t^{2k-2};q)_{\infty}(q^{\frac{1+k}{2}}t^{2-2k};q)_{\infty}]^{N-k}},
\end{align}
where $N-k$ pairs have $j-i=k$.
For $j=i+1$ the factor $(1;q)_{\infty}$ vanishes.
This zero corresponds to the zero mode in step 3, and we remove it by replacing $(1;q)_{\infty}$ with $(1;q)'_{\infty}=(q)_{\infty}$.
\item Strip off the split Fermi multiplets $\prod_{(J,\lambda)}\prod_{\kappa}F(q^{\frac{1+\kappa}{2}}t^{-2\kappa}\bar{u}^{w_{\lambda}}x^{f_{\lambda}};q)$,
where $\kappa$ runs from $-(J-\frac12)$ to $J-\frac12$ in steps of one.
The spin-$0$ components contribute no Fermi multiplets, while each fundamental hypermultiplet leads to $N-1$ Fermi multiplets.
\end{enumerate}
The prescription gives
\begin{align}
\label{Nahm_single}
&\mathbb{II}_{\mathrm{Nahm}'}^{U(N),R}
\nonumber\\*
&=\sum_{m_1\le\cdots\le m_N}\underbrace{(-q^{\frac12})^{m^Tk_{\rm eff}m}u^{\sum_i(k_{\rm eff}m)_i}(q^{\frac12}t^{-2})^{\sum_i\mu_i(k_{\rm eff}m)_i}\prod_{\lambda\in R}x^{w_{\lambda}(m)f_{\lambda}}z^{-\sum_im_i}}_{\textrm{bare monopole operators}}
\nonumber\displaybreak[0]\\
&\times \underbrace{\frac{1}{[(q)_{\infty}(q^{\frac12}t^{-2};q)_{\infty}]^N}{\prod_{i\neq j}}'\frac{1}{(q^{1+m_i-m_j+\frac{i-j}{2}}t^{-2(i-j)};q)_{\infty}(q^{\frac12+m_i-m_j+\frac{i-j}{2}}t^{-2(1+i-j)};q)_{\infty}}}_{\textrm{vector multiplet ($\mathcal{D}'$)}}
\nonumber\displaybreak[0]\\
&\times \underbrace{\prod_{\lambda\in R}(q^{\frac34\pm w_{\lambda}(m)\pm\frac12w_{\lambda}(\mu)}t^{-1\mp2w_{\lambda}(\mu)}(\bar{u}^{w_{\lambda}}x^{f_{\lambda}})^{\pm};q)_{\infty}}_{\textrm{hypermultiplets ($D'$)}}\;
\underbrace{\prod_{(J,\lambda)}\prod_{\kappa}\frac{1}{F(q^{\frac{1+\kappa}{2}}t^{-2\kappa}\bar{u}^{w_{\lambda}}x^{f_{\lambda}};q)}}_{(\textrm{split Fermi multiplets})^{-1}}
\nonumber\\*
&\times \underbrace{\prod_{k=1}^{N-1}[(q^{\frac{k}{2}}t^{-2k};q)_{\infty}(q^{1-\frac{k}{2}}t^{2k};q)_{\infty}(q^{\frac{1-k}{2}}t^{2k-2};q)'_{\infty}(q^{\frac{1+k}{2}}t^{2-2k};q)_{\infty}]^{N-k}}_{(\textrm{2d twisted hypermultiplets})^{-1}},
\end{align}
where the prime indicates that the second factor $(q^{-n_i};q)_{\infty}$ for $j=i+1$ is replaced by $(q^{-n_i};q)'_{\infty}$ as in step 3.
For $U(N)$ SQCD with $N_f$ flavors, (\ref{keff_SQCD}) leads to the factor $(q^{\frac12}t^{-2})^{(N_f-2N)\sum_i\mu_im_i}$,
which is trivial only for the balanced node.
We call (\ref{Nahm_single}) the expanded form, since it follows the Higgsing step by step.

\subsubsection{Formula}
\label{sec_UN_simple}
The factors of the expanded form (\ref{Nahm_single}) are not separately meaningful,
and the Nahm pole half-index takes a simpler form if we apply the prescription to the form (\ref{D_simple}) of the Dirichlet half-index.
At the specialization (\ref{specialization}) we have $u_j/u_i=(q^{\frac12}t^{-2})^{j-i}$,
and the weight $\lambda$ has the fugacity $\bar{u}^{w_{\lambda}}x^{f_{\lambda}}(q^{\frac12}t^{-2})^{w_{\lambda}(\mu)}$.
In the summand of (\ref{D_simple}) the factor $1/(q^{\frac12}t^2u_{i+1}/u_i;q)_{m_{i+1}-m_i}$ becomes $1/(q;q)_{m_{i+1}-m_i}$,
which vanishes for $m_{i+1}<m_i$.
Hence only the dominant sectors $m_1\le\cdots\le m_N$ survive, and their summands stay finite.
In the prefactor the factors $(q^{\frac12}t^{-2}u_i/u_{i+1};q)_{\infty}$ become $(1;q)_{\infty}$,
which we replace with $(q)_{\infty}$ as in steps 3 and 4.
Then the factors of the vector multiplet divided by the Goldstone modes (\ref{2dtHM_UN}) give the factor of the Slodowy slice (\ref{Slodowy}),
and the factors of the hypermultiplets divided by the split Fermi multiplets give the top components (\ref{top_component}).
Since (\ref{D_simple}) agrees with (\ref{Dhalf}) in each sector, the result is equal to the expanded form (\ref{Nahm_single}), and we find
\begin{align}
\label{Nahm_single_simple}
&\mathbb{II}_{\mathrm{Nahm}'}^{U(N),R}
\nonumber\\*
&=\underbrace{\prod_{(J,\lambda)}(q^{\frac34+\frac{J}{2}}t^{-1-2J}(\bar{u}^{w_{\lambda}}x^{f_{\lambda}})^{\pm};q)_{\infty}}_{\textrm{top components}}\;
\underbrace{\prod_{k=0}^{N-1}\frac{1}{(q^{1+\frac{k}{2}}t^{-2k};q)_{\infty}(q^{\frac12+\frac{k}{2}}t^{-2-2k};q)_{\infty}}}_{\textrm{Slodowy slice}}
\nonumber\\
&\times\sum_{m_1\le\cdots\le m_N}\quad\underbrace{z^{-\sum_im_i}}_{\textrm{topological charge}}
\nonumber\\
&\times\underbrace{\prod_{1\le i<j\le N}(q^{\frac12}t^{-2})^{m_i-m_j}\frac{(q^{\frac12+\frac{j-i}{2}}t^{-2-2(j-i)};q)_{m_j-m_i}(q^{1+\frac{j-i}{2}}t^{-2(j-i)};q)_{m_j-m_i}}{(q^{\frac{j-i}{2}}t^{-2(j-i)};q)_{m_j-m_i}(q^{\frac12+\frac{j-i}{2}}t^{2-2(j-i)};q)_{m_j-m_i}}}_{\textrm{vector multiplet ($\mathrm{Nahm}'$)}}
\nonumber\\*
&\times\underbrace{\prod_{\lambda\in R}(q^{\frac14}t^{-1})^{w_{\lambda}(m)}\frac{(q^{\frac14+\frac{w_{\lambda}(\mu)}{2}}t^{1-2w_{\lambda}(\mu)}\bar{u}^{w_{\lambda}}x^{f_{\lambda}};q)_{w_{\lambda}(m)}}{(q^{\frac34+\frac{w_{\lambda}(\mu)}{2}}t^{-1-2w_{\lambda}(\mu)}\bar{u}^{w_{\lambda}}x^{f_{\lambda}};q)_{w_{\lambda}(m)}}}_{\textrm{hypermultiplets ($D'$)}}.
\end{align}
Here $(J,\lambda)$ runs over the spin-$J$ multiplets of $R$.
By (\ref{spins_UN}) each fundamental hypermultiplet gives a single multiplet with $J=\frac{N-1}{2}$,
and the adjoint hypermultiplet gives $N$ multiplets with $J=0,1,\cdots,N-1$.
Comparing with (\ref{D_simple}), the topological charge, the vector multiplet and the hypermultiplets give the same factors at the specialization.
Only the prefactor, i.e. the contribution of the sector $m=0$, is changed by the Higgsing.
The Slodowy slice comes from the boundary currents and the moment maps of the vector multiplet.
As in (\ref{Slodowy}), only the lowest weight component of each multiplet of spin $k=0,\cdots,N-1$ survives,
with $q^{1+\frac{k}{2}}t^{-2k}$ and $q^{\frac12+\frac{k}{2}}t^{-2-2k}$ respectively.
Their charges are shifted by $k$ as for the hypermultiplets.

The vector multiplet factor in (\ref{Nahm_single_simple}) has a simple interpretation in terms of the Macdonald polynomials \cite{MR1354144}.
For symmetric Laurent polynomials $f(x)$ and $g(x)$ of $x=(x_1,\cdots,x_N)$, the Macdonald inner product is
\begin{align}
\label{Mac_inner}
\langle f,g\rangle&=\frac{1}{N!}\oint\prod_{i=1}^N\frac{dx_i}{2\pi ix_i}\prod_{i\neq j}\frac{(\frac{x_i}{x_j};\mathfrak{q})_{\infty}}{(\mathfrak{t}\frac{x_i}{x_j};\mathfrak{q})_{\infty}}\,f(x)g(x^{-1}),
\end{align}
where the contour is the unit torus.
For a partition $(m_N,\cdots,m_1)$ with $m_N\ge\cdots\ge m_1\ge 0$,
the Macdonald polynomial $P_m(x;\mathfrak{q},\mathfrak{t})$ is the symmetric polynomial
whose leading term is the sum of the monomials $x_1^{m_{\sigma(1)}}\cdots x_N^{m_{\sigma(N)}}$ over the distinct permutations $\sigma$ of $m$.
The other terms are such sums for the partitions lower than $m$ in the dominance order,
and they are fixed by the orthogonality $\langle P_m,P_{m'}\rangle=0$ for $m\neq m'$.
We label $P_m$ by the dominant flux $m$ and define $P_m$ with negative fluxes by $P_{m+(c,\cdots,c)}=(x_1\cdots x_N)^cP_m$ for $c\in\mathbb{Z}$.
Then $\langle P_m,P_m\rangle$ is the norm of $P_m$, and $\langle 1,1\rangle$ is that of $P_{(0,\cdots,0)}=1$.
We set $\mathfrak{q}=q$ and $\mathfrak{t}=q^{\frac12}t^{-2}$,
which are the parameters of the Macdonald inner product for the Neumann half-indices in 4d $\mathcal{N}=4$ SYM theories \cite{Hatsuda:2025yzp}.
The value of $P_m$ at $x=(q^{\frac12}t^{-2})^{\rho}$, i.e. $x_i=(q^{\frac12}t^{-2})^{\mu_i}$, is its principal specialization.
The formulas for the principal specialization \cite{MR1354144} and for the normalized norm \cite{Hatsuda:2025yzp} give
\begin{align}
\label{Mac_weight}
&\frac{P_m((q^{\frac12}t^{-2})^{\rho};q,q^{\frac12}t^{-2})^2}{\langle P_m,P_m\rangle/\langle 1,1\rangle}
\nonumber\\*
&=\prod_{1\le i<j\le N}(q^{\frac12}t^{-2})^{m_i-m_j}
\frac{(q^{\frac12+\frac{j-i}{2}}t^{-2-2(j-i)};q)_{m_j-m_i}(q^{1+\frac{j-i}{2}}t^{-2(j-i)};q)_{m_j-m_i}}{(q^{\frac{j-i}{2}}t^{-2(j-i)};q)_{m_j-m_i}(q^{\frac12+\frac{j-i}{2}}t^{2-2(j-i)};q)_{m_j-m_i}},
\end{align}
whose right-hand side is the vector multiplet factor in (\ref{Nahm_single_simple}).
Hence each boundary monopole sector is weighted by the squared principal specialization of the Macdonald polynomial divided by its norm.
We call (\ref{Mac_weight}) the Macdonald weight.
The point $x=(q^{\frac12}t^{-2})^{\rho}$ of the principal specialization is the specialization (\ref{specialization}) of the boundary fugacities
required by the Higgsing.
For example, the Macdonald polynomial of two variables is \cite{MR1354144}
\begin{align}
\label{Mac_U2}
P_{(m_1,m_2)}(x;\mathfrak{q},\mathfrak{t})&=(x_1x_2)^{m_1}\frac{(\mathfrak{q};\mathfrak{q})_n}{(\mathfrak{t};\mathfrak{q})_n}
\sum_{k=0}^{n}\frac{(\mathfrak{t};\mathfrak{q})_k(\mathfrak{t};\mathfrak{q})_{n-k}}{(\mathfrak{q};\mathfrak{q})_k(\mathfrak{q};\mathfrak{q})_{n-k}}x_1^kx_2^{n-k},
\end{align}
with $n=m_2-m_1$.
For $\mathfrak{q}=q$ and $\mathfrak{t}=q^{\frac12}t^{-2}$, its principal specialization and its normalized norm are
\begin{align}
P_{(m_1,m_2)}((q^{\frac12}t^{-2})^{\rho};q,q^{\frac12}t^{-2})&=(q^{\frac12}t^{-2})^{-\frac{n}{2}}\frac{(qt^{-4};q)_n}{(q^{\frac12}t^{-2};q)_n},
\\*
\frac{\langle P_{(m_1,m_2)},P_{(m_1,m_2)}\rangle}{\langle 1,1\rangle}&=\frac{(qt^{-4};q)_n(q;q)_n}{(q^{\frac12}t^{-2};q)_n(q^{\frac32}t^{-2};q)_n},
\end{align}
and the ratio (\ref{Mac_weight}) gives the vector multiplet factor $(q^{\frac12}t^{-2})^{-n}\frac{(qt^{-4};q)_n(q^{\frac32}t^{-2};q)_n}{(q^{\frac12}t^{-2};q)_n(q;q)_n}$ of (\ref{Nahm_single_simple}) for $N=2$.

The formula (\ref{Nahm_single_simple}) extends to quiver gauge theories with $G=\prod_IU(N_I)$ and hypermultiplets in a representation $R$.
As in all the examples in this paper, we impose the principal Nahm pole boundary conditions on the non-Abelian nodes
and the Dirichlet boundary conditions on the Abelian nodes.
We apply the prescription in section \ref{sec_proposal} at each Nahm pole node $U(N_I)$ with the specialization
\begin{align}
\label{specialization_gen}
u^{(I)}_i&=u^{(I)}q^{\frac{\mu^{(I)}_i}{2}}t^{-2\mu^{(I)}_i}, & \mu^{(I)}_i&=i-\frac{N_I+1}{2}, & i&=1,\cdots,N_I.
\end{align}
For a weight $\lambda=(w_{\lambda},f_{\lambda})$ of $R$ we write $w_{\lambda}(\mu)=\sum_{I,i}w^{(I)}_{\lambda,i}\mu^{(I)}_i$, and $w_{\lambda}(m)$ similarly.
Then the weight $\lambda$ has the fugacity $\bar{u}^{w_{\lambda}}x^{f_{\lambda}}(q^{\frac12}t^{-2})^{w_{\lambda}(\mu)}$,
where $\bar{u}^{w_{\lambda}}$ is the fugacity under $\prod_IU(1)_{u^{(I)}}$ (see section \ref{sec_B}).
When a node adjacent to a Nahm pole node obeys the Dirichlet boundary condition $\mathcal{D}'$, its boundary fugacities are not specialized,
and the bifundamental hypermultiplet is regarded as fundamental hypermultiplets of the Nahm pole node.
For example, the $(i,a)$ component of the bifundamental representation of two adjacent nodes $U(N_I)$ and $U(N_{I+1})$
has $\bar{u}^{w_{\lambda}}=u^{(I)}/u^{(I+1)}$ and $w_{\lambda}(\mu)=\mu^{(I)}_i-\mu^{(I+1)}_a$.
Since $\mu^{(I)}_i$ and $\mu^{(I+1)}_a$ run over the weights of the spins $\frac{N_I-1}{2}$ and $\frac{N_{I+1}-1}{2}$,
the bifundamental representation decomposes into the spin-$J$ multiplets with
\begin{align}
\label{spins_bifund}
J&=\frac{|N_I-N_{I+1}|}{2},\ \frac{|N_I-N_{I+1}|}{2}+1,\ \cdots,\ \frac{N_I+N_{I+1}}{2}-1,
\end{align}
as in section \ref{sec_dualrules}.
We find
\begin{align}
\label{Nahm_quiver_simple}
&\mathbb{II}_{\mathrm{Nahm}'}^{\prod_IU(N_I),R}
\nonumber\\*
&=\underbrace{\prod_{(J,\lambda)}(q^{\frac34+\frac{J}{2}}t^{-1-2J}(\bar{u}^{w_{\lambda}}x^{f_{\lambda}})^{\pm};q)_{\infty}}_{\textrm{top components}}\;
\underbrace{\prod_I\prod_{k=0}^{N_I-1}\frac{1}{(q^{1+\frac{k}{2}}t^{-2k};q)_{\infty}(q^{\frac12+\frac{k}{2}}t^{-2-2k};q)_{\infty}}}_{\textrm{Slodowy slices}}
\nonumber\\
&\times\sum_{m}\underbrace{\prod_Iz_I^{-\sum_im^{(I)}_i}}_{\textrm{topological charges}}
\nonumber\\
&\times\underbrace{\prod_I\prod_{1\le i<j\le N_I}(q^{\frac12}t^{-2})^{m^{(I)}_i-m^{(I)}_j}\frac{(q^{\frac12+\frac{j-i}{2}}t^{-2-2(j-i)};q)_{m^{(I)}_j-m^{(I)}_i}(q^{1+\frac{j-i}{2}}t^{-2(j-i)};q)_{m^{(I)}_j-m^{(I)}_i}}{(q^{\frac{j-i}{2}}t^{-2(j-i)};q)_{m^{(I)}_j-m^{(I)}_i}(q^{\frac12+\frac{j-i}{2}}t^{2-2(j-i)};q)_{m^{(I)}_j-m^{(I)}_i}}}_{\textrm{vector multiplets ($\mathrm{Nahm}'$)}}
\nonumber\\*
&\times\underbrace{\prod_{\lambda\in R}(q^{\frac14}t^{-1})^{w_{\lambda}(m)}\frac{(q^{\frac14+\frac{w_{\lambda}(\mu)}{2}}t^{1-2w_{\lambda}(\mu)}\bar{u}^{w_{\lambda}}x^{f_{\lambda}};q)_{w_{\lambda}(m)}}{(q^{\frac34+\frac{w_{\lambda}(\mu)}{2}}t^{-1-2w_{\lambda}(\mu)}\bar{u}^{w_{\lambda}}x^{f_{\lambda}};q)_{w_{\lambda}(m)}}}_{\textrm{hypermultiplets ($D'$)}},
\end{align}
where the sum runs over the fluxes $m^{(I)}_1\le\cdots\le m^{(I)}_{N_I}$ of all the nodes,
and $(J,\lambda)$ runs over the spin-$J$ multiplets of $R$, e.g. (\ref{spins_UN}) with $N=N_I$ and (\ref{spins_bifund}).
For a single node it reduces to (\ref{Nahm_single_simple}), and it follows from (\ref{D_simple}) in the same way.
The vector multiplet factor of each Nahm pole node is the Macdonald weight (\ref{Mac_weight}) of $U(N_I)$, and that of an Abelian node is $1$.
In sections \ref{sec_SQCD}, \ref{sec_TSUN} and \ref{sec_ADHM} we write (\ref{Nahm_single_simple}) and (\ref{Nahm_quiver_simple}) explicitly for all the examples.

The same argument applies to more general gauge groups and representations of the hypermultiplets with principal Nahm poles,
and we write the result in terms of the root system.
We abbreviate the specialization as $u=\bar{u}(q^{\frac12}t^{-2})^{\rho}$ with $\rho=\oplus_I\rho_I$.
The hypermultiplets decompose into the spin-$J$ multiplets $(J,\lambda)$ under $\rho(\mathfrak{su}(2))$,
and for a fixed $\bar{u}^{w_{\lambda}}x^{f_{\lambda}}$ the spins are read off from the values of $w_{\lambda}(\mu)$, which run from $-J$ to $J$ in each multiplet.
Let $\Delta^+_{\rho}$ be the set of the positive roots $\alpha$ of the Nahm pole nodes,
and let $\alpha(m)$ and $u^{\alpha}$ be its pairings with the flux $m$ and the fugacity $u$.
We write $\alpha=\sum_an_a\alpha_a$ with the simple roots $\alpha_a$ and non-negative integers $n_a$, and define the height of $\alpha$ by
\begin{align}
\label{ht_def}
\mathrm{ht}\,\alpha&=\sum_an_a,
\end{align}
i.e. the number of the simple roots contained in $\alpha$.
For example, $\alpha=e_j-e_i$ with $1\le i<j\le N_I$ for $U(N_I)$ has $\mathrm{ht}\,\alpha=j-i$, $\alpha(m)=m^{(I)}_j-m^{(I)}_i$ and $u^{\alpha}=u^{(I)}_j/u^{(I)}_i$.
In this convention the components $\varphi_{i,i+1}$ of the moment map, which contain the vev $\rho(\sigma^+)$, lie along the negative simple roots $e_i-e_{i+1}$.
At the specialization we have $u^{\alpha}=(q^{\frac12}t^{-2})^{\mathrm{ht}\,\alpha}$.
The Goldstone modes are the 2d twisted hypermultiplets with fugacities $q^{-\frac14}tu^{\alpha}$ for $\alpha\in\Delta^+_{\rho}$,
and the split Fermi multiplets are given by step 5 of the prescription for each spin-$J$ multiplet.
The Higgsing leads to the factorization
\begin{align}
\label{Nahm_limit}
\mathbb{II}^{\mathcal{T}}_{\mathcal{D}',D'}(u)
&\simeq \mathbb{II}^{\mathcal{T}}_{\mathrm{Nahm}'}(\bar{u})\,
\underbrace{\prod_{\alpha\in\Delta^+_{\rho}}I^{\textrm{2d tHM}}(t,q^{-\frac14}tu^{\alpha};q)}_{\textrm{2d twisted hypermultiplets}}
\nonumber\\*
&\times \underbrace{\prod_{(J,\lambda)}\prod_{\kappa}F(q^{\frac{1+\kappa}{2}}t^{-2\kappa}\bar{u}^{w_{\lambda}}x^{f_{\lambda}};q)}_{\textrm{split Fermi multiplets}},
\end{align}
as $u\rightarrow \bar{u}(q^{\frac12}t^{-2})^{\rho}$.
Both sides have poles of the same order from the simple roots, and $\simeq$ means that their leading coefficients agree.
We propose that the Nahm pole half-index is given by
\begin{align}
\label{Nahm_simple}
&\mathbb{II}^{\mathcal{T}}_{\mathrm{Nahm}'}(\bar{u})
\nonumber\\*
&=\underbrace{\prod_{(J,\lambda)}(q^{\frac34+\frac{J}{2}}t^{-1-2J}(\bar{u}^{w_{\lambda}}x^{f_{\lambda}})^{\pm};q)_{\infty}}_{\textrm{top components}}\;
\underbrace{\prod_{k}\frac{1}{(q^{1+\frac{k}{2}}t^{-2k};q)_{\infty}(q^{\frac12+\frac{k}{2}}t^{-2-2k};q)_{\infty}}}_{\textrm{Slodowy slices}}
\nonumber\\
&\times\sum_{m\in\Lambda^+_{\rho}}\underbrace{\prod_Iz_I^{-\sum_im^{(I)}_i}}_{\textrm{topological charges}}\;
\underbrace{\prod_{\alpha\in\Delta^+_{\rho}}(q^{\frac12}t^{-2})^{-\alpha(m)}
\frac{(q^{\frac12+\frac{\mathrm{ht}\alpha}{2}}t^{-2-2\mathrm{ht}\alpha};q)_{\alpha(m)}(q^{1+\frac{\mathrm{ht}\alpha}{2}}t^{-2\mathrm{ht}\alpha};q)_{\alpha(m)}}
{(q^{\frac{\mathrm{ht}\alpha}{2}}t^{-2\mathrm{ht}\alpha};q)_{\alpha(m)}(q^{\frac12+\frac{\mathrm{ht}\alpha}{2}}t^{2-2\mathrm{ht}\alpha};q)_{\alpha(m)}}}_{\textrm{vector multiplets ($\mathrm{Nahm}'$)}}
\nonumber\\*
&\times\underbrace{\prod_{\lambda}(q^{\frac14}t^{-1})^{w_{\lambda}(m)}
\frac{(q^{\frac14+\frac{w_{\lambda}(\mu)}{2}}t^{1-2w_{\lambda}(\mu)}\bar{u}^{w_{\lambda}}x^{f_{\lambda}};q)_{w_{\lambda}(m)}}
{(q^{\frac34+\frac{w_{\lambda}(\mu)}{2}}t^{-1-2w_{\lambda}(\mu)}\bar{u}^{w_{\lambda}}x^{f_{\lambda}};q)_{w_{\lambda}(m)}}}_{\textrm{hypermultiplets ($D'$)}},
\end{align}
where $k$ runs over the spins in the decomposition of the adjoint representation of the Lie algebra $\mathfrak{g}$ of $G$ under $\rho$,
and $\Lambda^+_{\rho}$ is the set of the cocharacters $m$ which are dominant at the Nahm pole nodes, i.e. the dominant coweights.
For $G=\prod_IU(N_I)$ it is (\ref{Nahm_quiver_simple}), which follows from (\ref{D_simple}) as above, so that (\ref{Nahm_limit}) holds by construction.
For $G=USp(2N)$ we have checked (\ref{Nahm_limit}) in examples (see Appendix \ref{app_checks}).
As for $U(N)$, the prefactor is the contribution of the sector $m=0$.
For $m\neq 0$ the shifts $q^{\pm w_{\lambda}(m)}$ spoil the cancellation between the hypermultiplets and the split Fermi multiplets,
which gives rise to the non-trivial sums over the fluxes.

All the factors in the sum are finite products,
and neither the zero modes nor the decoupled multiplets have to be stripped off.
Each summand is a rational function of $q^{\frac12}$, $q^{\frac14}t^{-1}$ and the fugacities,
so that (\ref{Nahm_simple}) is a multiple bilateral basic hypergeometric series \cite{MR2128719},
namely a $q$-hypergeometric series whose summation indices, the fluxes, run over all integers.
Together with the prefactor, the Nahm pole half-index depends on $t$ only through $q^{\frac14}t^{-1}$, the square root of $q^{\frac12}t^{-2}$,
as does the specialization (\ref{specialization_gen}).
The same holds for the dual Neumann half-indices in all the examples below,
so that the dualities are the identities of functions of $q^{\frac12}$ and $q^{\frac14}t^{-1}$ or $q^{\frac14}t$.

The formula only requires the gauge group, the Nahm pole, the weights of the hypermultiplets and their decomposition into the spin-$J$ multiplets.
For $G=\prod_IU(N_I)$ and $G=USp(2N)$ the dualities in sections \ref{sec_SQCD}--\ref{sec_USpADHM} test the proposed formula (\ref{Nahm_simple}).
For the principal embedding, the spins $k$ in the Slodowy slices are the exponents of $\mathfrak{g}$,
i.e. $d-1$ for the degrees $d$ of the Casimir invariants,
as for the Nahm pole half-indices of 4d $\mathcal{N}=4$ SYM theories \cite{Hatsuda:2024lcc,Hatsuda:2025yzp,Hatsuda:2026ysv}.
They are $k=0,1,\cdots,N-1$ for $U(N)$ and $k=1,3,\cdots,2N-1$ for $USp(2N)$.

For large fluxes the summand of (\ref{Nahm_simple}) behaves as $(q^{\frac12}t^{-2})^{\Delta(m)}$ up to a bounded factor,
with $\Delta(m)$ in (\ref{monopole_dim}), where $|\alpha(m)|=\alpha(m)$ in the dominant sectors.
Hence only finitely many sectors contribute at each order in $q^{\frac14}$ for good and ugly theories,
although (\ref{Nahm_simple}) contains no Gaussian factor.

\subsubsection{Boundary anomalies}
\label{sec_Nahm_anom}
Since the anomaly polynomials of decoupled sectors add up, the factorization (\ref{Nahm_limit}) implies that
the boundary anomaly (\ref{anom_DD}) of the Dirichlet boundary condition, restricted to the boundary symmetry preserved by the Nahm pole (see section \ref{sec_B}),
is the sum of the anomalies of the Nahm pole boundary condition, the Goldstone modes and the split Fermi multiplets.
In this restriction the vector multiplet does not contribute to (\ref{anom_DD}) since the roots are orthogonal to the boundary symmetry,
and the $2J+1$ components of a spin-$J$ multiplet of the hypermultiplets carry the same charges.
The Goldstone modes are neutral,
and the $2J$ split Fermi multiplets of the spin-$J$ multiplet contribute $2J$ times the square of its charges.
Hence the boundary anomaly of the Nahm pole boundary condition is
\begin{align}
\label{Nahm_anom}
\mathcal{A}^{\mathcal{T}}_{\mathrm{Nahm}'}
&=\underbrace{\sum_{(J,\lambda)}(w_{\lambda}\cdot\bar{\mathbf{u}}+f_{\lambda}\cdot\mathbf{x})^2}_{\textrm{top components}}
\underbrace{-2\sum_IN_I\,\mathbf{u}^{(I)}\mathbf{z}_I}_{\textrm{FI couplings}},
\end{align}
where $w_{\lambda}\cdot\bar{\mathbf{u}}=\sum_I(\sum_iw^{(I)}_{\lambda,i})\mathbf{u}^{(I)}$.
Each spin-$J$ multiplet thus contributes only once, through its top components, which leads to the following.
\begin{quote}
\textit{The boundary anomaly of the Nahm pole boundary condition receives contributions only from the top components and the FI couplings.}
\end{quote}

The terms with the field strengths $\mathbf{h}$ and $\mathbf{c}$ of the R-symmetries in section \ref{sec_anomaly}, which are omitted in (\ref{Nahm_anom}), change at the Nahm pole.
There $C$ is replaced by the generator $C_{\rho}$ in (\ref{C_rho}) of the twisted $U(1)_C$.
Then the fermions of the top component of a spin-$J$ multiplet have $C_{\rho}=\pm(1+2J)$,
and those of the Slodowy slice of spin $k$ have $H=\pm1$ and $C_{\rho}=\pm(1+2k)$.
Hence the top component contributes $(1+2J)^2\mathbf{c}^2$ in addition to its term in (\ref{Nahm_anom}),
and the Slodowy slice of spin $k$ contributes $-\mathbf{h}^2-(1+2k)^2\mathbf{c}^2$, with $\mathbf{h}$ and $\mathbf{c}$ exchanged for the twisted multiplets.

The R-charges can also be read off from the half-indices.
The contributions above and in section \ref{sec_anomaly} are the sums of
\begin{align}
\label{anom_R}
\pm\frac12\Bigl(\mathbf{X}+\bigl(2a-1+\tfrac{b}{2}\bigr)\mathbf{h}+\bigl(2a-1-\tfrac{b}{2}\bigr)\mathbf{c}\Bigr)^2
\end{align}
over the factors $(q^at^bX;q)_{\infty}^{\pm1}$ in the half-indices,
where $X$ is a monomial in the fugacities other than $q$ and $t$, and $\mathbf{X}$ is the sum of their field strengths weighted by the exponents.
By (\ref{DEF_hind}) the coefficients of $\mathbf{h}$ and $\mathbf{c}$ are the charges $H$ and $C$ for which $q^{a-\frac12}t^b=q^{\frac{H+C}{4}}t^{H-C}$.
For example, the Fermi multiplet $F(q^{\frac12}(q^{\frac12}t^2)^{\kappa}X;q)$ in section \ref{sec_USp_fw_N} gives $(\mathbf{X}+2\kappa\mathbf{h})^2$, so that its fermion has $H=2\kappa$.
In sections \ref{sec_SQCD}--\ref{sec_USpADHM} we compare the anomalies of the dual boundary conditions without these terms.
With these terms the anomalies still agree, up to the same mixed terms between the flavor and topological symmetries,
in the dualities for $U(N)$ SQCD with $2N$ flavors, $T[SU(N)]$ and the $U(N)$ and $USp(2N)$ ADHM theories, which we have checked explicitly in examples with $N\le3$.

\section{\texorpdfstring{$U(N)$ SQCD with $N_f$ flavors}{U(N) SQCD with Nf flavors}}
\label{sec_SQCD}
Consider 3d $\mathcal{N}=4$ $U(N)$ gauge theory with $2N$ fundamental hypermultiplets, i.e. $U(N)$ SQCD with $2N$ flavors, $(N)-[2N]$.
The gauge node is balanced, and the topological $U(1)$ symmetry is enhanced to $SU(2)$ \cite{Gaiotto:2008ak}.
The mirror theory is the linear quiver gauge theory \cite{Intriligator:1996ex,Hanany:1996ie}
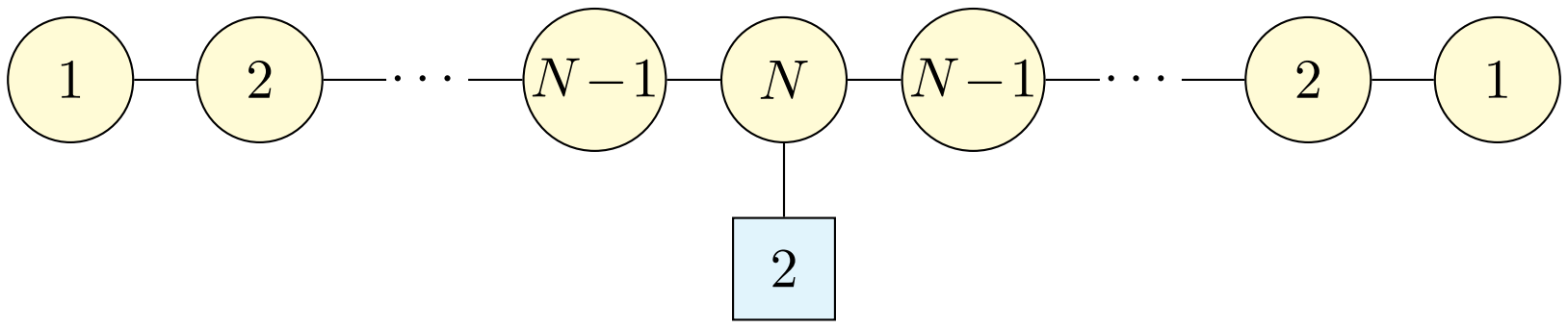
\begin{align}
\label{SQCD_mirror}
\begin{tikzpicture}[quiver]
\node[gauge](a1) at (0,0){$1$}; \node[gauge](a2) at (1.3,0){$2$}; \node[qdots](a3) at (2.45,0){$\cdots$};
\node[gauge](a4) at (3.6,0){$N\!-\!1$}; \node[gauge](a5) at (4.9,0){$N$}; \node[gauge](a6) at (6.2,0){$N\!-\!1$};
\node[qdots](a7) at (7.35,0){$\cdots$}; \node[gauge](a8) at (8.5,0){$2$}; \node[gauge](a9) at (9.8,0){$1$};
\node[flavor](f) at (4.9,-1.3){$2$};
\draw (a1)--(a2)--(a3)--(a4)--(a5)--(a6)--(a7)--(a8)--(a9); \draw (a5)--(f);
\end{tikzpicture}
\end{align}
with $2N-1$ nodes $U(r_I)$, $r_I=\min(I,2N-I)$, and two flavors at the $U(N)$ node
(see also \cite{Okazaki:2019ony} for tests of mirror symmetry using supersymmetric indices).
We denote the gauge fugacities of SQCD by $s_i$ and the flavor fugacities by $x_{\alpha}$, $\alpha=1,\cdots,2N$,
which we take to be $U(2N)$ fugacities without imposing $\prod_{\alpha}x_{\alpha}=1$.
As in \cite{Okazaki:2019bok}, the hypermultiplets have fugacities $s_ix_{\alpha}$.
The fugacities $z_1$, $z_2$ for the $SU(2)$ flavor symmetry of the mirror quiver
are the topological fugacities of SQCD (see section \ref{sec_conv}).
We denote the boundary fugacities for the node $U(r_I)$ of the mirror quiver by $u^{(I)}$.

\begin{figure}
\centering
\scalebox{0.85}{
\begin{tikzpicture}
\fill[yellow!25] (-2.8,0) rectangle (2.8,2.6);
\fill[yellow!50] (-1.2,0) rectangle (1.2,2.6);
\draw[thick,solid,black] (-3.6,0) -- (3.6,0) node[right] {\textrm{NS5}$'$};
\foreach \x in {-2.8,-1.2,1.2,2.8} \draw[thick,dashed,black] (\x,-0.6) -- (\x,2.9);
\node at (0,3.15) {\textrm{4 D5}};
\foreach \x in {-0.75,0.75} \draw[thick,solid,black] (\x,-0.6) -- (\x,2.9);
\node at (-2.0,1.3) {\textrm{1}};
\node at (0,1.3) {\textrm{2}};
\node at (2.0,1.3) {\textrm{1}};
\fill[black] (-1.2,0) circle (2pt) node[below left] {$\Gamma_A$};
\fill[black] (-0.75,0) circle (2pt) node[above right] {$\eta_+$};
\fill[black] (0.75,0) circle (2pt) node[above left] {$\eta_-$};
\fill[black] (1.2,0) circle (2pt) node[below right] {$\Gamma_B$};
\node at (0,-1.2) {(a) $U(2)$ SQCD, $(\mathcal{N}',N'+\Gamma+\eta)$};
\end{tikzpicture}
}
\hspace{0.5cm}
\scalebox{0.85}{
\begin{tikzpicture}
\fill[yellow!25] (-2.8,0) rectangle (2.8,2.6);
\fill[yellow!50] (-1.2,0) rectangle (1.2,2.6);
\draw[thick,dashed,black] (-3.6,0) -- (3.6,0) node[right] {\textrm{D5}$'$};
\foreach \x in {-2.8,-1.2,1.2,2.8} \draw[thick,solid,black] (\x,-0.6) -- (\x,2.9);
\node at (0,3.15) {\textrm{4 NS5}};
\foreach \x in {-0.75,0.75} \draw[thick,dashed,black] (\x,-0.6) -- (\x,2.9);
\node at (-2.0,1.3) {\textrm{1}};
\node at (0,1.3) {\textrm{2}};
\node at (2.0,1.3) {\textrm{1}};
\node at (-2.0,-0.3) {$\mathcal{D}$};
\node at (0,-0.3) {\textrm{Nahm}};
\node at (2.0,-0.3) {$\mathcal{D}$};
\node at (0,-1.2) {(b) mirror, $(\mathcal{D},\mathrm{Nahm},\mathcal{D})$};
\end{tikzpicture}
}
\\[0.5cm]
\scalebox{0.85}{
\begin{tikzpicture}
\fill[yellow!50] (-2.5,0) rectangle (2.5,2.6);
\draw[thick,dashed,black] (-3.3,0) -- (3.3,0) node[right] {\textrm{D5}$'$};
\draw[thick,solid,black] (-2.5,-0.6) -- (-2.5,2.9) node[above] {\textrm{NS5}};
\draw[thick,solid,black] (2.5,-0.6) -- (2.5,2.9) node[above] {\textrm{NS5}};
\foreach \x in {-1.8,-0.9,0.9,1.8} \draw[thick,dashed,black] (\x,-0.6) -- (\x,2.9);
\node at (0,3.15) {\textrm{4 D5}};
\node at (0,1.3) {\textrm{2 D3}};
\node at (0,-0.3) {$\mathrm{Nahm}'$};
\node at (0,-1.2) {(c) $U(2)$ SQCD, $\mathrm{Nahm}'$};
\end{tikzpicture}
}
\hspace{0.5cm}
\scalebox{0.85}{
\begin{tikzpicture}
\fill[yellow!50] (-2.5,0) rectangle (2.5,2.6);
\draw[thick,solid,black] (-3.3,0) -- (3.3,0) node[right] {\textrm{NS5}$'$};
\draw[thick,dashed,black] (-2.5,-0.6) -- (-2.5,2.9) node[above] {\textrm{D5}};
\draw[thick,dashed,black] (2.5,-0.6) -- (2.5,2.9) node[above] {\textrm{D5}};
\foreach \x in {-1.8,-0.9,0.9,1.8} \draw[thick,solid,black] (\x,-0.6) -- (\x,2.9);
\node at (0,3.15) {\textrm{4 NS5}};
\node at (0,1.3) {\textrm{2 D3}};
\fill[black] (-1.8,0) circle (2pt) node[below left] {$\tilde{\eta}_1$};
\fill[black] (-0.9,0) circle (2pt) node[below left] {$\tilde{\eta}_2$};
\fill[black] (0.9,0) circle (2pt) node[below right] {$\tilde{\eta}_3$};
\fill[black] (1.8,0) circle (2pt) node[below right] {$\tilde{\eta}_4$};
\node at (0,-1.2) {(d) mirror, $(\mathcal{N},N+\tilde{\eta})$};
\end{tikzpicture}
}
\caption{\textrm{
The brane configurations for $U(2)$ SQCD with four flavors (left) and its mirror (right).
}}
\label{fig_SQCD}
\end{figure}
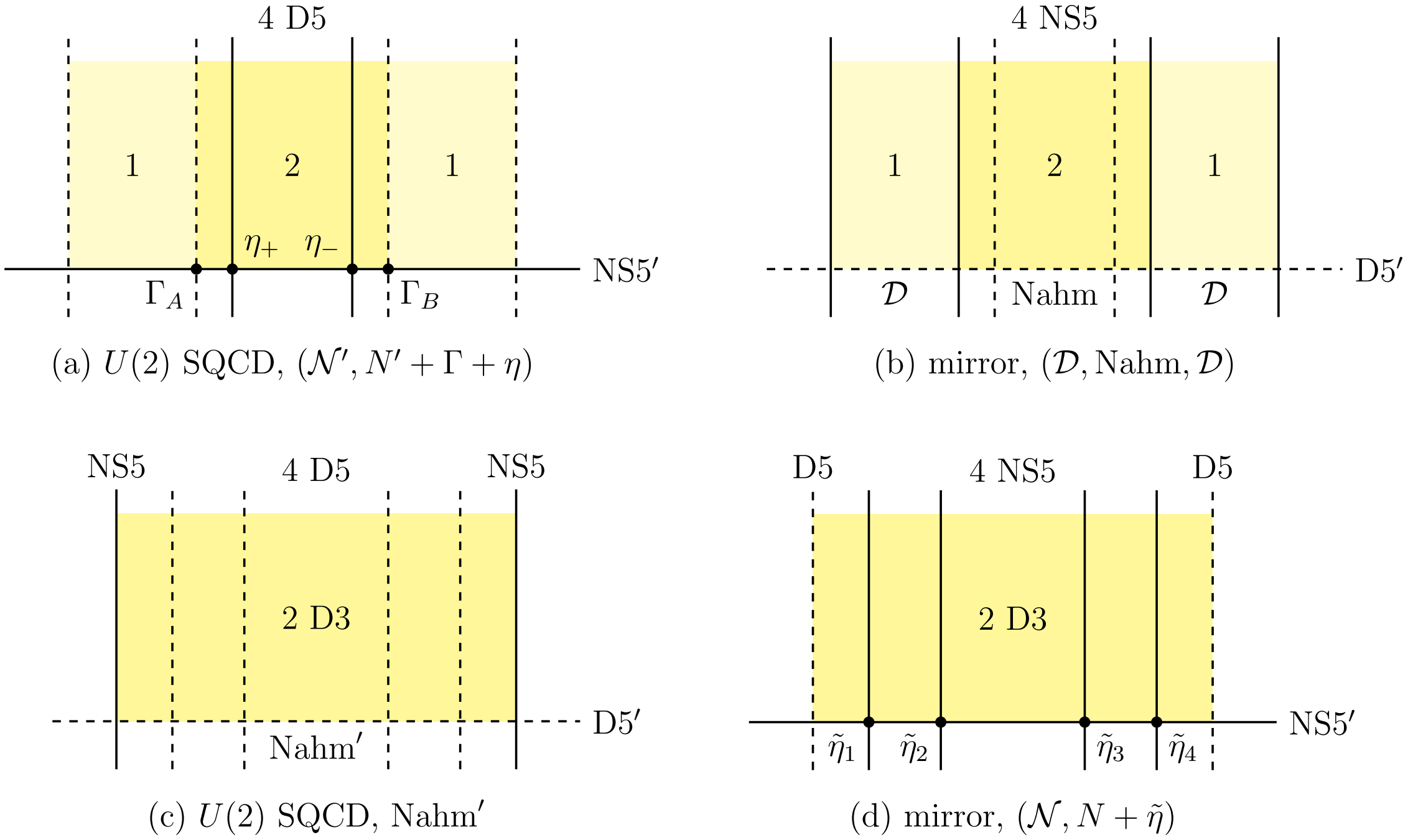

In the brane construction \cite{Hanany:1996ie}, $N$ D3-branes are suspended between two NS5-branes and intersect with $2N$ D5-branes.
The boundary is introduced by an NS5$'$- or D5$'$-brane on which the D3-branes end.
The two choices lead to two types of dualities, as illustrated in Figure \ref{fig_SQCD} for $N=2$.
In (b) the two D5-branes are moved inside the $U(2)$ segment by the Hanany-Witten transitions,
and (a) is drawn as the S-dual of (b).
Likewise (d) is drawn as the S-dual of (c), so that in (d) the D3-branes end on the two D5-branes.

\subsection{\texorpdfstring{$(\mathcal{N}',N'+\Gamma+\eta)$ and $(\mathcal{D},\mathrm{Nahm}^{2N-3},\mathcal{D})$}{(N', N'+Gamma+eta) and (D, Nahm, ..., Nahm, D)}}
\label{sec_SQCD_fw}
When the D3-branes end on the NS5$'$-brane, SQCD obeys the Neumann boundary condition $(\mathcal{N}',N')$.
Under S-duality the NS5$'$-brane maps to the D5$'$-brane.
For the mirror quiver, the $r_I$ D3-branes of the node $U(r_I)$ end on a single D5$'$-brane,
so that the node obeys the Nahm pole boundary condition for $r_I\ge 2$ and the Dirichlet boundary condition for $r_I=1$.
We thus find the boundary condition $(\mathcal{D},\mathrm{Nahm}^{2N-3},\mathcal{D})$ for the mirror quiver.
Since the mirror quiver consists of the twisted multiplets, we apply the prescription in section \ref{sec_proposal} at each Nahm pole node with $t\rightarrow t^{-1}$.

\subsubsection{\texorpdfstring{$N=2$, $N_f=4$}{N=2, Nf=4}}
\label{sec_SQCD_fwN2}
Consider $U(2)$ SQCD with four flavors.
The mirror quiver is $(1)-(2)-(1)$ with two flavors $[2]$ at the $U(2)$ node.

\paragraph{$U(2)$ SQCD}
As explained in section \ref{sec_dualrules}, for the Neumann boundary condition $(\mathcal{N}',N')$ the non-Abelian part of the boundary gauge anomaly cancels since $N_f=2N$,
and the remaining part $-2(\mathbf{s}_1+\mathbf{s}_2)^2$ is cancelled by the two cross-determinant Fermi multiplets $\eta_{\pm}$
(see (\ref{U2_4_Neu_anom}) for the full anomaly).
Together with two gauge-neutral Fermi multiplets $\Gamma_A$ and $\Gamma_B$, we find the Neumann half-index
\begin{align}
\label{U2_4_Neu}
\mathbb{II}^{U(2)+4}_{\mathcal{N}',N'+\Gamma+\eta}
&=\frac{1}{2}\oint \prod_{i=1}^2\frac{ds_i}{2\pi is_i}
\underbrace{(q)_{\infty}^2(q^{\frac12}t^2;q)_{\infty}^2\prod_{i\neq j}\left(\frac{s_i}{s_j};q\right)_{\infty}\left(q^{\frac12}t^2\frac{s_i}{s_j};q\right)_{\infty}}_{\textrm{vector multiplet ($\mathcal{N}'$)}}
\nonumber\\*
&\times \underbrace{\prod_{i=1}^2\prod_{\alpha=1}^4\frac{1}{(q^{\frac14}ts_i^{\pm}x_{\alpha}^{\pm};q)_{\infty}}}_{\textrm{hypermultiplets ($N'$)}}\;
\underbrace{F\left(q^{\frac12}s_1s_2x_1x_2\frac{z_1}{u^{(2)}};q\right)}_{\eta_+}
\underbrace{F\left(q^{\frac12}s_1s_2x_3x_4\frac{u^{(2)}}{z_2};q\right)}_{\eta_-}
\nonumber\\*
&\times \underbrace{F\left(q^{\frac12}\frac{x_2}{x_1}\frac{u^{(1)}}{u^{(2)}};q\right)}_{\Gamma_A}
\underbrace{F\left(q^{\frac12}\frac{x_3}{x_4}\frac{u^{(2)}}{u^{(3)}};q\right)}_{\Gamma_B}.
\end{align}
The Fermi multiplets $\eta_{\pm}$ carry the same gauge and flavor charges as the baryons $\epsilon^{ij}H_{i1}H_{j2}$ and $\epsilon^{ij}H_{i3}H_{j4}$,
where $H_{i\alpha}$ is the chiral field of the hypermultiplets with color $i$ and flavor $\alpha$.
In the brane configuration of Figure \ref{fig_SQCD} (a) they arise at the junctions of the two NS5-branes with the NS5$'$-brane \cite{Hanany:2018hlz},
and their flavor charges come from the D3-branes ending on the D5-branes on the outer sides (see section \ref{sec_dualrules}).
Their mixed anomaly $2(\mathbf{s}_1+\mathbf{s}_2)(\mathbf{z}_1-\mathbf{z}_2)$ cancels the FI term.
This is the non-Abelian generalization of the boundary Fermi multiplets $\Gamma^{(1)}$ and $\Gamma^{(N_f)}$ for SQED$_{N_f}$ in \cite{Okazaki:2019bok}.
The gauge-neutral Fermi multiplets $\Gamma_A$ and $\Gamma_B$ arise at the intersections of the second and third D5-branes with the NS5$'$-brane,
through which the D3-branes emanating from the first and fourth D5-branes pass.

\paragraph{Mirror quiver}
For the Nahm pole boundary condition, the formula (\ref{Nahm_quiver_simple}) with $t\rightarrow t^{-1}$ gives
\begin{align}
\label{U2_4_mirror_simple}
&\mathbb{II}^{\wtilde{(1)-(2)-(1)}}_{\mathcal{D},\mathrm{Nahm},\mathcal{D}}
\nonumber\\*
&=\underbrace{\left(qt^{2}\left(\frac{u^{(1)}}{u^{(2)}}\right)^{\pm};q\right)_{\infty}\left(qt^{2}\left(\frac{u^{(2)}}{u^{(3)}}\right)^{\pm};q\right)_{\infty}\prod_{a=1}^2\left(qt^{2}\left(\frac{u^{(2)}}{z_a}\right)^{\pm};q\right)_{\infty}}_{\textrm{top components}}
\nonumber\displaybreak[1]\\
&\times\underbrace{\frac{1}{(q)_{\infty}^3(q^{\frac12}t^{2};q)_{\infty}^3(q^{\frac32}t^{2};q)_{\infty}(qt^{4};q)_{\infty}}}_{\textrm{Slodowy slices}}
\nonumber\displaybreak[1]\\
&\times\sum_{m^{(1)},m^{(3)}\in\mathbb{Z}}\sum_{m^{(2)}_1\le m^{(2)}_2}
\underbrace{\left(\frac{x_2}{x_1}\right)^{m^{(1)}}\left(\frac{x_3}{x_2}\right)^{m^{(2)}_1+m^{(2)}_2}\left(\frac{x_4}{x_3}\right)^{m^{(3)}}}_{\textrm{topological charges}}
\nonumber\displaybreak[1]\\
&\times\underbrace{(q^{\frac12}t^{2})^{-n}\frac{(qt^{4};q)_n(q^{\frac32}t^{2};q)_n}{(q^{\frac12}t^{2};q)_n(q;q)_n}}_{\textrm{vector multiplet ($\mathrm{Nahm}$)}}\;
\underbrace{\prod_{a=1}^2(q^{\frac14}t)^{m^{(2)}_1+m^{(2)}_2}\frac{(t^{-2}\frac{u^{(2)}}{z_a};q)_{m^{(2)}_1}(q^{\frac12}\frac{u^{(2)}}{z_a};q)_{m^{(2)}_2}}{(q^{\frac12}\frac{u^{(2)}}{z_a};q)_{m^{(2)}_1}(qt^{2} \frac{u^{(2)}}{z_a};q)_{m^{(2)}_2}}}_{\textrm{fundamental hypermultiplets ($D$)}}
\nonumber\\
&\times\underbrace{(q^{\frac14}t)^{2m^{(1)}-m^{(2)}_1-m^{(2)}_2}\frac{(q^{\frac12}\frac{u^{(1)}}{u^{(2)}};q)_{m^{(1)}-m^{(2)}_1}(t^{-2}\frac{u^{(1)}}{u^{(2)}};q)_{m^{(1)}-m^{(2)}_2}}{(qt^{2} \frac{u^{(1)}}{u^{(2)}};q)_{m^{(1)}-m^{(2)}_1}(q^{\frac12}\frac{u^{(1)}}{u^{(2)}};q)_{m^{(1)}-m^{(2)}_2}}}_{\textrm{bifundamental hypermultiplet ($D$)}}
\nonumber\\*
&\times\underbrace{(q^{\frac14}t)^{m^{(2)}_1+m^{(2)}_2-2m^{(3)}}\frac{(t^{-2}\frac{u^{(2)}}{u^{(3)}};q)_{m^{(2)}_1-m^{(3)}}(q^{\frac12}\frac{u^{(2)}}{u^{(3)}};q)_{m^{(2)}_2-m^{(3)}}}{(q^{\frac12}\frac{u^{(2)}}{u^{(3)}};q)_{m^{(2)}_1-m^{(3)}}(qt^{2} \frac{u^{(2)}}{u^{(3)}};q)_{m^{(2)}_2-m^{(3)}}}}_{\textrm{bifundamental hypermultiplet ($D$)}},
\end{align}
where $n=m^{(2)}_2-m^{(2)}_1$.
The top components have $J=\frac12$ for the four doublets,
namely the two twisted bifundamental and the two twisted fundamental hypermultiplets of the $U(2)$ node.

\paragraph{Duality}
We propose that
\begin{align}
\label{U2_4_dual}
\mathbb{II}^{U(2)+4}_{\mathcal{N}',N'+\Gamma+\eta}&=\mathbb{II}^{\wtilde{(1)-(2)-(1)}}_{\mathcal{D},\mathrm{Nahm},\mathcal{D}}.
\end{align}
The identity singles out the flavor charge of the Fermi multiplet $\eta_+$ against its inverse (see Appendix \ref{app_checks}).

As a consistency check, we compare the boundary anomalies.
For the Neumann boundary condition (\ref{U2_4_Neu}) the boundary anomaly is
\begin{align}
\label{U2_4_Neu_anom}
&\mathcal{A}^{U(2)+4}_{\mathcal{N}',N'+\Gamma+\eta}
\nonumber\\*
&=\underbrace{-\sum_{i=1}^2\sum_{\alpha=1}^4(\mathbf{s}_i+\mathbf{x}_{\alpha})^2}_{\textrm{hypermultiplets ($N'$)}}
\quad\underbrace{+2(\mathbf{s}_1-\mathbf{s}_2)^2}_{\textrm{vector multiplet ($\mathcal{N}'$)}}
\quad\underbrace{-2(\mathbf{s}_1+\mathbf{s}_2)(\mathbf{z}_1-\mathbf{z}_2)}_{\textrm{FI coupling}}
\nonumber\\
&+\underbrace{(\mathbf{s}_1+\mathbf{s}_2+\mathbf{x}_1+\mathbf{x}_2+\mathbf{z}_1-\mathbf{u}^{(2)})^2}_{\eta_+}
+\underbrace{(\mathbf{s}_1+\mathbf{s}_2+\mathbf{x}_3+\mathbf{x}_4+\mathbf{u}^{(2)}-\mathbf{z}_2)^2}_{\eta_-}
\nonumber\\*
&+\underbrace{(\mathbf{u}^{(1)}-\mathbf{u}^{(2)}+\mathbf{x}_2-\mathbf{x}_1)^2}_{\Gamma_A}
+\underbrace{(\mathbf{u}^{(2)}-\mathbf{u}^{(3)}+\mathbf{x}_3-\mathbf{x}_4)^2}_{\Gamma_B}.
\end{align}
All the terms with the gauge field strengths cancel.
By (\ref{Nahm_anom}), the boundary anomaly for the Nahm pole side is
\begin{align}
&\mathcal{A}^{\wtilde{(1)-(2)-(1)}}_{\mathcal{D},\mathrm{Nahm},\mathcal{D}}
\nonumber\\*
&=\underbrace{(\mathbf{u}^{(1)}-\mathbf{u}^{(2)})^2+(\mathbf{u}^{(2)}-\mathbf{u}^{(3)})^2+\sum_{a=1}^2(\mathbf{u}^{(2)}-\mathbf{z}_a)^2}_{\textrm{top components}}
\nonumber\\
&\underbrace{-2\mathbf{u}^{(1)}(\mathbf{x}_1-\mathbf{x}_2)-4\mathbf{u}^{(2)}(\mathbf{x}_2-\mathbf{x}_3)-2\mathbf{u}^{(3)}(\mathbf{x}_3-\mathbf{x}_4)}_{\textrm{FI couplings}}
\nonumber\\*
&=\mathcal{A}^{U(2)+4}_{\mathcal{N}',N'+\Gamma+\eta}-2\mathbf{z}_1(\mathbf{x}_1+\mathbf{x}_2)+2\mathbf{z}_2(\mathbf{x}_3+\mathbf{x}_4).
\end{align}
Hence the anomalies of the two boundary conditions agree up to the mixed terms between the flavor symmetries of $x_{\alpha}$ and $z_a$.
They do not involve the boundary symmetries $U(1)_{u^{(I)}}$ and can be absorbed into background mixed Chern-Simons terms.
On the Neumann side $U(1)_{u^{(I)}}$ act only on the boundary Fermi multiplets.
For instance, the mixed term $-4\mathbf{u}^{(2)}(\mathbf{x}_2-\mathbf{x}_3)$ arises as the sum of the contributions from the four Fermi multiplets,
while the terms $\mp2\mathbf{u}^{(2)}(\mathbf{s}_1+\mathbf{s}_2)$ from the Fermi multiplets $\eta_{\pm}$ cancel.

\subsubsection{\texorpdfstring{General $N$, $N_f=2N$}{General N, Nf=2N}}
\label{sec_SQCD_fwN}
\paragraph{$U(N)$ SQCD}
As for $N=2$, the remaining part $-2(\sum_i\mathbf{s}_i)^2$ of the boundary gauge anomaly for $(\mathcal{N}',N')$
is cancelled by the two cross-determinant Fermi multiplets $\eta_{\pm}$
(see (\ref{SQCD_fw_Neu_anom}) for the full anomaly).
We propose that the Neumann boundary condition $(\mathcal{N}',N'+\Gamma+\eta)$ involves the Fermi multiplets $\eta_{\pm}$
and $N(N-1)$ gauge-neutral Fermi multiplets $\Gamma$.
The Neumann half-index reads
\begin{align}
\label{SQCD_fw_Neu}
&\mathbb{II}^{U(N)+2N}_{\mathcal{N}',N'+\Gamma+\eta}
\nonumber\\*
&=\frac{1}{N!}\oint\prod_{i=1}^N\frac{ds_i}{2\pi is_i}
\underbrace{(q)_{\infty}^N(q^{\frac12}t^2;q)_{\infty}^N\prod_{i\neq j}\left(\frac{s_i}{s_j};q\right)_{\infty}\left(q^{\frac12}t^2\frac{s_i}{s_j};q\right)_{\infty}}_{\textrm{vector multiplet ($\mathcal{N}'$)}}
\underbrace{\prod_{i=1}^N\prod_{\alpha=1}^{2N}\frac{1}{(q^{\frac14}ts_i^{\pm}x_{\alpha}^{\pm};q)_{\infty}}}_{\textrm{hypermultiplets ($N'$)}}
\nonumber\\
&\times \underbrace{F\left(q^{\frac12}\prod_{i=1}^Ns_i\,\frac{z_1}{u^{(N)}}\prod_{\alpha=1}^Nx_{\alpha};q\right)}_{\eta_+}
\underbrace{F\left(q^{\frac12}\prod_{i=1}^Ns_i\,\frac{u^{(N)}}{z_2}\prod_{\alpha=N+1}^{2N}x_{\alpha};q\right)}_{\eta_-}
\nonumber\\*
&\times \underbrace{\prod_{I=1}^{N-1}\prod_{j=1}^{I}F\left(q^{\frac12}\frac{u^{(I)}}{u^{(I+1)}}\frac{x_{I+1}}{x_j};q\right)
\prod_{I=N}^{2N-2}\prod_{j=I+2}^{2N}F\left(q^{\frac12}\frac{u^{(I)}}{u^{(I+1)}}\frac{x_{I+1}}{x_j};q\right)}_{\Gamma}.
\end{align}
For $N=2$ it is (\ref{U2_4_Neu}).

\paragraph{Mirror quiver}
Let the fluxes for the node $U(r_I)$ be $m^{(I)}=(m^{(I)}_1\le\cdots\le m^{(I)}_{r_I})$ and $\mu^{(I)}_i=i-\frac{r_I+1}{2}$.
The formula (\ref{Nahm_quiver_simple}) with $t\rightarrow t^{-1}$ gives
\begin{align}
\label{SQCD_fw_simple}
&\mathbb{II}^{\wtilde{(1)-(2)-\cdots-(N)-\cdots-(2)-(1)}}_{\mathcal{D},\mathrm{Nahm}^{2N-3},\mathcal{D}}
\nonumber\\*
&=\underbrace{\prod_{a=1}^2\left(q^{\frac12+\frac{N}{4}}t^{N}\left(\frac{u^{(N)}}{z_a}\right)^{\pm};q\right)_{\infty}
\prod_{I=1}^{2N-2}\prod_{J=|j_I-j_{I+1}|}^{j_I+j_{I+1}}\left(q^{\frac34+\frac{J}{2}}t^{1+2J}\left(\frac{u^{(I)}}{u^{(I+1)}}\right)^{\pm};q\right)_{\infty}}_{\textrm{top components}}
\nonumber\\
&\times\underbrace{\prod_{I=1}^{2N-1}\prod_{k=0}^{r_I-1}\frac{1}{(q^{1+\frac{k}{2}}t^{2k};q)_{\infty}(q^{\frac12+\frac{k}{2}}t^{2+2k};q)_{\infty}}}_{\textrm{Slodowy slices}}
\sum_{\{m^{(I)}\}}\underbrace{\prod_{I=1}^{2N-1}\left(\frac{x_{I+1}}{x_I}\right)^{\sum_im^{(I)}_i}}_{\textrm{topological charges}}
\nonumber\\
&\times\underbrace{\prod_{I=1}^{2N-1}\prod_{1\le i<j\le r_I}(q^{\frac12}t^2)^{m^{(I)}_i-m^{(I)}_j}\frac{(q^{\frac12+\frac{j-i}{2}}t^{2+2(j-i)};q)_{m^{(I)}_j-m^{(I)}_i}(q^{1+\frac{j-i}{2}}t^{2(j-i)};q)_{m^{(I)}_j-m^{(I)}_i}}{(q^{\frac{j-i}{2}}t^{2(j-i)};q)_{m^{(I)}_j-m^{(I)}_i}(q^{\frac12+\frac{j-i}{2}}t^{2(j-i)-2};q)_{m^{(I)}_j-m^{(I)}_i}}}_{\textrm{vector multiplets ($\mathrm{Nahm}$)}}
\nonumber\\
&\times\underbrace{\prod_{a=1}^2\prod_{i=1}^{N}(q^{\frac14}t)^{m^{(N)}_i}\frac{(q^{\frac14+\frac{\mu^{(N)}_i}{2}}t^{2\mu^{(N)}_i-1}\frac{u^{(N)}}{z_a};q)_{m^{(N)}_i}}{(q^{\frac34+\frac{\mu^{(N)}_i}{2}}t^{1+2\mu^{(N)}_i}\frac{u^{(N)}}{z_a};q)_{m^{(N)}_i}}}_{\textrm{fundamental hypermultiplets ($D$)}}
\nonumber\\*
&\times\underbrace{\prod_{I=1}^{2N-2}\prod_{i=1}^{r_I}\prod_{j=1}^{r_{I+1}}(q^{\frac14}t)^{m^{(I)}_i-m^{(I+1)}_j}\frac{(q^{\frac14+\frac{\mu^{(I)}_i-\mu^{(I+1)}_j}{2}}t^{2(\mu^{(I)}_i-\mu^{(I+1)}_j)-1}\frac{u^{(I)}}{u^{(I+1)}};q)_{m^{(I)}_i-m^{(I+1)}_j}}{(q^{\frac34+\frac{\mu^{(I)}_i-\mu^{(I+1)}_j}{2}}t^{1+2(\mu^{(I)}_i-\mu^{(I+1)}_j)}\frac{u^{(I)}}{u^{(I+1)}};q)_{m^{(I)}_i-m^{(I+1)}_j}}}_{\textrm{bifundamental hypermultiplets ($D$)}},
\end{align}
where the top components of the twisted hypermultiplets have
$J=\frac{N-1}{2}$ for the two fundamental ones of the node $U(N)$
and $J$ from $|j_I-j_{I+1}|$ to $j_I+j_{I+1}$ for the bifundamental ones of $U(r_I)\times U(r_{I+1})$,
which are the tensor products of the spins $j_I=\frac{r_I-1}{2}$ and $j_{I+1}$ as in (\ref{spins_bifund}).
For $N=2$ it reduces to (\ref{U2_4_mirror_simple}).

\paragraph{Duality}
We propose that
\begin{align}
\label{SQCD_fw_dual}
\mathbb{II}^{U(N)+2N}_{\mathcal{N}',N'+\Gamma+\eta}&=\mathbb{II}^{\wtilde{(1)-(2)-\cdots-(N)-\cdots-(2)-(1)}}_{\mathcal{D},\mathrm{Nahm}^{2N-3},\mathcal{D}},
\end{align}
where the left-hand side is (\ref{SQCD_fw_Neu}).
For $N=1$ it reduces to the duality between the Neumann boundary condition for SQED$_2$ and the Dirichlet boundary condition for its mirror in \cite{Okazaki:2019bok},
and for $N=2$ it is (\ref{U2_4_dual}).
For $N=3$ and $N=4$ the identity singles out the charges $\{\frac{x_N}{x_{N-2}},\frac{x_N}{x_{N-1}}\}\frac{u^{(N-1)}}{u^{(N)}}$ of two of the Fermi multiplets $\Gamma$
against $\{\frac{x_N^2}{x_{N-2}x_{N-1}},1\}\frac{u^{(N-1)}}{u^{(N)}}$, which give the same anomaly terms with $\mathbf{u}^{(I)}$ but leave a pure flavor term,
and for $N=3$ the determinant charge of the Fermi multiplet $\eta_-$ against its inverse (see Appendix \ref{app_checks}).

As a consistency check, we compare the boundary anomalies.
For the Neumann boundary condition (\ref{SQCD_fw_Neu}) the boundary anomaly is
\begin{align}
\label{SQCD_fw_Neu_anom}
&\mathcal{A}^{U(N)+2N}_{\mathcal{N}',N'+\Gamma+\eta}
\nonumber\\*
&=\underbrace{-\sum_{i=1}^N\sum_{\alpha=1}^{2N}(\mathbf{s}_i+\mathbf{x}_{\alpha})^2}_{\textrm{hypermultiplets ($N'$)}}
\quad\underbrace{+\sum_{i\neq j}(\mathbf{s}_i-\mathbf{s}_j)^2}_{\textrm{vector multiplet ($\mathcal{N}'$)}}
\quad\underbrace{-2\mathbf{S}(\mathbf{z}_1-\mathbf{z}_2)}_{\textrm{FI coupling}}
\nonumber\\
&+\underbrace{(\mathbf{S}+\mathbf{X}_++\mathbf{z}_1-\mathbf{u}^{(N)})^2}_{\eta_+}
+\underbrace{(\mathbf{S}+\mathbf{X}_-+\mathbf{u}^{(N)}-\mathbf{z}_2)^2}_{\eta_-}
\nonumber\\
&+\underbrace{\sum_{I=1}^{N-1}\sum_{j=1}^{I}(\mathbf{u}^{(I)}-\mathbf{u}^{(I+1)}+\mathbf{x}_{I+1}-\mathbf{x}_j)^2}_{\Gamma}
\nonumber\\*
&+\underbrace{\sum_{I=N}^{2N-2}\sum_{j=I+2}^{2N}(\mathbf{u}^{(I)}-\mathbf{u}^{(I+1)}+\mathbf{x}_{I+1}-\mathbf{x}_j)^2}_{\Gamma},
\end{align}
where $\mathbf{S}=\sum_i\mathbf{s}_i$, $\mathbf{X}_+=\sum_{\alpha=1}^N\mathbf{x}_{\alpha}$ and $\mathbf{X}_-=\sum_{\alpha=N+1}^{2N}\mathbf{x}_{\alpha}$.
All the terms with the gauge field strengths cancel.
By (\ref{Nahm_anom}), the boundary anomaly for the Nahm pole side is
\begin{align}
\label{SQCD_fw_anom}
&\mathcal{A}^{\wtilde{(1)-(2)-\cdots-(N)-\cdots-(2)-(1)}}_{\mathcal{D},\mathrm{Nahm}^{2N-3},\mathcal{D}}
\nonumber\\*
&=\underbrace{\sum_{I=1}^{2N-2}\min(r_I,r_{I+1})(\mathbf{u}^{(I)}-\mathbf{u}^{(I+1)})^2+\sum_{a=1}^2(\mathbf{u}^{(N)}-\mathbf{z}_a)^2}_{\textrm{top components}}
\quad\underbrace{-2\sum_{I=1}^{2N-1}r_I\,\mathbf{u}^{(I)}(\mathbf{x}_I-\mathbf{x}_{I+1})}_{\textrm{FI couplings}}
\nonumber\\*
&=\mathcal{A}^{U(N)+2N}_{\mathcal{N}',N'+\Gamma+\eta}-2\mathbf{z}_1\mathbf{X}_++2\mathbf{z}_2\mathbf{X}_-.
\end{align}
As for $N=2$, the anomalies agree up to the mixed terms between $x_{\alpha}$ and $z_a$.
In particular, the gauge-neutral Fermi multiplets $\Gamma$ reproduce the terms with $\min(r_I,r_{I+1})$.

\subsection{\texorpdfstring{$\mathrm{Nahm}'$ and $(\mathcal{N},N+\tilde{\eta})$}{Nahm' and (N, N+eta)}}
\label{sec_SQCD_rv}
When the $N$ D3-branes end on a single D5$'$-brane,
SQCD obeys the Nahm pole boundary condition for the vector multiplet and the Dirichlet boundary condition $D'$ for the hypermultiplets.
Under S-duality the D5$'$-brane maps to the NS5$'$-brane
and the mirror quiver obeys the Neumann boundary condition $(\mathcal{N},N)$ together with boundary Fermi multiplets.
In this case we apply (\ref{Nahm_single_simple}) to SQCD directly.

\subsubsection{\texorpdfstring{$N=2$, $N_f=4$}{N=2, Nf=4}}
\label{sec_SQCD_rvN2}
Consider $U(2)$ SQCD with four flavors.

\paragraph{$U(2)$ SQCD}
For the Dirichlet boundary condition $(\mathcal{D}',D')$ we have $k_{\rm eff}=2\mathbb{J}_2$.
With $n=m_2-m_1$, the formula (\ref{Nahm_single_simple}) with $z=z_1/z_2$ and $J=\frac12$ for the four doublets with $\bar{u}^{w_{\lambda}}x^{f_{\lambda}}=ux_{\alpha}$ gives
\begin{align}
\label{U2_4_simple}
\mathbb{II}^{U(2)+4}_{\mathrm{Nahm}'}
&=\underbrace{\prod_{\alpha=1}^4(qt^{-2}(ux_{\alpha})^{\pm};q)_{\infty}}_{\textrm{top components}}\;
\underbrace{\frac{1}{(q)_{\infty}(q^{\frac12}t^{-2};q)_{\infty}(q^{\frac32}t^{-2};q)_{\infty}(qt^{-4};q)_{\infty}}}_{\textrm{Slodowy slice}}
\nonumber\\*
&\times\sum_{m_1\le m_2}\underbrace{\left(\frac{z_2}{z_1}\right)^{m_1+m_2}}_{\textrm{topological charge}}\;
\underbrace{(q^{\frac12}t^{-2})^{-n}\frac{(qt^{-4};q)_n(q^{\frac32}t^{-2};q)_n}{(q^{\frac12}t^{-2};q)_n(q;q)_n}}_{\textrm{vector multiplet ($\mathrm{Nahm}'$)}}
\nonumber\\*
&\times\underbrace{\prod_{\alpha=1}^4(q^{\frac14}t^{-1})^{m_1+m_2}\frac{(t^{2}ux_{\alpha};q)_{m_1}(q^{\frac12}ux_{\alpha};q)_{m_2}}{(q^{\frac12}ux_{\alpha};q)_{m_1}(qt^{-2} ux_{\alpha};q)_{m_2}}}_{\textrm{fundamental hypermultiplets ($D'$)}}.
\end{align}
It is the $N=2$ case of (\ref{SQCD_rv_simple}) below.

\paragraph{Mirror quiver}
The mirror quiver is $(1)-(2)-(1)$ with two flavors $[2]$ at the $U(2)$ node,
and we denote its gauge fugacities by $s^{(1)}$, $s^{(2)}_i$ and $s^{(3)}$.
As in section \ref{sec_dualrules}, for the Neumann boundary condition $(\mathcal{N},N)$ the remaining part
$-(\mathbf{s}^{(1)})^2-(\mathbf{s}^{(2)}_1+\mathbf{s}^{(2)}_2-\mathbf{s}^{(1)})^2-(\mathbf{s}^{(3)}-\mathbf{s}^{(2)}_1-\mathbf{s}^{(2)}_2)^2-(\mathbf{s}^{(3)})^2$
of the boundary gauge anomaly
is cancelled by four cross-determinant Fermi multiplets $\tilde{\eta}_I$ (see (\ref{U2_4_rv_Neu_anom}) for the full anomaly).
Their global charges are fixed by the cancellation of the mixed anomalies with the FI terms and the matching with the anomaly on the Nahm pole side.
The Neumann half-index of the mirror quiver reads
\begin{align}
\label{U2_4_rv_Neu}
&\mathbb{II}^{\wtilde{(1)-(2)-(1)}}_{\mathcal{N},N+\tilde{\eta}}
\nonumber\\*
&=\frac{1}{2}\oint\prod_{I,i}\frac{ds^{(I)}_i}{2\pi is^{(I)}_i}
\underbrace{[(q)_{\infty}(q^{\frac12}t^{-2};q)_{\infty}]^4\prod_{i\neq j}\left(\frac{s^{(2)}_i}{s^{(2)}_j};q\right)_{\infty}\left(q^{\frac12}t^{-2}\frac{s^{(2)}_i}{s^{(2)}_j};q\right)_{\infty}}_{\textrm{vector multiplets ($\mathcal{N}$)}}
\nonumber\\
&\times \underbrace{\prod_{i=1}^2\frac{1}{(q^{\frac14}t^{-1}(s^{(1)}/s^{(2)}_i)^{\pm};q)_{\infty}(q^{\frac14}t^{-1}(s^{(2)}_i/s^{(3)})^{\pm};q)_{\infty}}}_{\textrm{bifundamental hypermultiplets ($N$)}}\;
\underbrace{\prod_{i=1}^2\prod_{a=1}^2\frac{1}{(q^{\frac14}t^{-1}(s^{(2)}_i/z_a)^{\pm};q)_{\infty}}}_{\textrm{fundamental hypermultiplets ($N$)}}
\nonumber\\
&\times \underbrace{F\left(q^{\frac12}\frac{1}{s^{(1)}}\frac{z_1}{ux_1};q\right)}_{\tilde{\eta}_1}
\underbrace{F\left(q^{\frac12}\frac{s^{(1)}}{s^{(2)}_1s^{(2)}_2}\frac{z_1}{ux_2};q\right)}_{\tilde{\eta}_2}
\nonumber\\*
&\times \underbrace{F\left(q^{\frac12}\frac{s^{(2)}_1s^{(2)}_2}{s^{(3)}}\frac{1}{ux_3z_2};q\right)}_{\tilde{\eta}_3}
\underbrace{F\left(q^{\frac12}s^{(3)}\frac{1}{ux_4z_2};q\right)}_{\tilde{\eta}_4}.
\end{align}
The Fermi multiplets $\tilde{\eta}_I$ arise at the junctions of the four NS5-branes with the NS5$'$-brane in Figure \ref{fig_SQCD} (d) \cite{Hanany:2018hlz},
and their charges under $z_1$ and $z_2$ come from the D3-branes ending on the two D5-branes (see section \ref{sec_dualrules}).
They correspond to the top components $(qt^{-2}(ux_{\alpha})^{\pm};q)_{\infty}$ of the four doublets of $U(2)$ at the Nahm pole.
Unlike section \ref{sec_SQCD_fw}, no gauge-neutral Fermi multiplet is needed,
since no D3-brane passes through the D5-branes in Figure \ref{fig_SQCD} (d).

\paragraph{Duality}
We propose that
\begin{align}
\label{U2_4_rv_dual}
\mathbb{II}^{U(2)+4}_{\mathrm{Nahm}'}&=\mathbb{II}^{\wtilde{(1)-(2)-(1)}}_{\mathcal{N},N+\tilde{\eta}}.
\end{align}

As a consistency check, we compare the boundary anomalies.
For the Neumann boundary condition (\ref{U2_4_rv_Neu}) of the mirror quiver the boundary anomaly is
\begin{align}
\label{U2_4_rv_Neu_anom}
&\mathcal{A}^{\wtilde{(1)-(2)-(1)}}_{\mathcal{N},N+\tilde{\eta}}
\nonumber\\*
&=\underbrace{2(\mathbf{s}^{(2)}_1-\mathbf{s}^{(2)}_2)^2}_{\textrm{vector multiplet ($\mathcal{N}$)}}\quad
\underbrace{-\sum_{i=1}^2\bigl[(\mathbf{s}^{(1)}-\mathbf{s}^{(2)}_i)^2+(\mathbf{s}^{(2)}_i-\mathbf{s}^{(3)})^2\bigr]}_{\textrm{bifundamental hypermultiplets ($N$)}}\quad
\underbrace{-\sum_{i=1}^2\sum_{a=1}^2(\mathbf{s}^{(2)}_i-\mathbf{z}_a)^2}_{\textrm{fundamental hypermultiplets ($N$)}}
\nonumber\\
&\underbrace{-2\mathbf{s}^{(1)}(\mathbf{x}_1-\mathbf{x}_2)-2(\mathbf{s}^{(2)}_1+\mathbf{s}^{(2)}_2)(\mathbf{x}_2-\mathbf{x}_3)-2\mathbf{s}^{(3)}(\mathbf{x}_3-\mathbf{x}_4)}_{\textrm{FI couplings}}
\nonumber\\
&+\underbrace{(\mathbf{z}_1-\mathbf{s}^{(1)}-\mathbf{u}-\mathbf{x}_1)^2}_{\tilde{\eta}_1}
+\underbrace{(\mathbf{s}^{(1)}-\mathbf{s}^{(2)}_1-\mathbf{s}^{(2)}_2+\mathbf{z}_1-\mathbf{u}-\mathbf{x}_2)^2}_{\tilde{\eta}_2}
\nonumber\\*
&+\underbrace{(\mathbf{s}^{(2)}_1+\mathbf{s}^{(2)}_2-\mathbf{s}^{(3)}-\mathbf{u}-\mathbf{x}_3-\mathbf{z}_2)^2}_{\tilde{\eta}_3}
+\underbrace{(\mathbf{s}^{(3)}-\mathbf{u}-\mathbf{x}_4-\mathbf{z}_2)^2}_{\tilde{\eta}_4}.
\end{align}
All the terms with the gauge field strengths cancel.
In particular, the FI couplings of the mirror quiver cancel the mixed terms of the Fermi multiplets $\tilde{\eta}_I$ between the gauge group and $x_{\alpha}$.
By (\ref{Nahm_anom}), the boundary anomaly for the Nahm pole side is
\begin{align}
\mathcal{A}^{U(2)+4}_{\mathrm{Nahm}'}&=\underbrace{\sum_{\alpha=1}^4(\mathbf{u}+\mathbf{x}_{\alpha})^2}_{\textrm{top components}}\underbrace{-4\mathbf{u}(\mathbf{z}_1-\mathbf{z}_2)}_{\textrm{FI coupling}}
\nonumber\\*
&=\mathcal{A}^{\wtilde{(1)-(2)-(1)}}_{\mathcal{N},N+\tilde{\eta}}+2\mathbf{z}_1(\mathbf{x}_1+\mathbf{x}_2)-2\mathbf{z}_2(\mathbf{x}_3+\mathbf{x}_4).
\end{align}
As in section \ref{sec_SQCD_fwN2}, the anomalies agree up to the mixed terms between $x_{\alpha}$ and $z_a$, which now appear with the opposite signs.
Hence in both dualities $\mathcal{A}^{U(2)+4}-\mathcal{A}^{\wtilde{(1)-(2)-(1)}}=2\mathbf{z}_1(\mathbf{x}_1+\mathbf{x}_2)-2\mathbf{z}_2(\mathbf{x}_3+\mathbf{x}_4)$,
as expected from a single background Chern-Simons term relating the two bulk theories.
The boundary symmetry $U(1)_u$ acts only on the Fermi multiplets $\tilde{\eta}_I$ on the Neumann side,
and the FI coupling of the Nahm pole side arises from their charges under $z_1$ and $z_2$.

\subsubsection{\texorpdfstring{General $N$, $N_f=2N$}{General N, Nf=2N}}
\label{sec_SQCD_rvN}
\paragraph{$U(N)$ SQCD}
The formula (\ref{Nahm_single_simple}) with $J=\frac{N-1}{2}$ in (\ref{spins_UN}) gives
\begin{align}
\label{SQCD_rv_simple}
&\mathbb{II}^{U(N)+2N}_{\mathrm{Nahm}'}
\nonumber\\*
&=\underbrace{\prod_{\alpha=1}^{2N}(q^{\frac12+\frac{N}{4}}t^{-N}(ux_{\alpha})^{\pm};q)_{\infty}}_{\textrm{top components}}\;
\underbrace{\prod_{k=0}^{N-1}\frac{1}{(q^{1+\frac{k}{2}}t^{-2k};q)_{\infty}(q^{\frac12+\frac{k}{2}}t^{-2-2k};q)_{\infty}}}_{\textrm{Slodowy slice}}\;
\sum_{m_1\le\cdots\le m_N}\underbrace{\left(\frac{z_2}{z_1}\right)^{\sum_im_i}}_{\makebox[0pt]{\scriptsize topological charge}}
\nonumber\displaybreak[1]\\
&\times\underbrace{\prod_{1\le i<j\le N}(q^{\frac12}t^{-2})^{m_i-m_j}\frac{(q^{\frac12+\frac{j-i}{2}}t^{-2-2(j-i)};q)_{m_j-m_i}(q^{1+\frac{j-i}{2}}t^{-2(j-i)};q)_{m_j-m_i}}{(q^{\frac{j-i}{2}}t^{-2(j-i)};q)_{m_j-m_i}(q^{\frac12+\frac{j-i}{2}}t^{2-2(j-i)};q)_{m_j-m_i}}}_{\textrm{vector multiplet ($\mathrm{Nahm}'$)}}
\nonumber\\*
&\times \underbrace{\prod_{\alpha=1}^{2N}\prod_{i=1}^N(q^{\frac14}t^{-1})^{m_i}\frac{(q^{\frac14+\frac{\mu_i}{2}}t^{1-2\mu_i}ux_{\alpha};q)_{m_i}}{(q^{\frac34+\frac{\mu_i}{2}}t^{-1-2\mu_i}ux_{\alpha};q)_{m_i}}}_{\textrm{fundamental hypermultiplets ($D'$)}},
\end{align}
where $\mu_i=i-\frac{N+1}{2}$.

\paragraph{Mirror quiver}
As for $N=2$, the remaining part of the boundary gauge anomaly for $(\mathcal{N},N)$ is $-\sum_{I=1}^{2N}(\sum_i\mathbf{s}^{(I)}_i-\sum_i\mathbf{s}^{(I-1)}_i)^2$ with $\mathbf{s}^{(0)}=\mathbf{s}^{(2N)}=0$
(see (\ref{SQCD_rv_Neu_anom}) for the full anomaly).
Hence we need $2N$ cross-determinant Fermi multiplets $\tilde{\eta}_I$ with the gauge fugacities $\prod_is^{(I-1)}_i/\prod_is^{(I)}_i$.
Generalizing the case $N=2$, we find the Neumann half-index of the mirror quiver
\begin{align}
\label{SQCD_rv_Neu}
&\mathbb{II}^{\wtilde{(1)-(2)-\cdots-(N)-\cdots-(2)-(1)}}_{\mathcal{N},N+\tilde{\eta}}
\nonumber\\*
&=\frac{1}{\prod_Ir_I!}\oint\prod_{I,i}\frac{ds^{(I)}_i}{2\pi is^{(I)}_i}
\underbrace{[(q)_{\infty}(q^{\frac12}t^{-2};q)_{\infty}]^{N^2}\prod_I\prod_{i\neq j}\left(\frac{s^{(I)}_i}{s^{(I)}_j};q\right)_{\infty}\left(q^{\frac12}t^{-2}\frac{s^{(I)}_i}{s^{(I)}_j};q\right)_{\infty}}_{\textrm{vector multiplets ($\mathcal{N}$)}}
\nonumber\\
&\times \underbrace{\prod_{I=1}^{2N-2}\prod_{i,j}\frac{1}{(q^{\frac14}t^{-1}(s^{(I)}_i/s^{(I+1)}_j)^{\pm};q)_{\infty}}}_{\textrm{bifundamental hypermultiplets ($N$)}}\;
\underbrace{\prod_{i=1}^N\prod_{a=1}^2\frac{1}{(q^{\frac14}t^{-1}(s^{(N)}_i/z_a)^{\pm};q)_{\infty}}}_{\textrm{fundamental hypermultiplets ($N$)}}
\nonumber\\*
&\times \underbrace{\prod_{I=1}^{N}F\left(q^{\frac12}\frac{\prod_is^{(I-1)}_i}{\prod_is^{(I)}_i}\frac{z_1}{ux_I};q\right)
\prod_{I=N+1}^{2N}F\left(q^{\frac12}\frac{\prod_is^{(I-1)}_i}{\prod_is^{(I)}_i}\frac{1}{ux_Iz_2};q\right)}_{\tilde{\eta}_I},
\end{align}
where $\prod_is^{(0)}_i=\prod_is^{(2N)}_i=1$.
As for $N=2$, the Fermi multiplet $\tilde{\eta}_I$ arises at the junction of the $I$-th NS5-brane with the NS5$'$-brane,
which separates the nodes $U(r_{I-1})$ and $U(r_I)$ when all the D3-branes of the mirror quiver are suspended between the NS5-branes,
and corresponds to the top component of the $I$-th fundamental hypermultiplet of SQCD at the Nahm pole.

\paragraph{Duality}
We propose that
\begin{align}
\label{SQCD_rv_dual}
\mathbb{II}^{U(N)+2N}_{\mathrm{Nahm}'}&=\mathbb{II}^{\wtilde{(1)-(2)-\cdots-(N)-\cdots-(2)-(1)}}_{\mathcal{N},N+\tilde{\eta}}.
\end{align}
For $N=1$ it reduces to the duality between the Dirichlet boundary condition for SQED$_2$ and the Neumann boundary condition for its mirror in \cite{Okazaki:2019bok},
and for $N=2$ it is (\ref{U2_4_rv_dual}).
The identity singles out the topological charge $(z_2/z_1)^{\sum_im_i}$ in (\ref{SQCD_rv_simple}) against $(z_1/z_2)^{\sum_im_i}$
and requires stripping off the 2d twisted hypermultiplets (see Appendix \ref{app_checks}).

As a consistency check, we compare the boundary anomalies.
For the Neumann boundary condition (\ref{SQCD_rv_Neu}) of the mirror quiver the boundary anomaly is
\begin{align}
\label{SQCD_rv_Neu_anom}
&\mathcal{A}^{\wtilde{(1)-(2)-\cdots-(N)-\cdots-(2)-(1)}}_{\mathcal{N},N+\tilde{\eta}}
\nonumber\\*
&=\underbrace{\sum_{I=1}^{2N-1}\sum_{i\neq j}(\mathbf{s}^{(I)}_i-\mathbf{s}^{(I)}_j)^2}_{\textrm{vector multiplets ($\mathcal{N}$)}}\quad
\underbrace{-\sum_{I=1}^{2N-2}\sum_{i,j}(\mathbf{s}^{(I)}_i-\mathbf{s}^{(I+1)}_j)^2}_{\textrm{bifundamental hypermultiplets ($N$)}}\quad
\underbrace{-\sum_{i=1}^N\sum_{a=1}^2(\mathbf{s}^{(N)}_i-\mathbf{z}_a)^2}_{\textrm{fundamental hypermultiplets ($N$)}}
\nonumber\\
&\underbrace{-2\sum_{I=1}^{2N-1}\mathbf{S}_I(\mathbf{x}_I-\mathbf{x}_{I+1})}_{\textrm{FI couplings}}
\nonumber\\*
&+\underbrace{\sum_{I=1}^{N}(\mathbf{S}_{I-1}-\mathbf{S}_I+\mathbf{z}_1-\mathbf{u}-\mathbf{x}_I)^2
+\sum_{I=N+1}^{2N}(\mathbf{S}_{I-1}-\mathbf{S}_I-\mathbf{u}-\mathbf{x}_I-\mathbf{z}_2)^2}_{\tilde{\eta}_I},
\end{align}
where $\mathbf{S}_I=\sum_i\mathbf{s}^{(I)}_i$ with $\mathbf{S}_0=\mathbf{S}_{2N}=0$.
All the terms with the gauge field strengths cancel.
By (\ref{Nahm_anom}), the boundary anomaly for the Nahm pole side is
\begin{align}
\label{SQCD_rv_anom}
\mathcal{A}^{U(N)+2N}_{\mathrm{Nahm}'}&=\underbrace{\sum_{\alpha=1}^{2N}(\mathbf{u}+\mathbf{x}_{\alpha})^2}_{\textrm{top components}}\underbrace{-2N\mathbf{u}(\mathbf{z}_1-\mathbf{z}_2)}_{\textrm{FI coupling}}
\nonumber\\*
&=\mathcal{A}^{\wtilde{(1)-(2)-\cdots-(N)-\cdots-(2)-(1)}}_{\mathcal{N},N+\tilde{\eta}}+2\mathbf{z}_1\mathbf{X}_+-2\mathbf{z}_2\mathbf{X}_-,
\end{align}
where $\mathbf{X}_{\pm}$ are as in section \ref{sec_SQCD_fwN}.
As for $N=2$, the anomalies agree up to the mixed terms between $x_{\alpha}$ and $z_a$.

\subsection{Unbalanced SQCD}
\label{sec_SQCD_nonbal}
For $U(N)$ SQCD with $N_f>2N$ flavors the mirror quiver is
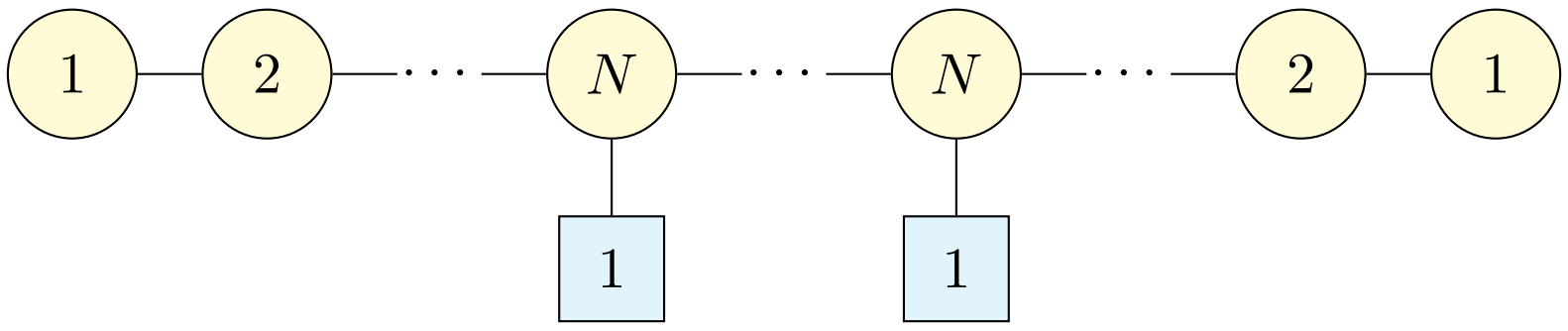
\begin{align}
\label{SQCD_nonbal_quiver}
\begin{tikzpicture}[quiver]
\node[gauge](a1) at (0,0){$1$}; \node[gauge](a2) at (1.3,0){$2$}; \node[qdots](a3) at (2.45,0){$\cdots$};
\node[gauge](a4) at (3.6,0){$N$}; \node[qdots](a5) at (4.75,0){$\cdots$}; \node[gauge](a6) at (5.9,0){$N$};
\node[qdots](a7) at (7.05,0){$\cdots$}; \node[gauge](a8) at (8.2,0){$2$}; \node[gauge](a9) at (9.5,0){$1$};
\node[flavor](f1) at (3.6,-1.3){$1$}; \node[flavor](f2) at (5.9,-1.3){$1$};
\draw (a1)--(a2)--(a3)--(a4)--(a5)--(a6)--(a7)--(a8)--(a9); \draw (a4)--(f1); \draw (a6)--(f2);
\end{tikzpicture}
\end{align}
with $N_f-1$ nodes $U(r_I)$, $r_I=\min(I,N,N_f-I)$, where one flavor with fugacity $z_1$ is attached to the $N$-th node and the other with $z_2$ to the $(N_f-N)$-th node.
The nodes from the $N$-th to the $(N_f-N)$-th form a plateau of $U(N)$ nodes.
All the nodes of the mirror quiver are balanced, while the node of SQCD is unbalanced.
As in the balanced case, we find the two types of dualities.
We consider the following cases (see Table \ref{Table_checks} for the checks),
\begin{align}
\label{SQCD_nonbal_list}
\begin{array}{c|c}
\textrm{SQCD}&\textrm{mirror}\\ \hline
U(2)+5&(1)-(2)-(2)-(1)\\
U(2)+6&(1)-(2)-(2)-(2)-(1)\\
U(2)+7&(1)-(2)-(2)-(2)-(2)-(1)\\
U(3)+7&(1)-(2)-(3)-(3)-(2)-(1)\\
U(3)+8&(1)-(2)-(3)-(3)-(3)-(2)-(1)\\
\end{array}.
\end{align}

\subsubsection{\texorpdfstring{$(\mathcal{N}',N'+\Gamma+\eta)$ and $(\mathcal{D},\mathrm{Nahm}^{N_f-3},\mathcal{D})$}{(N', N'+Gamma+eta) and (D, Nahm, ..., Nahm, D)}}
\label{sec_SQCD_nonbal_fw}
\paragraph{$U(N)$ SQCD}
For the Neumann boundary condition $(\mathcal{N}',N')$ the boundary gauge anomaly is
$-(N_f-2N)\sum_i\mathbf{s}_i^2-2(\sum_i\mathbf{s}_i)^2$,
up to the mixed terms with the flavor and topological symmetries (see (\ref{SQCD_nonbal_fw_Neu_anom}) for the full anomaly).
Unlike the balanced case, the non-Abelian part does not cancel.
The term $-(N_f-2N)\sum_i\mathbf{s}_i^2$ is cancelled by $N_f-2N$ Fermi multiplets $\Gamma_I$ in the fundamental representation,
and the remaining part $-2(\sum_i\mathbf{s}_i)^2$ is cancelled by the two cross-determinant Fermi multiplets $\eta_{\pm}$ as before.
Together with $N(N-1)$ gauge-neutral Fermi multiplets $\Gamma$, we propose the Neumann half-index
\begin{align}
\label{SQCD_nonbal_fw_Neu}
&\mathbb{II}^{U(N)+N_f}_{\mathcal{N}',N'+\Gamma+\eta}
\nonumber\\*
&=\frac{1}{N!}\oint \prod_{i=1}^N\frac{ds_i}{2\pi is_i}
\underbrace{(q)_{\infty}^N(q^{\frac12}t^2;q)_{\infty}^N\prod_{i\neq j}\left(\frac{s_i}{s_j};q\right)_{\infty}\left(q^{\frac12}t^2\frac{s_i}{s_j};q\right)_{\infty}}_{\textrm{vector multiplet ($\mathcal{N}'$)}}
\underbrace{\prod_{i=1}^N\prod_{\alpha=1}^{N_f}\frac{1}{(q^{\frac14}ts_i^{\pm}x_{\alpha}^{\pm};q)_{\infty}}}_{\textrm{hypermultiplets ($N'$)}}
\nonumber\\
&\times \underbrace{F\left(q^{\frac12}\prod_{i=1}^Ns_i\,\frac{z_1}{u^{(N)}}\prod_{\alpha=1}^Nx_{\alpha};q\right)}_{\eta_+}
\underbrace{F\left(q^{\frac12}\prod_{i=1}^Ns_i\,\frac{u^{(N_f-N)}}{z_2}\prod_{\alpha=N_f-N+1}^{N_f}x_{\alpha};q\right)}_{\eta_-}
\nonumber\\
&\times \underbrace{\prod_{I=N}^{N_f-N-1}\prod_{i=1}^NF\left(q^{\frac12}s_i\,x_{I+1}\frac{u^{(I)}}{u^{(I+1)}};q\right)}_{\Gamma_I}
\nonumber\\*
&\times \underbrace{\prod_{I=1}^{N-1}\prod_{j=1}^{I}F\left(q^{\frac12}\frac{u^{(I)}}{u^{(I+1)}}\frac{x_{I+1}}{x_j};q\right)
\prod_{I=N_f-N}^{N_f-2}\prod_{j=I+2}^{N_f}F\left(q^{\frac12}\frac{u^{(I)}}{u^{(I+1)}}\frac{x_{I+1}}{x_j};q\right)}_{\Gamma}.
\end{align}
For $N_f=2N$ there is no Fermi multiplet $\Gamma_I$ and (\ref{SQCD_nonbal_fw_Neu}) reduces to (\ref{SQCD_fw_Neu}).
The fundamental Fermi multiplets $\Gamma_I$ are the non-Abelian generalization of the Fermi multiplets $\Gamma^{(2)},\cdots,\Gamma^{(N_f-1)}$ for SQED$_{N_f}$ in \cite{Okazaki:2019bok}.
The fugacity of the Fermi multiplet $\Gamma_I$ contains $x_{I+1}$ and $u^{(I)}/u^{(I+1)}$,
so that the Fermi multiplet $\Gamma_I$ is associated with the $(I+1)$-th D5-brane.
In the configuration S-dual to that of the mirror quiver, the gauge D3-branes pass through the $(I+1)$-th D5-brane for $N\le I\le N_f-N-1$,
and the Fermi multiplet $\Gamma_I$ arises from the D3-D5 strings at its intersection with the NS5$'$-brane (see section \ref{sec_dualrules}).
It corresponds to the $N$ top components of the bifundamental hypermultiplet between the $U(N)$ nodes $I$ and $I+1$ of the plateau.

\paragraph{Mirror quiver}
As in section \ref{sec_SQCD_fw}, the $r_I$ D3-branes of the node $U(r_I)$ end on a single D5$'$-brane.
With the notation of section \ref{sec_SQCD_fwN}, the formula (\ref{Nahm_quiver_simple}) with $t\rightarrow t^{-1}$ gives
\begin{align}
\label{SQCD_nonbal_fw_simple}
&\mathbb{II}^{\wtilde{(1)-\cdots-(N)-\cdots-(N)-\cdots-(1)}}_{\mathcal{D},\mathrm{Nahm}^{N_f-3},\mathcal{D}}
\nonumber\\*
&=\underbrace{\left(q^{\frac12+\frac{N}{4}}t^{N}\left(\frac{u^{(N)}}{z_1}\right)^{\pm};q\right)_{\infty}
\left(q^{\frac12+\frac{N}{4}}t^{N}\left(\frac{u^{(N_f-N)}}{z_2}\right)^{\pm};q\right)_{\infty}}_{\textrm{top components}}
\nonumber\displaybreak[1]\\
&\times\underbrace{\prod_{I=1}^{N_f-2}\prod_{J=|j_I-j_{I+1}|}^{j_I+j_{I+1}}\left(q^{\frac34+\frac{J}{2}}t^{1+2J}\left(\frac{u^{(I)}}{u^{(I+1)}}\right)^{\pm};q\right)_{\infty}}_{\textrm{top components}}
\nonumber\displaybreak[1]\\
&\times\underbrace{\prod_{I=1}^{N_f-1}\prod_{k=0}^{r_I-1}\frac{1}{(q^{1+\frac{k}{2}}t^{2k};q)_{\infty}(q^{\frac12+\frac{k}{2}}t^{2+2k};q)_{\infty}}}_{\textrm{Slodowy slices}}
\sum_{\{m^{(I)}\}}\underbrace{\prod_{I=1}^{N_f-1}\left(\frac{x_{I+1}}{x_I}\right)^{\sum_im^{(I)}_i}}_{\textrm{topological charges}}
\nonumber\displaybreak[1]\\
&\times\underbrace{\prod_{I=1}^{N_f-1}\prod_{1\le i<j\le r_I}(q^{\frac12}t^2)^{m^{(I)}_i-m^{(I)}_j}\frac{(q^{\frac12+\frac{j-i}{2}}t^{2+2(j-i)};q)_{m^{(I)}_j-m^{(I)}_i}(q^{1+\frac{j-i}{2}}t^{2(j-i)};q)_{m^{(I)}_j-m^{(I)}_i}}{(q^{\frac{j-i}{2}}t^{2(j-i)};q)_{m^{(I)}_j-m^{(I)}_i}(q^{\frac12+\frac{j-i}{2}}t^{2(j-i)-2};q)_{m^{(I)}_j-m^{(I)}_i}}}_{\textrm{vector multiplets ($\mathrm{Nahm}$)}}
\nonumber\displaybreak[1]\\
&\times\underbrace{\prod_{i=1}^{N}(q^{\frac14}t)^{m^{(N)}_i}\frac{(q^{\frac14+\frac{\mu^{(N)}_i}{2}}t^{2\mu^{(N)}_i-1}\frac{u^{(N)}}{z_1};q)_{m^{(N)}_i}}{(q^{\frac34+\frac{\mu^{(N)}_i}{2}}t^{1+2\mu^{(N)}_i}\frac{u^{(N)}}{z_1};q)_{m^{(N)}_i}}}_{\textrm{fundamental hypermultiplets ($D$)}}
\nonumber\\
&\times\underbrace{\prod_{i=1}^{N}(q^{\frac14}t)^{m^{(N_f-N)}_i}\frac{(q^{\frac14+\frac{\mu^{(N_f-N)}_i}{2}}t^{2\mu^{(N_f-N)}_i-1}\frac{u^{(N_f-N)}}{z_2};q)_{m^{(N_f-N)}_i}}{(q^{\frac34+\frac{\mu^{(N_f-N)}_i}{2}}t^{1+2\mu^{(N_f-N)}_i}\frac{u^{(N_f-N)}}{z_2};q)_{m^{(N_f-N)}_i}}}_{\textrm{fundamental hypermultiplets ($D$)}}
\nonumber\\*
&\times\underbrace{\prod_{I=1}^{N_f-2}\prod_{i=1}^{r_I}\prod_{j=1}^{r_{I+1}}(q^{\frac14}t)^{m^{(I)}_i-m^{(I+1)}_j}\frac{(q^{\frac14+\frac{\mu^{(I)}_i-\mu^{(I+1)}_j}{2}}t^{2(\mu^{(I)}_i-\mu^{(I+1)}_j)-1}\frac{u^{(I)}}{u^{(I+1)}};q)_{m^{(I)}_i-m^{(I+1)}_j}}{(q^{\frac34+\frac{\mu^{(I)}_i-\mu^{(I+1)}_j}{2}}t^{1+2(\mu^{(I)}_i-\mu^{(I+1)}_j)}\frac{u^{(I)}}{u^{(I+1)}};q)_{m^{(I)}_i-m^{(I+1)}_j}}}_{\textrm{bifundamental hypermultiplets ($D$)}}.
\end{align}
The spins of the top components are as in (\ref{SQCD_fw_simple}).
In particular, they have $J=0,1,\cdots,N-1$ for the bifundamental ones between two nodes $U(N)$ of the plateau.

\paragraph{Duality}
We propose that
\begin{align}
\label{SQCD_nonbal_fw_dual}
\mathbb{II}^{U(N)+N_f}_{\mathcal{N}',N'+\Gamma+\eta}
&=\mathbb{II}^{\wtilde{(1)-\cdots-(N)-\cdots-(N)-\cdots-(1)}}_{\mathcal{D},\mathrm{Nahm}^{N_f-3},\mathcal{D}}.
\end{align}
For $N=1$ all the nodes are $U(1)$ with the Dirichlet boundary conditions
and it reduces to the duality between the Neumann boundary condition for SQED$_{N_f}$ and the Dirichlet boundary condition for its mirror in \cite{Okazaki:2019bok}.
The identity singles out the charges of the Fermi multiplets $\Gamma_I$ against those with $u^{(I)}/u^{(I+1)}$ inverted or with $x_I$ in place of $x_{I+1}$ (see Appendix \ref{app_checks}),
which also spoil the cancellation of the terms with the gauge field strengths.

As a consistency check, we compare the boundary anomalies.
For the Neumann boundary condition (\ref{SQCD_nonbal_fw_Neu}) the boundary anomaly is
\begin{align}
\label{SQCD_nonbal_fw_Neu_anom}
&\mathcal{A}^{U(N)+N_f}_{\mathcal{N}',N'+\Gamma+\eta}
\nonumber\\*
&=\underbrace{-\sum_{i=1}^N\sum_{\alpha=1}^{N_f}(\mathbf{s}_i+\mathbf{x}_{\alpha})^2}_{\textrm{hypermultiplets ($N'$)}}
\quad\underbrace{+\sum_{i\neq j}(\mathbf{s}_i-\mathbf{s}_j)^2}_{\textrm{vector multiplet ($\mathcal{N}'$)}}
\quad\underbrace{-2\mathbf{S}(\mathbf{z}_1-\mathbf{z}_2)}_{\textrm{FI coupling}}
\nonumber\\
&+\underbrace{\sum_{I=N}^{N_f-N-1}\sum_{i=1}^N(\mathbf{s}_i+\mathbf{x}_{I+1}+\mathbf{u}^{(I)}-\mathbf{u}^{(I+1)})^2}_{\Gamma_I}
\nonumber\\
&+\underbrace{\sum_{I=1}^{N-1}\sum_{j=1}^{I}(\mathbf{u}^{(I)}-\mathbf{u}^{(I+1)}+\mathbf{x}_{I+1}-\mathbf{x}_j)^2}_{\Gamma}
+\underbrace{\sum_{I=N_f-N}^{N_f-2}\sum_{j=I+2}^{N_f}(\mathbf{u}^{(I)}-\mathbf{u}^{(I+1)}+\mathbf{x}_{I+1}-\mathbf{x}_j)^2}_{\Gamma}
\nonumber\\*
&+\underbrace{(\mathbf{S}+\mathbf{X}_++\mathbf{z}_1-\mathbf{u}^{(N)})^2}_{\eta_+}
+\underbrace{(\mathbf{S}+\mathbf{X}_-+\mathbf{u}^{(N_f-N)}-\mathbf{z}_2)^2}_{\eta_-},
\end{align}
where $\mathbf{S}=\sum_i\mathbf{s}_i$, $\mathbf{X}_+=\sum_{\alpha=1}^N\mathbf{x}_{\alpha}$ and $\mathbf{X}_-=\sum_{\alpha=N_f-N+1}^{N_f}\mathbf{x}_{\alpha}$.
All the terms with the gauge field strengths $\mathbf{s}_i$ cancel.
By (\ref{Nahm_anom}), the boundary anomaly for the Nahm pole side is
\begin{align}
&\mathcal{A}^{\wtilde{(1)-\cdots-(N)-\cdots-(N)-\cdots-(1)}}_{\mathcal{D},\mathrm{Nahm}^{N_f-3},\mathcal{D}}
\nonumber\\*
&=\underbrace{\sum_{I=1}^{N_f-2}\min(r_I,r_{I+1})(\mathbf{u}^{(I)}-\mathbf{u}^{(I+1)})^2
+(\mathbf{u}^{(N)}-\mathbf{z}_1)^2+(\mathbf{u}^{(N_f-N)}-\mathbf{z}_2)^2}_{\textrm{top components}}
\nonumber\\
&\underbrace{-2\sum_{I=1}^{N_f-1}r_I\,\mathbf{u}^{(I)}(\mathbf{x}_I-\mathbf{x}_{I+1})}_{\textrm{FI couplings}}
\nonumber\\*
&=\mathcal{A}^{U(N)+N_f}_{\mathcal{N}',N'+\Gamma+\eta}-2\mathbf{z}_1\mathbf{X}_++2\mathbf{z}_2\mathbf{X}_-.
\end{align}
As for $N_f=2N$, the anomalies agree up to the mixed terms between $x_{\alpha}$ and $z_a$.
The terms $\min(r_I,r_{I+1})(\mathbf{u}^{(I)}-\mathbf{u}^{(I+1)})^2$ are reproduced by the Fermi multiplets $\Gamma_I$ for the bifundamental hypermultiplets between the nodes of the plateau
and by the Fermi multiplets $\Gamma$ for the other bifundamental hypermultiplets.

\subsubsection{\texorpdfstring{$\mathrm{Nahm}'$ and $(\mathcal{N},N+\tilde{\eta})$}{Nahm' and (N, N+eta)}}
\label{sec_SQCD_nonbal_rv}
\paragraph{$U(N)$ SQCD}
Since the gauge node is unbalanced, the expanded form (\ref{Nahm_single}) contains the factor $(q^{\frac12}t^{-2})^{(N_f-2N)\sum_i\mu_im_i}$ noted below it
and the sign $(-1)^{(N_f-2N)\sum_im_i^2}$.
They are absorbed into the hypermultiplet factors, and the formula (\ref{Nahm_single_simple}) gives
\begin{align}
\label{SQCD_nonbal_rv_simple}
&\mathbb{II}^{U(N)+N_f}_{\mathrm{Nahm}'}
\nonumber\\*
&=\underbrace{\prod_{\alpha=1}^{N_f}(q^{\frac12+\frac{N}{4}}t^{-N}(ux_{\alpha})^{\pm};q)_{\infty}}_{\textrm{top components}}\;
\underbrace{\prod_{k=0}^{N-1}\frac{1}{(q^{1+\frac{k}{2}}t^{-2k};q)_{\infty}(q^{\frac12+\frac{k}{2}}t^{-2-2k};q)_{\infty}}}_{\textrm{Slodowy slice}}\;
\sum_{m_1\le\cdots\le m_N}\underbrace{\left(\frac{z_2}{z_1}\right)^{\sum_im_i}}_{\makebox[0pt]{\scriptsize topological charge}}
\nonumber\\
&\times\underbrace{\prod_{1\le i<j\le N}(q^{\frac12}t^{-2})^{m_i-m_j}\frac{(q^{\frac12+\frac{j-i}{2}}t^{-2-2(j-i)};q)_{m_j-m_i}(q^{1+\frac{j-i}{2}}t^{-2(j-i)};q)_{m_j-m_i}}{(q^{\frac{j-i}{2}}t^{-2(j-i)};q)_{m_j-m_i}(q^{\frac12+\frac{j-i}{2}}t^{2-2(j-i)};q)_{m_j-m_i}}}_{\textrm{vector multiplet ($\mathrm{Nahm}'$)}}
\nonumber\\*
&\times \underbrace{\prod_{\alpha=1}^{N_f}\prod_{i=1}^N(q^{\frac14}t^{-1})^{m_i}\frac{(q^{\frac14+\frac{\mu_i}{2}}t^{1-2\mu_i}ux_{\alpha};q)_{m_i}}{(q^{\frac34+\frac{\mu_i}{2}}t^{-1-2\mu_i}ux_{\alpha};q)_{m_i}}}_{\textrm{fundamental hypermultiplets ($D'$)}},
\end{align}
where $\mu_i=i-\frac{N+1}{2}$.
For $N_f=2N$ it reduces to (\ref{SQCD_rv_simple}).

\paragraph{Mirror quiver}
As for $N_f=2N$, for the Neumann boundary condition $(\mathcal{N},N)$ of the mirror quiver (\ref{SQCD_nonbal_quiver})
the remaining part of the boundary gauge anomaly is cancelled by $N_f$ cross-determinant Fermi multiplets $\tilde{\eta}_I$
with the gauge fugacities $\prod_is^{(I-1)}_i/\prod_is^{(I)}_i$ (see (\ref{SQCD_nonbal_rv_Neu_anom}) for the full anomaly).
Generalizing (\ref{SQCD_rv_Neu}), we find the Neumann half-index
\begin{align}
\label{SQCD_nonbal_rv_Neu}
&\mathbb{II}^{\wtilde{(1)-\cdots-(N)-\cdots-(N)-\cdots-(1)}}_{\mathcal{N},N+\tilde{\eta}}
\nonumber\\*
&=\frac{1}{\prod_Ir_I!}\oint\prod_{I,i}\frac{ds^{(I)}_i}{2\pi is^{(I)}_i}
\underbrace{[(q)_{\infty}(q^{\frac12}t^{-2};q)_{\infty}]^{\sum_Ir_I}\prod_I\prod_{i\neq j}\left(\frac{s^{(I)}_i}{s^{(I)}_j};q\right)_{\infty}\left(q^{\frac12}t^{-2}\frac{s^{(I)}_i}{s^{(I)}_j};q\right)_{\infty}}_{\textrm{vector multiplets ($\mathcal{N}$)}}
\nonumber\\
&\times \underbrace{\prod_{I=1}^{N_f-2}\prod_{i,j}\frac{1}{(q^{\frac14}t^{-1}(s^{(I)}_i/s^{(I+1)}_j)^{\pm};q)_{\infty}}}_{\textrm{bifundamental hypermultiplets ($N$)}}
\underbrace{\prod_{i=1}^N\frac{1}{(q^{\frac14}t^{-1}(s^{(N)}_i/z_1)^{\pm};q)_{\infty}(q^{\frac14}t^{-1}(s^{(N_f-N)}_i/z_2)^{\pm};q)_{\infty}}}_{\textrm{fundamental hypermultiplets ($N$)}}
\nonumber\\
&\times \underbrace{\prod_{I=1}^{N}F\left(q^{\frac12}\frac{\prod_is^{(I-1)}_i}{\prod_is^{(I)}_i}\frac{z_1}{ux_I};q\right)
\prod_{I=N+1}^{N_f-N}F\left(q^{\frac12}\frac{\prod_is^{(I-1)}_i}{\prod_is^{(I)}_i}\frac{1}{ux_I};q\right)}_{\tilde{\eta}_I}
\nonumber\\*
&\times \underbrace{\prod_{I=N_f-N+1}^{N_f}F\left(q^{\frac12}\frac{\prod_is^{(I-1)}_i}{\prod_is^{(I)}_i}\frac{1}{ux_Iz_2};q\right)}_{\tilde{\eta}_I},
\end{align}
where $\prod_is^{(0)}_i=\prod_is^{(N_f)}_i=1$.
For $N_f=2N$ it reduces to (\ref{SQCD_rv_Neu}).

\paragraph{Duality}
We propose that
\begin{align}
\label{SQCD_nonbal_rv_dual}
\mathbb{II}^{U(N)+N_f}_{\mathrm{Nahm}'}&=\mathbb{II}^{\wtilde{(1)-\cdots-(N)-\cdots-(N)-\cdots-(1)}}_{\mathcal{N},N+\tilde{\eta}}.
\end{align}

As a consistency check, we compare the boundary anomalies.
For the Neumann boundary condition (\ref{SQCD_nonbal_rv_Neu}) of the mirror quiver the boundary anomaly is
\begin{align}
\label{SQCD_nonbal_rv_Neu_anom}
&\mathcal{A}^{\wtilde{(1)-\cdots-(N)-\cdots-(N)-\cdots-(1)}}_{\mathcal{N},N+\tilde{\eta}}
\nonumber\\*
&=\underbrace{\sum_{I=1}^{N_f-1}\sum_{i\neq j}(\mathbf{s}^{(I)}_i-\mathbf{s}^{(I)}_j)^2}_{\textrm{vector multiplets ($\mathcal{N}$)}}\quad
\underbrace{-\sum_{I=1}^{N_f-2}\sum_{i,j}(\mathbf{s}^{(I)}_i-\mathbf{s}^{(I+1)}_j)^2}_{\textrm{bifundamental hypermultiplets ($N$)}}
\nonumber\\
&\underbrace{-\sum_{i=1}^N\bigl[(\mathbf{s}^{(N)}_i-\mathbf{z}_1)^2+(\mathbf{s}^{(N_f-N)}_i-\mathbf{z}_2)^2\bigr]}_{\textrm{fundamental hypermultiplets ($N$)}}\quad
\underbrace{-2\sum_{I=1}^{N_f-1}\mathbf{S}_I(\mathbf{x}_I-\mathbf{x}_{I+1})}_{\textrm{FI couplings}}
\nonumber\\
&+\underbrace{\sum_{I=1}^{N}(\mathbf{S}_{I-1}-\mathbf{S}_I+\mathbf{z}_1-\mathbf{u}-\mathbf{x}_I)^2
+\sum_{I=N+1}^{N_f-N}(\mathbf{S}_{I-1}-\mathbf{S}_I-\mathbf{u}-\mathbf{x}_I)^2}_{\tilde{\eta}_I}
\nonumber\\*
&+\underbrace{\sum_{I=N_f-N+1}^{N_f}(\mathbf{S}_{I-1}-\mathbf{S}_I-\mathbf{u}-\mathbf{x}_I-\mathbf{z}_2)^2}_{\tilde{\eta}_I},
\end{align}
where $\mathbf{S}_I=\sum_i\mathbf{s}^{(I)}_i$ with $\mathbf{S}_0=\mathbf{S}_{N_f}=0$.
All the terms with the gauge field strengths cancel.
By (\ref{Nahm_anom}), the boundary anomaly for the Nahm pole side is
\begin{align}
\label{SQCD_nonbal_rv_anom}
\mathcal{A}^{U(N)+N_f}_{\mathrm{Nahm}'}&=\underbrace{\sum_{\alpha=1}^{N_f}(\mathbf{u}+\mathbf{x}_{\alpha})^2}_{\textrm{top components}}\underbrace{-2N\mathbf{u}(\mathbf{z}_1-\mathbf{z}_2)}_{\textrm{FI coupling}}
\nonumber\\*
&=\mathcal{A}^{\wtilde{(1)-\cdots-(N)-\cdots-(N)-\cdots-(1)}}_{\mathcal{N},N+\tilde{\eta}}+2\mathbf{z}_1\mathbf{X}_+-2\mathbf{z}_2\mathbf{X}_-,
\end{align}
where $\mathbf{X}_{\pm}$ are as in section \ref{sec_SQCD_nonbal_fw}.
As for $N_f=2N$, the anomalies agree up to the mixed terms between $x_{\alpha}$ and $z_a$.

\section{\texorpdfstring{$T[SU(N)]$}{T[SU(N)]}}
\label{sec_TSUN}
Next consider $T[SU(N)]$, the linear quiver gauge theory
\begin{align}
\label{TSUN_quiver}
\begin{tikzpicture}[quiver]
\node[gauge](a1) at (0,0){$1$}; \node[gauge](a2) at (1.3,0){$2$}; \node[qdots](a3) at (2.45,0){$\cdots$};
\node[gauge](a4) at (3.6,0){$N\!-\!1$}; \node[flavor](f) at (4.95,0){$N$};
\draw (a1)--(a2)--(a3)--(a4)--(f);
\end{tikzpicture}
\end{align}
All the nodes are balanced and the theory is self-mirror \cite{Gaiotto:2008ak}
(see also \cite{Okazaki:2019ony} for tests of mirror symmetry using supersymmetric indices).
We denote the gauge fugacities for the node $U(I)$ by $s^{(I)}_i$, $i=1,\cdots,I$.
The fugacities for the $SU(N)$ flavor symmetry are $x_{\alpha}$, $\alpha=1,\cdots,N$ with $\prod_{\alpha}x_{\alpha}=1$,
and the hypermultiplets at the node $U(N-1)$ have fugacities $s^{(N-1)}_ix_{\alpha}$ as in section \ref{sec_SQCD}.
The fugacities $z_a$, $a=1,\cdots,N$ for the topological symmetry of $T[SU(N)]$
are the flavor fugacities of the mirror $T[SU(N)]$ (see section \ref{sec_conv}).
We do not need to impose $\prod_az_a=1$.
In our conventions the FI coupling of the node $U(I)$ of $T[SU(N)]$ is $+2\sum_i\mathbf{s}^{(I)}_i(\mathbf{z}_I-\mathbf{z}_{I+1})$,
whose sign is opposite to that for $U(N)$ SQCD in section \ref{sec_SQCD}.

\subsection{\texorpdfstring{$(\mathcal{N}',N'+\Gamma+\eta)$ and $(\mathcal{D},\mathrm{Nahm}^{N-2})$}{(N', N'+Gamma+eta) and (D, Nahm, ..., Nahm)}}
\label{sec_TSUN_fw}
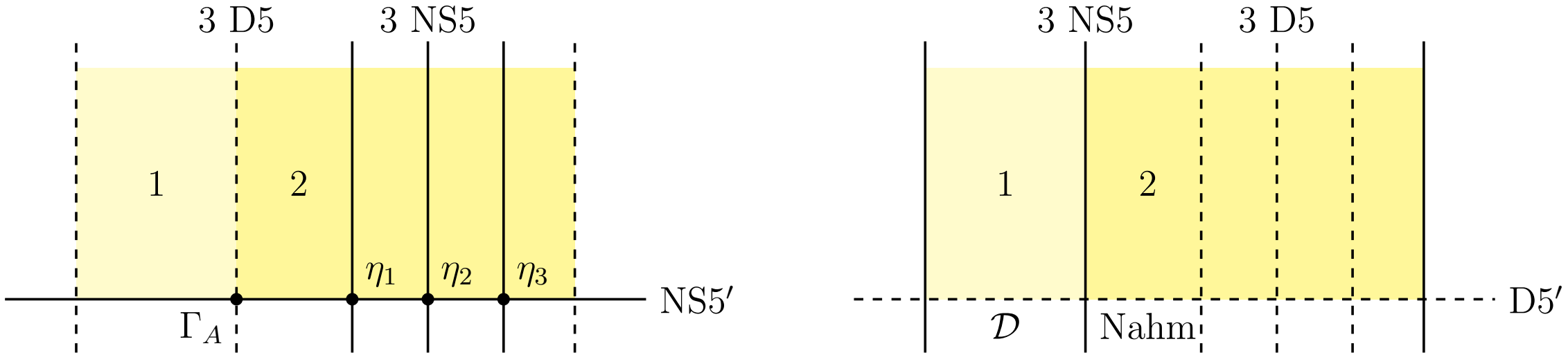
\begin{figure}
\centering
\scalebox{0.85}{
\begin{tikzpicture}
\fill[yellow!25] (-2.6,0) rectangle (-0.8,2.6);
\fill[yellow!50] (-0.8,0) rectangle (3.0,2.6);
\draw[thick,solid,black] (-3.4,0) -- (3.8,0) node[right] {\textrm{NS5}$'$};
\foreach \x in {-2.6,-0.8,3.0} \draw[thick,dashed,black] (\x,-0.6) -- (\x,2.9);
\foreach \x in {0.5,1.35,2.2} \draw[thick,solid,black] (\x,-0.6) -- (\x,2.9);
\node at (-0.8,3.15) {\textrm{3 D5}};
\node at (1.35,3.15) {\textrm{3 NS5}};
\node at (-1.7,1.3) {\textrm{1}};
\node at (-0.1,1.3) {\textrm{2}};
\fill[black] (-0.8,0) circle (2pt) node[below left] {$\Gamma_A$};
\fill[black] (0.5,0) circle (2pt) node[above right] {$\eta_1$};
\fill[black] (1.35,0) circle (2pt) node[above right] {$\eta_2$};
\fill[black] (2.2,0) circle (2pt) node[above right] {$\eta_3$};
\node at (0,-1.2) {(a) $T[SU(3)]$, $(\mathcal{N}',N'+\Gamma+\eta)$};
\end{tikzpicture}
}
\hspace{0.5cm}
\scalebox{0.85}{
\begin{tikzpicture}
\fill[yellow!25] (-2.6,0) rectangle (-0.8,2.6);
\fill[yellow!50] (-0.8,0) rectangle (3.0,2.6);
\draw[thick,dashed,black] (-3.4,0) -- (3.8,0) node[right] {\textrm{D5}$'$};
\foreach \x in {-2.6,-0.8,3.0} \draw[thick,solid,black] (\x,-0.6) -- (\x,2.9);
\foreach \x in {0.5,1.35,2.2} \draw[thick,dashed,black] (\x,-0.6) -- (\x,2.9);
\node at (-0.8,3.15) {\textrm{3 NS5}};
\node at (1.35,3.15) {\textrm{3 D5}};
\node at (-1.7,1.3) {\textrm{1}};
\node at (-0.1,1.3) {\textrm{2}};
\node at (-1.7,-0.3) {$\mathcal{D}$};
\node at (-0.1,-0.3) {\textrm{Nahm}};
\node at (0,-1.2) {(b) mirror, $(\mathcal{D},\mathrm{Nahm})$};
\end{tikzpicture}
}
\caption{\textrm{
The brane configurations for $T[SU(3)]$ (a) and its mirror (b).
}}
\label{fig_TSUN}
\end{figure}

When the D3-branes end on the NS5$'$-brane, $T[SU(N)]$ obeys the Neumann boundary condition $(\mathcal{N}',N')$.
In the S-dual configuration, for each node $U(I)$ of the mirror the $I$ D3-branes end on a single D5$'$-brane.
We thus find the boundary condition $(\mathcal{D},\mathrm{Nahm}^{N-2})$,
i.e. the Dirichlet boundary condition for the $U(1)$ node and the Nahm pole boundary conditions for the other nodes.
The brane configurations for $N=3$ are illustrated in Figure \ref{fig_TSUN}.
In (b) the D5-branes are moved inside the $U(2)$ segment by the Hanany-Witten transitions, and (a) is drawn as the S-dual of (b).

\subsubsection{\texorpdfstring{$N=3$}{N=3}}
\label{sec_TSUN_N3}
Consider $T[SU(3)]$, the quiver $(1)-(2)-[3]$ with the gauge fugacities $s^{(1)}$ and $s^{(2)}_i$, $i=1,2$.
Its mirror is again $T[SU(3)]$.

\paragraph{$T[SU(3)]$}
As in section \ref{sec_dualrules}, for the Neumann boundary condition $(\mathcal{N}',N')$ the remaining part
$-(\mathbf{s}^{(1)})^2-(\mathbf{s}^{(2)}_1+\mathbf{s}^{(2)}_2-\mathbf{s}^{(1)})^2-(\mathbf{s}^{(2)}_1+\mathbf{s}^{(2)}_2)^2$
of the boundary gauge anomaly
(see (\ref{TSU3_Neu_anom}) for the full anomaly) is cancelled by three cross-determinant Fermi multiplets $\eta_1$, $\eta_2$ and $\eta_3$
with the gauge fugacities $s^{(1)}$, $s^{(2)}_1s^{(2)}_2/s^{(1)}$ and $1/(s^{(2)}_1s^{(2)}_2)$.
Their mixed anomalies with the boundary symmetry $U(1)_{u^{(2)}}$ cancel only when they carry the same charge under it,
since their gauge charges add up to zero.
The term $(\mathbf{u}^{(1)}-\mathbf{u}^{(2)})^2$ in the anomaly (\ref{TSU3_Nahm_anom}) for the Nahm pole side below
requires a gauge-neutral Fermi multiplet $\Gamma_A$.
We find the Neumann half-index
\begin{align}
\label{TSU3_Neu}
&\mathbb{II}^{T[SU(3)]}_{\mathcal{N}',N'+\Gamma+\eta}
\nonumber\\*
&=\frac{1}{2}\oint\frac{ds^{(1)}}{2\pi is^{(1)}}\prod_{i=1}^2\frac{ds^{(2)}_i}{2\pi is^{(2)}_i}
\underbrace{[(q)_{\infty}(q^{\frac12}t^2;q)_{\infty}]^3\prod_{i\neq j}\left(\frac{s^{(2)}_i}{s^{(2)}_j};q\right)_{\infty}\left(q^{\frac12}t^2\frac{s^{(2)}_i}{s^{(2)}_j};q\right)_{\infty}}_{\textrm{vector multiplets ($\mathcal{N}'$)}}
\nonumber\displaybreak[1]\\
&\times \underbrace{\prod_{i=1}^2\frac{1}{(q^{\frac14}t(s^{(1)}/s^{(2)}_i)^{\pm};q)_{\infty}}}_{\textrm{bifundamental hypermultiplet ($N'$)}}\;
\underbrace{\prod_{i=1}^2\prod_{\alpha=1}^3\frac{1}{(q^{\frac14}t(s^{(2)}_i)^{\pm}x_{\alpha}^{\pm};q)_{\infty}}}_{\textrm{fundamental hypermultiplets ($N'$)}}\;
\underbrace{F\left(q^{\frac12}\frac{u^{(1)}}{u^{(2)}}\frac{x_2}{x_1};q\right)}_{\Gamma_A}
\nonumber\\*
&\times \underbrace{F\left(q^{\frac12}\frac{s^{(1)}u^{(2)}x_3}{z_1};q\right)}_{\eta_1}\underbrace{F\left(q^{\frac12}\frac{s^{(2)}_1s^{(2)}_2}{s^{(1)}}\frac{u^{(2)}x_3}{z_2};q\right)}_{\eta_2}
\underbrace{F\left(q^{\frac12}\frac{1}{s^{(2)}_1s^{(2)}_2}\frac{u^{(2)}x_3}{z_3};q\right)}_{\eta_3}.
\end{align}
The Fermi multiplet $\eta_I$ arises at the junction of the $I$-th NS5-brane with the NS5$'$-brane in Figure \ref{fig_TSUN} (a) \cite{Hanany:2018hlz}.
When all the D3-branes of $T[SU(3)]$ are suspended between the NS5-branes, this NS5-brane separates the nodes $U(I-1)$ and $U(I)$.
The common factor $u^{(2)}$ of the Fermi multiplets $\eta_I$ reflects that the three NS5-branes are located on the segment of the NS5$'$-brane
between the last two D5-branes.
The mixed anomalies of the Fermi multiplets $\eta_I$ between the gauge group and the topological symmetries amount to
$-2\mathbf{s}^{(1)}(\mathbf{z}_1-\mathbf{z}_2)-2(\mathbf{s}^{(2)}_1+\mathbf{s}^{(2)}_2)(\mathbf{z}_2-\mathbf{z}_3)$,
which is cancelled by the FI terms.
The gauge-neutral Fermi multiplet $\Gamma_A$ arises at the intersection of the second D5-brane with the NS5$'$-brane,
through which the D3-brane emanating from the first D5-brane passes (see section \ref{sec_dualrules}).
It takes the same form as the Fermi multiplet $\Gamma_A$ for $U(2)$ SQCD in (\ref{U2_4_Neu}).

\paragraph{Mirror $T[SU(3)]$}
The formula (\ref{Nahm_quiver_simple}) with $t\rightarrow t^{-1}$ and $J=\frac12$
for the bifundamental hypermultiplet of $U(1)\times U(2)$ and the three fundamental ones of $U(2)$ gives
\begin{align}
\label{TSU3_simple}
&\mathbb{II}^{\wtilde{T[SU(3)]}}_{\mathcal{D},\mathrm{Nahm}}
\nonumber\\*
&=\underbrace{\left(qt^{2}\left(\frac{u^{(1)}}{u^{(2)}}\right)^{\pm};q\right)_{\infty}\prod_{a=1}^3\left(qt^{2}\left(\frac{u^{(2)}}{z_a}\right)^{\pm};q\right)_{\infty}}_{\textrm{top components}}\;
\underbrace{\frac{1}{(q)_{\infty}^2(q^{\frac12}t^{2};q)_{\infty}^2(q^{\frac32}t^{2};q)_{\infty}(qt^{4};q)_{\infty}}}_{\textrm{Slodowy slices}}
\nonumber\displaybreak[1]\\
&\times \sum_{m^{(1)}\in\mathbb{Z}}\sum_{m^{(2)}_1\le m^{(2)}_2}\underbrace{\left(\frac{x_2}{x_1}\right)^{m^{(1)}}\left(\frac{x_3}{x_2}\right)^{m^{(2)}_1+m^{(2)}_2}}_{\textrm{topological charges}}\;
\underbrace{(q^{\frac12}t^{2})^{-n}\frac{(qt^{4};q)_{n}(q^{\frac32}t^{2};q)_{n}}{(q^{\frac12}t^{2};q)_{n}(q;q)_{n}}}_{\textrm{vector multiplet ($\mathrm{Nahm}$)}}
\nonumber\displaybreak[1]\\
&\times \underbrace{(q^{\frac14}t)^{2m^{(1)}-m^{(2)}_1-m^{(2)}_2}\frac{(q^{\frac12}\frac{u^{(1)}}{u^{(2)}};q)_{m^{(1)}-m^{(2)}_1}(t^{-2}\frac{u^{(1)}}{u^{(2)}};q)_{m^{(1)}-m^{(2)}_2}}{(qt^{2} \frac{u^{(1)}}{u^{(2)}};q)_{m^{(1)}-m^{(2)}_1}(q^{\frac12}\frac{u^{(1)}}{u^{(2)}};q)_{m^{(1)}-m^{(2)}_2}}}_{\textrm{bifundamental hypermultiplet ($D$)}}
\nonumber\\*
&\times \underbrace{\prod_{a=1}^3(q^{\frac14}t)^{m^{(2)}_1+m^{(2)}_2}\frac{(t^{-2}\frac{u^{(2)}}{z_a};q)_{m^{(2)}_1}(q^{\frac12}\frac{u^{(2)}}{z_a};q)_{m^{(2)}_2}}{(q^{\frac12}\frac{u^{(2)}}{z_a};q)_{m^{(2)}_1}(qt^{2} \frac{u^{(2)}}{z_a};q)_{m^{(2)}_2}}}_{\textrm{fundamental hypermultiplets ($D$)}},
\end{align}
where $m^{(1)}$ is the flux for the $U(1)$ node, $(m^{(2)}_1,m^{(2)}_2)$ are those for the $U(2)$ node and $n=m^{(2)}_2-m^{(2)}_1$.

\paragraph{Duality}
We propose that
\begin{align}
\label{TSU3_dual}
\mathbb{II}^{T[SU(3)]}_{\mathcal{N}',N'+\Gamma+\eta}&=\mathbb{II}^{\wtilde{T[SU(3)]}}_{\mathcal{D},\mathrm{Nahm}}.
\end{align}
The Fermi multiplets $\eta_I$ in (\ref{TSU3_Neu}) are singled out among the assignments
which differ in the choice of the $U(1)_{u^{(2)}}$ charges, of the flavor fugacity $x_{\alpha}$ and of the orientations of $z_I$.
For instance, the identity distinguishes the Fermi multiplets $\eta_I$ from those with $z_I^{-1}$ in place of $z_I$ (see Appendix \ref{app_checks}).

As a consistency check, we compare the boundary anomalies.
For the Neumann boundary condition (\ref{TSU3_Neu}) the boundary anomaly is
\begin{align}
\label{TSU3_Neu_anom}
&\mathcal{A}^{T[SU(3)]}_{\mathcal{N}',N'+\Gamma+\eta}
\nonumber\\*
&=\underbrace{2(\mathbf{s}^{(2)}_1-\mathbf{s}^{(2)}_2)^2}_{\textrm{vector multiplet ($\mathcal{N}'$)}}\quad
\underbrace{-\sum_{i=1}^2(\mathbf{s}^{(1)}-\mathbf{s}^{(2)}_i)^2}_{\textrm{bifundamental hypermultiplet ($N'$)}}\quad
\underbrace{-\sum_{i=1}^2\sum_{\alpha=1}^3(\mathbf{s}^{(2)}_i+\mathbf{x}_{\alpha})^2}_{\textrm{fundamental hypermultiplets ($N'$)}}
\nonumber\\
&\underbrace{+2\mathbf{s}^{(1)}(\mathbf{z}_1-\mathbf{z}_2)+2(\mathbf{s}^{(2)}_1+\mathbf{s}^{(2)}_2)(\mathbf{z}_2-\mathbf{z}_3)}_{\textrm{FI couplings}}
+\underbrace{(\mathbf{u}^{(1)}-\mathbf{u}^{(2)}+\mathbf{x}_2-\mathbf{x}_1)^2}_{\Gamma_A}
\nonumber\\
&+\underbrace{(\mathbf{s}^{(1)}+\mathbf{u}^{(2)}+\mathbf{x}_3-\mathbf{z}_1)^2}_{\eta_1}
+\underbrace{(\mathbf{s}^{(2)}_1+\mathbf{s}^{(2)}_2-\mathbf{s}^{(1)}+\mathbf{u}^{(2)}+\mathbf{x}_3-\mathbf{z}_2)^2}_{\eta_2}
\nonumber\\*
&+\underbrace{(-\mathbf{s}^{(2)}_1-\mathbf{s}^{(2)}_2+\mathbf{u}^{(2)}+\mathbf{x}_3-\mathbf{z}_3)^2}_{\eta_3}.
\end{align}
With $\sum_{\alpha}\mathbf{x}_{\alpha}=0$, all the terms with the gauge field strengths cancel.
By (\ref{Nahm_anom}), the boundary anomaly for the Nahm pole side is
\begin{align}
\label{TSU3_Nahm_anom}
&\mathcal{A}^{\wtilde{T[SU(3)]}}_{\mathcal{D},\mathrm{Nahm}}
\nonumber\\*
&=\underbrace{(\mathbf{u}^{(1)}-\mathbf{u}^{(2)})^2+\sum_{a=1}^3(\mathbf{u}^{(2)}-\mathbf{z}_a)^2}_{\textrm{top components}}
\underbrace{-2\mathbf{u}^{(1)}(\mathbf{x}_1-\mathbf{x}_2)-4\mathbf{u}^{(2)}(\mathbf{x}_2-\mathbf{x}_3)}_{\textrm{FI couplings}}
\nonumber\\*
&=\mathcal{A}^{T[SU(3)]}_{\mathcal{N}',N'+\Gamma+\eta}+2\mathbf{x}_3(\mathbf{z}_1+\mathbf{z}_2+\mathbf{z}_3).
\end{align}
Hence the anomalies agree up to the mixed term $2\mathbf{x}_3(\mathbf{z}_1+\mathbf{z}_2+\mathbf{z}_3)$,
which can be absorbed into a background mixed Chern-Simons term and vanishes for $\prod_az_a=1$.

\subsubsection{\texorpdfstring{General $N$}{General N}}
\label{sec_TSUN_N}
\paragraph{$T[SU(N)]$}
As for $N=3$, for the Neumann boundary condition $(\mathcal{N}',N')$
the remaining part $-\sum_{I=1}^N(\sum_i\mathbf{s}^{(I)}_i-\sum_i\mathbf{s}^{(I-1)}_i)^2$ with $\mathbf{s}^{(0)}=\mathbf{s}^{(N)}=0$ of the boundary gauge anomaly
is cancelled by $N$ cross-determinant Fermi multiplets $\eta_I$ with the gauge fugacities $\prod_is^{(I)}_i/\prod_is^{(I-1)}_i$, $I=1,\cdots,N$.
They carry the same charge under $U(1)_{u^{(N-1)}}$.
The terms $\sum_{I=1}^{N-2}I(\mathbf{u}^{(I)}-\mathbf{u}^{(I+1)})^2$ in the anomaly (\ref{TSUN_Nahm_anom}) for the Nahm pole side
require $\frac{(N-1)(N-2)}{2}$ gauge-neutral Fermi multiplets $\Gamma$.
We propose the Neumann half-index
\begin{align}
\label{TSUN_Neu}
&\mathbb{II}^{T[SU(N)]}_{\mathcal{N}',N'+\Gamma+\eta}
\nonumber\\*
&=\frac{1}{\prod_{I=1}^{N-1}I!}\oint\prod_{I,i}\frac{ds^{(I)}_i}{2\pi is^{(I)}_i}
\underbrace{[(q)_{\infty}(q^{\frac12}t^2;q)_{\infty}]^{\frac{N(N-1)}{2}}\prod_{I=1}^{N-1}\prod_{i\neq j}\left(\frac{s^{(I)}_i}{s^{(I)}_j};q\right)_{\infty}\left(q^{\frac12}t^2\frac{s^{(I)}_i}{s^{(I)}_j};q\right)_{\infty}}_{\textrm{vector multiplets ($\mathcal{N}'$)}}
\nonumber\\
&\times \underbrace{\prod_{I=1}^{N-2}\prod_{i=1}^I\prod_{j=1}^{I+1}\frac{1}{(q^{\frac14}t(s^{(I)}_i/s^{(I+1)}_j)^{\pm};q)_{\infty}}}_{\textrm{bifundamental hypermultiplets ($N'$)}}\;
\underbrace{\prod_{i=1}^{N-1}\prod_{\alpha=1}^N\frac{1}{(q^{\frac14}t(s^{(N-1)}_i)^{\pm}x_{\alpha}^{\pm};q)_{\infty}}}_{\textrm{fundamental hypermultiplets ($N'$)}}
\nonumber\\*
&\times \underbrace{\prod_{I=1}^NF\left(q^{\frac12}\frac{\prod_is^{(I)}_i}{\prod_is^{(I-1)}_i}\frac{u^{(N-1)}x_N}{z_I};q\right)}_{\eta_I}
\underbrace{\prod_{I=1}^{N-2}\prod_{j=1}^IF\left(q^{\frac12}\frac{u^{(I)}}{u^{(I+1)}}\frac{x_{I+1}}{x_j};q\right)}_{\Gamma},
\end{align}
where $\prod_is^{(0)}_i=\prod_is^{(N)}_i=1$.
For $N=3$ it is (\ref{TSU3_Neu}).
As for $N=3$, the Fermi multiplets $\eta_I$ arise at the junctions of the NS5-branes with the NS5$'$-brane,
and their mixed anomalies with the topological symmetries are cancelled by the FI terms.
The gauge-neutral Fermi multiplets $\Gamma$ arise as in section \ref{sec_dualrules}.
The $I$ D3-branes emanating from the first $I$ D5-branes pass through the $(I+1)$-th D5-brane,
which separates the segments of the NS5$'$-brane with $U(1)_{u^{(I)}}$ and $U(1)_{u^{(I+1)}}$.
This gives the charges $\frac{u^{(I)}}{u^{(I+1)}}\frac{x_{I+1}}{x_j}$, $j=1,\cdots,I$, in (\ref{TSUN_Neu}).
Similarly, the factor $u^{(N-1)}x_N$ of the Fermi multiplets $\eta_I$ reflects that the NS5-branes are located on the last segment of the NS5$'$-brane,
and the D3-branes there end on the last D5-brane.
They take the same form as those attached to the left half of the mirror quiver of $U(N)$ SQCD in (\ref{SQCD_fw_Neu}).

\paragraph{Mirror $T[SU(N)]$}
The formula (\ref{Nahm_quiver_simple}) with $t\rightarrow t^{-1}$ gives
\begin{align}
\label{TSUN_simple}
&\mathbb{II}^{\wtilde{T[SU(N)]}}_{\mathcal{D},\mathrm{Nahm}^{N-2}}
\nonumber\\*
&=\underbrace{\prod_{a=1}^N\left(q^{\frac14+\frac{N}{4}}t^{N-1}\left(\frac{u^{(N-1)}}{z_a}\right)^{\pm};q\right)_{\infty}
\prod_{I=1}^{N-2}\prod_{J=\frac12}^{I-\frac12}\left(q^{\frac34+\frac{J}{2}}t^{1+2J}\left(\frac{u^{(I)}}{u^{(I+1)}}\right)^{\pm};q\right)_{\infty}}_{\textrm{top components}}
\nonumber\displaybreak[1]\\
&\times\underbrace{\prod_{I=1}^{N-1}\prod_{k=0}^{I-1}\frac{1}{(q^{1+\frac{k}{2}}t^{2k};q)_{\infty}(q^{\frac12+\frac{k}{2}}t^{2+2k};q)_{\infty}}}_{\textrm{Slodowy slices}}
\sum_{\{m^{(I)}\}}\underbrace{\prod_{I=1}^{N-1}\left(\frac{x_{I+1}}{x_I}\right)^{\sum_im^{(I)}_i}}_{\textrm{topological charges}}
\nonumber\displaybreak[1]\\
&\times\underbrace{\prod_{I=1}^{N-1}\prod_{1\le i<j\le I}(q^{\frac12}t^2)^{m^{(I)}_i-m^{(I)}_j}\frac{(q^{\frac12+\frac{j-i}{2}}t^{2+2(j-i)};q)_{m^{(I)}_j-m^{(I)}_i}(q^{1+\frac{j-i}{2}}t^{2(j-i)};q)_{m^{(I)}_j-m^{(I)}_i}}{(q^{\frac{j-i}{2}}t^{2(j-i)};q)_{m^{(I)}_j-m^{(I)}_i}(q^{\frac12+\frac{j-i}{2}}t^{2(j-i)-2};q)_{m^{(I)}_j-m^{(I)}_i}}}_{\textrm{vector multiplets ($\mathrm{Nahm}$)}}
\nonumber\displaybreak[1]\\
&\times\underbrace{\prod_{a=1}^N\prod_{i=1}^{N-1}(q^{\frac14}t)^{m^{(N-1)}_i}\frac{(q^{\frac14+\frac{\mu^{(N-1)}_i}{2}}t^{2\mu^{(N-1)}_i-1}\frac{u^{(N-1)}}{z_a};q)_{m^{(N-1)}_i}}{(q^{\frac34+\frac{\mu^{(N-1)}_i}{2}}t^{1+2\mu^{(N-1)}_i}\frac{u^{(N-1)}}{z_a};q)_{m^{(N-1)}_i}}}_{\textrm{fundamental hypermultiplets ($D$)}}
\nonumber\\*
&\times\underbrace{\prod_{I=1}^{N-2}\prod_{i=1}^{I}\prod_{j=1}^{I+1}(q^{\frac14}t)^{m^{(I)}_i-m^{(I+1)}_j}\frac{(q^{\frac14+\frac{\mu^{(I)}_i-\mu^{(I+1)}_j}{2}}t^{2(\mu^{(I)}_i-\mu^{(I+1)}_j)-1}\frac{u^{(I)}}{u^{(I+1)}};q)_{m^{(I)}_i-m^{(I+1)}_j}}{(q^{\frac34+\frac{\mu^{(I)}_i-\mu^{(I+1)}_j}{2}}t^{1+2(\mu^{(I)}_i-\mu^{(I+1)}_j)}\frac{u^{(I)}}{u^{(I+1)}};q)_{m^{(I)}_i-m^{(I+1)}_j}}}_{\textrm{bifundamental hypermultiplets ($D$)}},
\end{align}
where $\mu^{(I)}_i=i-\frac{I+1}{2}$ and the fluxes are restricted to $m^{(I)}_1\le\cdots\le m^{(I)}_I$.
The top components of the bifundamental hypermultiplet of $U(I)\times U(I+1)$ have $J=\frac12,\frac32,\cdots,I-\frac12$
from $\frac{I-1}{2}\otimes\frac{I}{2}$, and those of the $N$ fundamental ones of $U(N-1)$ have $J=\frac{N-2}{2}$.
For $N=3$ it is (\ref{TSU3_simple}).

\paragraph{Duality}
We propose that
\begin{align}
\label{TSUN_dual}
\mathbb{II}^{T[SU(N)]}_{\mathcal{N}',N'+\Gamma+\eta}&=\mathbb{II}^{\wtilde{T[SU(N)]}}_{\mathcal{D},\mathrm{Nahm}^{N-2}}.
\end{align}
For $N=2$ it is the duality for $T[SU(2)]$ in \cite{Okazaki:2019bok}
under the identification $x_1=x^{-1}$, $x_2=x$, $z_1=z$, $z_2=z^{-1}$ and $u^{(1)}=u$.
It follows from the minimal mirror identity (\ref{minimal_mirror}) and hence from Ramanujan's ${}_1\psi_1$ summation formula.
For $N=3$ it is (\ref{TSU3_dual}).
For $N\ge 4$ the Fermi multiplets $\eta_I$ singled out for $N=3$ work without any modification.

As a consistency check, we compare the boundary anomalies.
For the Neumann boundary condition (\ref{TSUN_Neu}) the boundary anomaly is
\begin{align}
&\mathcal{A}^{T[SU(N)]}_{\mathcal{N}',N'+\Gamma+\eta}
\nonumber\\*
&=\underbrace{\sum_{I=1}^{N-1}\sum_{i\neq j}(\mathbf{s}^{(I)}_i-\mathbf{s}^{(I)}_j)^2}_{\textrm{vector multiplets ($\mathcal{N}'$)}}\quad
\underbrace{-\sum_{I=1}^{N-2}\sum_{i,j}(\mathbf{s}^{(I)}_i-\mathbf{s}^{(I+1)}_j)^2}_{\textrm{bifundamental hypermultiplets ($N'$)}}\quad
\underbrace{-\sum_{i=1}^{N-1}\sum_{\alpha=1}^N(\mathbf{s}^{(N-1)}_i+\mathbf{x}_{\alpha})^2}_{\textrm{fundamental hypermultiplets ($N'$)}}
\nonumber\\
&\underbrace{+2\sum_{I=1}^{N-1}\mathbf{S}_I(\mathbf{z}_I-\mathbf{z}_{I+1})}_{\textrm{FI couplings}}
+\underbrace{\sum_{I=1}^{N-2}\sum_{j=1}^{I}(\mathbf{u}^{(I)}-\mathbf{u}^{(I+1)}+\mathbf{x}_{I+1}-\mathbf{x}_j)^2}_{\Gamma}
\nonumber\\*
&+\underbrace{\sum_{I=1}^{N}(\mathbf{S}_I-\mathbf{S}_{I-1}+\mathbf{u}^{(N-1)}+\mathbf{x}_N-\mathbf{z}_I)^2}_{\eta_I},
\end{align}
where $\mathbf{S}_I=\sum_i\mathbf{s}^{(I)}_i$ with $\mathbf{S}_0=\mathbf{S}_N=0$.
As for $N=3$, all the terms with the gauge field strengths cancel.
By (\ref{Nahm_anom}), the boundary anomaly for the Nahm pole side is
\begin{align}
\label{TSUN_Nahm_anom}
\mathcal{A}^{\wtilde{T[SU(N)]}}_{\mathcal{D},\mathrm{Nahm}^{N-2}}
&=\underbrace{\sum_{I=1}^{N-2}I(\mathbf{u}^{(I)}-\mathbf{u}^{(I+1)})^2+\sum_{a=1}^N(\mathbf{u}^{(N-1)}-\mathbf{z}_a)^2}_{\textrm{top components}}
\quad\underbrace{-2\sum_{I=1}^{N-1}I\,\mathbf{u}^{(I)}(\mathbf{x}_I-\mathbf{x}_{I+1})}_{\textrm{FI couplings}}
\nonumber\\*
&=\mathcal{A}^{T[SU(N)]}_{\mathcal{N}',N'+\Gamma+\eta}+2\mathbf{x}_N\sum_{a=1}^N\mathbf{z}_a,
\end{align}
where we have used $\sum_{\alpha}\mathbf{x}_{\alpha}=0$.
As for $N=3$, the anomalies agree up to the mixed term which vanishes for $\prod_az_a=1$.
The $I$ gauge-neutral Fermi multiplets with fugacities $q^{\frac12}\frac{u^{(I)}}{u^{(I+1)}}\frac{x_{I+1}}{x_j}$ give $I(\mathbf{u}^{(I)}-\mathbf{u}^{(I+1)})^2$,
and the Fermi multiplets $\eta_I$ give $N(\mathbf{u}^{(N-1)})^2$.
In the Neumann description, $u^{(N-1)}$ is the common flavor charge of the gauge-charged Fermi multiplets $\eta_I$,
whereas $u^{(I)}$ with $I<N-1$ act only on the gauge-neutral Fermi multiplets.

\section{\texorpdfstring{$U(N)$ ADHM theory}{U(N) ADHM theory}}
\label{sec_ADHM}
The $U(N)$ ADHM theory with $l$ flavors consists of the $U(N)$ vector multiplet, an adjoint hypermultiplet and $l$ fundamental hypermultiplets.
It describes $N$ M2-branes probing $\mathbb{C}^2\times\mathbb{C}^2/\mathbb{Z}_l$ and is realized by $N$ D3-branes on a circle
with a single NS5-brane and $l$ D5-branes.
Its mirror is the necklace quiver gauge theory $U(N)^l$
with $l$ nodes $U(N)^{(I)}$, $I=1,\cdots,l$ \cite{deBoer:1996mp,Porrati:1996xi,deBoer:1996ck}
(see also \cite{Hayashi:2022ldo} for tests of mirror symmetry using supersymmetric indices).
It has the twisted bifundamental hypermultiplets between the adjacent nodes $(I,I+1)$ with $I+l\equiv I$
and one twisted fundamental hypermultiplet at the node $I=1$,
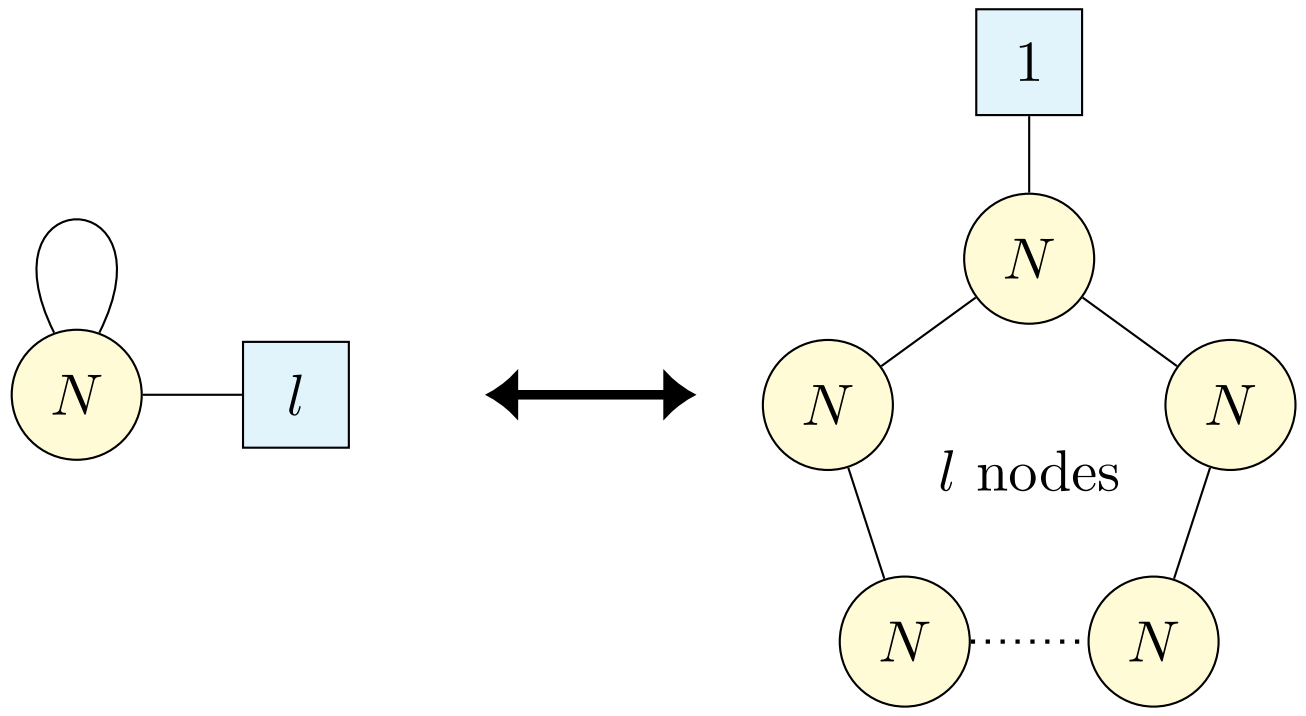
\begin{align}
\label{ADHM_quiver}
\begin{tikzpicture}[quiver]
\node[gauge](a) at (0,0){$N$}; \node[flavor](b) at (1.45,0){$l$}; \draw (a)--(b);
\draw (a.110) .. controls +(-0.5,1.0) and +(0.5,1.0) .. (a.70);
\draw[mirror] (2.7,0)--(4.1,0);
\begin{scope}[shift={(6.3,-0.5)}]
\foreach \i/\a in {1/90,2/162,3/234,4/306,5/18} \node[gauge](n\i) at (\a:1.4){$N$};
\node[flavor](f) at (0,2.7){$1$};
\draw (n1)--(n2)--(n3); \draw (n4)--(n5)--(n1); \draw[dotted,thick] (n3)--(n4); \draw (n1)--(f);
\node[font=\small] at (0,0) {$l$ nodes};
\end{scope}
\end{tikzpicture}
\end{align}
where the loop is the adjoint hypermultiplet and the dotted line stands for the omitted nodes of the necklace.
For $l=1$ the bifundamental hypermultiplet becomes the adjoint hypermultiplet and the theory is self-mirror.

\begin{figure}
\centering
\scalebox{0.85}{
\begin{tikzpicture}
\fill[yellow!40] (-3,0) rectangle (3,2.6);
\draw[thick,blue,->] (-3,-0.6) -- (-3,2.9);
\draw[thick,blue,->] (3,-0.6) -- (3,2.9);
\draw[thick,solid,black] (-3.4,0) -- (3.4,0) node[right] {\textrm{NS5}$'$};
\draw[thick,solid,black] (0.3,-0.6) -- (0.3,2.9) node[above] {\textrm{NS5}};
\draw[thick,dashed,black] (-1.5,-0.6) -- (-1.5,2.9) node[above] {\textrm{D5}};
\draw[thick,dashed,black] (1.5,-0.6) -- (1.5,2.9) node[above] {\textrm{D5}};
\node at (-2.25,1.3) {$N$ \textrm{D3}};
\fill[black] (0.3,0) circle (2pt) node[below right] {$\eta$};
\fill[black] (-1.5,0) circle (2pt) node[below left] {$\Gamma_{2,1}$};
\fill[black] (1.5,0) circle (2pt) node[below right] {$\Gamma_{1,2}$};
\node at (0,-1.2) {(a) $U(N)$ ADHM, $(\mathcal{N}',N',N'+\Gamma+\eta)$};
\end{tikzpicture}
}
\hspace{0.3cm}
\scalebox{0.85}{
\begin{tikzpicture}
\fill[yellow!40] (-3,0) rectangle (3,2.6);
\draw[thick,blue,->] (-3,-0.6) -- (-3,2.9);
\draw[thick,blue,->] (3,-0.6) -- (3,2.9);
\draw[thick,dashed,black] (-3.4,0) -- (3.4,0) node[right] {\textrm{D5}$'$};
\draw[thick,dashed,black] (0.3,-0.6) -- (0.3,2.9) node[above] {\textrm{D5}};
\draw[thick,solid,black] (-1.5,-0.6) -- (-1.5,2.9) node[above] {\textrm{NS5}};
\draw[thick,solid,black] (1.5,-0.6) -- (1.5,2.9) node[above] {\textrm{NS5}};
\node at (-2.25,1.3) {$N$ \textrm{D3}};
\node at (-0.6,-0.3) {\textrm{Nahm}};
\node at (-2.25,-0.3) {\textrm{Nahm}};
\node at (2.25,-0.3) {\textrm{Nahm}};
\node at (0,-1.2) {(b) necklace quiver $U(N)^2$, $\mathrm{Nahm}^2$};
\end{tikzpicture}
}
\\[0.5cm]
\scalebox{0.85}{
\begin{tikzpicture}
\fill[yellow!40] (-3,0) rectangle (3,2.6);
\draw[thick,blue,->] (-3,-0.6) -- (-3,2.9);
\draw[thick,blue,->] (3,-0.6) -- (3,2.9);
\draw[thick,dashed,black] (-3.4,0) -- (3.4,0) node[right] {\textrm{D5}$'$};
\draw[thick,solid,black] (0,-0.6) -- (0,2.9) node[above] {\textrm{NS5}};
\draw[thick,dashed,black] (-1.5,-0.6) -- (-1.5,2.9) node[above] {\textrm{D5}};
\draw[thick,dashed,black] (1.5,-0.6) -- (1.5,2.9) node[above] {\textrm{D5}};
\node at (-2.25,1.3) {$N$ \textrm{D3}};
\node at (-0.75,-0.3) {$\mathrm{Nahm}'$};
\node at (0,-1.2) {(c) $U(N)$ ADHM, $\mathrm{Nahm}'$};
\end{tikzpicture}
}
\hspace{0.3cm}
\scalebox{0.85}{
\begin{tikzpicture}
\fill[yellow!40] (-3,0) rectangle (3,2.6);
\draw[thick,blue,->] (-3,-0.6) -- (-3,2.9);
\draw[thick,blue,->] (3,-0.6) -- (3,2.9);
\draw[thick,solid,black] (-3.4,0) -- (3.4,0) node[right] {\textrm{NS5}$'$};
\draw[thick,dashed,black] (0,-0.6) -- (0,2.9) node[above] {\textrm{D5}};
\draw[thick,solid,black] (-1.5,-0.6) -- (-1.5,2.9) node[above] {\textrm{NS5}};
\draw[thick,solid,black] (1.5,-0.6) -- (1.5,2.9) node[above] {\textrm{NS5}};
\node at (-2.25,1.3) {$N$ \textrm{D3}};
\fill[black] (-1.5,0) circle (2pt) node[below left] {$\tilde{\eta}_{2,1}$};
\fill[black] (1.5,0) circle (2pt) node[below right] {$\tilde{\eta}_{1,2}$};
\fill[black] (0,0) circle (2pt) node[below right] {$\tilde{\Gamma}$};
\node at (0,-1.2) {(d) necklace quiver $U(N)^2$, $(\mathcal{N},N,N+\tilde{\Gamma}+\tilde{\eta})$};
\end{tikzpicture}
}
\caption{\textrm{
The brane configurations for the $U(N)$ ADHM theory with $l=2$ flavors (left) and its mirror (right).
The direction $x^6$ is periodic and the blue lines are identified.
}}
\label{fig_ADHM}
\end{figure}
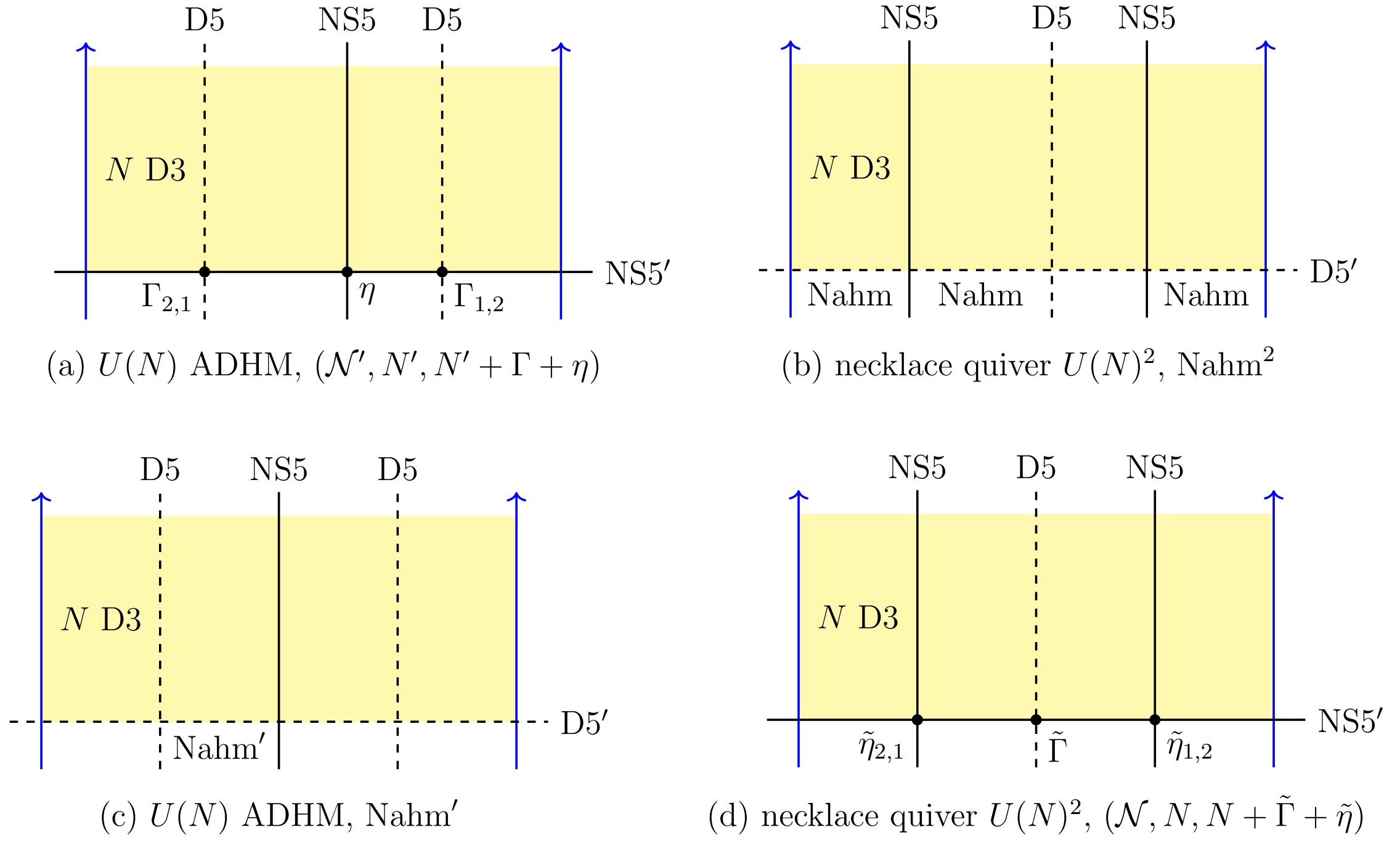

The $\mathcal{N}=(0,4)$ boundary conditions of the ADHM theories and their Abelian dualities have been studied in \cite{Okazaki:2025rjl}.
As for SQCD, the boundary is introduced by an NS5$'$- or D5$'$-brane on which the D3-branes end,
and the two choices lead to two types of dualities, as illustrated in Figure \ref{fig_ADHM} for $l=2$.
For $l=1$ they are related by mirror symmetry (section \ref{sec_ADHM_l1}), while for $l\ge 2$ they are different (section \ref{sec_ADHM_l2}).
In Figure \ref{fig_ADHM}, (a) and (d) are drawn as the S-duals of (b) and (c).

We follow the conventions in \cite{Okazaki:2025rjl}.
We denote the fugacities for the flavor symmetry $U(1)_x$ of the adjoint hypermultiplet, the topological symmetry $U(1)_z$
and the $SU(l)$ flavor symmetry of the fundamental hypermultiplets by $x$, $z$ and $y_I$ with $\prod_Iy_I=1$.
As in the previous sections, the fundamental hypermultiplets have fugacities $s_iy_I$.
Under mirror symmetry $x$ is exchanged with $z$ and $t$ with $t^{-1}$,
so that $x$ and $y_I$ are the fugacities for the topological symmetries of the necklace quiver.
On the Neumann side the segments of the NS5$'$-brane between adjacent D5-branes support the boundary symmetries $U(1)_{v_I}$, $I=1,\cdots,l$,
where the $I$-th D5-brane separates the segments $I$ and $I+1$ and $v_{l+1}\equiv v_1$.
On the Nahm pole side we denote the boundary fugacity for the $I$-th node by $u^{(I)}$, and by $u$ for a single node.

\subsection{\texorpdfstring{Self-mirror case $l=1$}{Self-mirror case l=1}}
\label{sec_ADHM_l1}
When the D3-branes end on the NS5$'$-brane, the ADHM theory obeys the Neumann boundary condition $(\mathcal{N}',N',N'+\Gamma+\eta)$.
In the S-dual configuration the $N$ D3-branes end on a single D5$'$-brane,
so that the dual boundary condition is the principal Nahm pole boundary condition.
For $N=1$ it reduces to the Dirichlet boundary condition.
For $l=1$ the twisted bifundamental hypermultiplet of the necklace quiver becomes the twisted adjoint hypermultiplet,
and the mirror is the ADHM theory itself with the twisted multiplets.
For $l=1$ we have $y_1=1$.
Since the theory is self-mirror, the other type of duality between the Nahm pole boundary condition $\mathrm{Nahm}'$ of the ADHM theory
and the Neumann boundary condition $(\mathcal{N},N,N+\tilde{\Gamma}+\tilde{\eta})$ of its mirror
follows from the duality in this section by $t\rightarrow t^{-1}$ and $x\leftrightarrow z$.
Hence the examples below cover both types of dualities.

\subsubsection{\texorpdfstring{$N=2$}{N=2}}
\label{sec_ADHM_l1N2}
Consider the $U(2)$ ADHM theory with one flavor.

\paragraph{$U(2)$ ADHM theory}
For the Neumann boundary condition $(\mathcal{N}',N',N')$ the adjoint hypermultiplet cancels the non-Abelian part of the boundary gauge anomaly
of the vector multiplet, and the fundamental hypermultiplet leaves the gauge anomaly $-\mathrm{Tr}(\mathbf{s}^2)$.
It is cancelled by the fundamental Fermi multiplet $\Gamma_{1,1}$
arising from the D3-D5 strings at the intersection of the D5-brane with the NS5$'$-brane \cite{Hanany:2018hlz}.
In addition, the junction of the NS5-brane and the NS5$'$-brane supports the cross-determinant Fermi multiplet $\eta$.
Since the quiver is circular, it is gauge-neutral (see section \ref{sec_dualrules}) and has fugacity $q^{\frac12}v_1/x$ \cite{Okazaki:2025rjl}.
The Neumann half-index is \cite{Okazaki:2025rjl}
\begin{align}
\label{ADHM2_Neu}
\mathbb{II}^{U(2)\,\textrm{ADHM-}[1]}_{\mathcal{N}',N',N'+\Gamma+\eta}
&=\frac{1}{2}\oint\prod_{i=1}^2\frac{ds_i}{2\pi is_i}
\underbrace{(q)_{\infty}^2(q^{\frac12}t^2;q)_{\infty}^2\prod_{i\neq j}\left(\frac{s_i}{s_j};q\right)_{\infty}\left(q^{\frac12}t^2\frac{s_i}{s_j};q\right)_{\infty}}_{\textrm{vector multiplet ($\mathcal{N}'$)}}
\nonumber\\*
&\times \underbrace{\frac{1}{(q^{\frac14}tx^{\pm};q)_{\infty}^2\prod_{i\neq j}(q^{\frac14}tx^{\pm}\frac{s_i}{s_j};q)_{\infty}}}_{\textrm{adjoint hypermultiplet ($N'$)}}\;
\underbrace{\prod_{i=1}^2\frac{1}{(q^{\frac14}ts_i^{\pm};q)_{\infty}}}_{\textrm{fundamental hypermultiplet ($N'$)}}
\nonumber\\*
&\times \underbrace{\prod_{i=1}^2F\left(q^{\frac12}\frac{s_i}{z};q\right)}_{\Gamma_{1,1}}\;\underbrace{F(q^{\frac12}v_1x^{-1};q)}_{\eta}.
\end{align}
The Fermi multiplets $\Gamma_{1,1}$ and $\eta$ correspond to the top components
of the twisted adjoint hypermultiplet and of the twisted fundamental hypermultiplet on the Nahm pole side (see section \ref{sec_dualrules}).

\paragraph{Mirror ADHM theory}
For the Dirichlet boundary condition $(\mathcal{D},D,D)$ of the mirror \cite{Okazaki:2025rjl}
the twisted adjoint hypermultiplet cancels the twisted vector multiplet in $k_{\rm eff}$,
and the twisted fundamental hypermultiplet gives $k_{\rm eff}=\mathbf{1}_2$.
The node is unbalanced, so that the Dirichlet half-index (\ref{Dhalf}) contains the sign $(-1)^{m_1+m_2}$,
in agreement with \cite{Okazaki:2025rjl}.
The formula (\ref{Nahm_quiver_simple}) with $t\rightarrow t^{-1}$ gives
\begin{align}
\label{ADHM2_Nahm_simple}
&\mathbb{II}^{\wtilde{U(2)\,\textrm{ADHM-}[1]}}_{\mathrm{Nahm}}
\nonumber\\*
&=\underbrace{(q^{\frac34}tz^{\pm};q)_{\infty}(q^{\frac54}t^{3}z^{\pm};q)_{\infty}(qt^{2} u^{\pm};q)_{\infty}}_{\textrm{top components}}\;
\underbrace{\frac{1}{(q)_{\infty}(q^{\frac32}t^{2};q)_{\infty}(q^{\frac12}t^{2};q)_{\infty}(qt^{4};q)_{\infty}}}_{\textrm{Slodowy slice}}
\nonumber\\
&\times\sum_{m_1\le m_2}\quad\underbrace{x^{-m_1-m_2}}_{\textrm{topological charge}}
\nonumber\\*
&\times\underbrace{\frac{(qt^{4};q)_n(q^{\frac32}t^{2};q)_n(q^{\frac34}tz^{\pm};q)_n}{(q^{\frac12}t^{2};q)_n(q;q)_n(q^{\frac54}t^{3}z^{\pm};q)_n}}_{\textrm{vector multiplet ($\mathrm{Nahm}$) and adjoint hypermultiplet ($D$)}}\;
\underbrace{(q^{\frac14}t)^{m_1+m_2}\frac{(t^{-2}u;q)_{m_1}(q^{\frac12}u;q)_{m_2}}{(q^{\frac12}u;q)_{m_1}(qt^{2} u;q)_{m_2}}}_{\textrm{fundamental hypermultiplet ($D$)}},
\end{align}
where $n=m_2-m_1$.
The top components have $J=0$ and $J=1$ for the adjoint hypermultiplet and $J=\frac12$ for the fundamental hypermultiplet.
The sign $(-1)^{m_1+m_2}$ is absorbed into the fundamental hypermultiplet.

\paragraph{Duality}
We propose that
\begin{align}
\label{ADHM2_dual}
\mathbb{II}^{U(2)\,\textrm{ADHM-}[1]}_{\mathcal{N}',N',N'+\Gamma+\eta}(t,x,z,v_1;q)
&=\mathbb{II}^{\wtilde{U(2)\,\textrm{ADHM-}[1]}}_{\mathrm{Nahm}}(t,x,z,u=v_1x;q).
\end{align}
The identification $u=v_1x$ is read off from the brane configuration,
in which the Fermi multiplet $\eta$ is dressed by the flavor fugacity $x$ of the adjoint hypermultiplet (see section \ref{sec_dualrules})
and has fugacity $q^{\frac12}ux^{-2}=q^{\frac12}v_1/x$.
The identity singles out the identification $u=v_1x$ against $u=v_1$ and $u=v_1x^2$,
and requires stripping off the split Fermi multiplets of the adjoint and of the fundamental hypermultiplets (see Appendix \ref{app_checks}).

As a consistency check, we compare the boundary anomalies.
For the Neumann boundary condition (\ref{ADHM2_Neu}) the boundary anomaly is
\begin{align}
&\mathcal{A}^{U(2)\,\textrm{ADHM-}[1]}_{\mathcal{N}',N',N'+\Gamma+\eta}
\nonumber\\*
&=\underbrace{2(\mathbf{s}_1-\mathbf{s}_2)^2}_{\textrm{vector multiplet ($\mathcal{N}'$)}}\quad
\underbrace{-\sum_{i,j=1}^2(\mathbf{s}_i-\mathbf{s}_j+\mathbf{x})^2}_{\textrm{adjoint hypermultiplet ($N'$)}}\quad
\underbrace{-\sum_{i=1}^2\mathbf{s}_i^2}_{\textrm{fundamental hypermultiplet ($N'$)}}
\nonumber\\*
&\underbrace{+2(\mathbf{s}_1+\mathbf{s}_2)\mathbf{z}}_{\textrm{FI coupling}}\quad
\underbrace{+\sum_{i=1}^2(\mathbf{s}_i-\mathbf{z})^2}_{\Gamma_{1,1}}\quad
\underbrace{+(\mathbf{v}_1-\mathbf{x})^2}_{\eta}.
\end{align}
All the terms with the gauge field strengths cancel.
By (\ref{Nahm_anom}), the boundary anomaly for the Nahm pole side is
\begin{align}
\label{ADHM2_u_anom}
\mathcal{A}^{\wtilde{U(2)\,\textrm{ADHM-}[1]}}_{\mathrm{Nahm}}&=\underbrace{2\mathbf{z}^2+\mathbf{u}^2}_{\textrm{top components}}\;\underbrace{-4\mathbf{x}\mathbf{u}}_{\textrm{FI coupling}}
\nonumber\\*
&=2\mathbf{z}^2-4\mathbf{x}^2+(\mathbf{v}_1-\mathbf{x})^2
=\mathcal{A}^{U(2)\,\textrm{ADHM-}[1]}_{\mathcal{N}',N',N'+\Gamma+\eta},
\end{align}
where the second equality holds for $\mathbf{u}=\mathbf{v}_1+\mathbf{x}$.
Hence the anomalies of the two boundary conditions agree exactly.
In the Neumann anomaly, $2\mathbf{z}^2$ comes from the Fermi multiplet $\Gamma_{1,1}$, $-4\mathbf{x}^2$ from the adjoint hypermultiplet and $(\mathbf{v}_1-\mathbf{x})^2$ from the Fermi multiplet $\eta$.
Conversely, letting $\mathbf{u}=\mathbf{v}_1+c\,\mathbf{x}$, the coefficient of $\mathbf{v}_1\mathbf{x}$ fixes $c=1$,
and then the coefficient $1-4=-3$ of $\mathbf{x}^2$ also agrees.
Hence the anomaly matching also fixes the identification $u=v_1x$.

The Fermi multiplet $\eta$ is essential for the duality.
Suppose that we drop it and take the Neumann boundary condition $(\mathcal{N}',N',N'+\Gamma)$
as a naive candidate for the dual of the Nahm pole boundary condition.
Then its half-index does not agree with the Nahm pole half-index (see Appendix \ref{app_checks}).
Being gauge-neutral, the Fermi multiplet $\eta$ is required only by the 't Hooft anomaly (\ref{ADHM2_u_anom}) for $U(1)_u$.

\subsubsection{\texorpdfstring{General $N$}{General N}}
\label{sec_ADHM_l1N}
\paragraph{$U(N)$ ADHM theory}
As for $N=2$, the fundamental hypermultiplet leaves the gauge anomaly $-\mathrm{Tr}(\mathbf{s}^2)$,
which is cancelled by the fundamental Fermi multiplet $\Gamma_{1,1}$,
and the junction of the NS5-brane and the NS5$'$-brane supports the gauge-neutral Fermi multiplet $\eta$.
The Neumann half-index is \cite{Okazaki:2025rjl}
\begin{align}
\label{ADHM1_Neu}
\mathbb{II}^{U(N)\,\textrm{ADHM-}[1]}_{\mathcal{N}',N',N'+\Gamma+\eta}
&=\frac{1}{N!}\oint\prod_{i=1}^N\frac{ds_i}{2\pi is_i}
\underbrace{(q)_{\infty}^N(q^{\frac12}t^2;q)_{\infty}^N\prod_{i\neq j}\left(\frac{s_i}{s_j};q\right)_{\infty}\left(q^{\frac12}t^2\frac{s_i}{s_j};q\right)_{\infty}}_{\textrm{vector multiplet ($\mathcal{N}'$)}}
\nonumber\\*
&\times \underbrace{\frac{1}{(q^{\frac14}tx^{\pm};q)_{\infty}^N\prod_{i\neq j}(q^{\frac14}tx^{\pm}\frac{s_i}{s_j};q)_{\infty}}}_{\textrm{adjoint hypermultiplet ($N'$)}}\;
\underbrace{\prod_{i=1}^N\frac{1}{(q^{\frac14}ts_i^{\pm};q)_{\infty}}}_{\textrm{fundamental hypermultiplet ($N'$)}}
\nonumber\\*
&\times \underbrace{\prod_{i=1}^NF\left(q^{\frac12}\frac{s_i}{z};q\right)}_{\Gamma_{1,1}}\;\underbrace{F(q^{\frac12}v_1x^{-1};q)}_{\eta}.
\end{align}
For $N=2$ it is (\ref{ADHM2_Neu}).

\paragraph{Mirror ADHM theory}
As for $N=2$, we have $k_{\rm eff}=\mathbf{1}_N$ and the sign $(-1)^{\sum_im_i}$ in the Dirichlet half-index (\ref{Dhalf}).
The formula (\ref{Nahm_quiver_simple}) with $t\rightarrow t^{-1}$ gives
\begin{align}
\label{ADHM1_Nahm_simple}
&\mathbb{II}^{\wtilde{U(N)\,\textrm{ADHM-}[1]}}_{\mathrm{Nahm}}
\nonumber\\*
&=\underbrace{\prod_{J=0}^{N-1}(q^{\frac34+\frac{J}{2}}t^{1+2J}z^{\pm};q)_{\infty}\,(q^{\frac12+\frac{N}{4}}t^{N}u^{\pm};q)_{\infty}}_{\textrm{top components}}
\nonumber\\
&\times\underbrace{\prod_{k=0}^{N-1}\frac{1}{(q^{1+\frac{k}{2}}t^{2k};q)_{\infty}(q^{\frac12+\frac{k}{2}}t^{2+2k};q)_{\infty}}}_{\textrm{Slodowy slice}}
\sum_{m_1\le\cdots\le m_N}\quad\underbrace{x^{-\sum_im_i}}_{\textrm{topological charge}}
\nonumber\\
&\times\underbrace{\prod_{i<j}\frac{(q^{\frac12+\frac{j-i}{2}}t^{2+2(j-i)};q)_{m_j-m_i}(q^{1+\frac{j-i}{2}}t^{2(j-i)};q)_{m_j-m_i}(q^{\frac14+\frac{j-i}{2}}t^{2(j-i)-1}z^{\pm};q)_{m_j-m_i}}
{(q^{\frac{j-i}{2}}t^{2(j-i)};q)_{m_j-m_i}(q^{\frac12+\frac{j-i}{2}}t^{2(j-i)-2};q)_{m_j-m_i}(q^{\frac34+\frac{j-i}{2}}t^{1+2(j-i)}z^{\pm};q)_{m_j-m_i}}}_{\textrm{vector multiplet ($\mathrm{Nahm}$) and adjoint hypermultiplet ($D$)}}
\nonumber\\*
&\times \underbrace{\prod_{i=1}^N(q^{\frac14}t)^{m_i}\frac{(q^{\frac14+\frac{\mu_i}{2}}t^{2\mu_i-1}u;q)_{m_i}}{(q^{\frac34+\frac{\mu_i}{2}}t^{1+2\mu_i}u;q)_{m_i}}}_{\textrm{fundamental hypermultiplet ($D$)}},
\end{align}
where $\mu_i=i-\frac{N+1}{2}$.
The top components have $J=0,1,\cdots,N-1$ for the adjoint hypermultiplet and $J=\frac{N-1}{2}$ for the fundamental hypermultiplet.
The factor $(q^{\frac12}t^{2})^{m_i-m_j}$ of the vector multiplet is cancelled by the adjoint hypermultiplet,
and the sign $(-1)^{\sum_im_i}$ is absorbed into the fundamental hypermultiplet.
For $N=2$ it is (\ref{ADHM2_Nahm_simple}).

\paragraph{Duality}
We propose that
\begin{align}
\label{ADHM1_dual}
\mathbb{II}^{U(N)\,\textrm{ADHM-}[1]}_{\mathcal{N}',N',N'+\Gamma+\eta}(t,x,z,v_1;q)
&=\mathbb{II}^{\wtilde{U(N)\,\textrm{ADHM-}[1]}}_{\mathrm{Nahm}}(t,x,z,u=v_1x^{N-1};q).
\end{align}
For $N=1$ it reduces to the duality in \cite{Okazaki:2025rjl}, and for $N=2$ it is (\ref{ADHM2_dual}).
The factor $x^{N-1}$ in the identification reflects the mixing of $U(1)_u$ with $U(1)_x$ through the Fermi multiplet $\eta$,
whose fugacity is $q^{\frac12}ux^{-N}=q^{\frac12}v_1/x$ as in the brane configuration.
By the self-mirror property, (\ref{ADHM1_dual}) becomes the $l=1$ case of (\ref{ADHM_rv_dual}) in section \ref{sec_ADHM_rv} with $v_1=uz^{-(N-1)}$.

As a consistency check, we compare the boundary anomalies.
For the Neumann boundary condition (\ref{ADHM1_Neu}) the boundary anomaly is
\begin{align}
&\mathcal{A}^{U(N)\,\textrm{ADHM-}[1]}_{\mathcal{N}',N',N'+\Gamma+\eta}
\nonumber\\*
&=\underbrace{\sum_{i\neq j}(\mathbf{s}_i-\mathbf{s}_j)^2}_{\textrm{vector multiplet ($\mathcal{N}'$)}}\quad
\underbrace{-\sum_{i,j=1}^N(\mathbf{s}_i-\mathbf{s}_j+\mathbf{x})^2}_{\textrm{adjoint hypermultiplet ($N'$)}}\quad
\underbrace{-\sum_{i=1}^N\mathbf{s}_i^2}_{\textrm{fundamental hypermultiplet ($N'$)}}
\nonumber\\*
&\underbrace{+2\mathbf{S}\mathbf{z}}_{\textrm{FI coupling}}\quad
\underbrace{+\sum_{i=1}^N(\mathbf{s}_i-\mathbf{z})^2}_{\Gamma_{1,1}}\quad
\underbrace{+(\mathbf{v}_1-\mathbf{x})^2}_{\eta},
\end{align}
where $\mathbf{S}=\sum_i\mathbf{s}_i$.
All the terms with the gauge field strengths cancel.
By (\ref{Nahm_anom}), the boundary anomaly for the Nahm pole side is
\begin{align}
\label{ADHM1_Nahm_anom}
\mathcal{A}^{\wtilde{U(N)\,\textrm{ADHM-}[1]}}_{\mathrm{Nahm}}&=\underbrace{N\mathbf{z}^2+\mathbf{u}^2}_{\textrm{top components}}\;\underbrace{-2N\mathbf{x}\mathbf{u}}_{\textrm{FI coupling}}
\nonumber\\*
&=N\mathbf{z}^2-N^2\mathbf{x}^2+(\mathbf{v}_1-\mathbf{x})^2
=\mathcal{A}^{U(N)\,\textrm{ADHM-}[1]}_{\mathcal{N}',N',N'+\Gamma+\eta},
\end{align}
where the second equality holds for $\mathbf{u}=\mathbf{v}_1+(N-1)\mathbf{x}$.
As for $N=2$, the anomalies agree exactly.
Conversely, letting $\mathbf{u}=\mathbf{v}_1+c\,\mathbf{x}$, the coefficient of $\mathbf{v}_1\mathbf{x}$ fixes $c=N-1$,
and then the coefficient $(N-1)^2-2N(N-1)=-(N^2-1)$ of $\mathbf{x}^2$ also agrees.
Hence the anomaly matching also fixes the identification $u=v_1x^{N-1}$.

\subsection{\texorpdfstring{The case $l\ge 2$}{The case l>=2}}
\label{sec_ADHM_l2}
For each of the two dualities we first consider the $U(2)$ ADHM theory with two flavors, whose brane configurations are shown in Figure \ref{fig_ADHM},
and then general $N$ and $l$.

\subsubsection{\texorpdfstring{$(\mathcal{N}',N',N'+\Gamma+\eta)$ and $\mathrm{Nahm}^l$}{(N',N',N'+Gamma+eta) and Nahm, ..., Nahm}}
\label{sec_ADHM_fw}
We consider the Neumann boundary condition $(\mathcal{N}',N',N'+\Gamma+\eta)$ of the ADHM theory.
As for $l=1$, in the S-dual configuration the $N$ D3-branes of each node $U(N)^{(I)}$ of the necklace quiver end on a single D5$'$-brane
(see Figure \ref{fig_ADHM} (a) and (b)).
Hence all the nodes obey the principal Nahm pole boundary conditions.

\paragraph{$U(2)$ ADHM theory with two flavors}
For the Neumann boundary condition $(\mathcal{N}',N',N')$ the fundamental hypermultiplets leave the gauge anomaly $-2\,\mathrm{Tr}(\mathbf{s}^2)$.
It is cancelled by the two fundamental Fermi multiplets $\Gamma_{1,2}$ and $\Gamma_{2,1}$ at the intersections of the two D5-branes with the NS5$'$-brane.
The Fermi multiplet $\Gamma_{I,I+1}$ carries the charges $(+1,-1)$ under $U(1)_{v_I}\times U(1)_{v_{I+1}}$.
As for $l=1$, the junction of the NS5-brane and the NS5$'$-brane supports the gauge-neutral Fermi multiplet $\eta$.
The Neumann half-index is \cite{Okazaki:2025rjl}
\begin{align}
\label{ADHM22_Neu}
&\mathbb{II}^{U(2)\,\textrm{ADHM-}[2]}_{\mathcal{N}',N',N'+\Gamma+\eta}
\nonumber\\*
&=\frac{1}{2}\oint\prod_{i=1}^2\frac{ds_i}{2\pi is_i}
\underbrace{(q)_{\infty}^2(q^{\frac12}t^2;q)_{\infty}^2\prod_{i\neq j}\left(\frac{s_i}{s_j};q\right)_{\infty}\left(q^{\frac12}t^2\frac{s_i}{s_j};q\right)_{\infty}}_{\textrm{vector multiplet ($\mathcal{N}'$)}}
\nonumber\displaybreak[1]\\
&\times\underbrace{\frac{1}{(q^{\frac14}tx^{\pm};q)_{\infty}^2\prod_{i\neq j}(q^{\frac14}tx^{\pm}\frac{s_i}{s_j};q)_{\infty}}}_{\textrm{adjoint hypermultiplet ($N'$)}}\;
\underbrace{\prod_{i=1}^2\prod_{I=1}^2\frac{1}{(q^{\frac14}ts_i^{\pm}y_I^{\pm};q)_{\infty}}}_{\textrm{fundamental hypermultiplets ($N'$)}}
\nonumber\\*
&\times \underbrace{\prod_{i=1}^2F\left(q^{\frac12}s_i\frac{v_1y_1}{v_2z};q\right)}_{\Gamma_{1,2}}\;
\underbrace{\prod_{i=1}^2F\left(q^{\frac12}s_i\frac{v_2y_2}{v_1z};q\right)}_{\Gamma_{2,1}}\;
\underbrace{F(q^{\frac12}v_1x^{-1};q)}_{\eta}.
\end{align}
The Fermi multiplets $\Gamma_{I,I+1}$ and $\eta$ correspond to the top components
of the twisted bifundamental hypermultiplets and of the twisted fundamental hypermultiplet on the Nahm pole side.

\paragraph{Mirror quiver}
The mirror is the necklace quiver $U(2)^{(1)}\times U(2)^{(2)}$
with two twisted bifundamental hypermultiplets and a twisted fundamental hypermultiplet at the node $I=1$.
For the Dirichlet boundary condition $(\mathcal{D},D,D)$ \cite{Okazaki:2025rjl} we have
$m^Tk_{\rm eff}m=\sum_i(m^{(1)}_i)^2+2(\sum_im^{(1)}_i-\sum_im^{(2)}_i)^2$, which is the case $N=l=2$ of (\ref{ADHM_keff}) below.
The node $I=1$ is unbalanced, so that the Dirichlet half-index (\ref{Dhalf}) contains the sign $(-1)^{m^{(1)}_1+m^{(1)}_2}$.
The formula (\ref{Nahm_quiver_simple}) with $t\rightarrow t^{-1}$ gives
\begin{align}
\label{ADHM22_Nahm_simple}
&\mathbb{II}^{\wtilde{U(2)^2}}_{\mathrm{Nahm}^2}
\nonumber\\*
&=\underbrace{(qt^2(u^{(1)})^{\pm};q)_{\infty}\prod_{I=1}^2\left(q^{\frac34}t\left(\frac{u^{(I)}}{u^{(I+1)}z}\right)^{\pm};q\right)_{\infty}\left(q^{\frac54}t^3\left(\frac{u^{(I)}}{u^{(I+1)}z}\right)^{\pm};q\right)_{\infty}}_{\textrm{top components}}
\nonumber\displaybreak[1]\\
&\times\underbrace{\frac{1}{[(q)_{\infty}(q^{\frac12}t^{2};q)_{\infty}(q^{\frac32}t^{2};q)_{\infty}(qt^{4};q)_{\infty}]^2}}_{\textrm{Slodowy slices}}
\sum_{m^{(1)}_1\le m^{(1)}_2}\ \sum_{m^{(2)}_1\le m^{(2)}_2}
\underbrace{x^{-\sum_im^{(1)}_i}\prod_{I=1}^2y_I^{\sum_i(m^{(I)}_i-m^{(I+1)}_i)}}_{\textrm{topological charges}}
\nonumber\displaybreak[1]\\
&\times\underbrace{\prod_{I=1}^2(q^{\frac12}t^{2})^{-n^{(I)}}\frac{(qt^{4};q)_{n^{(I)}}(q^{\frac32}t^{2};q)_{n^{(I)}}}{(q^{\frac12}t^{2};q)_{n^{(I)}}(q;q)_{n^{(I)}}}}_{\textrm{vector multiplets ($\mathrm{Nahm}$)}}\;
\underbrace{(q^{\frac14}t)^{m^{(1)}_1+m^{(1)}_2}\frac{(t^{-2}u^{(1)};q)_{m^{(1)}_1}(q^{\frac12}u^{(1)};q)_{m^{(1)}_2}}{(q^{\frac12}u^{(1)};q)_{m^{(1)}_1}(qt^{2}u^{(1)};q)_{m^{(1)}_2}}}_{\textrm{fundamental hypermultiplet ($D$)}}
\nonumber\\*
&\times\underbrace{\prod_{I=1}^2\prod_{i,j=1}^2(q^{\frac14}t)^{m^{(I)}_i-m^{(I+1)}_j}\frac{(q^{\frac14+\frac{i-j}{2}}t^{2(i-j)-1}\frac{u^{(I)}}{u^{(I+1)}z};q)_{m^{(I)}_i-m^{(I+1)}_j}}{(q^{\frac34+\frac{i-j}{2}}t^{1+2(i-j)}\frac{u^{(I)}}{u^{(I+1)}z};q)_{m^{(I)}_i-m^{(I+1)}_j}}}_{\textrm{bifundamental hypermultiplets ($D$)}},
\end{align}
where $n^{(I)}=m^{(I)}_2-m^{(I)}_1$, $u^{(3)}\equiv u^{(1)}$ and $m^{(3)}\equiv m^{(1)}$.
The top components have $J=0$ and $J=1$ for each bifundamental hypermultiplet and $J=\frac12$ for the fundamental hypermultiplet.

\paragraph{Duality}
We propose that
\begin{align}
\label{ADHM22_dual}
\mathbb{II}^{U(2)\,\textrm{ADHM-}[2]}_{\mathcal{N}',N',N'+\Gamma+\eta}(t,x,z,v_I,y_I;q)
&=\mathbb{II}^{\wtilde{U(2)^2}}_{\mathrm{Nahm}^2}(t,x,z,u^{(I)}=v_Ix,y_I;q).
\end{align}
As for $l=1$, the factor $x$ in the identification comes from the Fermi multiplet $\eta$.
The identity singles out the identification $u^{(I)}=v_Ix$ against $u^{(I)}=v_I$ and $u^{(I)}=v_Ix^2$,
and requires stripping off the split Fermi multiplets of the bifundamental and of the fundamental hypermultiplets.
As for $l=1$, it also requires the Fermi multiplet $\eta$ (see Appendix \ref{app_checks}).

As a consistency check, we compare the boundary anomalies.
For the Neumann boundary condition (\ref{ADHM22_Neu}) the boundary anomaly is
\begin{align}
&\mathcal{A}^{U(2)\,\textrm{ADHM-}[2]}_{\mathcal{N}',N',N'+\Gamma+\eta}
\nonumber\\*
&=\underbrace{2(\mathbf{s}_1-\mathbf{s}_2)^2}_{\textrm{vector multiplet ($\mathcal{N}'$)}}\quad
\underbrace{-\sum_{i,j=1}^2(\mathbf{s}_i-\mathbf{s}_j+\mathbf{x})^2}_{\textrm{adjoint hypermultiplet ($N'$)}}\quad
\underbrace{-\sum_{i=1}^2\sum_{I=1}^2(\mathbf{s}_i+\mathbf{y}_I)^2}_{\textrm{fundamental hypermultiplets ($N'$)}}\quad
\underbrace{+4\mathbf{S}\mathbf{z}}_{\textrm{FI coupling}}
\nonumber\\*
&+\underbrace{\sum_{i=1}^2(\mathbf{s}_i+\mathbf{v}_1-\mathbf{v}_2+\mathbf{y}_1-\mathbf{z})^2}_{\Gamma_{1,2}}
+\underbrace{\sum_{i=1}^2(\mathbf{s}_i+\mathbf{v}_2-\mathbf{v}_1+\mathbf{y}_2-\mathbf{z})^2}_{\Gamma_{2,1}}
\quad\underbrace{+(\mathbf{v}_1-\mathbf{x})^2}_{\eta},
\end{align}
where $\mathbf{S}=\mathbf{s}_1+\mathbf{s}_2$.
With $\mathbf{y}_1+\mathbf{y}_2=0$, all the terms with the gauge field strengths cancel.
By (\ref{Nahm_anom}), the boundary anomaly for the Nahm pole side is
\begin{align}
\label{ADHM22_Nahm_anom}
&\mathcal{A}^{\wtilde{U(2)^2}}_{\mathrm{Nahm}^2}
\nonumber\\*
&=\underbrace{2\sum_{I=1}^2(\mathbf{u}^{(I)}-\mathbf{u}^{(I+1)}-\mathbf{z})^2+(\mathbf{u}^{(1)})^2}_{\textrm{top components}}
\underbrace{-4\mathbf{u}^{(1)}\mathbf{x}+4(\mathbf{u}^{(1)}-\mathbf{u}^{(2)})(\mathbf{y}_1-\mathbf{y}_2)}_{\textrm{FI couplings}}
\nonumber\\
&=4(\mathbf{v}_1-\mathbf{v}_2)^2+4(\mathbf{v}_1-\mathbf{v}_2)(\mathbf{y}_1-\mathbf{y}_2)+4\mathbf{z}^2-4\mathbf{x}^2+(\mathbf{v}_1-\mathbf{x})^2
\nonumber\\*
&=\mathcal{A}^{U(2)\,\textrm{ADHM-}[2]}_{\mathcal{N}',N',N'+\Gamma+\eta},
\end{align}
where the second equality holds for $\mathbf{u}^{(I)}=\mathbf{v}_I+\mathbf{x}$.
As for $l=1$, the anomalies agree exactly.
Conversely, the terms with $\mathbf{u}^{(1)}-\mathbf{u}^{(2)}$ are reproduced by the Fermi multiplets $\Gamma_{I,I+1}$ only for $\mathbf{u}^{(1)}-\mathbf{u}^{(2)}=\mathbf{v}_1-\mathbf{v}_2$,
and the terms $(\mathbf{u}^{(1)})^2-4\mathbf{u}^{(1)}\mathbf{x}$ by the Fermi multiplet $\eta$ and the adjoint hypermultiplet only for $\mathbf{u}^{(1)}=\mathbf{v}_1+\mathbf{x}$.
Hence the anomaly matching also fixes the identification $u^{(I)}=v_Ix$.

\paragraph{$U(N)$ ADHM theory with $l$ flavors}
As for $N=l=2$, the fundamental hypermultiplets leave the gauge anomaly $-l\,\mathrm{Tr}(\mathbf{s}^2)$,
which is cancelled by the $l$ fundamental Fermi multiplets $\Gamma_{I,I+1}$,
and the junction of the NS5-brane and the NS5$'$-brane supports the Fermi multiplet $\eta$.
The Neumann half-index is \cite{Okazaki:2025rjl}
\begin{align}
\label{ADHM_Neu}
\mathbb{II}^{U(N)\,\textrm{ADHM-}[l]}_{\mathcal{N}',N',N'+\Gamma+\eta}
&=\frac{1}{N!}\oint\prod_{i=1}^N\frac{ds_i}{2\pi is_i}
\underbrace{(q)_{\infty}^N(q^{\frac12}t^2;q)_{\infty}^N\prod_{i\neq j}\left(\frac{s_i}{s_j};q\right)_{\infty}\left(q^{\frac12}t^2\frac{s_i}{s_j};q\right)_{\infty}}_{\textrm{vector multiplet ($\mathcal{N}'$)}}
\nonumber\\*
&\times \underbrace{\frac{1}{(q^{\frac14}tx^{\pm};q)_{\infty}^N\prod_{i\neq j}(q^{\frac14}tx^{\pm}\frac{s_i}{s_j};q)_{\infty}}}_{\textrm{adjoint hypermultiplet ($N'$)}}\;
\underbrace{\prod_{i=1}^N\prod_{I=1}^l\frac{1}{(q^{\frac14}ts_i^{\pm}y_I^{\pm};q)_{\infty}}}_{\textrm{fundamental hypermultiplets ($N'$)}}
\nonumber\\*
&\times \underbrace{\prod_{i=1}^N\prod_{I=1}^lF\left(q^{\frac12}s_i\frac{v_Iy_I}{v_{I+1}z};q\right)}_{\Gamma_{I,I+1}}\;\underbrace{F(q^{\frac12}v_1x^{-1};q)}_{\eta},
\end{align}
with $v_{l+1}\equiv v_1$.
For $l=1$ it is (\ref{ADHM1_Neu}), and for $N=l=2$ it is (\ref{ADHM22_Neu}).

\paragraph{Mirror quiver}
For the Dirichlet boundary condition $(\mathcal{D},D,D)$ of the necklace quiver \cite{Okazaki:2025rjl},
at each node the two twisted bifundamental hypermultiplets contribute $2N\mathbf{1}_N$ to $k_{\rm eff}$,
which cancels the contribution $-2N\mathbf{1}_N$ of the twisted vector multiplet,
while the node $I=1$ receives $\mathbf{1}_N$ from the twisted fundamental hypermultiplet.
We find
\begin{align}
\label{ADHM_keff_matrix}
(k_{\rm eff})_{II}&=2\mathbb{J}_N+\delta_{I1}\mathbf{1}_N, \qquad (k_{\rm eff})_{I,I\pm1}=-\mathbb{J}_N,
\\*
\label{ADHM_keff}
m^Tk_{\rm eff}m&=\sum_{i=1}^N(m^{(1)}_i)^2+\sum_{I=1}^l\Bigl(\sum_i(m^{(I)}_i-m^{(I+1)}_i)\Bigr)^2,
\end{align}
where $m^{(l+1)}\equiv m^{(1)}$.
The node $I=1$ is unbalanced, so that the Dirichlet half-index (\ref{Dhalf}) contains the sign $(-1)^{m^Tk_{\rm eff}m}=(-1)^{\sum_im^{(1)}_i}$.
The formula (\ref{Nahm_quiver_simple}) with $t\rightarrow t^{-1}$ gives
\begin{align}
\label{ADHM_Nahm_simple}
&\mathbb{II}^{\wtilde{U(N)^l}}_{\mathrm{Nahm}^l}
\nonumber\\*
&=\underbrace{(q^{\frac12+\frac{N}{4}}t^{N}(u^{(1)})^{\pm};q)_{\infty}
\prod_{I=1}^l\prod_{J=0}^{N-1}\left(q^{\frac34+\frac{J}{2}}t^{1+2J}\left(\frac{u^{(I)}}{u^{(I+1)}z}\right)^{\pm};q\right)_{\infty}}_{\textrm{top components}}
\nonumber\\
&\times\underbrace{\prod_{k=0}^{N-1}\frac{1}{[(q^{1+\frac{k}{2}}t^{2k};q)_{\infty}(q^{\frac12+\frac{k}{2}}t^{2+2k};q)_{\infty}]^l}}_{\textrm{Slodowy slices}}\;
\sum_{m^{(1)},\cdots,m^{(l)}}\underbrace{x^{-\sum_im^{(1)}_i}\prod_{I=1}^ly_I^{\sum_i(m^{(I)}_i-m^{(I+1)}_i)}}_{\textrm{topological charges}}
\nonumber\\
&\times \underbrace{\prod_{I=1}^l\prod_{1\le i<j\le N}(q^{\frac12}t^2)^{m^{(I)}_i-m^{(I)}_j}\frac{(q^{\frac12+\frac{j-i}{2}}t^{2+2(j-i)};q)_{m^{(I)}_j-m^{(I)}_i}(q^{1+\frac{j-i}{2}}t^{2(j-i)};q)_{m^{(I)}_j-m^{(I)}_i}}{(q^{\frac{j-i}{2}}t^{2(j-i)};q)_{m^{(I)}_j-m^{(I)}_i}(q^{\frac12+\frac{j-i}{2}}t^{2(j-i)-2};q)_{m^{(I)}_j-m^{(I)}_i}}}_{\textrm{vector multiplets ($\mathrm{Nahm}$)}}
\nonumber\\
&\times \underbrace{\prod_{i=1}^N(q^{\frac14}t)^{m^{(1)}_i}\frac{(q^{\frac14+\frac{\mu_i}{2}}t^{2\mu_i-1}u^{(1)};q)_{m^{(1)}_i}}{(q^{\frac34+\frac{\mu_i}{2}}t^{1+2\mu_i}u^{(1)};q)_{m^{(1)}_i}}}_{\textrm{fundamental hypermultiplet ($D$)}}
\nonumber\\*
&\times \underbrace{\prod_{I=1}^l\prod_{i,j=1}^N(q^{\frac14}t)^{m^{(I)}_i-m^{(I+1)}_j}\frac{(q^{\frac14+\frac{i-j}{2}}t^{2(i-j)-1}\frac{u^{(I)}}{u^{(I+1)}z};q)_{m^{(I)}_i-m^{(I+1)}_j}}{(q^{\frac34+\frac{i-j}{2}}t^{1+2(i-j)}\frac{u^{(I)}}{u^{(I+1)}z};q)_{m^{(I)}_i-m^{(I+1)}_j}}}_{\textrm{bifundamental hypermultiplets ($D$)}},
\end{align}
with $\mu_i=i-\frac{N+1}{2}$, $u^{(l+1)}\equiv u^{(1)}$ and $m^{(l+1)}\equiv m^{(1)}$.
The top component of the twisted fundamental hypermultiplet has $J=\frac{N-1}{2}$
and those of each twisted bifundamental hypermultiplet have $J=0,1,\cdots,N-1$ from $\frac{N-1}{2}\otimes\frac{N-1}{2}$.
The sign $(-1)^{\sum_im^{(1)}_i}$ for the unbalanced node is absorbed into the hypermultiplet factors.
For $N=l=2$ it is (\ref{ADHM22_Nahm_simple}).
For $l=1$ the product over $i,j$ is the contribution of the adjoint hypermultiplet including its diagonal components,
and (\ref{ADHM_Nahm_simple}) reduces to (\ref{ADHM1_Nahm_simple}).

\paragraph{Duality}
We propose that
\begin{align}
\label{ADHM_dual}
\mathbb{II}^{U(N)\,\textrm{ADHM-}[l]}_{\mathcal{N}',N',N'+\Gamma+\eta}(t,x,z,v_I,y_I;q)
&=\mathbb{II}^{\wtilde{U(N)^l}}_{\mathrm{Nahm}^l}(t,x,z,u^{(I)}=v_Ix^{N-1},y_I;q).
\end{align}
For $N=1$ we have $u^{(I)}=v_I$ and (\ref{ADHM_dual}) reduces to the dualities in \cite{Okazaki:2025rjl}.
For $l=1$ and $N=l=2$ it is (\ref{ADHM1_dual}) and (\ref{ADHM22_dual}).
In the identification $u^{(I)}=v_Ix^{N-1}$ the ratios $u^{(I)}/u^{(I+1)}=v_I/v_{I+1}$ appear in the fugacities of the Fermi multiplets $\Gamma_{I,I+1}$,
while the common factor $x^{N-1}$ comes from the Fermi multiplet $\eta$ as for $l=1$.
As for $N=l=2$, the identity singles out the identification $u^{(I)}=v_Ix^{N-1}$ against $u^{(I)}=v_Ix^{N-2}$ and $u^{(I)}=v_Ix^{N}$,
and also against those with powers of $y_I$ mixed into $u^{(I)}$ (see Appendix \ref{app_checks}).

As a consistency check, we compare the boundary anomalies.
For the Neumann boundary condition (\ref{ADHM_Neu}) the boundary anomaly is \cite{Okazaki:2025rjl}
\begin{align}
&\mathcal{A}^{U(N)\,\textrm{ADHM-}[l]}_{\mathcal{N}',N',N'+\Gamma+\eta}
\nonumber\\*
&=\underbrace{\sum_{i\neq j}(\mathbf{s}_i-\mathbf{s}_j)^2}_{\textrm{vector multiplet ($\mathcal{N}'$)}}\quad
\underbrace{-\sum_{i,j=1}^N(\mathbf{s}_i-\mathbf{s}_j+\mathbf{x})^2}_{\textrm{adjoint hypermultiplet ($N'$)}}\quad
\underbrace{-\sum_{i=1}^N\sum_{I=1}^l(\mathbf{s}_i+\mathbf{y}_I)^2}_{\textrm{fundamental hypermultiplets ($N'$)}}\quad
\underbrace{+2l\mathbf{S}\mathbf{z}}_{\textrm{FI coupling}}
\nonumber\\*
&+\underbrace{\sum_{I=1}^l\sum_{i=1}^N(\mathbf{s}_i+\mathbf{v}_I-\mathbf{v}_{I+1}+\mathbf{y}_I-\mathbf{z})^2}_{\Gamma_{I,I+1}}
\quad\underbrace{+(\mathbf{v}_1-\mathbf{x})^2}_{\eta},
\end{align}
where $\mathbf{S}=\sum_i\mathbf{s}_i$ and $\mathbf{v}_{l+1}\equiv\mathbf{v}_1$.
With $\sum_I\mathbf{y}_I=0$, all the terms with the gauge field strengths cancel.\footnote{
In the anomaly of \cite{Okazaki:2025rjl}, the Fermi multiplet $\Gamma_{I,I+1}$ is assigned the opposite charge under $U(1)_z$, and the FI coupling has the opposite sign.
The total anomaly is the same, since it depends on $\mathbf{z}$ only through $lN\mathbf{z}^2$.
}
By (\ref{Nahm_anom}), the boundary anomaly for the Nahm pole side is
\begin{align}
\label{ADHM_Nahm_anom}
\mathcal{A}^{\wtilde{U(N)^l}}_{\mathrm{Nahm}^l}
&=\underbrace{N\sum_{I=1}^l(\mathbf{u}^{(I)}-\mathbf{u}^{(I+1)}-\mathbf{z})^2+(\mathbf{u}^{(1)})^2}_{\textrm{top components}}
\quad\underbrace{-2N\mathbf{u}^{(1)}\mathbf{x}+2N\sum_{I=1}^l(\mathbf{u}^{(I)}-\mathbf{u}^{(I+1)})\mathbf{y}_I}_{\textrm{FI couplings}}
\nonumber\\
&=N\sum_{I=1}^l(\mathbf{v}_I-\mathbf{v}_{I+1})^2+2N\sum_{I=1}^l(\mathbf{v}_I-\mathbf{v}_{I+1})\mathbf{y}_I+lN\mathbf{z}^2-N^2\mathbf{x}^2+(\mathbf{v}_1-\mathbf{x})^2
\nonumber\\*
&=\mathcal{A}^{U(N)\,\textrm{ADHM-}[l]}_{\mathcal{N}',N',N'+\Gamma+\eta},
\end{align}
where the second equality holds for $\mathbf{u}^{(I)}=\mathbf{v}_I+(N-1)\mathbf{x}$.
As for $l=1$, the anomalies agree exactly.
Conversely, the terms with $\mathbf{u}^{(I)}-\mathbf{u}^{(I+1)}$ agree only for $\mathbf{u}^{(I)}-\mathbf{u}^{(I+1)}=\mathbf{v}_I-\mathbf{v}_{I+1}$,
and the terms with $\mathbf{u}^{(1)}$ and $\mathbf{x}$ agree only for $\mathbf{u}^{(1)}=\mathbf{v}_1+(N-1)\mathbf{x}$ as for $l=1$.
Hence the anomaly matching also fixes the identification $u^{(I)}=v_Ix^{N-1}$.

\subsubsection{\texorpdfstring{$\mathrm{Nahm}'$ and $(\mathcal{N},N,N+\tilde{\Gamma}+\tilde{\eta})$}{Nahm' and (N,N,N+Gamma+eta)}}
\label{sec_ADHM_rv}
When the $N$ D3-branes end on a single D5$'$-brane,
the ADHM theory obeys the Nahm pole boundary condition for the vector multiplet
and the Dirichlet boundary condition $D'$ for the adjoint and fundamental hypermultiplets.
Under S-duality the D5$'$-brane maps to the NS5$'$-brane,
and all the nodes of the necklace quiver obey the Neumann boundary condition $(\mathcal{N},N,N)$ (see Figure \ref{fig_ADHM} (c) and (d)).
As for SQCD in section \ref{sec_SQCD_rv}, we apply the formula (\ref{Nahm_single_simple}) directly.

\paragraph{$U(2)$ ADHM theory with two flavors}
The adjoint hypermultiplet cancels the contribution of the vector multiplet to $k_{\rm eff}$
for the Dirichlet boundary condition $(\mathcal{D}',D',D')$ \cite{Okazaki:2025rjl},
so that $k_{\rm eff}=2\,\mathbf{1}_2$ and the sign $(-1)^{2\sum_im_i^2}$ in the Dirichlet half-index (\ref{Dhalf}) is trivial.
With $n=m_2-m_1$, the formula (\ref{Nahm_single_simple}) with $J=0,1$ for the adjoint hypermultiplet
and $J=\frac12$ for the two fundamental hypermultiplets in (\ref{spins_UN}) gives
\begin{align}
\label{ADHM2_rv_simple}
&\mathbb{II}^{U(2)\,\textrm{ADHM-}[2]}_{\mathrm{Nahm}'}
\nonumber\\*
&=\underbrace{(q^{\frac34}t^{-1}x^{\pm};q)_{\infty}(q^{\frac54}t^{-3}x^{\pm};q)_{\infty}\prod_{I=1}^2(qt^{-2}(uy_I)^{\pm};q)_{\infty}}_{\textrm{top components}}
\nonumber\\
&\times\underbrace{\frac{1}{(q)_{\infty}(q^{\frac32}t^{-2};q)_{\infty}(q^{\frac12}t^{-2};q)_{\infty}(qt^{-4};q)_{\infty}}}_{\textrm{Slodowy slice}}
\nonumber\\
&\times\sum_{m_1\le m_2}\quad\underbrace{z^{-2(m_1+m_2)}}_{\textrm{topological charge}}\;
\underbrace{\frac{(qt^{-4};q)_n(q^{\frac32}t^{-2};q)_n(q^{\frac34}t^{-1}x^{\pm};q)_n}{(q^{\frac12}t^{-2};q)_n(q;q)_n(q^{\frac54}t^{-3}x^{\pm};q)_n}}_{\textrm{vector multiplet ($\mathrm{Nahm}'$) and adjoint hypermultiplet ($D'$)}}
\nonumber\\*
&\times\underbrace{\prod_{I=1}^2(q^{\frac14}t^{-1})^{m_1+m_2}\frac{(t^{2}uy_I;q)_{m_1}(q^{\frac12}uy_I;q)_{m_2}}{(q^{\frac12}uy_I;q)_{m_1}(qt^{-2} uy_I;q)_{m_2}}}_{\textrm{fundamental hypermultiplets ($D'$)}}.
\end{align}
The topological fugacity appears as $z^{-2(m_1+m_2)}$.
This reflects that the monopole of unit flux is mapped to the closed loop of the two twisted bifundamental hypermultiplets of the necklace quiver,
each of which carries charge $-1$ under $U(1)_z$.
It is the $N=2$ and $l=2$ case of (\ref{ADHM_simple}) below.

\paragraph{Mirror quiver}
We denote the gauge fugacities for the two nodes of the necklace quiver by $s^{(1)}_i$ and $s^{(2)}_i$.
For the Neumann boundary condition $(\mathcal{N},N,N)$ the twisted bifundamental hypermultiplets cancel the non-Abelian part of the boundary gauge anomaly
of the twisted vector multiplets.
The remaining part is $-2(\sum_i\mathbf{s}^{(1)}_i-\sum_i\mathbf{s}^{(2)}_i)^2-\sum_{i=1}^2(\mathbf{s}^{(1)}_i)^2$,
where the last term comes from the twisted fundamental hypermultiplet (see (\ref{ADHM2_rv_Neu_anom}) for the full anomaly).
It is cancelled by the two cross-determinant Fermi multiplets $\tilde{\eta}_{1,2}$ and $\tilde{\eta}_{2,1}$
at the junctions of the two NS5-branes with the NS5$'$-brane in Figure \ref{fig_ADHM} (d)
and the fundamental Fermi multiplet $\tilde{\Gamma}$ at the intersection of the D5-brane with the NS5$'$-brane, as in \cite{Hanany:2018hlz,Okazaki:2025rjl}.
The Neumann half-index of the mirror quiver reads
\begin{align}
\label{ADHM2_rv_Neu}
&\mathbb{II}^{\wtilde{U(2)^2}}_{\mathcal{N},N,N+\tilde{\Gamma}+\tilde{\eta}}
\nonumber\\*
&=\frac{1}{4}\oint\prod_{I=1}^2\prod_{i=1}^2\frac{ds^{(I)}_i}{2\pi is^{(I)}_i}
\underbrace{[(q)_{\infty}(q^{\frac12}t^{-2};q)_{\infty}]^{4}\prod_{I=1}^2\prod_{i\neq j}\left(\frac{s^{(I)}_i}{s^{(I)}_j};q\right)_{\infty}\left(q^{\frac12}t^{-2}\frac{s^{(I)}_i}{s^{(I)}_j};q\right)_{\infty}}_{\textrm{vector multiplets ($\mathcal{N}$)}}
\nonumber\\
&\times \underbrace{\prod_{i,j=1}^2\frac{1}{(q^{\frac14}t^{-1}(s^{(1)}_i/s^{(2)}_jz)^{\pm};q)_{\infty}(q^{\frac14}t^{-1}(s^{(2)}_i/s^{(1)}_jz)^{\pm};q)_{\infty}}}_{\textrm{bifundamental hypermultiplets ($N$)}}\;
\underbrace{\prod_{i=1}^2\frac{1}{(q^{\frac14}t^{-1}(s^{(1)}_i)^{\pm};q)_{\infty}}}_{\textrm{fundamental hypermultiplet ($N$)}}
\nonumber\\*
&\times \underbrace{\prod_{i=1}^2F\left(q^{\frac12}\frac{s^{(1)}_i}{x};q\right)}_{\tilde{\Gamma}}\;
\underbrace{F\left(q^{\frac12}\frac{s^{(1)}_1s^{(1)}_2}{s^{(2)}_1s^{(2)}_2}\,y_1\frac{u}{z^2};q\right)}_{\tilde{\eta}_{1,2}}\;
\underbrace{F\left(q^{\frac12}\frac{s^{(2)}_1s^{(2)}_2}{s^{(1)}_1s^{(1)}_2}\,y_2\frac{u}{z^2};q\right)}_{\tilde{\eta}_{2,1}}.
\end{align}
The Fermi multiplets $\tilde{\Gamma}$ and $\tilde{\eta}_{I,I+1}$ correspond to the two top components of the adjoint hypermultiplet
and to the top components of the two fundamental hypermultiplets of the ADHM theory at the Nahm pole respectively (see section \ref{sec_dualrules}).
The factor $z^{-2}$ in the Fermi multiplets $\tilde{\eta}_{I,I+1}$ comes from the flavor fugacity $z$ of the twisted bifundamental hypermultiplets as in section \ref{sec_dualrules}.

\paragraph{Duality}
We propose that
\begin{align}
\label{ADHM2_rv_dual}
\mathbb{II}^{U(2)\,\textrm{ADHM-}[2]}_{\mathrm{Nahm}'}
&=\mathbb{II}^{\wtilde{U(2)^2}}_{\mathcal{N},N,N+\tilde{\Gamma}+\tilde{\eta}}.
\end{align}
The identity singles out $z^{-2}$ in the Fermi multiplets $\tilde{\eta}_{I,I+1}$ against $z^{-1}$ and $z^{-3}$,
and the topological charge $z^{-2(m_1+m_2)}$ in (\ref{ADHM2_rv_simple}) against $z^{-(m_1+m_2)}$ (see Appendix \ref{app_checks}).

As a consistency check, we compare the boundary anomalies.
For the Neumann boundary condition (\ref{ADHM2_rv_Neu}) of the mirror quiver the boundary anomaly is
\begin{align}
\label{ADHM2_rv_Neu_anom}
&\mathcal{A}^{\wtilde{U(2)^2}}_{\mathcal{N},N,N+\tilde{\Gamma}+\tilde{\eta}}
\nonumber\\*
&=\underbrace{\sum_{I=1}^2\sum_{i\neq j}(\mathbf{s}^{(I)}_i-\mathbf{s}^{(I)}_j)^2}_{\textrm{vector multiplets ($\mathcal{N}$)}}\quad
\underbrace{-\sum_{I=1}^2\sum_{i,j=1}^2(\mathbf{s}^{(I)}_i-\mathbf{s}^{(I+1)}_j-\mathbf{z})^2}_{\textrm{bifundamental hypermultiplets ($N$)}}\quad
\underbrace{-\sum_{i=1}^2(\mathbf{s}^{(1)}_i)^2}_{\textrm{fundamental hypermultiplet ($N$)}}
\nonumber\\
&\underbrace{+2\mathbf{S}_1\mathbf{x}-2(\mathbf{S}_1-\mathbf{S}_2)(\mathbf{y}_1-\mathbf{y}_2)}_{\textrm{FI couplings}}
\quad\underbrace{+\sum_{i=1}^2(\mathbf{s}^{(1)}_i-\mathbf{x})^2}_{\tilde{\Gamma}}
\nonumber\\*
&+\underbrace{(\mathbf{S}_1-\mathbf{S}_2+\mathbf{y}_1+\mathbf{u}-2\mathbf{z})^2}_{\tilde{\eta}_{1,2}}
+\underbrace{(\mathbf{S}_2-\mathbf{S}_1+\mathbf{y}_2+\mathbf{u}-2\mathbf{z})^2}_{\tilde{\eta}_{2,1}},
\end{align}
where $\mathbf{S}_I=\sum_i\mathbf{s}^{(I)}_i$ and $\mathbf{s}^{(3)}\equiv\mathbf{s}^{(1)}$.
All the terms with the gauge field strengths cancel.
By (\ref{Nahm_anom}), the boundary anomaly for the Nahm pole side is
\begin{align}
\mathcal{A}^{U(2)\,\textrm{ADHM-}[2]}_{\mathrm{Nahm}'}&=\underbrace{\sum_{I=1}^2(\mathbf{u}+\mathbf{y}_I)^2+2\mathbf{x}^2}_{\textrm{top components}}\underbrace{-8\mathbf{u}\mathbf{z}}_{\textrm{FI coupling}}
\nonumber\\*
&=\sum_{I=1}^2(\mathbf{y}_I+\mathbf{u}-2\mathbf{z})^2+2\mathbf{x}^2-8\mathbf{z}^2
=\mathcal{A}^{\wtilde{U(2)^2}}_{\mathcal{N},N,N+\tilde{\Gamma}+\tilde{\eta}},
\end{align}
where we have used $\mathbf{y}_1+\mathbf{y}_2=0$.
Hence the anomalies agree exactly.
In the Neumann anomaly, the three terms come from the Fermi multiplets $\tilde{\eta}_{I,I+1}$, $\tilde{\Gamma}$ and the bifundamental hypermultiplets.
The factor $z^{-2}$ in the Fermi multiplets $\tilde{\eta}_{I,I+1}$ is fixed by both the $\mathbf{u}\mathbf{z}$ and $\mathbf{z}^2$ terms.

\paragraph{$U(N)$ ADHM theory with $l$ flavors}
The adjoint hypermultiplet cancels the contribution of the vector multiplet to $k_{\rm eff}$
for the Dirichlet boundary condition $(\mathcal{D}',D',D')$ \cite{Okazaki:2025rjl},
so that $k_{\rm eff}=l\,\mathbf{1}_N$.
Hence the sign $(-1)^{l\sum_im_i^2}$ appears in the Dirichlet half-index (\ref{Dhalf}).
In the formula (\ref{Nahm_single_simple}),
the components $(i,j)$ and $(j,i)$ of the adjoint hypermultiplet with $i<j$ have the weights $\pm(e_i-e_j)$ and $w_{\lambda}(\mu)=\mp(j-i)$ (see section \ref{sec_Higgsing}),
while the Cartan components contribute only to the prefactor.
Combining the components $(i,j)$ and $(j,i)$, we find
\begin{align}
\label{ADHM_simple}
&\mathbb{II}^{U(N)\,\textrm{ADHM-}[l]}_{\mathrm{Nahm}'}
\nonumber\\*
&=\underbrace{\prod_{J=0}^{N-1}(q^{\frac34+\frac{J}{2}}t^{-1-2J}x^{\pm};q)_{\infty}\prod_{I=1}^{l}(q^{\frac12+\frac{N}{4}}t^{-N}(uy_I)^{\pm};q)_{\infty}}_{\textrm{top components}}
\nonumber\\
&\times\underbrace{\prod_{k=0}^{N-1}\frac{1}{(q^{1+\frac{k}{2}}t^{-2k};q)_{\infty}(q^{\frac12+\frac{k}{2}}t^{-2-2k};q)_{\infty}}}_{\textrm{Slodowy slice}}\;
\sum_{m_1\le\cdots\le m_N}\quad\underbrace{z^{-l\sum_im_i}}_{\textrm{topological charge}}
\nonumber\\
&\times\underbrace{\prod_{i<j}\frac{(q^{\frac12+\frac{j-i}{2}}t^{-2-2(j-i)};q)_{m_j-m_i}(q^{1+\frac{j-i}{2}}t^{-2(j-i)};q)_{m_j-m_i}(q^{\frac14+\frac{j-i}{2}}t^{1-2(j-i)}x^{\pm};q)_{m_j-m_i}}
{(q^{\frac{j-i}{2}}t^{-2(j-i)};q)_{m_j-m_i}(q^{\frac12+\frac{j-i}{2}}t^{2-2(j-i)};q)_{m_j-m_i}(q^{\frac34+\frac{j-i}{2}}t^{-1-2(j-i)}x^{\pm};q)_{m_j-m_i}}}_{\textrm{vector multiplet ($\mathrm{Nahm}'$) and adjoint hypermultiplet ($D'$)}}
\nonumber\\*
&\times \underbrace{\prod_{I=1}^{l}\prod_{i=1}^N(q^{\frac14}t^{-1})^{m_i}\frac{(q^{\frac14+\frac{\mu_i}{2}}t^{1-2\mu_i}uy_I;q)_{m_i}}{(q^{\frac34+\frac{\mu_i}{2}}t^{-1-2\mu_i}uy_I;q)_{m_i}}}_{\textrm{fundamental hypermultiplets ($D'$)}}.
\end{align}
The top components have $J=0,1,\cdots,N-1$ for the adjoint hypermultiplet and $J=\frac{N-1}{2}$ for the fundamental hypermultiplets.
As for $N=2$ and $l=2$, the topological fugacity appears as $z^{-l\sum_im_i}$.
The factor $(q^{\frac12}t^{-2})^{m_i-m_j}$ of the Macdonald weight is cancelled by the adjoint hypermultiplet,
and the sign $(-1)^{l\sum_im_i^2}$ is absorbed into the fundamental hypermultiplets.
The invariance under $x\rightarrow x^{-1}$ is manifest.
At $x=q^{\frac14}t$ the adjoint hypermultiplet cancels the vector multiplet both in the prefactor and in the weights for $i<j$.
For $N=2$ and $l=2$ it is (\ref{ADHM2_rv_simple}).
For $l=1$ it coincides with (\ref{ADHM1_Nahm_simple}) under $x\leftrightarrow z$ and $t\rightarrow t^{-1}$.

\paragraph{Mirror quiver}
For the Neumann boundary condition $(\mathcal{N},N,N)$ of the necklace quiver the non-Abelian part of the boundary gauge anomaly cancels as for $N=l=2$,
and the remaining part is $-\sum_{I=1}^l(\sum_i\mathbf{s}^{(I)}_i-\sum_i\mathbf{s}^{(I+1)}_i)^2-\sum_{i=1}^N(\mathbf{s}^{(1)}_i)^2$
with $\mathbf{s}^{(l+1)}\equiv\mathbf{s}^{(1)}$ (see (\ref{ADHM_rv_Neu_anom}) for the full anomaly).
It is cancelled by the $l$ cross-determinant Fermi multiplets $\tilde{\eta}_{I,I+1}$
and the fundamental Fermi multiplet $\tilde{\Gamma}$,
which correspond to the top components of the $l$ fundamental hypermultiplets
and to the $N$ top components of the adjoint hypermultiplet of the ADHM theory at the Nahm pole.
The Neumann half-index of the mirror quiver reads
\begin{align}
\label{ADHM_rv_Neu}
&\mathbb{II}^{\wtilde{U(N)^l}}_{\mathcal{N},N,N+\tilde{\Gamma}+\tilde{\eta}}
\nonumber\\*
&=\prod_{I=1}^l\left[\frac{1}{N!}\oint\prod_{i=1}^N\frac{ds^{(I)}_i}{2\pi is^{(I)}_i}\right]
\underbrace{[(q)_{\infty}(q^{\frac12}t^{-2};q)_{\infty}]^{lN}\prod_{I=1}^l\prod_{i\neq j}\left(\frac{s^{(I)}_i}{s^{(I)}_j};q\right)_{\infty}\left(q^{\frac12}t^{-2}\frac{s^{(I)}_i}{s^{(I)}_j};q\right)_{\infty}}_{\textrm{vector multiplets ($\mathcal{N}$)}}
\nonumber\\
&\times \underbrace{\prod_{I=1}^l\prod_{i,j=1}^N\frac{1}{(q^{\frac14}t^{-1}(s^{(I)}_i/s^{(I+1)}_jz)^{\pm};q)_{\infty}}}_{\textrm{bifundamental hypermultiplets ($N$)}}\;
\underbrace{\prod_{i=1}^N\frac{1}{(q^{\frac14}t^{-1}(s^{(1)}_i)^{\pm};q)_{\infty}}}_{\textrm{fundamental hypermultiplet ($N$)}}
\nonumber\\*
&\times \underbrace{\prod_{i=1}^NF\left(q^{\frac12}\frac{s^{(1)}_i}{x};q\right)}_{\tilde{\Gamma}}
\underbrace{\prod_{I=1}^lF\left(q^{\frac12}\frac{\prod_is^{(I)}_i}{\prod_is^{(I+1)}_i}\,y_I\frac{u}{z^N};q\right)}_{\tilde{\eta}_{I,I+1}},
\end{align}
with $s^{(l+1)}_i\equiv s^{(1)}_i$.
For $N=2$ and $l=2$ it is (\ref{ADHM2_rv_Neu}).
For $l=1$, the Fermi multiplet $\tilde{\eta}_{1,1}=F(q^{\frac12}uz^{-N};q)$ is gauge-neutral.

\paragraph{Duality}
We propose that
\begin{align}
\label{ADHM_rv_dual}
\mathbb{II}^{U(N)\,\textrm{ADHM-}[l]}_{\mathrm{Nahm}'}
&=\mathbb{II}^{\wtilde{U(N)^l}}_{\mathcal{N},N,N+\tilde{\Gamma}+\tilde{\eta}}.
\end{align}
For $l=1$ it is equivalent to (\ref{ADHM1_dual}) as shown in section \ref{sec_ADHM_l1N}.
For $N=1$ it reduces to the duality for the Abelian ADHM theory in \cite{Okazaki:2025rjl} up to the conventions of the fugacities,
and for $N=2$ and $l=2$ it is (\ref{ADHM2_rv_dual}).
As for $N=2$ and $l=2$, the identity singles out $z^{-N}$ in the Fermi multiplets $\tilde{\eta}_{I,I+1}$ against $z^{-N\pm1}$
and $z^{-l\sum_im_i}$ in (\ref{ADHM_simple}) against $z^{-\sum_im_i}$.
It also requires the factor $(q^{\frac12}t^{-2})^{l\sum_i\mu_im_i}$ in the expanded form (\ref{Nahm_single})
and stripping off the split Fermi multiplets of the adjoint and of the fundamental hypermultiplets.
For $l=3$ it moreover singles out $y_I$ in the Fermi multiplets $\tilde{\eta}_{I,I+1}$ against $y_I^{-1}$ and requires the sign $(-1)^{l\sum_im_i^2}$,
while for $l=2$ the half-indices are unchanged under these modifications (see Appendix \ref{app_checks}).

As a consistency check, we compare the boundary anomalies.
For the Neumann boundary condition (\ref{ADHM_rv_Neu}) of the mirror quiver the boundary anomaly is
\begin{align}
\label{ADHM_rv_Neu_anom}
&\mathcal{A}^{\wtilde{U(N)^l}}_{\mathcal{N},N,N+\tilde{\Gamma}+\tilde{\eta}}
\nonumber\\*
&=\underbrace{\sum_{I=1}^l\sum_{i\neq j}(\mathbf{s}^{(I)}_i-\mathbf{s}^{(I)}_j)^2}_{\textrm{vector multiplets ($\mathcal{N}$)}}\quad
\underbrace{-\sum_{I=1}^l\sum_{i,j=1}^N(\mathbf{s}^{(I)}_i-\mathbf{s}^{(I+1)}_j-\mathbf{z})^2}_{\textrm{bifundamental hypermultiplets ($N$)}}\quad
\underbrace{-\sum_{i=1}^N(\mathbf{s}^{(1)}_i)^2}_{\textrm{fundamental hypermultiplet ($N$)}}
\nonumber\\
&\underbrace{+2\mathbf{S}_1\mathbf{x}-2\sum_{I=1}^l(\mathbf{S}_I-\mathbf{S}_{I+1})\mathbf{y}_I}_{\textrm{FI couplings}}
\quad\underbrace{+\sum_{i=1}^N(\mathbf{s}^{(1)}_i-\mathbf{x})^2}_{\tilde{\Gamma}}
\nonumber\\*
&+\underbrace{\sum_{I=1}^l(\mathbf{S}_I-\mathbf{S}_{I+1}+\mathbf{y}_I+\mathbf{u}-N\mathbf{z})^2}_{\tilde{\eta}_{I,I+1}},
\end{align}
where $\mathbf{S}_I=\sum_i\mathbf{s}^{(I)}_i$ and $\mathbf{s}^{(l+1)}\equiv\mathbf{s}^{(1)}$.
All the terms with the gauge field strengths cancel.
In particular, the FI couplings of the necklace quiver cancel the mixed terms of the Fermi multiplets $\tilde{\Gamma}$ and $\tilde{\eta}_{I,I+1}$ between the gauge group and the symmetries of $x$ and $y_I$.\footnote{
In \cite{Okazaki:2025rjl} the Fermi multiplets $\tilde{\Gamma}$ and $\tilde{\eta}_{I,I+1}$ have the opposite charges under $U(1)_x$ and $U(1)_{y_I}$ relative to their gauge charges,
and the FI couplings have the opposite signs.
The half-index (\ref{ADHM_rv_Neu}) does not change under $x\rightarrow x^{-1}$, since the dual half-index depends on $x$ only through $x^{\pm}$.
}
By (\ref{Nahm_anom}), the boundary anomaly for the Nahm pole side is
\begin{align}
\mathcal{A}^{U(N)\,\textrm{ADHM-}[l]}_{\mathrm{Nahm}'}&=\underbrace{\sum_{I=1}^l(\mathbf{u}+\mathbf{y}_I)^2+N\mathbf{x}^2}_{\textrm{top components}}\underbrace{-2lN\mathbf{u}\mathbf{z}}_{\textrm{FI coupling}}
\nonumber\\*
&=\sum_{I=1}^l(\mathbf{y}_I+\mathbf{u}-N\mathbf{z})^2+N\mathbf{x}^2-lN^2\mathbf{z}^2
=\mathcal{A}^{\wtilde{U(N)^l}}_{\mathcal{N},N,N+\tilde{\Gamma}+\tilde{\eta}},
\end{align}
where we have used $\sum_I\mathbf{y}_I=0$.
As for $N=l=2$, the anomalies agree exactly.
The coefficient $N$ of $\mathbf{z}$ in the Fermi multiplets $\tilde{\eta}_{I,I+1}$ is fixed by both the $\mathbf{u}\mathbf{z}$ and $\mathbf{z}^2$ terms,
in the same way as the factor $x^{N-1}$ in (\ref{ADHM_dual}).

\section{\texorpdfstring{$USp(2N)$ ADHM theory}{USp(2N) ADHM theory}}
\label{sec_USpADHM}
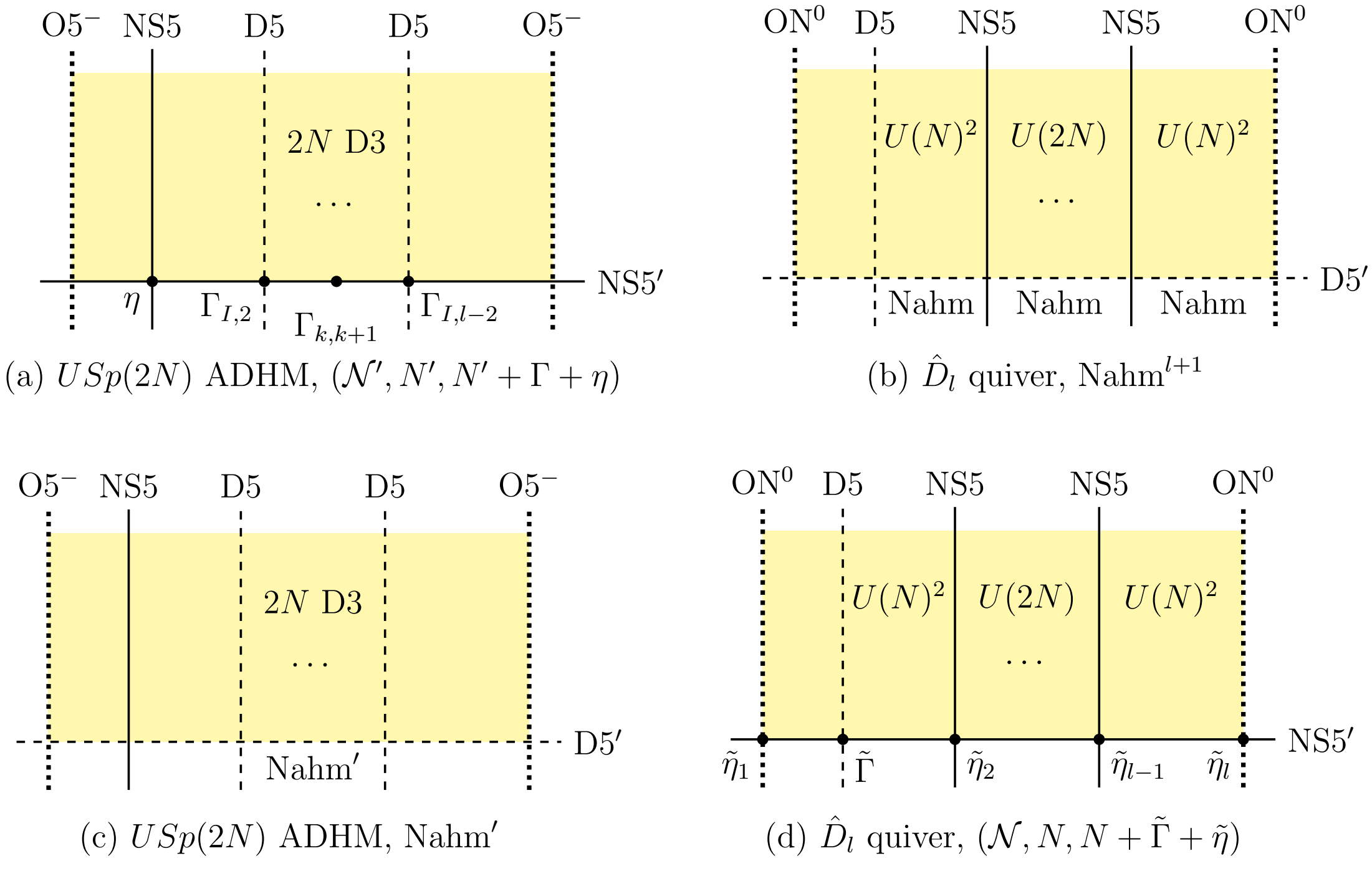
\begin{figure}
\centering
\scalebox{0.85}{
\begin{tikzpicture}
\fill[yellow!40] (-3,0) rectangle (3,2.6);
\draw[line width=1.6pt,dotted] (-3,-0.6) -- (-3,2.9) node[above] {\textrm{O5}$^-$};
\draw[line width=1.6pt,dotted] (3,-0.6) -- (3,2.9) node[above] {\textrm{O5}$^-$};
\draw[thick,solid,black] (-3.4,0) -- (3.4,0) node[right] {\textrm{NS5}$'$};
\draw[thick,solid,black] (-2.0,-0.6) -- (-2.0,2.9) node[above] {\textrm{NS5}};
\draw[thick,dashed,black] (-0.6,-0.6) -- (-0.6,2.9) node[above] {\textrm{D5}};
\draw[thick,dashed,black] (1.2,-0.6) -- (1.2,2.9) node[above] {\textrm{D5}};
\node at (0.3,1.75) {$2N$ \textrm{D3}};
\node at (0.3,0.95) {$\cdots$};
\fill[black] (-2.0,0) circle (2pt) node[below left] {$\eta$};
\fill[black] (-0.6,0) circle (2pt) node[below left] {$\Gamma_{I,2}$};
\fill[black] (0.3,0) circle (2pt) node[below,yshift=-7pt] {$\Gamma_{k,k+1}$};
\fill[black] (1.2,0) circle (2pt) node[below right] {$\Gamma_{I,l-2}$};
\node at (0,-1.2) {(a) $USp(2N)$ ADHM, $(\mathcal{N}',N',N'+\Gamma+\eta)$};
\end{tikzpicture}
}
\hspace{0.3cm}
\scalebox{0.85}{
\begin{tikzpicture}
\fill[yellow!40] (-3,0) rectangle (3,2.6);
\draw[line width=1.6pt,dotted] (-3,-0.6) -- (-3,2.9) node[above] {\textrm{ON}$^0$};
\draw[line width=1.6pt,dotted] (3,-0.6) -- (3,2.9) node[above] {\textrm{ON}$^0$};
\draw[thick,dashed,black] (-3.4,0) -- (3.4,0) node[right] {\textrm{D5}$'$};
\draw[thick,dashed,black] (-2.0,-0.6) -- (-2.0,2.9) node[above] {\textrm{D5}};
\draw[thick,solid,black] (-0.6,-0.6) -- (-0.6,2.9) node[above] {\textrm{NS5}};
\draw[thick,solid,black] (1.2,-0.6) -- (1.2,2.9) node[above] {\textrm{NS5}};
\node at (-1.3,1.75) {$U(N)^2$};
\node at (0.3,1.75) {$U(2N)$};
\node at (0.3,0.95) {$\cdots$};
\node at (2.1,1.75) {$U(N)^2$};
\node at (-1.3,-0.3) {\textrm{Nahm}};
\node at (0.3,-0.3) {\textrm{Nahm}};
\node at (2.1,-0.3) {\textrm{Nahm}};
\node at (0,-1.2) {(b) $\hat{D}_l$ quiver, $\mathrm{Nahm}^{l+1}$};
\end{tikzpicture}
}
\\[0.5cm]
\scalebox{0.85}{
\begin{tikzpicture}
\fill[yellow!40] (-3,0) rectangle (3,2.6);
\draw[line width=1.6pt,dotted] (-3,-0.6) -- (-3,2.9) node[above] {\textrm{O5}$^-$};
\draw[line width=1.6pt,dotted] (3,-0.6) -- (3,2.9) node[above] {\textrm{O5}$^-$};
\draw[thick,dashed,black] (-3.4,0) -- (3.4,0) node[right] {\textrm{D5}$'$};
\draw[thick,solid,black] (-2.0,-0.6) -- (-2.0,2.9) node[above] {\textrm{NS5}};
\draw[thick,dashed,black] (-0.6,-0.6) -- (-0.6,2.9) node[above] {\textrm{D5}};
\draw[thick,dashed,black] (1.2,-0.6) -- (1.2,2.9) node[above] {\textrm{D5}};
\node at (0.3,1.75) {$2N$ \textrm{D3}};
\node at (0.3,0.95) {$\cdots$};
\node at (0.3,-0.3) {$\mathrm{Nahm}'$};
\node at (0,-1.2) {(c) $USp(2N)$ ADHM, $\mathrm{Nahm}'$};
\end{tikzpicture}
}
\hspace{0.3cm}
\scalebox{0.85}{
\begin{tikzpicture}
\fill[yellow!40] (-3,0) rectangle (3,2.6);
\draw[line width=1.6pt,dotted] (-3,-0.6) -- (-3,2.9) node[above] {\textrm{ON}$^0$};
\draw[line width=1.6pt,dotted] (3,-0.6) -- (3,2.9) node[above] {\textrm{ON}$^0$};
\draw[thick,solid,black] (-3.4,0) -- (3.4,0) node[right] {\textrm{NS5}$'$};
\draw[thick,dashed,black] (-2.0,-0.6) -- (-2.0,2.9) node[above] {\textrm{D5}};
\draw[thick,solid,black] (-0.6,-0.6) -- (-0.6,2.9) node[above] {\textrm{NS5}};
\draw[thick,solid,black] (1.2,-0.6) -- (1.2,2.9) node[above] {\textrm{NS5}};
\node at (-1.3,1.75) {$U(N)^2$};
\node at (0.3,1.75) {$U(2N)$};
\node at (0.3,0.95) {$\cdots$};
\node at (2.1,1.75) {$U(N)^2$};
\fill[black] (-3,0) circle (2pt) node[below left] {$\tilde{\eta}_1$};
\fill[black] (-2.0,0) circle (2pt) node[below right] {$\tilde{\Gamma}$};
\fill[black] (-0.6,0) circle (2pt) node[below right] {$\tilde{\eta}_2$};
\fill[black] (1.2,0) circle (2pt) node[below right] {$\tilde{\eta}_{l-1}$};
\fill[black] (3,0) circle (2pt) node[below left] {$\tilde{\eta}_l$};
\node at (0,-1.2) {(d) $\hat{D}_l$ quiver, $(\mathcal{N},N,N+\tilde{\Gamma}+\tilde{\eta})$};
\end{tikzpicture}
}
\caption{\textrm{
The brane configurations for the $USp(2N)$ ADHM theory (left) and its mirror (right).
The dotted lines are the O5$^-$-planes with a D5-brane on top and their S-duals, the ON$^0$-planes.
The NS5-brane (left) and the D5-brane (right) lie on these planes and are drawn apart.
}}
\label{fig_USpADHM}
\end{figure}

The $USp(2N)$ ADHM theory with $l$ flavors consists of the $USp(2N)$ vector multiplet, a hypermultiplet in the rank-two antisymmetric representation
and $l$ fundamental hypermultiplets, i.e. $2l$ fundamental half-hypermultiplets.
As in \cite{Hayashi:2022ldo}, the label $[2l]$ in $USp(2N)\,\textrm{ADHM-}[2l]$ counts the half-hypermultiplets.
The antisymmetric representation is reducible and contains a singlet.
We denote the fugacities for the flavor symmetry $U(1)_x$ of the antisymmetric hypermultiplet
and for the Cartan subalgebra of the $SO(2l)$ flavor symmetry of the fundamental hypermultiplets by $x$ and $y_a$, $a=1,\cdots,l$.
The theory describes $N$ M2-branes probing $\mathbb{C}^2\times\mathbb{C}^2/\hat{D}_{l-2}$,
where $\hat{D}_{l-2}$ is the binary dihedral group of order $4(l-2)$ \cite{Hayashi:2022ldo}.
It is realized by $2N$ D3-branes in the covering space suspended between two O5$^-$-planes,
with an NS5-brane on one of the O5$^-$-planes and $l$ D5-branes \cite{Kapustin:1998fa,Hanany:1999sj,Hayashi:2022ldo}.
The boundary is introduced by an NS5$'$- or D5$'$-brane on which the D3-branes end, as illustrated in Figure \ref{fig_USpADHM},
where (a) and (d) are drawn as the S-duals of (b) and (c).
Its mirror is the quiver gauge theory of the affine $D_l$ Dynkin diagram with one flavor \cite{deBoer:1996mp,Porrati:1996xi,Kapustin:1998fa}
(see also \cite{Dey:2014dwa}, and \cite{Hayashi:2022ldo} for tests of mirror symmetry using supersymmetric indices),
\begin{align}
\label{USp_quiver}
\begin{tikzpicture}[quiver]
\node[gauge,font=\scriptsize,inner sep=0.3pt](a) at (0,0){$USp(2N)$}; \node[flavor,font=\scriptsize](b) at (1.9,0){$SO(2l)$}; \draw (a)--(b);
\draw (a.115) .. controls +(-0.5,1.0) and +(0.5,1.0) .. (a.65);
\draw[mirror] (3.3,0)--(4.7,0);
\begin{scope}[shift={(7.2,0)}]
\node[gauge](n0) at (-1.05,0.9){$N$}; \node[gauge](n1) at (-1.05,-0.9){$N$};
\node[gauge](n2) at (0,0){$2N$}; \node[qdots](d) at (1.15,0){$\cdots$}; \node[gauge](n3) at (2.3,0){$2N$};
\node[gauge](n4) at (3.35,0.9){$N$}; \node[gauge](n5) at (3.35,-0.9){$N$};
\node[flavor](f) at (-2.4,0.9){$1$};
\draw (f)--(n0)--(n2)--(d)--(n3)--(n4); \draw (n1)--(n2); \draw (n3)--(n5);
\node[font=\scriptsize] at (-1.05,1.55) {$\alpha_0$}; \node[font=\scriptsize] at (-1.05,-1.55) {$\alpha_1$};
\node[font=\scriptsize] at (0,-0.68) {$\alpha_2$}; \node[font=\scriptsize] at (2.3,-0.68) {$\alpha_{l-2}$};
\node[font=\scriptsize] at (3.35,1.55) {$\alpha_{l-1}$}; \node[font=\scriptsize] at (3.35,-1.55) {$\alpha_l$};
\end{scope}
\end{tikzpicture}
\end{align}
where the loop is the antisymmetric hypermultiplet.
The nodes are labeled by the simple roots of the affine Lie algebra of type $D_l$,
\begin{align}
\label{Dl_roots}
\alpha_0&=-e_1-e_2, & \alpha_I&=e_I-e_{I+1}\quad (1\le I\le l-1), & \alpha_l&=e_{l-1}+e_l,
\end{align}
where $e_a$ is the orthonormal basis of $\mathbb{R}^l$.
Two nodes $\alpha_I$ and $\alpha_J$ are connected by a twisted bifundamental hypermultiplet when $\alpha_I\cdot\alpha_J=-1$.
The nodes $\alpha_0$, $\alpha_1$, $\alpha_{l-1}$ and $\alpha_l$ are $U(N)$ and the others are $U(2N)$.
The twisted fundamental hypermultiplet is attached to the node $\alpha_0$, and all the other nodes are balanced.
For $l=3$ the quiver is the necklace quiver $U(N)^4$ with one flavor in section \ref{sec_ADHM},
so that the $USp(2N)$ ADHM theory with three flavors and the $U(N)$ ADHM theory with four flavors have the same mirror \cite{Kapustin:1998fa,Hayashi:2022ldo}.
In the brane configuration we move one D5-brane onto each O5$^-$-plane.
S-duality maps an O5$^-$-plane with a D5-brane on top to an ON$^0$-plane \cite{Kutasov:1995te,Sen:1996na},
the remaining $l-2$ D5-branes to NS5-branes, and the NS5-brane to a D5-brane on one of the ON$^0$-planes \cite{Kapustin:1998fa,Hayashi:2022ldo}.
Near an ON$^0$-plane the orbifold projection breaks the gauge group $U(2N)$ of the D3-branes to $U(N)\times U(N)$ \cite{Kapustin:1998fa},
which gives the two end nodes.

\subsection{\texorpdfstring{$(\mathcal{N}',N',N'+\Gamma+\eta)$ and $\mathrm{Nahm}^{l+1}$}{(N',N',N'+Gamma+eta) and Nahm, ..., Nahm}}
\label{sec_USp_fw}
When the D3-branes end on an NS5$'$-brane (Figure \ref{fig_USpADHM} (a)),
the $USp(2N)$ ADHM theory obeys the Neumann boundary condition $(\mathcal{N}',N',N')$.
Its boundary gauge anomaly is $(8-2l)\sum_i\mathbf{s}_i^2$ (see (\ref{USp_fw_anom})).
For $l\ge4$ it is cancelled by $l-4$ fundamental Fermi multiplets.
For $l=3$ it has the sign of the contribution of a Fermi multiplet and cannot be cancelled by boundary Fermi multiplets,
while the Nahm pole boundary condition of the mirror necklace quiver $U(N)^4$ is dual to the Neumann boundary condition
of the $U(N)$ ADHM theory with four flavors by (\ref{ADHM_dual}) with $z=1$.
In the following we take $l\ge4$.
In the S-dual configuration the D3-branes end on a D5$'$-brane,
and all the nodes of the mirror quiver obey the Nahm pole boundary condition (Figure \ref{fig_USpADHM} (b)).
On the Neumann side we denote the fugacities for the boundary symmetries by $v_I$, $I=0,\cdots,l$, labeled by the nodes of the mirror quiver.
The Fermi multiplet $\Gamma_{I,J}$ carries the charges $(+1,-1)$ under $U(1)_{v_I}\times U(1)_{v_J}$, and the Fermi multiplet $\eta$ the charge $+1$ under $U(1)_{v_0}$.

\subsubsection{\texorpdfstring{$N=1$ and $l=4$}{N=1 and l=4}}
\label{sec_USp_fw_N1}
\paragraph{$USp(2)$ ADHM theory with four flavors}
For $N=1$ and $l=4$ the boundary gauge anomaly vanishes, and all the boundary Fermi multiplets are gauge-neutral.
They are the Fermi multiplets $\Gamma_{I,2}$ for the end nodes $\alpha_I$, $I=0,1,3,4$, and the Fermi multiplet $\eta$.
They correspond to the top components of the twisted bifundamental hypermultiplets and of the twisted fundamental hypermultiplet of the mirror quiver at the Nahm poles.
The Neumann half-index reads
\begin{align}
\label{USp2_fw_Neu}
&\mathbb{II}^{USp(2)\,\textrm{ADHM-}[8]}_{\mathcal{N}',N',N'+\Gamma+\eta}
=\frac12\oint\frac{ds}{2\pi is}
\underbrace{(q)_{\infty}(q^{\frac12}t^2;q)_{\infty}(s^{\pm2};q)_{\infty}(q^{\frac12}t^2s^{\pm2};q)_{\infty}}_{\textrm{vector multiplet ($\mathcal{N}'$)}}\;
\underbrace{\frac{1}{(q^{\frac14}tx^{\pm};q)_{\infty}}}_{\textrm{singlet hypermultiplet ($N'$)}}
\nonumber\\*
&\times\underbrace{\prod_{a=1}^4\frac{1}{(q^{\frac14}ts^{\pm}y_a^{\pm};q)_{\infty}(q^{\frac14}ts^{\pm}y_a^{\mp};q)_{\infty}}}_{\textrm{fundamental hypermultiplets ($N'$)}}\;
\underbrace{F\left(q^{\frac12}\frac{v_0}{v_2y_1y_2};q\right)}_{\Gamma_{0,2}}\;
\underbrace{F\left(q^{\frac12}\frac{v_1y_1}{v_2y_2};q\right)}_{\Gamma_{1,2}}\;
\underbrace{F\left(q^{\frac12}\frac{v_3y_3}{v_2y_4};q\right)}_{\Gamma_{3,2}}
\nonumber\\*
&\times\underbrace{F\left(q^{\frac12}\frac{v_4y_3y_4}{v_2};q\right)}_{\Gamma_{4,2}}\;
\underbrace{F(q^{\frac12}v_0x^{-1};q)}_{\eta}.
\end{align}
The $SO(8)$ fugacities in the Fermi multiplets $\Gamma_{I,2}$ are $\prod_ay_a^{(\alpha_I)_a}$.

\paragraph{Mirror quiver}
For $N=1$ the Nahm pole of a $U(1)$ node is trivial,
so that the four end nodes obey the Dirichlet boundary condition and the node $\alpha_2$ the Nahm pole boundary condition for $U(2)$.
The boundary symmetry is $U(1)_{u^{(I)}}$ for each node.
As for the necklace quiver, the twisted bifundamental hypermultiplets cancel the twisted vector multiplet in $k_{\rm eff}$ at the balanced node $\alpha_2$,
and the sign $(-1)^{m^{(0)}}$ for the node $\alpha_0$ is absorbed into the hypermultiplet factors.
The top components of the twisted bifundamental hypermultiplets have $J=\frac12$,
and that of the twisted fundamental hypermultiplet has $J=0$.
The formula (\ref{Nahm_quiver_simple}) with $t\rightarrow t^{-1}$ gives
\begin{align}
\label{USp2_fw_Nahm}
&\mathbb{II}^{\wtilde{U(1)^4\times U(2)}}_{\mathrm{Nahm}^{5}}
=\underbrace{(q^{\frac34}t(u^{(0)})^{\pm};q)_{\infty}\prod_{I=0,1,3,4}\left(qt^2\left(\frac{u^{(I)}}{u^{(2)}}\right)^{\pm};q\right)_{\infty}}_{\textrm{top components}}
\nonumber\\*
&\times\underbrace{\frac{1}{[(q)_{\infty}(q^{\frac12}t^2;q)_{\infty}]^{5}(q^{\frac32}t^2;q)_{\infty}(qt^4;q)_{\infty}}}_{\textrm{Slodowy slices}}
\nonumber\\
&\times\sum_{m}\underbrace{(xy_1y_2)^{-m^{(0)}}(y_1/y_2)^{m^{(1)}}(y_2/y_3)^{m^{(2)}_1+m^{(2)}_2}(y_3/y_4)^{m^{(3)}}(y_3y_4)^{m^{(4)}}}_{\textrm{topological charges}}
\nonumber\\
&\times\underbrace{(q^{\frac12}t^2)^{-n}\frac{(qt^4;q)_{n}(q^{\frac32}t^2;q)_{n}}{(q^{\frac12}t^2;q)_{n}(q;q)_{n}}}_{\textrm{vector multiplet ($\mathrm{Nahm}$)}}\;
\underbrace{(q^{\frac14}t)^{m^{(0)}}\frac{(q^{\frac14}t^{-1}u^{(0)};q)_{m^{(0)}}}{(q^{\frac34}tu^{(0)};q)_{m^{(0)}}}}_{\textrm{fundamental hypermultiplet ($D$)}}
\nonumber\\*
&\times\underbrace{\prod_{I=0,1,3,4}(q^{\frac14}t)^{2m^{(I)}-m^{(2)}_1-m^{(2)}_2}
\frac{(q^{\frac12}u^{(I)}/u^{(2)};q)_{m^{(I)}-m^{(2)}_1}(t^{-2}u^{(I)}/u^{(2)};q)_{m^{(I)}-m^{(2)}_2}}
{(qt^2u^{(I)}/u^{(2)};q)_{m^{(I)}-m^{(2)}_1}(q^{\frac12}u^{(I)}/u^{(2)};q)_{m^{(I)}-m^{(2)}_2}}}_{\textrm{bifundamental hypermultiplets ($D$)}},
\end{align}
where the sum runs over $m^{(I)}\in\mathbb{Z}$ for $I=0,1,3,4$ and $m^{(2)}_1\le m^{(2)}_2$, and $n=m^{(2)}_2-m^{(2)}_1$.
The topological charges of the node $\alpha_I$ are given by $\prod_ay_a^{(\alpha_I)_a}$, multiplied by $x^{-1}$ for $I=0$.

\paragraph{Duality}
We propose that
\begin{align}
\label{USp2_fw_dual}
\mathbb{II}^{USp(2)\,\textrm{ADHM-}[8]}_{\mathcal{N}',N',N'+\Gamma+\eta}(t,x,v_I,y_a;q)
&=\mathbb{II}^{\wtilde{U(1)^4\times U(2)}}_{\mathrm{Nahm}^{5}}(t,x,u^{(I)}=v_I,y_a;q).
\end{align}
The boundary anomalies agree, as we show for general $N$ and $l$ in (\ref{USp_fw_Nahm_anom}).

\subsubsection{\texorpdfstring{General $N$ and $l$}{General N and l}}
\label{sec_USp_fw_N}
\paragraph{$USp(2N)$ ADHM theory with $l$ flavors}
The boundary gauge anomaly is cancelled by the fundamental Fermi multiplets $\Gamma_{k,k+1}$ with $2\le k\le l-3$.
They correspond to the $2N$ top components of the twisted bifundamental hypermultiplets between the $U(2N)$ nodes $\alpha_k$ and $\alpha_{k+1}$.
The gauge-neutral Fermi multiplets are $\eta$ and $\Gamma_{I,J}$ for the four pairs $(I,J)=(0,2)$, $(1,2)$, $(l-1,l-2)$ and $(l,l-2)$ of an end node $\alpha_I$ and its neighbor $\alpha_J$.
For each pair there are $N$ Fermi multiplets $\Gamma_{I,J}$,
as many as the top components of the twisted bifundamental hypermultiplet between $U(N)$ and $U(2N)$.
They have the same flavor charges and are rotated by a $U(N)$ flavor symmetry.
We find that their fugacities are shifted by $(q^{\frac12}t^2)^{\kappa}$ along the principal embedding of $\mathfrak{su}(2)$ into $\mathfrak{u}(N)$,
i.e. the R-symmetry mixes with the Cartan subalgebra of $U(N)$.
For the Fermi multiplets $\Gamma_{0,2}$ the embedding is into $\mathfrak{u}(N-1)\subset\mathfrak{u}(N)$,
and the commutant $U(1)$ mixes with $U(1)_x$.
We write $y^{\alpha_I}=\prod_ay_a^{(\alpha_I)_a}$.
The Neumann half-index reads
\begin{align}
\label{USp_fw_Neu}
&\mathbb{II}^{USp(2N)\,\textrm{ADHM-}[2l]}_{\mathcal{N}',N',N'+\Gamma+\eta}
=\frac{1}{2^NN!}\oint\prod_{i=1}^N\frac{ds_i}{2\pi is_i}
\underbrace{(q)_{\infty}^N(q^{\frac12}t^2;q)_{\infty}^N\prod_{\alpha\in\Delta}(s^{\alpha};q)_{\infty}(q^{\frac12}t^2s^{\alpha};q)_{\infty}}_{\textrm{vector multiplet ($\mathcal{N}'$)}}
\nonumber\\*
&\times\underbrace{\frac{1}{(q^{\frac14}tx^{\pm};q)_{\infty}^N\prod_{\alpha\in\Delta_{\rm s}}(q^{\frac14}tx^{\pm}s^{\alpha};q)_{\infty}}}_{\textrm{antisymmetric hypermultiplet ($N'$)}}\;
\underbrace{\prod_{i=1}^N\prod_{a=1}^l\frac{1}{(q^{\frac14}ts_i^{\pm}y_a^{\pm};q)_{\infty}(q^{\frac14}ts_i^{\pm}y_a^{\mp};q)_{\infty}}}_{\textrm{fundamental hypermultiplets ($N'$)}}
\nonumber\\
&\times\underbrace{\prod_{k=2}^{l-3}\prod_{i=1}^NF\left(q^{\frac12}s_i\frac{v_k}{v_{k+1}y_{k+1}};q\right)F\left(q^{\frac12}s_i^{-1}\frac{v_k}{v_{k+1}y_{k+1}};q\right)}_{\Gamma_{k,k+1}}\;
\underbrace{F(q^{\frac12}v_0x^{-1};q)}_{\eta}
\nonumber\\
&\times\underbrace{\prod_{(I,J)\neq(0,2)}\prod_{\kappa}F\left(q^{\frac12}(q^{\frac12}t^2)^{\kappa}\frac{v_I}{v_J}y^{\alpha_I};q\right)}_{\Gamma_{I,J},\ (I,J)\neq(0,2)}
\nonumber\\*
&\times\underbrace{\prod_{\kappa'}F\left(q^{\frac12}(q^{\frac12}t^2)^{\kappa'}\frac{v_0}{v_2x}y^{\alpha_0};q\right)F\left(q^{\frac12}\frac{v_0x^{N-1}}{v_2}y^{\alpha_0};q\right)}_{\Gamma_{0,2}},
\end{align}
where $\Delta_{\rm s}$ is the set of the short roots $\pm e_i\pm e_j$ with $i<j$,
and $\kappa$ and $\kappa'$ run from $-\frac{N-1}{2}$ to $\frac{N-1}{2}$ and from $-\frac{N-2}{2}$ to $\frac{N-2}{2}$ in steps of one (the range of $\kappa'$ is empty for $N=1$).
For $N=1$ and $l=4$ it is (\ref{USp2_fw_Neu}).
For $(I,J)\neq(0,2)$ the Fermi multiplets $\Gamma_{I,J}$ take the form of the split Fermi multiplets in (\ref{Nahm_limit}) for spin $\frac{N}{2}$ with $t\rightarrow t^{-1}$.
Their components with $\kappa=-\frac{N-1}{2}$ give terms of order $q^{\frac{3-N}{4}}$,
which on the Nahm pole side arise from the fluxes of a single end node.
For $N\ge4$ this is a negative power of $q$ at fixed $t$, while it is $q^{\frac12}(q^{\frac14}t)^{1-N}$ in the variables $q^{\frac12}$ and $q^{\frac14}t$ of the duality (see section \ref{sec_UN_simple}).

\paragraph{Mirror quiver}
As for $N=1$ and $l=4$, the twisted bifundamental hypermultiplets cancel the twisted vector multiplets in $k_{\rm eff}$ at the balanced nodes,
and the sign $(-1)^{\sum_im^{(0)}_i}$ for the node $\alpha_0$ is absorbed into the hypermultiplet factors.
The top components of the twisted bifundamental hypermultiplets have $J=\frac{N}{2},\frac{N}{2}+1,\cdots,\frac{3N}{2}-1$ from $\frac{N-1}{2}\otimes\frac{2N-1}{2}$ between $U(N)$ and $U(2N)$
and $J=0,1,\cdots,2N-1$ between the $U(2N)$ nodes,
and that of the twisted fundamental hypermultiplet has $J=\frac{N-1}{2}$.
We write $\mu^{(I)}_i=i-\frac{N_I+1}{2}$ and $w_{ij}=\mu^{(I)}_i-\mu^{(J)}_j$ for the connected nodes $\alpha_I$ and $\alpha_J$.
The formula (\ref{Nahm_quiver_simple}) with $t\rightarrow t^{-1}$ gives
\begin{align}
\label{USp_fw_Nahm}
&\mathbb{II}^{\wtilde{\hat{D}_l}}_{\mathrm{Nahm}^{l+1}}
=\underbrace{(q^{\frac12+\frac{N}{4}}t^{N}(u^{(0)})^{\pm};q)_{\infty}
\prod_{(I,J)}\prod_{p=0}^{N-1}\left(q^{\frac34+\frac{N}{4}+\frac{p}{2}}t^{1+N+2p}\left(\frac{u^{(I)}}{u^{(J)}}\right)^{\pm};q\right)_{\infty}}_{\textrm{top components}}
\nonumber\\*
&\times\underbrace{\prod_{k=2}^{l-3}\prod_{p=0}^{2N-1}\left(q^{\frac34+\frac{p}{2}}t^{1+2p}\left(\frac{u^{(k)}}{u^{(k+1)}}\right)^{\pm};q\right)_{\infty}}_{\textrm{top components}}\;
\underbrace{\prod_{I=0}^l\prod_{n=0}^{N_I-1}\frac{1}{(q^{1+\frac{n}{2}}t^{2n};q)_{\infty}(q^{\frac12+\frac{n}{2}}t^{2+2n};q)_{\infty}}}_{\textrm{Slodowy slices}}
\nonumber\\
&\times\sum_{m}\underbrace{x^{-\sum_im^{(0)}_i}\prod_{I=0}^l(y^{\alpha_I})^{\sum_im^{(I)}_i}}_{\textrm{topological charges}}
\nonumber\\
&\times\underbrace{\prod_{I=0}^l\prod_{1\le i<j\le N_I}(q^{\frac12}t^2)^{m^{(I)}_i-m^{(I)}_j}
\frac{(q^{\frac12+\frac{j-i}{2}}t^{2+2(j-i)};q)_{m^{(I)}_j-m^{(I)}_i}(q^{1+\frac{j-i}{2}}t^{2(j-i)};q)_{m^{(I)}_j-m^{(I)}_i}}
{(q^{\frac{j-i}{2}}t^{2(j-i)};q)_{m^{(I)}_j-m^{(I)}_i}(q^{\frac12+\frac{j-i}{2}}t^{2(j-i)-2};q)_{m^{(I)}_j-m^{(I)}_i}}}_{\textrm{vector multiplets ($\mathrm{Nahm}$)}}
\nonumber\\
&\times\underbrace{\prod_{i=1}^N(q^{\frac14}t)^{m^{(0)}_i}\frac{(q^{\frac14+\frac{\mu^{(0)}_i}{2}}t^{2\mu^{(0)}_i-1}u^{(0)};q)_{m^{(0)}_i}}{(q^{\frac34+\frac{\mu^{(0)}_i}{2}}t^{1+2\mu^{(0)}_i}u^{(0)};q)_{m^{(0)}_i}}}_{\textrm{fundamental hypermultiplet ($D$)}}
\nonumber\\*
&\times\underbrace{\prod_{\alpha_I\cdot\alpha_J=-1}\prod_{i=1}^{N_I}\prod_{j=1}^{N_J}(q^{\frac14}t)^{m^{(I)}_i-m^{(J)}_j}
\frac{(q^{\frac14+\frac{w_{ij}}{2}}t^{2w_{ij}-1}\frac{u^{(I)}}{u^{(J)}};q)_{m^{(I)}_i-m^{(J)}_j}}{(q^{\frac34+\frac{w_{ij}}{2}}t^{1+2w_{ij}}\frac{u^{(I)}}{u^{(J)}};q)_{m^{(I)}_i-m^{(J)}_j}}}_{\textrm{bifundamental hypermultiplets ($D$)}},
\end{align}
where $(I,J)$ runs over the four pairs above, the sum runs over the fluxes $m^{(I)}_1\le\cdots\le m^{(I)}_{N_I}$ of all the nodes, and each pair of the connected nodes appears once in the last product.
For $N=1$ and $l=4$ it is (\ref{USp2_fw_Nahm}).

\paragraph{Duality}
We propose that
\begin{align}
\label{USp_fw_dual}
\mathbb{II}^{USp(2N)\,\textrm{ADHM-}[2l]}_{\mathcal{N}',N',N'+\Gamma+\eta}(t,x,v_I,y_a;q)
&=\mathbb{II}^{\wtilde{\hat{D}_l}}_{\mathrm{Nahm}^{l+1}}(t,x,u^{(I)}=v_Ix^{N-1},y_a;q).
\end{align}
For $N=1$ and $l=4$ it is (\ref{USp2_fw_dual}).
As in (\ref{ADHM_dual}), the common factor $x^{N-1}$ comes from the Fermi multiplet $\eta$.
The identity singles out the shifts of the fugacities of the Fermi multiplets $\Gamma_{I,J}$.
Without the shifts the two sides differ already at order $q^{\frac14}$ for $N=2$,
and for $N=3$ so do the shifts $x^{2\kappa}$ of all the Fermi multiplets $\Gamma_{0,2}$ in place of their splitting.

As a consistency check, we compare the boundary anomalies.
We write $\mathbf{b}_k=\mathbf{v}_k-\mathbf{v}_{k+1}-\mathbf{y}_{k+1}$ and $\mathbf{b}_I=\mathbf{v}_I-\mathbf{v}_J+\alpha_I\cdot\mathbf{y}$ for the charges of the Fermi multiplets $\Gamma_{k,k+1}$ and $\Gamma_{I,J}$.
For the Neumann boundary condition (\ref{USp_fw_Neu}) the boundary anomaly is
\begin{align}
\label{USp_fw_anom}
&\mathcal{A}^{USp(2N)\,\textrm{ADHM-}[2l]}_{\mathcal{N}',N',N'+\Gamma+\eta}
\nonumber\\*
&=\underbrace{\sum_{\alpha\in\Delta}(\alpha\cdot\mathbf{s})^2}_{\textrm{vector multiplet ($\mathcal{N}'$)}}\quad
\underbrace{-\sum_{\alpha\in\Delta_{\rm s}}(\alpha\cdot\mathbf{s}+\mathbf{x})^2-N\mathbf{x}^2}_{\textrm{antisymmetric hypermultiplet ($N'$)}}\quad
\underbrace{-\sum_{i=1}^N\sum_{a=1}^l\bigl[(\mathbf{s}_i+\mathbf{y}_a)^2+(-\mathbf{s}_i+\mathbf{y}_a)^2\bigr]}_{\textrm{fundamental hypermultiplets ($N'$)}}
\nonumber\\
&+\underbrace{\sum_{k=2}^{l-3}\sum_{i=1}^N\bigl[(\mathbf{s}_i+\mathbf{b}_k)^2+(-\mathbf{s}_i+\mathbf{b}_k)^2\bigr]}_{\Gamma_{k,k+1}}
+\underbrace{(\mathbf{v}_0-\mathbf{x})^2}_{\eta}
\nonumber\\*
&+\underbrace{N\sum_{(I,J)\neq(0,2)}\mathbf{b}_I^2+(N-1)(\mathbf{b}_0-\mathbf{x})^2+(\mathbf{b}_0+(N-1)\mathbf{x})^2}_{\Gamma_{I,J}}.
\end{align}
The shifts by $(q^{\frac12}t^2)^{\kappa}$ do not enter (\ref{USp_fw_anom}), which contains no field strengths of the R-symmetries.
The vector multiplet, the antisymmetric hypermultiplet, the fundamental hypermultiplets and the Fermi multiplets $\Gamma_{k,k+1}$
contribute $4N+4$, $-4(N-1)$, $-2l$ and $2(l-4)$ times $\sum_i\mathbf{s}_i^2$,
so that all the terms with the gauge field strengths cancel.
By (\ref{Nahm_anom}), the boundary anomaly for the Nahm pole side is
\begin{align}
\label{USp_fw_Nahm_anom}
\mathcal{A}^{\wtilde{\hat{D}_l}}_{\mathrm{Nahm}^{l+1}}
&=\underbrace{N\sum_{(I,J)}(\mathbf{u}^{(I)}-\mathbf{u}^{(J)})^2+2N\sum_{k=2}^{l-3}(\mathbf{u}^{(k)}-\mathbf{u}^{(k+1)})^2+(\mathbf{u}^{(0)})^2}_{\textrm{top components}}
\nonumber\\*
&\underbrace{-2N\mathbf{u}^{(0)}\mathbf{x}+2\sum_{I=0}^lN_I\mathbf{u}^{(I)}\,\alpha_I\cdot\mathbf{y}}_{\textrm{FI couplings}}
\nonumber\\*
&=\mathcal{A}^{USp(2N)\,\textrm{ADHM-}[2l]}_{\mathcal{N}',N',N'+\Gamma+\eta},
\end{align}
where the second equality holds for $\mathbf{u}^{(I)}=\mathbf{v}_I+(N-1)\mathbf{x}$ and uses $\sum_IN_I\alpha_I=0$.
Hence the anomalies agree exactly.
On both sides the terms in $\mathbf{x}^2$ add up to $(1-N^2)\mathbf{x}^2$.
On the Neumann side the antisymmetric hypermultiplet gives $-N(2N-1)\mathbf{x}^2$, the Fermi multiplet $\eta$ gives $\mathbf{x}^2$,
and the splitting of the Fermi multiplets $\Gamma_{0,2}$ supplies $N(N-1)\mathbf{x}^2$.

The shifts enter the terms with the field strengths of the R-symmetries in section \ref{sec_Nahm_anom}.
The Fermi multiplets $\Gamma_{I,J}$ carry $H=2\kappa$ and $H=2\kappa'$,
so that for each $(I,J)\neq(0,2)$ they contribute $4\sum_{\kappa}\kappa^2\,\mathbf{h}^2=\frac{N(N^2-1)}{3}\mathbf{h}^2$ in addition to (\ref{USp_fw_anom}),
and the Fermi multiplets $\Gamma_{0,2}$ contribute $4\sum_{\kappa'}\kappa'^2\,\mathbf{h}^2=\frac{N(N-1)(N-2)}{3}\mathbf{h}^2$.
With these contributions the anomalies agree exactly also in the terms with $\mathbf{h}$ and $\mathbf{c}$,
as for the duality (\ref{USp_rv_dual}) between the same bulk theories in section \ref{sec_USp_rv}.
Without the shifts they would differ by $\frac{N(N-1)(4N+1)}{3}\mathbf{h}^2$.
This cannot be absorbed into a background Chern-Simons term for the R-symmetries,
which would contribute in the same way to (\ref{USp_rv_dual}).
Hence the anomalies require the shifts, although they constrain only the sums of $\kappa^2$ and $\kappa'^2$.

\subsection{\texorpdfstring{$\mathrm{Nahm}'$ and $(\mathcal{N},N,N+\tilde{\Gamma}+\tilde{\eta})$}{Nahm' and (N,N,N+Gamma+eta)}}
\label{sec_USp_rv}
When the D3-branes end on a D5$'$-brane (Figure \ref{fig_USpADHM} (c)),
the $USp(2N)$ ADHM theory obeys the Nahm pole boundary condition for the vector multiplet
and the Dirichlet boundary condition $D'$ for the hypermultiplets.
For the principal embedding the commutant is the center $\mathbb{Z}_2$ of $USp(2N)_{\partial}$,
which is the boundary symmetry at the Nahm pole.
This agrees with the brane configuration, in which the O5$^-$-planes reduce the world-volume $U(1)$ of the D5$'$-brane to $O(1)\cong\mathbb{Z}_2$.
Its fugacity $\epsilon=\pm1$ enters only through $\epsilon y_a$,
since the center of $USp(2N)$ acts on the fundamental hypermultiplets in the same way as that of $SO(2l)$.
We set $\epsilon=1$.

In the S-dual configuration the D5$'$-brane maps to the NS5$'$-brane,
and all the nodes of the mirror quiver obey the Neumann boundary condition $(\mathcal{N},N,N)$ (Figure \ref{fig_USpADHM} (d)).
As for the $U(N)$ ADHM theory in section \ref{sec_ADHM_rv}, we apply the formula (\ref{Nahm_simple}) directly.

\subsubsection{\texorpdfstring{$N=1$ and $l=4$}{N=1 and l=4}}
\label{sec_USp_N1}
\paragraph{$USp(2)$ ADHM theory with four flavors}
For $N=1$ the gauge group is $USp(2)\cong SU(2)$, and the antisymmetric hypermultiplet is the singlet, i.e. a free hypermultiplet.
For $l=4$ the fundamental hypermultiplets have the flavor symmetry $SO(8)$.
For the Dirichlet boundary condition $(\mathcal{D}',D',D')$ the vector multiplet contributes $-8$ and the fundamental hypermultiplets contribute $8$ to $k_{\rm eff}$,
so that $k_{\rm eff}=0$ and the sign $(-1)^{m^Tk_{\rm eff}m}$ in the Dirichlet half-index (\ref{Dhalf}) is trivial.
The positive root $2e_1$ has the height $\mathrm{ht}\,(2e_1)=1$ in (\ref{ht_def}), and the specialization is $u_1=(q^{\frac12}t^{-2})^{\frac12}$.
Each fundamental hypermultiplet is the spin-$\frac12$ multiplet with the weights $\pm e_1$, the singlet has $J=0$, and the Slodowy slice has $k=1$.
The formula (\ref{Nahm_simple}) gives
\begin{align}
\label{USp2_simple}
&\mathbb{II}^{USp(2)\,\textrm{ADHM-}[8]}_{\mathrm{Nahm}'}
\nonumber\\*
&=\underbrace{(q^{\frac34}t^{-1}x^{\pm};q)_{\infty}\prod_{a=1}^{4}(qt^{-2}y_a^{\pm};q)_{\infty}}_{\textrm{top components}}\;
\underbrace{\frac{1}{(qt^{-4};q)_{\infty}(q^{\frac32}t^{-2};q)_{\infty}}}_{\textrm{Slodowy slice}}
\nonumber\\*
&\times\sum_{m=0}^{\infty}(q^{\frac12}t^{-2})^{2m}
\underbrace{\frac{(qt^{-4};q)_{2m}(q^{\frac32}t^{-2};q)_{2m}}{(q^{\frac12}t^{-2};q)_{2m}(q;q)_{2m}}}_{\textrm{vector multiplet ($\mathrm{Nahm}'$)}}\;
\underbrace{\prod_{a=1}^4\frac{(q^{\frac12}y_a^{\pm};q)_m}{(qt^{-2}y_a^{\pm};q)_m}}_{\textrm{fundamental hypermultiplets ($D'$)}}.
\end{align}
The free hypermultiplet contributes only its top component, which is its half-index for the Dirichlet boundary condition.
The factor $(q^{\frac12}t^{-2})^{2m}$ combines $(q^{\frac12}t^{-2})^{-2m}$ from the vector multiplet and $(q^{\frac12}t^{-2})^{4m}$ from the fundamental hypermultiplets.
The invariance under $y_a\rightarrow y_a^{-1}$ is manifest.

\paragraph{Mirror quiver}
For $N=1$ and $l=4$ the mirror quiver has the four nodes $\alpha_0$, $\alpha_1$, $\alpha_3$ and $\alpha_4$ of $U(1)$, all connected to the node $\alpha_2$ of $U(2)$.
We denote the gauge fugacities by $s^{(I)}$ for the $U(1)$ nodes and by $s^{(2)}_i$, $i=1,2$, for the $U(2)$ node, and write $S_2=s^{(2)}_1s^{(2)}_2$.
The $U(2)$ node is balanced, so that the non-Abelian part of the boundary gauge anomaly cancels.
The remaining part is cancelled by the fundamental Fermi multiplet $\tilde{\Gamma}$ and the cross-determinant Fermi multiplets $\tilde{\eta}_a$, $a=1,\cdots,4$.
They correspond to the top component of the singlet hypermultiplet and to the top components of the four fundamental hypermultiplets of the $USp(2)$ ADHM theory at the Nahm pole.
The Neumann half-index of the mirror quiver reads
\begin{align}
\label{USp2_rv_Neu}
&\mathbb{II}^{\wtilde{U(1)^4\times U(2)}}_{\mathcal{N},N,N+\tilde{\Gamma}+\tilde{\eta}}
=\oint\prod_{I=0,1,3,4}\frac{ds^{(I)}}{2\pi is^{(I)}}\;\frac12\oint\prod_{i=1}^2\frac{ds^{(2)}_i}{2\pi is^{(2)}_i}
\nonumber\\*
&\times\underbrace{[(q)_{\infty}(q^{\frac12}t^{-2};q)_{\infty}]^{6}\prod_{i\neq j}\left(\frac{s^{(2)}_i}{s^{(2)}_j};q\right)_{\infty}\left(q^{\frac12}t^{-2}\frac{s^{(2)}_i}{s^{(2)}_j};q\right)_{\infty}}_{\textrm{vector multiplets ($\mathcal{N}$)}}
\nonumber\\
&\times\underbrace{\prod_{I=0,1,3,4}\prod_{i=1}^2\frac{1}{(q^{\frac14}t^{-1}(s^{(I)}/s^{(2)}_i)^{\pm};q)_{\infty}}}_{\textrm{bifundamental hypermultiplets ($N$)}}\;
\underbrace{\frac{1}{(q^{\frac14}t^{-1}(s^{(0)})^{\pm};q)_{\infty}}}_{\textrm{fundamental hypermultiplet ($N$)}}\;
\underbrace{F\left(q^{\frac12}\frac{s^{(0)}}{x};q\right)}_{\tilde{\Gamma}}
\nonumber\\*
&\times\underbrace{F\left(q^{\frac12}y_1\frac{s^{(1)}}{s^{(0)}};q\right)}_{\tilde{\eta}_1}\;
\underbrace{F\left(q^{\frac12}y_2\frac{S_2}{s^{(0)}s^{(1)}};q\right)}_{\tilde{\eta}_2}\;
\underbrace{F\left(q^{\frac12}y_3\frac{s^{(3)}s^{(4)}}{S_2};q\right)}_{\tilde{\eta}_3}\;
\underbrace{F\left(q^{\frac12}y_4\frac{s^{(4)}}{s^{(3)}};q\right)}_{\tilde{\eta}_4}.
\end{align}
The charges of the Fermi multiplets $\tilde{\eta}_a$ are the components of the simple roots (\ref{Dl_roots}) (see section \ref{sec_USp_N}).

\paragraph{Duality}
We propose that
\begin{align}
\label{USp2_rv_dual}
\mathbb{II}^{USp(2)\,\textrm{ADHM-}[8]}_{\mathrm{Nahm}'}
&=\mathbb{II}^{\wtilde{U(1)^4\times U(2)}}_{\mathcal{N},N,N+\tilde{\Gamma}+\tilde{\eta}}.
\end{align}
By (\ref{Nahm_anom}), the boundary anomaly for the Nahm pole side is $\sum_a\mathbf{y}_a^2+\mathbf{x}^2$.
It agrees with that of the Neumann boundary condition, which we compute for general $N$ and $l$ in (\ref{USp_rv_Neu_anom}).

\subsubsection{\texorpdfstring{General $N$ and $l$}{General N and l}}
\label{sec_USp_N}
\paragraph{$USp(2N)$ ADHM theory with $l$ flavors}
The roots of $USp(2N)$ are $\pm e_i\pm e_j$ with $i<j$ and $\pm2e_i$, where $e_i$ are the weights of the fundamental representation.
For the Dirichlet boundary condition $(\mathcal{D}',D',D')$ the antisymmetric hypermultiplet cancels the contribution of the short roots $\pm e_i\pm e_j$ to $k_{\rm eff}$,
so that $k_{\rm eff}=(2l-8)\mathbf{1}_N$ and the sign $(-1)^{m^Tk_{\rm eff}m}$ in the Dirichlet half-index (\ref{Dhalf}) is trivial.
We take the positive roots $e_j-e_i$ and $e_i+e_j$ with $1\le i<j\le N$ and $2e_i$ with $1\le i\le N$.
The specialization $u_i=(q^{\frac12}t^{-2})^{\mu_i}$ with $\mu_i=i-\frac12$ gives $u^{\alpha}=(q^{\frac12}t^{-2})^{\mathrm{ht}\,\alpha}$,
where $\mathrm{ht}(e_j-e_i)=j-i$, $\mathrm{ht}(e_i+e_j)=i+j-1$ and $\mathrm{ht}(2e_i)=2i-1$.
The dominant coweights are $0\le m_1\le\cdots\le m_N$.
The fundamental representation has $w_{\lambda}(\mu)=\pm\mu_i$ and forms a single multiplet of spin $J=N-\frac12$.
The antisymmetric representation has the weights $\pm e_i\pm e_j$ with $i<j$ and $N$ zero weights,
and it decomposes into the spins $J=0,2,\cdots,2N-2$.
The spins in the Slodowy slice are $2k-1$ with $k=1,\cdots,N$.
Combining the weights $w_{\lambda}$ and $-w_{\lambda}$ of the hypermultiplets as in (\ref{ADHM_simple}), we find
\begin{align}
\label{USp_simple}
&\mathbb{II}^{USp(2N)\,\textrm{ADHM-}[2l]}_{\mathrm{Nahm}'}
\nonumber\\*
&=\underbrace{\prod_{k=0}^{N-1}(q^{\frac34+k}t^{-1-4k}x^{\pm};q)_{\infty}\prod_{a=1}^{l}(q^{\frac12+\frac{N}{2}}t^{-2N}y_a^{\pm};q)_{\infty}}_{\textrm{top components}}\;
\underbrace{\prod_{k=1}^{N}\frac{1}{(q^{k}t^{-4k};q)_{\infty}(q^{k+\frac12}t^{2-4k};q)_{\infty}}}_{\textrm{Slodowy slice}}
\nonumber\\
&\times\sum_{0\le m_1\le\cdots\le m_N}(q^{\frac12}t^{-2})^{(l-2)\sum_im_i}
\underbrace{\prod_{\alpha\in\Delta^+}\frac{(q^{\frac12+\frac{\mathrm{ht}\alpha}{2}}t^{-2-2\mathrm{ht}\alpha};q)_{\alpha(m)}(q^{1+\frac{\mathrm{ht}\alpha}{2}}t^{-2\mathrm{ht}\alpha};q)_{\alpha(m)}}
{(q^{\frac{\mathrm{ht}\alpha}{2}}t^{-2\mathrm{ht}\alpha};q)_{\alpha(m)}(q^{\frac12+\frac{\mathrm{ht}\alpha}{2}}t^{2-2\mathrm{ht}\alpha};q)_{\alpha(m)}}}_{\textrm{vector multiplet ($\mathrm{Nahm}'$)}}
\nonumber\\*
&\times\underbrace{\prod_{\alpha\in\Delta^+_{\rm s}}\frac{(q^{\frac14+\frac{\mathrm{ht}\alpha}{2}}t^{1-2\mathrm{ht}\alpha}x^{\pm};q)_{\alpha(m)}}{(q^{\frac34+\frac{\mathrm{ht}\alpha}{2}}t^{-1-2\mathrm{ht}\alpha}x^{\pm};q)_{\alpha(m)}}}_{\textrm{antisymmetric hypermultiplet ($D'$)}}\;
\underbrace{\prod_{a=1}^{l}\prod_{i=1}^N\frac{(q^{\frac{i}{2}}t^{2-2i}y_a^{\pm};q)_{m_i}}{(q^{\frac12+\frac{i}{2}}t^{-2i}y_a^{\pm};q)_{m_i}}}_{\textrm{fundamental hypermultiplets ($D'$)}},
\end{align}
where $\Delta^+$ is the set of the positive roots and $\Delta^+_{\rm s}$ that of the short positive roots $e_j-e_i$ and $e_i+e_j$.
The factor $(q^{\frac12}t^{-2})^{(l-2)\sum_im_i}$ collects the powers of $q^{\frac12}t^{-2}$ in (\ref{Nahm_simple}).
For the short roots they cancel between the vector multiplet and the antisymmetric hypermultiplet as in (\ref{ADHM_simple}),
while the long roots $2e_i$ and the fundamental hypermultiplets give $(q^{\frac12}t^{-2})^{-2m_i}$ and $(q^{\frac12}t^{-2})^{lm_i}$.
The zero weights of the antisymmetric representation contribute only to the prefactor.
Here the Dirichlet half-index is (\ref{Dhalf}) with the roots and weights of $USp(2N)$,
and the Goldstone modes and the split Fermi multiplets in (\ref{Nahm_limit}) are given by $\Delta^+$ and the spins above.
For $N=1$ and $l=4$ it is (\ref{USp2_simple}).
We have confirmed that (\ref{USp_simple}) agrees with the limit (\ref{Nahm_limit}) of the Dirichlet half-index for $l=3$ and $N\le3$ (see Appendix \ref{app_checks}).

\paragraph{Mirror quiver}
We denote the gauge fugacities for the node $\alpha_I$ of rank $N_I$ by $s^{(I)}_i$, $i=1,\cdots,N_I$,
and write $\mathbf{S}_I=\sum_i\mathbf{s}^{(I)}_i$.
For the Neumann boundary condition $(\mathcal{N},N,N)$ the non-Abelian part of the boundary gauge anomaly cancels at the balanced nodes as for the necklace quiver.
The remaining part is $-\sum_{I,J=0}^lC_{IJ}\mathbf{S}_I\mathbf{S}_J-\sum_{i=1}^N(\mathbf{s}^{(0)}_i)^2$,
where $C_{IJ}=\alpha_I\cdot\alpha_J$ is the Cartan matrix of the affine $D_l$ algebra (see (\ref{USp_rv_Neu_anom}) for the full anomaly).
It is cancelled by the fundamental Fermi multiplet $\tilde{\Gamma}$ and $l$ Fermi multiplets $\tilde{\eta}_a$, $a=1,\cdots,l$,
where the Fermi multiplet $\tilde{\eta}_a$ has the charge $(\alpha_I)_a$, the $a$-th component of $\alpha_I$, under the determinant of the node $\alpha_I$.
Indeed, $\sum_a(\alpha_I)_a(\alpha_J)_a=C_{IJ}$.
They correspond to the $N$ top components of the antisymmetric hypermultiplet
and to the top components of the $l$ fundamental hypermultiplets of the $USp(2N)$ ADHM theory at the Nahm pole.
The Neumann half-index of the mirror quiver reads
\begin{align}
\label{USp_rv_Neu}
&\mathbb{II}^{\wtilde{\hat{D}_l}}_{\mathcal{N},N,N+\tilde{\Gamma}+\tilde{\eta}}
\nonumber\\*
&=\prod_{I=0}^l\left[\frac{1}{N_I!}\oint\prod_{i=1}^{N_I}\frac{ds^{(I)}_i}{2\pi is^{(I)}_i}\right]
\underbrace{[(q)_{\infty}(q^{\frac12}t^{-2};q)_{\infty}]^{2(l-1)N}\prod_{I=0}^l\prod_{i\neq j}\left(\frac{s^{(I)}_i}{s^{(I)}_j};q\right)_{\infty}\left(q^{\frac12}t^{-2}\frac{s^{(I)}_i}{s^{(I)}_j};q\right)_{\infty}}_{\textrm{vector multiplets ($\mathcal{N}$)}}
\nonumber\\
&\times \underbrace{\prod_{\alpha_I\cdot\alpha_J=-1}\prod_{i=1}^{N_I}\prod_{j=1}^{N_J}\frac{1}{(q^{\frac14}t^{-1}(s^{(I)}_i/s^{(J)}_j)^{\pm};q)_{\infty}}}_{\textrm{bifundamental hypermultiplets ($N$)}}\;
\underbrace{\prod_{i=1}^N\frac{1}{(q^{\frac14}t^{-1}(s^{(0)}_i)^{\pm};q)_{\infty}}}_{\textrm{fundamental hypermultiplet ($N$)}}
\nonumber\\*
&\times \underbrace{\prod_{i=1}^NF\left(q^{\frac12}\frac{s^{(0)}_i}{x};q\right)}_{\tilde{\Gamma}}
\underbrace{\prod_{a=1}^lF\left(q^{\frac12}y_a\prod_{I=0}^l\Bigl(\prod_is^{(I)}_i\Bigr)^{(\alpha_I)_a};q\right)}_{\tilde{\eta}_a},
\end{align}
where the bifundamental hypermultiplets run over the pairs of the connected nodes.
For $N=1$ and $l=4$ it is (\ref{USp2_rv_Neu}).
Unlike the Fermi multiplets $\tilde{\eta}_{I,I+1}$ in (\ref{ADHM_rv_Neu}), the Fermi multiplets $\tilde{\eta}_a$ carry no $U(1)_u$ charge.

\paragraph{Duality}
We propose that
\begin{align}
\label{USp_rv_dual}
\mathbb{II}^{USp(2N)\,\textrm{ADHM-}[2l]}_{\mathrm{Nahm}'}
&=\mathbb{II}^{\wtilde{\hat{D}_l}}_{\mathcal{N},N,N+\tilde{\Gamma}+\tilde{\eta}}.
\end{align}
For $N=1$ and $l=4$ it is (\ref{USp2_rv_dual}).

As a consistency check, we compare the boundary anomalies.
For the Neumann boundary condition (\ref{USp_rv_Neu}) of the mirror quiver the boundary anomaly is
\begin{align}
\label{USp_rv_Neu_anom}
&\mathcal{A}^{\wtilde{\hat{D}_l}}_{\mathcal{N},N,N+\tilde{\Gamma}+\tilde{\eta}}
\nonumber\\*
&=\underbrace{\sum_{I=0}^l\sum_{i\neq j}(\mathbf{s}^{(I)}_i-\mathbf{s}^{(I)}_j)^2}_{\textrm{vector multiplets ($\mathcal{N}$)}}\quad
\underbrace{-\sum_{\alpha_I\cdot\alpha_J=-1}\sum_{i,j}(\mathbf{s}^{(I)}_i-\mathbf{s}^{(J)}_j)^2}_{\textrm{bifundamental hypermultiplets ($N$)}}\quad
\underbrace{-\sum_{i=1}^N(\mathbf{s}^{(0)}_i)^2}_{\textrm{fundamental hypermultiplet ($N$)}}
\nonumber\\*
&\underbrace{+2\mathbf{S}_0\mathbf{x}-2\sum_{I=0}^l\mathbf{S}_I\,\alpha_I\cdot\mathbf{y}}_{\textrm{FI couplings}}
\quad\underbrace{+\sum_{i=1}^N(\mathbf{s}^{(0)}_i-\mathbf{x})^2}_{\tilde{\Gamma}}
\quad+\underbrace{\sum_{a=1}^l\Bigl(\sum_{I=0}^l(\alpha_I)_a\mathbf{S}_I+\mathbf{y}_a\Bigr)^2}_{\tilde{\eta}_a},
\end{align}
where $\alpha_I\cdot\mathbf{y}=\sum_a(\alpha_I)_a\mathbf{y}_a$.
All the terms with the gauge field strengths cancel.
The FI couplings reflect the mirror map,
under which the topological fugacity for the node $\alpha_I$ is $\prod_ay_a^{(\alpha_I)_a}$, multiplied by $x^{-1}$ for $I=0$.
By (\ref{Nahm_anom}), the boundary anomaly for the Nahm pole side is
\begin{align}
\mathcal{A}^{USp(2N)\,\textrm{ADHM-}[2l]}_{\mathrm{Nahm}'}&=\underbrace{\sum_{a=1}^l\mathbf{y}_a^2+N\mathbf{x}^2}_{\textrm{top components}}
=\mathcal{A}^{\wtilde{\hat{D}_l}}_{\mathcal{N},N,N+\tilde{\Gamma}+\tilde{\eta}}.
\end{align}
Hence the anomalies agree exactly.
Given the mirror map, the anomalies thus fix the charges of the Fermi multiplets $\tilde{\eta}_a$ by $\sum_a(\alpha_I)_a(\alpha_J)_a=C_{IJ}$ and by the mixed terms.

\subsection{Brane configuration}
\label{sec_USp_brane}
For the duality (\ref{USp_fw_dual}) the NS5$'$-brane intersects the $l-2$ D5-branes in Figure \ref{fig_USpADHM} (a), which are S-dual to the NS5-branes in (b),
and the NS5-brane on the O5$^-$-plane, which supports the Fermi multiplet $\eta$.
The $l-4$ middle D5-branes in (a), S-dual to the NS5-branes between the segments of the $U(2N)$ nodes in (b), give the fundamental Fermi multiplets $\Gamma_{k,k+1}$ as in Figure \ref{fig_local} (b).
The two D5-branes next to the O5$^-$-planes correspond to the Fermi multiplets $\Gamma_{I,J}$ of the end nodes.
We leave the derivation of their gauge neutrality and of their shifted fugacities from the brane configuration for future work.

For the duality (\ref{USp_rv_dual}) the Fermi multiplets $\tilde{\eta}_a$ with $2\le a\le l-1$ follow from Figure \ref{fig_USpADHM} (d) by the rules in section \ref{sec_dualrules}.
We label the NS5-branes by $a=2,\cdots,l-1$ from left to right.
At the junction of the $a$-th NS5-brane and the NS5$'$-brane,
the Fermi multiplet $\tilde{\eta}_a$ carries the determinant of the gauge group on the right and the inverse determinant of that on the left.
For $a=2$ the gauge group on the left is $U(N)\times U(N)$ of the nodes $\alpha_0$ and $\alpha_1$,
and for $a=l-1$ that on the right is $U(N)\times U(N)$ of the nodes $\alpha_{l-1}$ and $\alpha_l$.
These charges are the components $(\alpha_I)_a$ of the simple roots (\ref{Dl_roots}).
The other two Fermi multiplets $\tilde{\eta}_1$ and $\tilde{\eta}_l$ couple to the two end nodes at an ON$^0$-plane with the opposite charges.
The rules in section \ref{sec_dualrules} do not directly apply to the ON$^0$-planes.
Since an ON$^0$-plane carries the charge of an NS5-brane,
we expect a cross-determinant Fermi multiplet at its junction with the NS5$'$-brane.
Then the Fermi multiplets $\tilde{\eta}_a$ correspond to the NS5-branes and the ON$^0$-planes,
i.e. to the $l$ D5-branes on the Nahm pole side, as do the top components of the $l$ fundamental hypermultiplets.
Given the other Fermi multiplets, the anomaly cancellation fixes the charges of the Fermi multiplets $\tilde{\eta}_1$ and $\tilde{\eta}_l$.
The D5-brane on the ON$^0$-plane gives the twisted fundamental hypermultiplet of the node $\alpha_0$
and the fundamental Fermi multiplet $\tilde{\Gamma}$ as in Figure \ref{fig_local} (b).

\subsection{\texorpdfstring{Relation to the $U(N)$ ADHM theory}{Relation to the U(N) ADHM theory}}
\label{sec_USp_ADHM4}
For $l=3$ the Neumann boundary conditions (\ref{ADHM_rv_Neu}) with $l=4$ and (\ref{USp_rv_Neu}) with $l=3$ are defined for the same necklace quiver.
We identify the nodes $\alpha_0$, $\alpha_2$, $\alpha_1$ and $\alpha_3$ with the nodes $I=1,2,3,4$ of the necklace quiver and set $z=1$.
Then the two boundary conditions differ only in the Fermi multiplets,
the four Fermi multiplets $\tilde{\eta}_{I,I+1}$ with the boundary $U(1)_u$ charge and the three Fermi multiplets $\tilde{\eta}_a$.
By the Jacobi triple product we find
\begin{align}
\label{theta_A3D3}
\frac12\sum_{\pm}\prod_{I=1}^4F(\pm q^{\frac12}X_Iu;q)\Big|_{u^2=-q^{-\frac12}}
&=\frac{F(q^{\frac12}X_1X_2;q)F(q^{\frac12}X_1X_3;q)F(q^{\frac12}X_1X_4;q)}{(q;q^2)_{\infty}^2}
\end{align}
for $X_1X_2X_3X_4=1$.
For $X_I=y_I\prod_is^{(I)}_i/\prod_is^{(I+1)}_i$ the left-hand side is the average of the contributions of the Fermi multiplets $\tilde{\eta}_{I,I+1}$ over $u=\pm iq^{-\frac14}$,
and the right-hand side is $(q;q^2)_{\infty}^{-2}$ times those of the Fermi multiplets $\tilde{\eta}_a$
with the $SO(6)$ fugacities $y_3y_4$, $y_2y_4$ and $y_1y_4$ given by the $SU(4)$ fugacities $y_I$ of the $U(N)$ ADHM theory.
Hence, given the duality (\ref{ADHM_rv_dual}), the duality (\ref{USp_rv_dual}) for $l=3$ is equivalent to
\begin{align}
\label{USp_ADHM4}
\mathbb{II}^{USp(2N)\,\textrm{ADHM-}[6]}_{\mathrm{Nahm}'}
&=(q;q^2)_{\infty}^2\,\frac12\sum_{\pm}\mathbb{II}^{U(N)\,\textrm{ADHM-}[4]}_{\mathrm{Nahm}'}\Big|_{u=\pm iq^{-\frac14},\,z=1}.
\end{align}
The average over $u\rightarrow-u$, which acts as $-\mathbf{1}_4\in SU(4)$ on the fundamental hypermultiplets,
restricts $SU(4)$ to $SO(6)=SU(4)/\mathbb{Z}_2$.

\section{Twist limits and boundary VOAs}
\label{sec_VOA}
In this section we study the twist limits of the half-indices.
In these limits the formula (\ref{Nahm_simple}) gives closed forms of the Nahm pole half-indices (section \ref{sec_VOA_closed}),
which we apply to $U(N)$ SQCD, $T[SU(N)]$ and the $U(N)$ and $USp(2N)$ ADHM theories.
We study the quasimodularity of their unflavored limits (section \ref{sec_VOA_qmod}).
We then discuss the boundary VOAs whose vacuum characters they are expected to be (section \ref{sec_VOA_general}).

\subsection{Twist limits}
\label{sec_VOA_limits}
We call the following limits of the fugacity $t$ the H-twist limit and the C-twist limit \cite{Gaiotto:2019jvo,Okazaki:2019bok},
\begin{align}
\label{twist_limits}
\textrm{H-twist limit:}&\quad t\rightarrow q^{\frac14}, &
\textrm{C-twist limit:}&\quad t\rightarrow q^{-\frac14}.
\end{align}
The Neumann boundary condition $(\mathcal{N}',N')$, with the boundary gauge anomaly cancelled, admits a deformation compatible with the H-twist,
and the Dirichlet boundary condition $(\mathcal{D}',D')$ is expected to admit one compatible with the C-twist \cite{Costello:2018fnz}.
With these deformations the boundary local operators form VOAs.
The Nahm pole boundary condition $(\mathrm{Nahm}',D')$, which arises from $(\mathcal{D}',D')$ by Higgsing, is expected to admit one compatible with the C-twist as well.
In the corresponding twist limit the half-index stays finite (see section \ref{sec_VOA_general}),
although for the Nahm pole boundary condition the individual factors in (\ref{Nahm_limit}) are singular.
We denote the twist limit of $\mathbb{II}^{\mathcal{T}}_{\mathcal{B}}$ by $\chi[\mathsf{V}^{\mathcal{T}}_{\mathcal{B}}]$,
anticipating that it is the vacuum character of the boundary VOA $\mathsf{V}^{\mathcal{T}}_{\mathcal{B}}$ for the boundary condition $\mathcal{B}$ in the theory $\mathcal{T}$
(see section \ref{sec_VOA_general}).

For the Dirichlet boundary condition $(\mathcal{D}',D')$ in the C-twist limit, the half-index (\ref{Dhalf}) becomes the vacuum character of the boundary VOA $\mathsf{V}^{\mathcal{T}}_{\mathcal{D}',D'}$ (see section \ref{sec_VOA_DS})
\begin{align}
\label{Dhalf_C}
\chi[\mathsf{V}^{\mathcal{T}}_{\mathcal{D}',D'}](u)
&=\frac{1}{(q)_{\infty}^{2\,\mathrm{rk}\,G}}\sum_{m}(-q^{\frac12})^{m^Tk_{\rm eff}m}u^{k_{\rm eff}m}\prod_{\lambda}x^{w_{\lambda}(m)f_{\lambda}}\prod_{I}z_I^{-\sum_im^{(I)}_i}
\nonumber\\*
&\times \prod_{\alpha\in\Delta}\frac{1}{(q^{1+\alpha(m)}u^{\alpha};q)_{\infty}^2}
\prod_{\lambda}(q^{1\pm w_{\lambda}(m)}(u^{w_{\lambda}}x^{f_{\lambda}})^{\pm};q)_{\infty}.
\end{align}

Since mirror symmetry exchanges $SU(2)_H$ and $SU(2)_C$, the H-twist of a theory is the C-twist of its mirror.
As we write the mirror theory in terms of the twisted multiplets (see section \ref{sec_conv}),
the two sides of each dual pair stay finite in the same twist limit of $t$.

\subsection{Closed-form formula}
\label{sec_VOA_closed}
\subsubsection{Nahm pole VOA character}
\label{sec_VOA_charformula}
The formula (\ref{Nahm_simple}) gives the C-twist limit $t\rightarrow q^{-\frac14}$ of the Nahm pole half-index immediately.
Let
\begin{align}
p^{(I)}_i&=m^{(I)}_i+\mu^{(I)}_i,
\end{align}
which are strictly increasing at each Nahm pole node and take values in $\mathbb{Z}+\frac{N_I-1}{2}$.
The Macdonald polynomials in (\ref{Mac_weight}) reduce to the Schur polynomials at $\mathfrak{t}=\mathfrak{q}$,
and the weight of the vector multiplet becomes $\prod_{\alpha\in\Delta^+_{\rho}}q^{-\alpha(m)}(1-q^{\alpha(p)})^2/(1-q^{\mathrm{ht}\alpha})^2$.
The factors $(1-q^{\mathrm{ht}\alpha})^{-2}$ combine with the Slodowy slice into $(q)_{\infty}^{-2\,\mathrm{rk}\,G}$.
The hypermultiplet factor becomes $q^{\frac{w_{\lambda}(m)}{2}}(1-q^{w_{\lambda}(\mu)}\bar{u}^{w_{\lambda}}x^{f_{\lambda}})/(1-q^{w_{\lambda}(p)}\bar{u}^{w_{\lambda}}x^{f_{\lambda}})$,
and for each spin-$J$ multiplet the numerators combine with the top component into $\theta(q^{-J}\bar{u}^{w_{\lambda}}x^{f_{\lambda}};q)$,
where $\theta(x;q)=(x;q)_{\infty}(qx^{-1};q)_{\infty}$ is the theta function.
Using $\sum_{\alpha\in\Delta^+_{\rho}}\alpha(m)=\sum_{\alpha\in\Delta^+_{\rho}}(\alpha(p)-\mathrm{ht}\,\alpha)$, $\sum_{\lambda}w_{\lambda}(\mu)=0$ and $\sum_i\mu^{(I)}_i=0$, we obtain
\begin{align}
\label{Nahm_closed}
\chi[\mathsf{V}^{\mathcal{T}}_{\mathrm{Nahm}'}](\bar{u})
&=(-1)^{|\Delta^+_{\rho}|}q^{\sum_{\alpha\in\Delta^+_{\rho}}\mathrm{ht}\alpha}\,
\frac{\prod_{(J,\lambda)}\theta(q^{-J}\bar{u}^{w_{\lambda}}x^{f_{\lambda}};q)}{(q)_{\infty}^{2\,\mathrm{rk}\,G}}
\nonumber\\*
&\times\sum_{p}\prod_Iz_I^{-|p^{(I)}|}\,q^{\frac12\sum_{\lambda}w_{\lambda}(p)}
\frac{\prod_{\alpha\in\Delta_{\rho}}(1-q^{\alpha(p)})}{\prod_{\lambda}(1-q^{w_{\lambda}(p)}\bar{u}^{w_{\lambda}}x^{f_{\lambda}})},
\end{align}
where the sum runs over the strictly increasing $p$, $|p^{(I)}|=\sum_ip^{(I)}_i$, $\mathrm{ht}\,\alpha$ is the height (\ref{ht_def}) and $\Delta_{\rho}$ is the set of roots of the Nahm pole nodes.
For a single $U(1)$ node with one hypermultiplet, (\ref{Nahm_closed}) reduces to Kronecker's formula \cite{zbMATH02706826,MR1723749,MR1106744}
\begin{align}
\label{Kronecker}
\sum_{n\in\mathbb{Z}}\frac{w^n}{1-aq^n}&=\frac{(q)_{\infty}^2\,\theta(aw;q)}{\theta(a;q)\,\theta(w;q)},
\end{align}
where the sum converges for $|q|<|w|<1$.
It is Ramanujan's ${}_1\psi_1$ summation formula (\ref{1psi1_Ramanujan}) with $b=aq$, after dividing by $1-a$.
It is the C-twist limit of the minimal mirror identity (\ref{minimal_mirror}) and the identity for $U(1)$ gauge theory with one hypermultiplet in \cite{Costello:2018fnz}.

\subsubsection{Jackson integral}
\label{sec_VOA_Jackson}
The sum in (\ref{Nahm_closed}) is a Jackson integral.
A Jackson integral is a $q$-analogue of an integral,
in which the integral is replaced by the sum over a lattice of points in geometric progression \cite{MR2128719}.
For a function $f(s)$ and $\xi\in\mathbb{C}^{\times}$ it is defined by
\begin{align}
\label{Jackson_def}
\int_0^{\xi\infty}f(s)\frac{d_qs}{s}&=(1-q)\sum_{p\in\mathbb{Z}}f(\xi q^p).
\end{align}
For $\xi=1$ it reduces to $\int_0^{\infty}f(s)\frac{ds}{s}$ as $q\rightarrow1$,
since the right-hand side is a Riemann sum over the points $s=q^p$ with the spacings $(1-q)s$.
With the variables $s^{(I)}_i$ for all the nodes and $z_I=q^{\zeta_I}$, (\ref{Nahm_closed}) reads
\begin{align}
\label{Nahm_Jackson}
\chi[\mathsf{V}^{\mathcal{T}}_{\mathrm{Nahm}'}](\bar{u})
&=\frac{(-1)^{|\Delta^+_{\rho}|}q^{\sum_{\alpha\in\Delta^+_{\rho}}\mathrm{ht}\alpha}}{|\mathrm{Weyl}(G)|(1-q)^{\mathrm{rk}\,G}}\,
\frac{\prod_{(J,\lambda)}\theta(q^{-J}\bar{u}^{w_{\lambda}}x^{f_{\lambda}};q)}{(q)_{\infty}^{2\,\mathrm{rk}\,G}}
\nonumber\\*
&\times\int_0^{\xi\infty}\prod_{I,i}\frac{d_qs^{(I)}_i}{s^{(I)}_i}\,(s^{(I)}_i)^{-\zeta_I}\,
\prod_{\lambda}(s^{w_{\lambda}})^{\frac12}\,
\frac{\prod_{\alpha\in\Delta_{\rho}}(1-s^{\alpha})}{\prod_{\lambda}(1-s^{w_{\lambda}}\bar{u}^{w_{\lambda}}x^{f_{\lambda}})},
\end{align}
where $\xi$ is $\xi_I=q^{\frac{N_I-1}{2}}$ for $s^{(I)}_i$, so that $s^{(I)}_i$ runs over the lattice $\xi_Iq^{\mathbb{Z}}$,
$s^{\alpha}$ and $s^{w_{\lambda}}$ are the corresponding monomials in $s^{(I)}_i$, and $|\mathrm{Weyl}(G)|$ is the order of the Weyl group of $G$, which is $\prod_IN_I!$ for $G=\prod_IU(N_I)$.

To see this, note that the points of the lattice are $s^{(I)}_i=q^{p^{(I)}_i}$ with $p^{(I)}_i\in\mathbb{Z}+\frac{N_I-1}{2}$.
At these points $s^{\alpha}=q^{\alpha(p)}$, $s^{w_{\lambda}}=q^{w_{\lambda}(p)}$ and $\prod_{I,i}(s^{(I)}_i)^{-\zeta_I}=\prod_Iz_I^{-|p^{(I)}|}$,
so that the integrand is the summand of (\ref{Nahm_closed}).
The summand is Weyl invariant and vanishes when two of the $p^{(I)}_i$ coincide,
so that the sum over the strictly increasing $p$ is $1/|\mathrm{Weyl}(G)|$ times the sum over all $p$.
This is the Weyl symmetry which appears for the other order of the limits in section \ref{sec_VOA_DS}.
The sum over all $p$ is the Jackson integral (\ref{Jackson_def}) in each variable divided by $(1-q)^{\mathrm{rk}\,G}$.

Thus (\ref{Nahm_Jackson}) is a $q$-analogue of an integral over the complexified maximal torus of $G$.
The points of the lattice are labeled by the fluxes of the boundary monopole sectors.
The factor $\prod_{\alpha}(1-s^{\alpha})$ is the Weyl denominator which appears in the Haar measure,
and the hypermultiplets give the poles at $s^{w_{\lambda}}=(\bar{u}^{w_{\lambda}}x^{f_{\lambda}})^{-1}$.
The shift by the Weyl vector turns the dominant coweights $m^{(I)}_1\le\cdots\le m^{(I)}_{N_I}$ into the strictly increasing $p^{(I)}_i$.
Since the integrand vanishes when two of them coincide, the sum takes the form of a partition function of $N_I$ fermions at each node.
The points of $\mathbb{Z}+\frac{N_I-1}{2}$ are the momenta of the one-particle states,
the fermions occupy different momenta $p^{(I)}_i$ by the Pauli exclusion principle,
and the summand is the Boltzmann weight of this configuration.
This leads to the free Fermi gas formulas below.

\subsubsection{Examples}
\label{sec_VOA_closed_examples}
We have checked the closed forms below against the twist limits of the half-indices (see Appendix \ref{app_checks}).
\paragraph{$U(N)$ SQCD}
\label{sec_VOA_SQCD}
We first consider the C-twist limit of (\ref{SQCD_rv_dual}).
On the SQCD side the closed form (\ref{Nahm_closed}) reads
\begin{align}
\label{SQCD_rv_closed}
\chi[\mathsf{V}^{U(N)+2N}_{\mathrm{Nahm}'}]
&=(-1)^{\frac{N(N-1)}{2}}q^{\frac{N(N^2-1)}{6}}\frac{\prod_{\alpha=1}^{2N}\theta(q^{-\frac{N-1}{2}}ux_{\alpha};q)}{(q)_{\infty}^{2N}}
\nonumber\\*
&\times\sum_{p_1<\cdots<p_N}\left(\frac{z_2}{z_1}\right)^{|p|}q^{N|p|}
\frac{\prod_{i\neq j}(1-q^{p_i-p_j})}{\prod_{i=1}^N\prod_{\alpha=1}^{2N}(1-q^{p_i}ux_{\alpha})},
\end{align}
where $p_i\in\mathbb{Z}+\frac{N-1}{2}$ and $|p|=\sum_ip_i$.
Each fundamental hypermultiplet is the spin-$\frac{N-1}{2}$ multiplet with $\bar{u}^{w_{\lambda}}x^{f_{\lambda}}=ux_{\alpha}$,
and (\ref{SQCD_rv_closed}) is also obtained from (\ref{SQCD_rv_simple}) at $t=q^{-\frac14}$.
Since the denominator only contains the $2N$ hypermultiplets, we can expand it in partial fractions and perform the sums by Kronecker's formula (\ref{Kronecker}).
The factors $q^{kp_i}$ from the Vandermonde determinant are then replaced by $(ux_{\alpha})^{-k}$,
and the sum localizes to the subsets $S\subset\{1,\cdots,2N\}$ of $N$ flavors,
\begin{align}
\label{SQCD_rv_Higgs}
\chi[\mathsf{V}^{U(N)+2N}_{\mathrm{Nahm}'}]
&=(-1)^{\frac{N(N-1)}{2}}q^{\frac{N(N^2-1)}{6}}\frac{(z_2/z_1)^{\frac{N(N-1)}{2}}}{\theta(z_1/z_2;q)^N}
\nonumber\\*
&\times\sum_{|S|=N}\prod_{\alpha\in S}\theta\Bigl(q^{\frac{N+1}{2}}\frac{z_2}{z_1}ux_{\alpha};q\Bigr)\prod_{\beta\notin S}\theta(q^{-\frac{N-1}{2}}ux_{\beta};q)
\prod_{\alpha\in S}\prod_{\beta\notin S}\frac{x_{\alpha}}{x_{\alpha}-x_{\beta}}.
\end{align}
The $\binom{2N}{N}$ terms are labeled by the choices of $N$ flavors.
The sum in (\ref{SQCD_rv_closed}) is the Jackson integral of type $A$ with $N$ variables in \cite{MR3831248}
with its parameter $t$ set to $q$, with the $2N$ pairs of parameters $(ux_{\alpha})^{-1}$ and $ux_{\alpha}$ and the power $(s_1\cdots s_N)^{\gamma}$ with $q^{\gamma}=qz_2/z_1$.
These values lie outside the range of its connection formula \cite{MR3831248},
and (\ref{SQCD_rv_Higgs}) follows instead from the partial fractions above.
Thus the C-twist limit of (\ref{SQCD_rv_dual}) expresses the twist limit of the Neumann half-index for the mirror quiver
as a finite sum of products of theta functions.

In the H-twist limit of (\ref{SQCD_fw_dual}), the closed form (\ref{Nahm_closed}) with $t\rightarrow t^{-1}$ applies to the mirror quiver $(1)-(2)-\cdots-(N)-\cdots-(2)-(1)$
with the ranks $r_I=\min(I,2N-I)$.
With the spins of the top components in (\ref{SQCD_fw_simple}) we obtain
\begin{align}
\label{SQCD_fw_closed}
&\chi[\mathsf{V}^{\wtilde{(1)-(2)-\cdots-(N)-\cdots-(2)-(1)}}_{\mathcal{D},\mathrm{Nahm}^{2N-3},\mathcal{D}}]
\nonumber\\*
&=\frac{(-1)^{\frac{N(N-1)(2N-1)}{6}}q^{\frac{N^2(N^2-1)}{12}}}{(q)_{\infty}^{2N^2}}
\prod_{a=1}^2\theta\Bigl(q^{-\frac{N-1}{2}}\frac{u^{(N)}}{z_a};q\Bigr)
\prod_{I=1}^{2N-2}\prod_{J=\frac12}^{\min(r_I,r_{I+1})-\frac12}\theta\Bigl(q^{-J}\frac{u^{(I)}}{u^{(I+1)}};q\Bigr)
\nonumber\\
&\times\sum_{p}\prod_{I=1}^{N}\Bigl(q\frac{x_{I+1}}{x_I}\Bigr)^{|p^{(I)}|}\prod_{I=N+1}^{2N-1}\Bigl(q^{-1}\frac{x_{I+1}}{x_I}\Bigr)^{|p^{(I)}|}
\prod_{I=1}^{2N-1}\prod_{i\neq j}(1-q^{p^{(I)}_i-p^{(I)}_j})
\nonumber\\*
&\times\prod_{i=1}^N\prod_{a=1}^2\frac{1}{1-q^{p^{(N)}_i}u^{(N)}/z_a}
\prod_{I=1}^{2N-2}\prod_{i=1}^{r_I}\prod_{j=1}^{r_{I+1}}\frac{1}{1-q^{p^{(I)}_i-p^{(I+1)}_j}u^{(I)}/u^{(I+1)}},
\end{align}
where $p^{(I)}_i\in\mathbb{Z}+\frac{r_I-1}{2}$ are strictly increasing in $i$.
Here $q^{\frac12\sum_{\lambda}w_{\lambda}(p)}=q^{\sum_{I\le N}|p^{(I)}|-\sum_{I>N}|p^{(I)}|}$ is combined with the topological charges.
By (\ref{SQCD_fw_dual}), (\ref{SQCD_fw_closed}) is also the H-twist limit of the Neumann half-index (\ref{SQCD_fw_Neu}) for $U(N)$ SQCD.
The unflavored characters are listed in Appendix \ref{app_vac}.

\paragraph{$T[SU(N)]$}
\label{sec_VOA_TSUN}
In the H-twist limit of (\ref{TSUN_dual}), the closed form (\ref{Nahm_closed}) with $t\rightarrow t^{-1}$ applies to the mirror $T[SU(N)]$, including its $U(1)$ node with the Dirichlet boundary condition.
With the spins of the top components in (\ref{TSUN_simple}) we obtain
\begin{align}
\label{TSUN_closed}
\chi[\mathsf{V}^{\wtilde{T[SU(N)]}}_{\mathcal{D},\mathrm{Nahm}^{N-2}}]
&=\frac{(-1)^{\binom{N}{3}}q^{\binom{N+1}{4}}}{(q)_{\infty}^{N(N-1)}}
\prod_{I=1}^{N-2}\prod_{J=\frac12}^{I-\frac12}\theta\Bigl(q^{-J}\frac{u^{(I)}}{u^{(I+1)}};q\Bigr)
\prod_{a=1}^N\theta\Bigl(q^{-\frac{N-2}{2}}\frac{u^{(N-1)}}{z_a};q\Bigr)
\nonumber\\*
&\times\sum_{p}\prod_{I=1}^{N-1}\Bigl(q\frac{x_{I+1}}{x_I}\Bigr)^{|p^{(I)}|}\prod_{i\neq j}(1-q^{p^{(I)}_i-p^{(I)}_j})
\prod_{i=1}^{N-1}\prod_{a=1}^N\frac{1}{1-q^{p^{(N-1)}_i}u^{(N-1)}/z_a}
\nonumber\\*
&\times\prod_{I=1}^{N-2}\prod_{i=1}^{I}\prod_{j=1}^{I+1}\frac{1}{1-q^{p^{(I)}_i-p^{(I+1)}_j}u^{(I)}/u^{(I+1)}},
\end{align}
where $p^{(I)}_i\in\mathbb{Z}+\frac{I-1}{2}$ are strictly increasing in $i$.
Here $q^{\frac12\sum_{\lambda}w_{\lambda}(p)}=q^{\sum_I|p^{(I)}|}$ is combined with the topological charges.
For $N=2$, (\ref{TSUN_closed}) is (\ref{SQCD_rv_closed}) with $N=1$, $ux_{\alpha}\rightarrow u^{(1)}/z_a$ and $z_2/z_1\rightarrow x_2/x_1$.
Then (\ref{SQCD_rv_Higgs}) gives the vacuum character as a sum of two products of theta functions,
\begin{align}
\label{TSU2_closed}
\chi[\mathsf{V}^{T[SU(2)]}_{\mathcal{N}',N'+\eta}]
&=\frac{1}{\theta(x_1/x_2;q)}\sum_{a\neq b}\frac{z_b}{z_b-z_a}\,
\theta\Bigl(q\frac{x_2}{x_1}\frac{u^{(1)}}{z_a};q\Bigr)\theta\Bigl(\frac{u^{(1)}}{z_b};q\Bigr),
\end{align}
which reduces to $(q)_{\infty}^2$ in the unflavored limit.

\paragraph{ADHM theory}
\label{sec_VOA_ADHM}
In the H-twist limit of (\ref{ADHM_dual}), the closed form (\ref{Nahm_closed}) with $t\rightarrow t^{-1}$ applies to the mirror necklace quiver $U(N)^l$.
Each twisted bifundamental hypermultiplet decomposes into the spins $J=0,\cdots,N-1$.
We take its fugacity to be $zu^{(I+1)}/u^{(I)}$, so that for $l=1$ it is the twisted adjoint hypermultiplet with fugacity $z$.
The twisted fundamental hypermultiplet with fugacity $u^{(1)}$ is the spin-$\frac{N-1}{2}$ multiplet.
We obtain
\begin{align}
\label{ADHM_closed}
\chi[\mathsf{V}^{\wtilde{U(N)^l}}_{\mathrm{Nahm}^l}]
&=(-1)^{\frac{lN(N-1)}{2}}q^{\frac{lN(N^2-1)}{6}}\frac{\prod_{I=1}^l\prod_{J=0}^{N-1}\theta(q^{-J}zu^{(I+1)}/u^{(I)};q)\,\theta(q^{-\frac{N-1}{2}}u^{(1)};q)}{(q)_{\infty}^{2lN}}
\nonumber\\*
&\times\sum_{p}\frac{(q^{\frac12}x^{-1})^{|p^{(1)}|}\prod_{I=1}^ly_I^{|p^{(I)}|-|p^{(I+1)}|}}{\prod_{i=1}^N(1-q^{p^{(1)}_i}u^{(1)})}
\prod_{I=1}^l\frac{\prod_{i\neq j}(1-q^{p^{(I)}_i-p^{(I)}_j})}{\prod_{i,j=1}^N(1-zq^{p^{(I+1)}_j-p^{(I)}_i}u^{(I+1)}/u^{(I)})},
\end{align}
where $p^{(I)}_i\in\mathbb{Z}+\frac{N-1}{2}$ are strictly increasing in $i$, $p^{(l+1)}\equiv p^{(1)}$ and $u^{(l+1)}\equiv u^{(1)}$.
Here $q^{\frac12\sum_{\lambda}w_{\lambda}(p)}=q^{\frac12|p^{(1)}|}$, since the weights of the twisted bifundamental hypermultiplets add up to zero.
The vector multiplets and the twisted bifundamental hypermultiplets combine into the Cauchy determinants
\begin{align}
\prod_{I=1}^l\frac{\prod_{i\neq j}(1-q^{p^{(I)}_i-p^{(I)}_j})}{\prod_{i,j}(1-zq^{p^{(I+1)}_j-p^{(I)}_i}u^{(I+1)}/u^{(I)})}
&=z^{-\frac{lN(N-1)}{2}}\prod_{I=1}^l\det_{i,j}\left(\frac{1}{1-zq^{p^{(I+1)}_j-p^{(I)}_i}u^{(I+1)}/u^{(I)}}\right),
\end{align}
and $\prod_{J=0}^{N-1}\theta(q^{-J}a;q)=(-1)^{\frac{N(N-1)}{2}}q^{-\frac{N(N^2-1)}{6}}a^{\frac{N(N-1)}{2}}\theta(a;q)^N$.
Thus (\ref{ADHM_closed}) becomes
\begin{align}
\label{ADHM_closed_Fermi}
\chi[\mathsf{V}^{\wtilde{U(N)^l}}_{\mathrm{Nahm}^l}]
&=\frac{\prod_{I=1}^l\theta(zu^{(I+1)}/u^{(I)};q)^N\,\theta(q^{-\frac{N-1}{2}}u^{(1)};q)}{(q)_{\infty}^{2lN}}
\nonumber\\*
&\times\sum_{p}\prod_{I=1}^l\det_{i,j}\left(\frac{(y_I/y_{I-1})^{p^{(I)}_i}}{1-zq^{p^{(I+1)}_j-p^{(I)}_i}u^{(I+1)}/u^{(I)}}\right)
\prod_{i=1}^N\frac{(q^{\frac12}x^{-1})^{p^{(1)}_i}}{1-q^{p^{(1)}_i}u^{(1)}}
\end{align}
with $y_0\equiv y_l$.
By the Cauchy-Binet formula \cite{MR1657129} the sums over $p^{(2)},\cdots,p^{(l)}$ give the products of the matrices,
so that (\ref{ADHM_closed_Fermi}) is the canonical partition function of $N$ fermions with the density matrix $\rho_1\rho_2\cdots\rho_l$, where
\begin{align}
\rho_1(p,p')&=\frac{(q^{\frac12}x^{-1}y_1/y_l)^p}{(1-q^pu^{(1)})(1-zq^{p'-p}u^{(2)}/u^{(1)})},
\\*
\rho_I(p,p')&=\frac{(y_I/y_{I-1})^p}{1-zq^{p'-p}u^{(I+1)}/u^{(I)}}\qquad(I\ge2).
\end{align}
For $z=1$ and $v_I=v_1$, i.e. $zu^{(I+1)}/u^{(I)}=1$, the $I$-th determinant has a pole of order $N$ only for $p^{(I+1)}=p^{(I)}$,
which is cancelled by $\theta(zu^{(I+1)}/u^{(I)};q)^N$.
Hence the sum localizes to $p^{(1)}=\cdots=p^{(l)}$, the fugacities $y_I$ cancel, and the fermions become free.
On the other side, for $z=1$ and $v_I=v_1$ the Fermi multiplets $\Gamma_{I,I+1}$ cancel the fundamental hypermultiplets,
and the Neumann half-index for the ADHM theory reduces to the factor from the Fermi multiplet $\eta$
times the Schur index of 4d $\mathcal{N}=4$ $U(N)$ SYM theory flavored by $x$ \cite{Okazaki:2025rjl}.
Hence the duality (\ref{ADHM_dual}) gives for any $l$
\begin{align}
\label{Schur_Fermi}
\mathcal{I}^{\textrm{Schur}}_{U(N)}(x;q)
&=\frac{\theta(q^{-\frac{N-1}{2}}u^{(1)};q)}{\theta(q^{\frac12}u^{(1)}x^{-N};q)}
\sum_{\substack{p_1<\cdots<p_N\\ p_i\in\mathbb{Z}+\frac{N-1}{2}}}\prod_{i=1}^N\frac{(q^{\frac12}x^{-1})^{p_i}}{1-q^{p_i}u^{(1)}},
\end{align}
where $u^{(1)}=v_1x^{N-1}$ and $\theta(q^{\frac12}u^{(1)}x^{-N};q)=F(q^{\frac12}v_1x^{-1};q)$ is the contribution of the Fermi multiplet $\eta$.
This is the free Fermi gas formula for the Schur index in \cite{Hatsuda:2022xdv}.\footnote{The parameter $\xi$ in \cite{Hatsuda:2022xdv} is $q^{-\frac12}x$,
and the auxiliary parameter $u$ there is $u^{(1)}$ for odd $N$ and $q^{\frac12}u^{(1)}$ for even $N$, up to the normalization of the theta functions.
The Schur index does not depend on $u^{(1)}$, since $U(1)_{u^{(1)}}$ acts only on the Fermi multiplet $\eta$ on the Neumann side.}
Since it is derived there from the contour integral, the H-twist limit of (\ref{ADHM_dual}) at $z=1$ and $v_I=v_1$ holds for all $N$ and $l$.

In the C-twist limit of (\ref{ADHM_rv_dual}), the closed form (\ref{Nahm_closed}) on the ADHM side and the same manipulation as for (\ref{ADHM_closed_Fermi}) give
\begin{align}
\label{ADHM_rv_closed}
\chi[\mathsf{V}^{U(N)\,\textrm{ADHM-}[l]}_{\mathrm{Nahm}'}]
&=\frac{\theta(x;q)^N\prod_{I=1}^l\theta(q^{-\frac{N-1}{2}}uy_I;q)}{(q)_{\infty}^{2N}}
\nonumber\\*
&\times\sum_{p_1<\cdots<p_N}\det_{i,j}\left(\frac{1}{1-xq^{p_j-p_i}}\right)\prod_{i=1}^N\frac{(q^{\frac12}z^{-1})^{lp_i}}{\prod_{I=1}^l(1-q^{p_i}uy_I)}
\end{align}
with $\prod_Iy_I=1$ and $p_i\in\mathbb{Z}+\frac{N-1}{2}$.
For $l=1$ it is (\ref{ADHM_closed_Fermi}) for $l=1$ with $x\leftrightarrow z$ and $u^{(1)}\rightarrow u$.
For $x=1$ the fermions become free, and the C-twist limit of (\ref{ADHM_rv_dual}) gives
\begin{align}
\label{necklace_Fermi}
\chi[\mathsf{V}^{\wtilde{U(N)^l}}_{\mathcal{N},N,N+\tilde{\Gamma}+\tilde{\eta}}]
&=\prod_{I=1}^l\theta(q^{-\frac{N-1}{2}}uy_I;q)
\sum_{\substack{p_1<\cdots<p_N\\ p_i\in\mathbb{Z}+\frac{N-1}{2}}}\prod_{i=1}^N\frac{(q^{\frac12}z^{-1})^{lp_i}}{\prod_{I=1}^l(1-q^{p_i}uy_I)}.
\end{align}
The $l$ fundamental hypermultiplets of the ADHM theory give the $l$ factors in the denominator,
and for $l=1$ (\ref{necklace_Fermi}) is (\ref{Schur_Fermi}) with $x\rightarrow z$ and $u^{(1)}\rightarrow uy_1$,
multiplied by the contribution $\theta(q^{\frac12}uy_1z^{-N};q)$ of the Fermi multiplet $\tilde{\eta}_{1,1}$.
At $x=1$ the Neumann side is the Schur index of the 4d $\mathcal{N}=2$ $SU(N)^l$ necklace quiver gauge theory
whose $U(1)$ flavor symmetries are gauged with the Fermi multiplets $\tilde{\eta}_{I,I+1}$ (see (\ref{necklace_4d})).
Hence (\ref{necklace_Fermi}) is a free Fermi gas formula for this gauged Schur index.\footnote{The flavored Schur index of the 4d $SU(N)^l$ necklace quiver gauge theory itself is recovered
from the Fourier coefficients of (\ref{necklace_Fermi}) with respect to $y_I$,
which give its components with fixed differences of the charges under the $U(1)$ flavor symmetries.
It agrees with the Fermi gas formula in \cite{Bourdier:2015sga}.
There the $SU(N)$ constraints lead to the sum over the sectors labeled by $n\in\mathbb{Z}^l$ with $n_l=0$,
while in (\ref{necklace_Fermi}) the Fermi multiplets $\tilde{\eta}_{I,I+1}$ replace it by the dependence on the fugacities $uy_I$.}
The unflavored characters are listed in Appendix \ref{app_vac}.

\paragraph{$USp(2N)$ ADHM theory}
\label{sec_VOA_USp}
In the C-twist limit of (\ref{USp_rv_dual}), the closed form (\ref{Nahm_closed}) applies to $G=USp(2N)$
with $|\Delta^+|=N^2$ and $\sum_{\alpha\in\Delta^+}\mathrm{ht}\,\alpha=\frac{N(N+1)(4N-1)}{6}$ for the height (\ref{ht_def}).
The variables $p_i=m_i+\mu_i$ are the positive half-integers $p_1<\cdots<p_N$,
and $q^{\frac12\sum_{\lambda}w_{\lambda}(p)}=1$, since the weights of the hypermultiplets come in pairs $\pm w_{\lambda}$.
Using $\theta(q^{-n}a;q)=(-1)^nq^{-\frac{n(n+1)}{2}}a^n\theta(a;q)$ for the top components, we obtain
\begin{align}
\label{USp_rv_closed}
\chi[\mathsf{V}^{USp(2N)\,\textrm{ADHM-}[2l]}_{\mathrm{Nahm}'}]
&=q^{-\frac{(l-2)N^2}{2}}x^{N(N-1)}\Bigl(\frac{\theta(x;q)}{(q)_{\infty}^2(1-x)}\Bigr)^N\prod_{a=1}^lF(q^{\frac12}y_a;q)
\nonumber\\*
&\times\sum_{p_1<\cdots<p_N}\prod_{\alpha\in\Delta_{\rm s}}\frac{1-q^{\alpha(p)}}{1-xq^{\alpha(p)}}
\prod_{i=1}^N\frac{q^{(l-2)p_i}(1-q^{2p_i})^2}{\prod_{a=1}^l(1-q^{p_i}y_a)(1-q^{p_i}y_a^{-1})},
\end{align}
where $p_i\in\mathbb{Z}_{\ge0}+\frac12$ and $\Delta_{\rm s}$ is the set of the short roots $\pm e_i\pm e_j$, i.e. the non-zero weights of the antisymmetric hypermultiplet.
The summand is invariant under the Weyl group of $USp(2N)$, which acts on $p$ by permutations and sign changes,
and it vanishes for $p_i=\pm p_j$.
Hence (\ref{USp_rv_closed}) takes the form (\ref{Nahm_Jackson}) with $|\mathrm{Weyl}(G)|=2^NN!$ and $\xi=q^{\frac12}$.
For $x=1$ the antisymmetric hypermultiplet cancels the vector multiplet along the short roots, and the factor in the parentheses becomes $1$.
Then the fermions become free, and the C-twist limit of (\ref{USp_rv_dual}) gives
\begin{align}
\label{USp_Fermi}
\chi[\mathsf{V}^{\wtilde{\hat{D}_l}}_{\mathcal{N},N,N+\tilde{\Gamma}+\tilde{\eta}}]
&=q^{-\frac{(l-2)N^2}{2}}\prod_{a=1}^lF(q^{\frac12}y_a;q)
\sum_{\substack{0<p_1<\cdots<p_N\\ p_i\in\mathbb{Z}+\frac12}}\prod_{i=1}^N\frac{q^{(l-2)p_i}(1-q^{2p_i})^2}{\prod_{a=1}^l(1-q^{p_i}y_a)(1-q^{p_i}y_a^{-1})}.
\end{align}
Unlike (\ref{necklace_Fermi}), the momenta $p_i$ of the fermions are positive because of the sign changes in the Weyl group,
and the long roots give the factors $(1-q^{2p_i})^2$.\footnote{A free Fermi gas description is also known for the $S^3$ partition functions of these theories, in which the fermions live on a half-line \cite{Assel:2015hsa}.}
At $x=1$ the Neumann side is the Schur index of the 4d $\mathcal{N}=2$ affine $D_l$ quiver gauge theory
whose $U(1)$ flavor symmetries are gauged with the Fermi multiplets $\tilde{\eta}_a$ (see (\ref{Dquiver_4d})).
Hence (\ref{USp_Fermi}) is a free Fermi gas formula for this gauged Schur index.
For the Schur indices of the 4d $\mathcal{N}=2$ affine $D_l$ quiver gauge theories themselves, the techniques of \cite{Bourdier:2015sga} give a sum over the sectors from the $SU(N)$ and $SU(2N)$ constraints as for the necklace quivers,
but each term lacks an interpretation as a free Fermi gas, and a closed form has been obtained only for their grand canonical version \cite{Bourdier:2017thesis}.
The unflavored characters are listed in Appendix \ref{app_vac}.

In the H-twist limit of (\ref{USp_fw_dual}), the closed form (\ref{Nahm_closed}) with $t\rightarrow t^{-1}$ applies to the affine $D_l$ quiver with $\mathrm{rk}\,G=2N(l-1)$.
We take the fugacities of the twisted bifundamental hypermultiplets to be $u^{(I)}/u^{(J)}$ for the four pairs $(I,J)$ in (\ref{USp_fw_Nahm}) and $u^{(k)}/u^{(k+1)}$ for $2\le k\le l-3$.
With the spins of the top components in (\ref{USp_fw_Nahm}) we obtain
\begin{align}
\label{USp_fw_closed}
\chi[\mathsf{V}^{\wtilde{\hat{D}_l}}_{\mathrm{Nahm}^{l+1}}]
&=\frac{(-1)^{|\Delta^+_{\rho}|}q^{\sum_{\alpha\in\Delta^+_{\rho}}\mathrm{ht}\alpha}}{(q)_{\infty}^{4N(l-1)}}\,
\theta(q^{-\frac{N-1}{2}}u^{(0)};q)
\prod_{(I,J)}\prod_{n=0}^{N-1}\theta\Bigl(q^{-\frac{N}{2}-n}\frac{u^{(I)}}{u^{(J)}};q\Bigr)
\nonumber\\*
&\times\prod_{k=2}^{l-3}\prod_{j=0}^{2N-1}\theta\Bigl(q^{-j}\frac{u^{(k)}}{u^{(k+1)}};q\Bigr)
\sum_{p}\Bigl(\frac{q^{\frac12}}{x}\Bigr)^{|p^{(0)}|}\prod_{I=0}^l(q^{n_I}y^{\alpha_I})^{|p^{(I)}|}
\nonumber\\*
&\times\frac{\prod_{I=0}^l\prod_{i\neq j}(1-q^{p^{(I)}_i-p^{(I)}_j})}{\prod_{i=1}^N(1-q^{p^{(0)}_i}u^{(0)})\prod_{(I,J)\in E}\prod_{i,j}(1-q^{p^{(I)}_i-p^{(J)}_j}u^{(I)}/u^{(J)})},
\end{align}
where $p^{(I)}_i\in\mathbb{Z}+\frac{N_I-1}{2}$ are strictly increasing in $i$,
and $E$ consists of the four pairs $(I,J)$ and the pairs $(k,k+1)$ with $2\le k\le l-3$.
Here $q^{\frac12\sum_{\lambda}w_{\lambda}(p)}=q^{\frac12|p^{(0)}|+\sum_In_I|p^{(I)}|}$ is combined with the topological charges,
where $n_I=N$ for the four end nodes, $n_{l-2}=-2N$ and $n_I=0$ otherwise.

For $v_2=\cdots=v_{l-2}$, i.e. $u^{(2)}=\cdots=u^{(l-2)}$, the theta functions in the second line vanish.
As for (\ref{ADHM_closed_Fermi}), the sum then localizes to $p^{(2)}=\cdots=p^{(l-2)}$.
Since $\alpha_2+\cdots+\alpha_{l-2}=e_2-e_{l-1}$, (\ref{USp_fw_closed}) reduces to that for $l=4$ with the $SO(8)$ fugacities $y_1$, $y_2$, $y_{l-1}$ and $y_l$.
On the Neumann side the Fermi multiplets $\Gamma_{k,k+1}$ cancel the fundamental hypermultiplets with $y_3,\cdots,y_{l-2}$.
For $l=4$ the H-twist limit of the Neumann half-index (\ref{USp_fw_Neu}) is the Schur index
\begin{align}
\label{Schur_USp}
\mathcal{I}^{\textrm{Schur}}_{USp(2N)}(x,y_a;q)
&=\frac{(q)_{\infty}^{2N}}{2^NN!}\oint\prod_{i=1}^N\frac{ds_i}{2\pi is_i}\,
\frac{\prod_{\alpha\in\Delta}(s^{\alpha};q)_{\infty}(qs^{\alpha};q)_{\infty}}{(q^{\frac12}x^{\pm};q)_{\infty}^N\prod_{\alpha\in\Delta_{\rm s}}(q^{\frac12}x^{\pm}s^{\alpha};q)_{\infty}}
\nonumber\\*
&\times\prod_{i=1}^N\prod_{a=1}^4\frac{1}{(q^{\frac12}s_i^{\pm}y_a^{\pm};q)_{\infty}(q^{\frac12}s_i^{\pm}y_a^{\mp};q)_{\infty}}
\end{align}
of 4d $\mathcal{N}=2$ $USp(2N)$ gauge theory with the antisymmetric hypermultiplet and four fundamental hypermultiplets,
times the contributions of the Fermi multiplets $\eta$ and $\Gamma_{I,2}$.
Here and below we omit the factor $q^{-\frac{c}{24}}$ in the Schur indices.
The shifted Fermi multiplets give $\theta(q^{\frac12+\kappa}X;q)$.
Hence the duality (\ref{USp_fw_dual}) gives
\begin{align}
\label{Schur_D4}
&\mathcal{I}^{\textrm{Schur}}_{USp(2N)}(x,y_a;q)\,\theta\Bigl(q^{\frac12}\frac{v_0}{x};q\Bigr)\prod_{I=1,3,4}\prod_{\kappa}\theta(q^{\frac12+\kappa}X_I;q)
\prod_{\kappa'}\theta\Bigl(q^{\frac12+\kappa'}\frac{X_0}{x};q\Bigr)\,\theta(q^{\frac12}X_0x^{N-1};q)
\nonumber\\*
&=\chi[\mathsf{V}^{\wtilde{\hat{D}_4}}_{\mathrm{Nahm}^{5}}](u^{(I)}=v_Ix^{N-1})
\end{align}
with $X_I=v_Iy^{\alpha_I}/v_2$.
As for (\ref{Schur_Fermi}), the Schur index does not depend on $v_I$, which act only on the Fermi multiplets on the Neumann side.
The antisymmetric representation of $USp(2N)$ contains a singlet,
so that $\mathcal{I}^{\textrm{Schur}}_{USp(2N)}$ is the Schur index of the rank-$N$ $D_4$ theory times that of a free hypermultiplet.
The rank-$N$ $D_4$ theory is the 4d $\mathcal{N}=2$ $USp(2N)$ gauge theory with the traceless antisymmetric hypermultiplet and four fundamental hypermultiplets,
for which $N=1$ is $SU(2)$ gauge theory with four flavors,
and for $N=2$ and $3$ its Schur index has been studied as a $q$-series and through modular differential equations \cite{Beem:2019tfp}.
By (\ref{Schur_D4}) and (\ref{USp_fw_closed}) the duality gives it for any $N$ as a Jackson integral.

\subsection{Quasimodularity}
\label{sec_VOA_qmod}
The closed form (\ref{Nahm_closed}) also shows the modular properties of the unflavored twist limits.
We first summarize our conventions for the Eisenstein series and the quasimodular forms.

\subsubsection{Eisenstein series}
\label{sec_qmod_eisenstein}
Let $q=e^{2\pi i\tau}$ with $\mathrm{Im}\,\tau>0$.
For $k\ge1$ the Eisenstein series of weight $2k$ is
\begin{align}
\label{Ek_def}
E_{2k}(\tau)&=1-\frac{4k}{B_{2k}}\sum_{n>0}\frac{n^{2k-1}q^n}{1-q^n},
\end{align}
where $B_{2k}$ are the Bernoulli numbers, $B_2=\frac16$, $B_4=-\frac{1}{30}$, $B_6=\frac{1}{42}$, $\cdots$.
We write $E_{2k}=E_{2k}(\tau)$ and $E_{2k}(q^a)=E_{2k}(a\tau)$, and similarly for $\eta$.
Let $\gamma=\bigl(\begin{smallmatrix}a&b\\c&d\end{smallmatrix}\bigr)\in SL(2,\mathbb{Z})$ act as $\gamma\tau=\frac{a\tau+b}{c\tau+d}$.
For $k\ge2$, $E_{2k}$ is a modular form of weight $2k$ for $SL(2,\mathbb{Z})$, i.e. $E_{2k}(\gamma\tau)=(c\tau+d)^{2k}E_{2k}(\tau)$,
and the ring of modular forms for $SL(2,\mathbb{Z})$ is generated by $E_4$ and $E_6$ \cite{MR1027834}.
For example, $E_8=E_4^2$, $E_{10}=E_4E_6$ and $E_{12}=(441E_4^3+250E_6^2)/691$.
The Eisenstein series $E_2$ is not modular,
\begin{align}
\label{E2_transf}
E_2(\gamma\tau)&=(c\tau+d)^2E_2(\tau)+\frac{6}{\pi i}c(c\tau+d).
\end{align}
A quasimodular form of weight $k$ and depth $s$ for $SL(2,\mathbb{Z})$ is a polynomial $\sum_{j=0}^sg_jE_2^j$ with $g_s\neq0$,
whose coefficients $g_j$ are modular forms of weight $k-2j$ \cite{MR1363056}.
By (\ref{E2_transf}) it transforms as
\begin{align}
\label{qmod_transf}
f(\gamma\tau)&=(c\tau+d)^k\sum_{j=0}^{s}f_j(\tau)\Bigl(\frac{c}{c\tau+d}\Bigr)^j,
\end{align}
where $f_0=f$ and $f_j$ are quasimodular forms of weight $k-2j$.
The ring of quasimodular forms is generated by $E_2$, $E_4$ and $E_6$, and it is closed under the derivative $D=q\frac{d}{dq}$,
which raises the weight by $2$ \cite{MR1363056}.
Indeed, Ramanujan's identities \cite{MR2280861} read
\begin{align}
\label{Ramanujan}
DE_2&=\frac{E_2^2-E_4}{12},&
DE_4&=\frac{E_2E_4-E_6}{3},&
DE_6&=\frac{E_2E_6-E_4^2}{2}.
\end{align}
The Dedekind eta function $\eta(\tau)=q^{\frac{1}{24}}(q)_{\infty}$ satisfies $\eta(\gamma\tau)=\epsilon(\gamma)(c\tau+d)^{\frac12}\eta(\tau)$ with $\epsilon(\gamma)^{24}=1$ \cite{MR1027834},
i.e. it is a modular form of weight $\frac12$ with a multiplier system.
It has neither zeros nor poles in the upper half plane.

The subgroup $\Gamma^0(2)$ of $SL(2,\mathbb{Z})$ consists of the $\gamma$ with even $b$.
For $\gamma\in\Gamma^0(2)$ we have $\frac{\gamma\tau}{2}=\gamma'\frac{\tau}{2}$ with $\gamma'=\bigl(\begin{smallmatrix}a&b/2\\2c&d\end{smallmatrix}\bigr)\in SL(2,\mathbb{Z})$
and $2c\cdot\frac{\tau}{2}+d=c\tau+d$.
Hence $\eta(q^{\frac12})$ and $E_{2k}(q^{\frac12})$ transform under $\Gamma^0(2)$
as $\eta(\tau)$ and $E_{2k}(\tau)$ under $SL(2,\mathbb{Z})$, where $c$ in (\ref{E2_transf}) is replaced by $2c$.
We write $\delta_{2k}=E_{2k}(q^{\frac12})-E_{2k}$.
Then $\delta_{2k}$ with $k\ge2$ and $E_2-\delta_2=2E_2-E_2(q^{\frac12})$ are modular forms of weight $2k$ and $2$ for $\Gamma^0(2)$.
Let $\widetilde{M}_k(\Gamma)$ be the space of quasimodular forms of weight $k$ for $\Gamma$,
i.e. the polynomials in $E_2$ whose coefficients are modular forms for $\Gamma$.
For $\Gamma=\Gamma^0(2)$ they are the polynomials in $E_2$, $E_4$, $E_2(q^{\frac12})$ and $E_4(q^{\frac12})$, and (\ref{qmod_transf}) holds for $\gamma\in\Gamma^0(2)$.

The sums in section \ref{sec_qmod_examples} reduce to polynomials in the double sums
\begin{align}
\label{double_sum}
\sum_{p>0}\sum_{d>0}p^kd^lq^{pd}&=-\frac{B_{2m}}{4m}D^{\min(k,l)}(E_{2m}-1),&
2m&=|k-l|+1,
\end{align}
with odd $k+l$, where $p$ and $d$ run over the positive integers.
This follows from (\ref{Ek_def}), since the sum is symmetric under $p\leftrightarrow d$,
$\sum_{p,d}p^kd^lq^{pd}=D^l\sum_{p,d}p^{k-l}q^{pd}$ for $k>l$, and $\sum_{p,d}p^{2m-1}q^{pd}=\sum_{n>0}n^{2m-1}q^n/(1-q^n)$.
If $p$ runs over $\mathbb{Z}_{\ge0}+\frac12$ instead, the double sum becomes $2^{-k}f(q^{\frac12})-f(q)$, where $f(q)$ is the right-hand side of (\ref{double_sum}),
since $\mathbb{Z}_{\ge0}+\frac12=\frac12\mathbb{Z}_{>0}\setminus\mathbb{Z}_{>0}$.
Hence the double sums are quasimodular forms for $SL(2,\mathbb{Z})$ and $\Gamma^0(2)$.
They have pure weight $k+l+1$ unless $l=0$, or $k=0$ and $p$ is an integer.

\subsubsection{Unflavored limit}
\label{sec_qmod_conj}
In the unflavored limit, in which all the fugacities other than $q$ are set to unity, (\ref{Nahm_closed}) reduces to
\begin{align}
\label{chi_unflavored}
\chi[\mathsf{V}^{\mathcal{T}}_{\mathrm{Nahm}'}]&=\pm q^{\frac{c}{24}}\,
\frac{\eta(q^{\frac12})^{2n_{\frac12}}}{\eta(q)^{2\,\mathrm{rk}\,G-2n_0+2n_{\frac12}}}\,\mathcal{Q}^{\mathcal{T}},
\\*
\label{QT_def}
\mathcal{Q}^{\mathcal{T}}&=\sum_{p\in P_0}\,\prod_{\alpha\in\Delta_{\rho}}(1-q^{\alpha(p)})\prod_{\lambda:\,w_{\lambda}(p)\neq0}\frac{q^{\frac12w_{\lambda}(p)}}{1-q^{w_{\lambda}(p)}}
\end{align}
where $c$ is the central charge of the boundary VOA, which we discuss in section \ref{sec_VOA_DS}.
Here $n_0$ and $n_{\frac12}$ are the numbers of the spin-$J$ multiplets with integer and half-integer $J$,
for which $\theta(q^{-J};q)$ in (\ref{Nahm_closed}) is equal to $\theta(1;q)$ and $\theta(q^{\frac12};q)$ up to monomials.
For the quiver gauge theories in this paper they are fixed by the ranks $N_I$ as follows.
A fundamental hypermultiplet of $U(N_I)$ has the weights $p^{(I)}_i\in\mathbb{Z}+\frac{N_I-1}{2}$
and forms a single spin-$\frac{N_I-1}{2}$ multiplet by (\ref{spins_UN}).
A bifundamental hypermultiplet of $U(N_I)\times U(N_K)$ has the weights $p^{(I)}_i-p^{(K)}_k\in\mathbb{Z}+\frac{N_I-N_K}{2}$
and forms $\min(N_I,N_K)$ spin-$J$ multiplets with $J\in\mathbb{Z}+\frac{N_I-N_K}{2}$ by (\ref{spins_bifund}).
The adjoint hypermultiplet of $U(N_I)$ is the case $K=I$ with $J=0,1,\cdots,N_I-1$.
Hence
\begin{align}
\label{n0_def}
n_0&=\sum_{\substack{\textrm{fund. of }U(N_I)\\N_I\ \textrm{odd}}}1
+\sum_{\substack{\textrm{bifund. of }U(N_I)\times U(N_K)\\N_I-N_K\ \textrm{even}}}\min(N_I,N_K),
\nonumber\\*
n_{\frac12}&=\sum_{\substack{\textrm{fund. of }U(N_I)\\N_I\ \textrm{even}}}1
+\sum_{\substack{\textrm{bifund. of }U(N_I)\times U(N_K)\\N_I-N_K\ \textrm{odd}}}\min(N_I,N_K).
\end{align}
Only the spin-$J$ multiplets with integer $J$ have weights which can vanish.
For unitary gauge groups the sum runs over the set $P_0$ of the strictly increasing $p$ such that
$p^{(I)}_i=0$ for some $i$ at each node with odd $N_I$ and fundamental hypermultiplets,
and the smaller one of the sets $\{p^{(I)}_i\}$ and $\{p^{(K)}_k\}$ is contained in the other for each bifundamental hypermultiplet with even $N_I-N_K$.
Exactly for these $p$, $n_0$ weights vanish, and their poles are cancelled by the zeros of $\theta(q^{-J}\bar{u}^{w_{\lambda}}x^{f_{\lambda}};q)$ at $\bar{u}^{w_{\lambda}}x^{f_{\lambda}}=1$ for integer $J$.
For the other $p$ fewer weights vanish, and the summands drop out.
For $USp(2N)$ gauge theories $P_0$ consists of all the strictly increasing $p$.
The prefactor is an eta-quotient of weight $n_0-\mathrm{rk}\,G$, since $\eta(q)$ and $\eta(q^{\frac12})$ have weight $\frac12$.

For all our examples we show the following in section \ref{sec_qmod_examples},
and we conjecture it for the quiver gauge theories of unitary gauge groups with fundamental, bifundamental and adjoint hypermultiplets
and for $USp(2N)$ gauge theories with fundamental hypermultiplets and at most one antisymmetric hypermultiplet.
\begin{quote}
\textit{The function $\mathcal{Q}^{\mathcal{T}}$ in (\ref{QT_def}) is a quasimodular form for $\Gamma^0(2)$ of mixed weight,}
\begin{align}
\label{qmod_weight}
\mathcal{Q}^{\mathcal{T}}&\in\bigoplus_{k=0}^{w^{\mathcal{T}}}\widetilde{M}_k(\Gamma^0(2)),&
w^{\mathcal{T}}&=\dim R-|\Delta_{\rho}|-n_0,
\end{align}
\textit{where $R$ is the representation of the hypermultiplets of $G$.}
\end{quote}
Then the unflavored character is, up to a power of $q$, an eta-quotient times a quasimodular form.
Since $|\Delta_{\rho}|+\mathrm{rk}\,G=\dim G$, its weight is at most $\dim R-\dim G$,
which is the quaternionic dimension of the Higgs branch of $\mathcal{T}$ for the theories in this paper.

The conjecture is suggested by the form of $\mathcal{Q}^{\mathcal{T}}$.
The factors $q^{\frac{n}{2}}/(1-q^n)$ are the coefficients of the Laurent series
\begin{align}
\label{qell_int}
\sum_{n\in\mathbb{Z}\setminus\{0\}}\frac{q^{\frac{n}{2}}}{1-q^n}\,y^n&=-y\frac{\partial}{\partial y}\log\theta(q^{\frac12}y;q),
\\*
\label{qell_half}
\sum_{n\in\mathbb{Z}+\frac12}\frac{q^{\frac{n}{2}}}{1-q^n}\,y^n&=-q^{\frac14}y^{-\frac12}\frac{(q)_{\infty}^2\,\theta(y;q)}{\theta(q^{\frac12};q)\,\theta(q^{\frac12}y;q)},
\end{align}
which converge for $|q|^{\frac12}<|y|<|q|^{-\frac12}$.
They follow from (\ref{Kronecker}) with $w=q^{\frac12}y$, at $a=q^{\frac12}$ for (\ref{qell_half}) and in the limit $a\rightarrow1$ after subtracting the term $n=0$ for (\ref{qell_int}).
With $y=e^{2\pi iz}$ the right-hand sides have weight $1$ under the action $(z,\tau)\rightarrow(\frac{z}{c\tau+d},\gamma\tau)$ of $\gamma\in\Gamma^0(2)$,
up to an additive term $-cz$ for (\ref{qell_int}) and a sign for (\ref{qell_half}).
Since the weights $w_{\lambda}(p)$ are linear in $p$, the sum over $p$ of the product of the $\dim R-n_0$ factors is a constant term of a product of these functions,
which suggests a quasimodular form of weight at most $\dim R-n_0$.
For example, $\sum_{m>0}q^{km}/(1-q^m)^{2k}$ is a linear combination of $1,E_2,\cdots,E_{2k}$.
The $|\Delta_{\rho}|$ factors $1-q^{\alpha(p)}$ have weight $-1$, since each of them removes one of these factors.
Indeed, $\alpha(p)$ is the difference $n-n'$ of two weights $n=w_{\lambda}(p)$ and $n'=w_{\lambda'}(p)$ of $R$, and
\begin{align}
\label{qell_root}
(1-q^{n-n'})\,\frac{q^{\frac{n}{2}}}{1-q^n}\,\frac{q^{\frac{n'}{2}}}{1-q^{n'}}
&=q^{\frac{n}{2}-n'}\,\frac{q^{\frac{n'}{2}}}{1-q^{n'}}-q^{-\frac{n'}{2}}\,\frac{q^{\frac{n}{2}}}{1-q^n}.
\end{align}
This suggests the bound (\ref{qmod_weight}), while the argument does not determine the group $\Gamma^0(2)$.

The quasimodularity is also expected from the dual Neumann boundary conditions and from the boundary VOAs.
In the unflavored twist limit the Neumann half-index is the constant term of an elliptic function of the gauge fugacities, since the boundary gauge anomaly is cancelled.
The Schur indices of 4d $\mathcal{N}=2$ superconformal gauge theories are of this form,
and their unflavored limits are quasimodular forms of mixed weight \cite{Pan:2021mrw,Beem:2021zvt}.
The associated variety\footnote{The associated variety of a vertex operator algebra is the affine Poisson variety given by its $C_2$ algebra \cite{MR2875849}.
A vertex operator algebra is quasi-lisse if its associated variety has finitely many symplectic leaves.}
of the boundary VOA for a Neumann boundary condition is expected to be the Higgs branch of the bulk theory \cite{Costello:2018swh}.
This is suggested by the classical limit of the BRST reduction, in which the symplectic bosons give the Hamiltonian reduction of the space of the hypermultiplets by the gauge group while the complex fermions do not contribute,
and it holds for the vertex operator superalgebra $\mathsf{V}_{S_N}$ of the ADHM theory with one flavor \cite{Arakawa:2023cki} and for the Abelian gauge theories \cite{Arakawa:2026nut} (see section \ref{sec_VOA_examples}).
Since the Higgs branch has finitely many symplectic leaves, the VOA is then quasi-lisse, and so is the Nahm pole boundary VOA through the duality.
Then its vacuum character satisfies a modular linear differential equation \cite{Arakawa:2016hkg}, which for half-integer weights is a twisted one for $\Gamma^0(2)$ \cite{Beem:2017ooy}.
The logarithmic solutions of such an equation transform into polynomials in $\tau$, as does an eta-quotient times a quasimodular form of positive depth by (\ref{qmod_transf}).

The eta-quotient in (\ref{chi_unflavored}) is a weakly holomorphic modular form for $\Gamma^0(2)$
with a multiplier system valued in the $24$th roots of unity.
By (\ref{qmod_transf}), the components of $\mathcal{Q}^{\mathcal{T}}$ of different weights acquire different powers of $c\tau+d$,
and the components of positive depth give polynomials in $c/(c\tau+d)$.
Hence the character is a modular form only if $\mathcal{Q}^{\mathcal{T}}$ has pure weight and depth zero.
For example, for the Abelian quivers with fundamental hypermultiplets, including $T[SU(2)]$, we have $\mathcal{Q}^{\mathcal{T}}=1$,
and the characters $(q)_{\infty}^{2(\dim R-\mathrm{rk}\,G)}$ are powers of $\eta(q)$ up to powers of $q$.
For the non-Abelian gauge theories in this paper $\mathcal{Q}^{\mathcal{T}}$ vanishes or has mixed weight or positive depth, and the non-vanishing characters are not modular.

The formula (\ref{chi_unflavored}) does not apply directly to the Dirichlet boundary conditions of non-Abelian gauge theories.
In the C-twist limit of (\ref{D_simple}) the vector multiplet gives $\prod_{\alpha\in\Delta}\theta(u^{\alpha};q)^{-1}$,
which has poles on the walls $u^{\alpha}=1$ of the fugacity $u$ for $G_{\partial}$,
and the unflavored limit does not exist unless the poles are cancelled as for the ADHM theory at $x=1$.
We comment on them in section \ref{sec_future}.

\subsubsection{Examples}
\label{sec_qmod_examples}
We now write $\mathcal{Q}^{\mathcal{T}}$ explicitly in terms of the Eisenstein series for all our examples.
In each case we first identify the eta-quotient and the sum $\mathcal{Q}^{\mathcal{T}}$, and then derive the result.

\paragraph{$U(N)$ SQCD}
For the Nahm pole boundary condition $\mathrm{Nahm}'$ of $U(N)$ SQCD with $2N$ flavors we have $\mathrm{rk}\,G=N$,
and the $2N$ fundamental hypermultiplets have the spin $\frac{N-1}{2}$.
Hence $n_0=2N$ for odd $N$ and $n_{\frac12}=2N$ for even $N$, and (\ref{chi_unflavored}) for (\ref{SQCD_rv_closed}) reads
\begin{align}
\label{SQCD_C_S}
\chi[\mathsf{V}^{U(N)+2N}_{\mathrm{Nahm}'}]&=\pm q^{\frac{c}{24}}
\begin{cases}
\eta(q)^{2N}\,\mathcal{Q}^{U(N)+2N}&(N\ \textrm{odd})\\
\frac{\eta(q^{1/2})^{4N}}{\eta(q)^{6N}}\,\mathcal{Q}^{U(N)+2N}&(N\ \textrm{even})\\
\end{cases},
\nonumber\\*
\mathcal{Q}^{U(N)+2N}&=\sum_{p_1<\cdots<p_N}\prod_{i\neq j}(1-q^{p_i-p_j})\prod_{i:\,p_i\neq0}\frac{q^{Np_i}}{(1-q^{p_i})^{2N}},
\end{align}
where $p_i\in\mathbb{Z}+\frac{N-1}{2}$, and one of $p_i$ is zero for odd $N$.
Let $N=2n$ or $N=2n+1$.
Then the bound (\ref{qmod_weight}) is $N^2+N-n_0=4n^2+2n$.
We find
\begin{align}
\label{SQCD_C_Hankel}
\mathcal{Q}^{U(N)+2N}&=(-1)^n\det_{0\le i,j\le n-1}(Z_{i+j+1})\det_{0\le i,j\le n-1}(Z_{i+j+1}+4Z_{i+j+2}),
\\*
\label{Zk_def}
Z_k&=\sum_{\substack{p\in\mathbb{Z}+\frac{N-1}{2}\\ p>0}}\frac{q^{kp}}{(1-q^p)^{2k}}.
\end{align}
For $N=2n$ and $N=2n+1$ the formula is the same, except that $p$ is a half-integer or an integer.
To derive (\ref{SQCD_C_Hankel}), we write $\prod_{i\neq j}(1-q^{p_i-p_j})=(-1)^{\frac{N(N-1)}{2}}\det_{i,b}(q^{m_bp_i})^2$ with $m_b=b-\frac{N+1}{2}$.
Then the Cauchy-Binet formula turns the $N$-fold sum into $(-1)^{\frac{N(N-1)}{2}}\det\mathcal{M}$ with the single sums
\begin{align}
\label{SQCD_CauchyBinet}
\mathcal{M}_{ab}&=\sum_{\substack{p\in\mathbb{Z}+\frac{N-1}{2}\\ p\neq0}}\frac{q^{(m_a+m_b+N)p}}{(1-q^p)^{2N}},
\end{align}
where for odd $N$ the terms with $p_i=0$ replace $\det\mathcal{M}$ by $\det(\mathcal{M}+\mathbb{J}_N)-\det\mathcal{M}$ with the matrix $\mathbb{J}_N$ of ones.
Since the summand is invariant under $p\rightarrow-p$, $\mathcal{M}$ becomes block diagonal in the basis
$(q^{\frac{p}{2}}+q^{-\frac{p}{2}})^{\epsilon}(q^{\frac{p}{2}}-q^{-\frac{p}{2}})^r$ with $\epsilon=0,1$, $r+\epsilon\le N-1$ and $r+\epsilon\equiv N-1$ mod $2$.
Since $q^{Np}/(1-q^p)^{2N}=(q^{\frac{p}{2}}-q^{-\frac{p}{2}})^{-2N}$ and $(q^{\frac{p}{2}}+q^{-\frac{p}{2}})^2=(q^{\frac{p}{2}}-q^{-\frac{p}{2}})^2+4$,
the entries of the two blocks are $2Z_k$ and $2(Z_k+4Z_{k+1})$, which gives (\ref{SQCD_C_Hankel}).
For odd $N$ the terms with $p_i=0$ only contribute to the entry of the even block with $r=\epsilon=0$,
and $\det(\mathcal{M}+\mathbb{J}_N)-\det\mathcal{M}$ is the product of its cofactor and the determinant of the odd block.

For integer $p$, $Z_k$ are the multiple $q$-zeta values of depth one with the argument $2k$ in \cite{okounkov2014hilbert}.
Since $q^{kp}/(1-q^p)^{2k}=\sum_{d>0}\binom{d+k-1}{2k-1}q^{pd}$ for $p>0$ and $\binom{d+k-1}{2k-1}$ is an odd polynomial in $d$,
they are linear combinations of the double sums (\ref{double_sum}) without powers of $p$, and hence of $1,E_2,\cdots,E_{2k}$,
\begin{align}
\label{Zk}
Z_1&=\frac{1-E_2}{24},
\\*
\label{Zk_2}
Z_2&=\frac{E_4+10E_2-11}{1440},
\\
\label{Zk_3}
Z_3&=\frac{191-168E_2-21E_4-2E_6}{120960},
\\*
\label{Zk_4}
Z_4&=\frac{3E_4^2+294E_4+40E_6+2160E_2-2497}{7257600}.
\end{align}
For half-integer $p$, $Z_k$ is replaced by $Z_k(q^{\frac12})-Z_k(q)$.
Hence $\mathcal{Q}^{U(N)+2N}$ is a quasimodular form of mixed weight at most $4n^2+2n$, in agreement with (\ref{qmod_weight}).
For $N\le5$ the component of this weight is non-zero.
For example, we find
\begin{align}
\label{SQCD_C_qmod2}
\mathcal{Q}^{U(2)+4}&=-\frac{1}{8640}\bigl(E_2-E_2(q^{\frac12})\bigr)\bigl(5E_2-5E_2(q^{\frac12})-E_4+E_4(q^{\frac12})\bigr),
\\*
\label{SQCD_C_qmod3}
\mathcal{Q}^{U(3)+6}&=\frac{1}{8640}(E_2-1)(E_4-5E_2+4).
\end{align}
For $N=4,5$, i.e. $n=2$, the two determinants in (\ref{SQCD_C_Hankel}) give the two factors of
\begin{align}
\label{SQCD_C_qmod4}
\mathcal{Q}^{U(4)+8}&=\frac{140\delta_2^2-35\delta_2\delta_4+10\delta_2\delta_6-7\delta_4^2}{14515200}
\nonumber\\*
&\times\frac{1}{4572288000}\bigl(3780\delta_2^2-1785\delta_2\delta_4+490\delta_2\delta_6-147\delta_4^2+70\delta_4\delta_6-20\delta_6^2
\nonumber\\
&\qquad-105\delta_2\delta_4^2+21\delta_4^3-210E_4\delta_2\delta_4+42E_4\delta_4^2\bigr),
\\
\label{SQCD_C_qmod5}
\mathcal{Q}^{U(5)+10}&=\frac{108-255E_2+49E_4+140E_2^2-10E_6-35E_2E_4-7E_4^2+10E_2E_6}{14515200}
\nonumber\\
&\times\frac{1}{4572288000}\bigl(2304-6160E_2+1988E_4+3780E_2^2-520E_6-1785E_2E_4
\nonumber\\*
&\qquad-63E_4^2+490E_2E_6-105E_2E_4^2+70E_4E_6+21E_4^3-20E_6^2\bigr),
\end{align}
Thus $\mathcal{Q}^{U(5)+10}$ is a quasimodular form for $SL(2,\mathbb{Z})$, while $\mathcal{Q}^{U(4)+8}$ is one for $\Gamma^0(2)$.

\paragraph{Mirror quiver}
For the mirror quiver of $U(N)$ SQCD in (\ref{SQCD_fw_closed}) we consider the Nahm pole boundary conditions $(\mathcal{D},\mathrm{Nahm}^{2N-3},\mathcal{D})$.
The ranks of the adjacent nodes differ by one, so that all the bifundamental hypermultiplets have half-integer spins.

For the quiver $(1)-(2)-\cdots-(N)-\cdots-(2)-(1)$ with the ranks $r_I=\min(I,2N-I)$ we have $\mathrm{rk}\,G=N^2$,
$n_0=2$ for odd $N$ and $n_0=0$ for even $N$, and $n_{\frac12}=N(N-1)+2-n_0$.
Hence (\ref{chi_unflavored}) reads
\begin{align}
\label{SQCD_H_S}
\chi[\mathsf{V}^{\wtilde{(1)-(2)-\cdots-(N)-\cdots-(2)-(1)}}_{\mathcal{D},\mathrm{Nahm}^{2N-3},\mathcal{D}}]&=\pm q^{\frac{c}{24}}\,\mathcal{Q}^{(1)-(2)-\cdots-(N)-\cdots-(2)-(1)}
\nonumber\\*
&\quad\times
\begin{cases}
\frac{\eta(q^{1/2})^{2N(N-1)}}{\eta(q)^{4N^2-2N-4}}&(N\ \textrm{odd})\\
\frac{\eta(q^{1/2})^{2N^2-2N+4}}{\eta(q)^{4N^2-2N+4}}&(N\ \textrm{even})\\
\end{cases},
\nonumber\\
\mathcal{Q}^{(1)-(2)-\cdots-(N)-\cdots-(2)-(1)}&=\sum_{p}\prod_{I=1}^{2N-1}\prod_{i\neq j}(1-q^{p^{(I)}_i-p^{(I)}_j})
\prod_{i:\,p^{(N)}_i\neq0}\frac{q^{p^{(N)}_i}}{(1-q^{p^{(N)}_i})^2}
\nonumber\\*
&\times\prod_{I=1}^{2N-2}\prod_{i,j}\frac{q^{\frac12(p^{(I)}_i-p^{(I+1)}_j)}}{1-q^{p^{(I)}_i-p^{(I+1)}_j}},
\end{align}
where $p^{(I)}_i\in\mathbb{Z}+\frac{r_I-1}{2}$ are strictly increasing in $i$, and one of $p^{(N)}_i$ is zero for odd $N$.
The bound (\ref{qmod_weight}) is $N^2+N-n_0$.
We find
\begin{align}
\label{SQCD_H_Hankel}
\mathcal{Q}^{(1)-(2)-\cdots-(N)-\cdots-(2)-(1)}&=\pm\frac{2^{2n}}{\bigl(\prod_{k=1}^{N-1}k!\bigr)^2}\det_{0\le i,j\le n-1}(a_{i+j+\epsilon})\det_{0\le i,j\le n-1}(a_{i+j+\epsilon+1}),
\\*
\label{ak_def}
a_k&=\sum_{\substack{p\in\mathbb{Z}+\frac{N-1}{2}\\ p>0}}\frac{p^{2k}q^p}{(1-q^p)^2},
\end{align}
where $N=2n+\epsilon$ with $\epsilon=0,1$.

To derive (\ref{SQCD_H_Hankel}), we use the following identity for a tail $U(1)-\cdots-U(n-1)$ attached to a node $U(n)$.
For the strictly increasing summation variables $p\in(\mathbb{Z}+\frac{n-1}{2})^n$ of the node $U(n)$,
the sum over the tail with the strictly increasing $p^{(I)}\in(\mathbb{Z}+\frac{I-1}{2})^I$ is
\begin{align}
\label{tail_identity}
\sum_{p^{(1)},\cdots,p^{(n-1)}}\prod_{I=1}^{n-1}\Bigl(\prod_{i\neq j}(1-q^{p^{(I)}_i-p^{(I)}_j})\prod_{i,j}\frac{q^{\frac12(p^{(I)}_i-p^{(I+1)}_j)}}{1-q^{p^{(I)}_i-p^{(I+1)}_j}}\Bigr)
&=\pm\frac{1}{\prod_{k=1}^{n-1}k!}\prod_{i<j}\frac{(p_j-p_i)\,q^{\frac12(p_j-p_i)}}{1-q^{p_j-p_i}}
\end{align}
with $p^{(n)}=p$.
We show (\ref{tail_identity}) by induction on $n$.
With $p'=p^{(n-1)}$, the induction step requires
\begin{align}
\label{tail_step}
\sum_{p'\in(\mathbb{Z}+\frac{n-2}{2})^{n-1}}\prod_{i<j}(p'_j-p'_i)\,\frac{\prod_iq^{p'_i}\prod_{i<j}(q^{p'_j}-q^{p'_i})\prod_{i<j}(q^{p_j}-q^{p_i})}{\prod_{i,j}(q^{p_j}-q^{p'_i})}&=\pm\prod_{i<j}(p_j-p_i),
\end{align}
where the sum over all $p'$ is $(n-1)!$ times the sum over the strictly increasing $p'$.
The fraction is the Cauchy determinant $\pm\det_{n\times n}$ whose $i$-th row is $(q^{p'_i}/(q^{p'_i}-q^{p_j}))_j$ for $i\le n-1$ and whose last row is $(1)_j$.
We write $\prod_{i<j}(p'_j-p'_i)$ as the operator $\prod_{i<j}(z_j\partial_{z_j}-z_i\partial_{z_i})$ acting on $\prod_iz_i^{p'_i}$.
Since $p'_i-p_j\in\mathbb{Z}+\frac12$, the sum of $z_i^{p'_i}$ times the $i$-th row over $p'_i$ is $(z_i^{p_j})_j\,g(z_i)$ with $g(z)=\sum_{t\in\mathbb{Z}+\frac12}z^t/(1-q^{-t})$,
which converges for $1<|z_i|<|q|^{-1}$.
The sum over all $p'$ converges for $|q|<|z_i|<|q|^{-1}$, since the $i$-th row approaches the last row as $p'_i\rightarrow-\infty$,
and hence it is the analytic continuation of $\pm\prod_ig(z_i)\det_{i,j}(z_i^{p_j})$ with $z_n=1$.
The left-hand side of (\ref{tail_step}) is the operator acting on this function at $z_i=1$.
As $z_i\rightarrow1$ we have $g(z_i)\simeq1/\log z_i$ and $\det_{i,j}(z_i^{p_j})\simeq\prod_{i<j}(\log z_j-\log z_i)(p_j-p_i)/\prod_{k=1}^{n-1}k!$, so that the poles cancel.
Since the function is antisymmetric in $\log z_1,\cdots,\log z_{n-1}$, it is $\prod_{i<j<n}(\log z_j-\log z_i)$ times a function regular at $z_i=1$.
The operator has the same degree, so that only the leading terms contribute at $z_i=1$, which gives $\pm\prod_{i<j}(p_j-p_i)$.
We have checked numerically that the sign in (\ref{tail_identity}) is $(-1)^{\frac{n(n-1)}{2}}$ for $n\le5$.

For the mirror quiver, by (\ref{tail_identity}) and $(1-q^x)(1-q^{-x})\,q^x/(1-q^x)^2=-1$,
the two tails and the vector multiplet of $U(N)$ give $\pm\prod_{i<j}(p^{(N)}_j-p^{(N)}_i)^2/(\prod_{k=1}^{N-1}k!)^2$.
Then the Cauchy-Binet formula gives
\begin{align}
\label{SQCD_H_det}
\mathcal{Q}^{(1)-(2)-\cdots-(N)-\cdots-(2)-(1)}&=\pm\frac{1}{\bigl(\prod_{k=1}^{N-1}k!\bigr)^2}\det_{0\le j,k\le N-1}\Bigl(\sum_{p\in\mathbb{Z}+\frac{N-1}{2}}\frac{p^{j+k}q^p}{(1-q^p)^2}\Bigr),
\end{align}
where for odd $N$ the sums omit $p=0$, and the terms with $p_i=0$ remove the row and the column $j=k=0$.
The entries vanish for odd $j+k$, and those for even $j+k$ are $2a_{\frac{j+k}{2}}$.
Hence the determinant is the product of the determinants for even $j,k$ and for odd $j,k$, which gives (\ref{SQCD_H_Hankel}).
Since $q^p/(1-q^p)^2=\sum_{d>0}dq^{pd}$, $a_k$ is the double sum (\ref{double_sum}) with $p^{2k}d$.
For integer $p$ we have $a_0=\frac{1-E_2}{24}$ and $a_k=-\frac{B_{2k}}{4k}DE_{2k}$ for $k\ge1$.
For half-integer $p$, $a_k$ is replaced by $2^{-2k}a_k(q^{\frac12})-a_k(q)$.
Since $a_0$ for half-integer $p$ and $DE_{2k}$ are quasimodular forms of pure weights $2$ and $2k+2$,
$\mathcal{Q}^{(1)-(2)-\cdots-(N)-\cdots-(2)-(1)}$ is a quasimodular form of pure weight $N^2+N-n_0$, in agreement with (\ref{qmod_weight}).
It is a form for $SL(2,\mathbb{Z})$ for odd $N$ and for $\Gamma^0(2)$ for even $N$.
For example, we find up to signs
\begin{align}
\label{SQCD_H_qmod2}
\mathcal{Q}^{(1)-(2)-(1)}&=\frac{1}{6912}\bigl(E_2-E_2(q^{\frac12})\bigr)\bigl(4E_2^2-4E_4-E_2(q^{\frac12})^2+E_4(q^{\frac12})\bigr),
\\*
\label{SQCD_H_qmod3}
\mathcal{Q}^{(1)-(2)-(3)-(2)-(1)}&=\frac{1}{207360}(E_2^2-E_4)(E_2E_4-E_6),
\end{align}
and for $N=5$
\begin{align}
\label{SQCD_H_qmod5}
&\mathcal{Q}^{(1)-(2)-(3)-(4)-(5)-(4)-(3)-(2)-(1)}
\nonumber\\*
&=\frac{1}{5184}\cdot\frac{25E_2^3E_6-39E_2^2E_4^2+3E_2E_4E_6+25E_4^3-14E_6^2}{7257600}
\nonumber\\*
&\times\frac{49E_2^2E_4^3-25E_2^2E_6^2-48E_2E_4^2E_6-25E_4^4+49E_4E_6^2}{25401600}.
\end{align}

\paragraph{Unbalanced SQCD and its mirror}
For the Nahm pole boundary condition $\mathrm{Nahm}'$ of $U(N)$ SQCD with $N_f>2N$ flavors,
the $N_f$ fundamental hypermultiplets have the spin $\frac{N-1}{2}$.
Hence $n_0=N_f$ for odd $N$ and $n_{\frac12}=N_f$ for even $N$, and (\ref{chi_unflavored}) for (\ref{SQCD_nonbal_rv_simple}) reads
\begin{align}
\label{SQCD_nonbal_C_S}
\chi[\mathsf{V}^{U(N)+N_f}_{\mathrm{Nahm}'}]&=\pm q^{\frac{c}{24}}
\begin{cases}
\eta(q)^{2N_f-2N}\,\mathcal{Q}^{U(N)+N_f}&(N\ \textrm{odd})\\
\frac{\eta(q^{1/2})^{2N_f}}{\eta(q)^{2N_f+2N}}\,\mathcal{Q}^{U(N)+N_f}&(N\ \textrm{even})\\
\end{cases},
\nonumber\\*
\mathcal{Q}^{U(N)+N_f}&=\sum_{p_1<\cdots<p_N}\prod_{i\neq j}(1-q^{p_i-p_j})\prod_{i:\,p_i\neq0}\frac{q^{\frac{N_f}{2}p_i}}{(1-q^{p_i})^{N_f}},
\end{align}
where $p_i\in\mathbb{Z}+\frac{N-1}{2}$, and one of $p_i$ is zero for odd $N$.
The bound (\ref{qmod_weight}) is $(N-1)(N_f-N)$ for odd $N$ and $N(N_f-N+1)$ for even $N$.
Let $N=2n$ or $N=2n+1$.
We find
\begin{align}
\label{SQCD_nonbal_even}
\mathcal{Q}^{U(N)+2N+2s}&=(-1)^n\det_{0\le i,j\le n-1}(Z_{i+j+s+1})\det_{0\le i,j\le n-1}(Z_{i+j+s+1}+4Z_{i+j+s+2}),
\\*
\label{SQCD_nonbal_odd}
\mathcal{Q}^{U(N)+2N+2s+1}&=\det_{0\le i,j\le n-1}(Y_{i+j+s+2})^2,
\end{align}
with $Z_k$ in (\ref{Zk_def}) and
\begin{align}
\label{SQCD_nonbal_Y}
Y_j&=\sum_{\substack{p\in\mathbb{Z}+\frac{N-1}{2}\\ p>0}}\frac{q^{(j-\frac12)p}(1+q^p)}{(1-q^p)^{2j}}.
\end{align}
For $s=0$, (\ref{SQCD_nonbal_even}) is (\ref{SQCD_C_Hankel}).
These follow as for $N_f=2N$, where $q^{(m_a+m_b+N)p}/(1-q^p)^{2N}$ in (\ref{SQCD_CauchyBinet}) is replaced by $q^{(m_a+m_b)p}(q^{-\frac{p}{2}}-q^{\frac{p}{2}})^{-N_f}$.
In the basis above the entries are the sums of $(q^{\frac{p}{2}}+q^{-\frac{p}{2}})^{\epsilon+\epsilon'}(q^{\frac{p}{2}}-q^{-\frac{p}{2}})^{k+k'-N_f}$,
which vanish unless $k+k'-N_f$ is even.
Hence $\mathcal{M}$ is block diagonal for even $N_f$, which gives (\ref{SQCD_nonbal_even}).
For odd $N_f$ it is block off-diagonal with the entries $2Y_j$, $j=\frac{N_f-k-k'}{2}$, which gives (\ref{SQCD_nonbal_odd}).
For integer $p$, $Y_j$ is a linear combination of the sums $\sum_{p>0}\sum_{n\in2\mathbb{Z}_{\ge0}+1}n^{2k-1}q^{\frac{pn}{2}}$ with $1\le k\le j$.
By (\ref{Ek_def}) it is a linear combination of $1$ and the modular forms $E_{2k}(q^{\frac12})-2^{2k-1}E_{2k}$ for $\Gamma^0(2)$.
For half-integer $p$, we find that $Y_j$ is $\eta(q)^8/\eta(q^{\frac12})^4$ times a polynomial of mixed weight at most $2j-2$ in the modular forms $2E_2-E_2(q^{\frac12})$ and $E_4$,
where $\eta(q)^{16}/\eta(q^{\frac12})^8=\delta_4/240$.
Hence $\mathcal{Q}^{U(N)+N_f}$ is a quasimodular form for $\Gamma^0(2)$, and for $SL(2,\mathbb{Z})$ if $N$ is odd and $N_f$ is even, in agreement with (\ref{qmod_weight}).
For odd $N_f$ it has depth zero.
For the cases in (\ref{SQCD_nonbal_list}) we find
\begin{align}
\label{SQCD_nonbal_25}
\mathcal{Q}^{U(2)+5}&=\frac{\delta_4\bigl(1-2E_2+E_2(q^{\frac12})\bigr)^2}{138240},
\\*
\label{SQCD_nonbal_26}
\mathcal{Q}^{U(2)+6}&=-\frac{(10\delta_2+\delta_4)(21\delta_2-\delta_6)}{21772800},
\\
\label{SQCD_nonbal_27}
\mathcal{Q}^{U(2)+7}&=\frac{\delta_4}{884736000}\Bigl(9-10\bigl(2E_2-E_2(q^{\frac12})\bigr)+5\bigl(2E_2-E_2(q^{\frac12})\bigr)^2-4E_4\Bigr)^2,
\\
\label{SQCD_nonbal_37}
\mathcal{Q}^{U(3)+7}&=\frac{\bigl(17-20E_2+10E_2(q^{\frac12})-8E_4+E_4(q^{\frac12})\bigr)^2}{33177600},
\\*
\label{SQCD_nonbal_38}
\mathcal{Q}^{U(3)+8}&=-\frac{(10E_2+E_4-11)(21E_2-E_6-20)}{21772800}.
\end{align}

For the Dirichlet and Nahm pole boundary conditions $(\mathcal{D},\mathrm{Nahm}^{N_f-3},\mathcal{D})$ of the mirror quiver (\ref{SQCD_nonbal_quiver}),
the bifundamental hypermultiplets between the $N_f-2N+1$ nodes of the plateau have integer spins.
In $P_0$ all the nodes of the plateau carry the same $p$.
The non-vanishing weights of these bifundamental hypermultiplets then cancel the vector multiplets of all but one node of the plateau, as for the necklace quiver.
Hence
\begin{align}
\label{SQCD_nonbal_H}
\mathcal{Q}^{(1)-\cdots-(N)-\cdots-(N)-\cdots-(1)}&=\mathcal{Q}^{(1)-(2)-\cdots-(N)-\cdots-(2)-(1)}
\end{align}
for any $N_f$.
The extra rank $N(N_f-2N)$ equals the extra $n_0$, so that the eta-quotient in (\ref{chi_unflavored}) is also the same as in (\ref{SQCD_H_S}),
and the characters coincide with those for $N_f=2N$.

\paragraph{$T[SU(N)]$}
For the Nahm pole boundary condition $(\mathcal{D},\mathrm{Nahm}^{N-2})$ of $T[SU(N)]$ in (\ref{TSUN_closed}),
the ranks of the adjacent nodes also differ by one.
We have $\mathrm{rk}\,G=\frac{N(N-1)}{2}$, $n_0=N$ for even $N$ and $n_0=0$ for odd $N$, and $n_{\frac12}=\frac{(N-1)(N-2)}{2}+N-n_0$.
Hence (\ref{chi_unflavored}) reads
\begin{align}
\label{TSUN_S}
\chi[\mathsf{V}^{\wtilde{T[SU(N)]}}_{\mathcal{D},\mathrm{Nahm}^{N-2}}]&=\pm q^{\frac{c}{24}}
\begin{cases}
\frac{\eta(q^{1/2})^{N^2-N+2}}{\eta(q)^{2N^2-2N+2}}\,\mathcal{Q}^{T[SU(N)]}&(N\ \textrm{odd})\\
\frac{\eta(q^{1/2})^{(N-1)(N-2)}}{\eta(q)^{2N^2-6N+2}}\,\mathcal{Q}^{T[SU(N)]}&(N\ \textrm{even})\\
\end{cases},
\nonumber\\*
\mathcal{Q}^{T[SU(N)]}&=\sum_{p}\prod_{I=1}^{N-1}\prod_{i\neq j}(1-q^{p^{(I)}_i-p^{(I)}_j})
\prod_{i:\,p^{(N-1)}_i\neq0}\frac{q^{\frac{N}{2}p^{(N-1)}_i}}{(1-q^{p^{(N-1)}_i})^N}
\nonumber\\*
&\times\prod_{I=1}^{N-2}\prod_{i,j}\frac{q^{\frac12(p^{(I)}_i-p^{(I+1)}_j)}}{1-q^{p^{(I)}_i-p^{(I+1)}_j}},
\end{align}
where $p^{(I)}_i\in\mathbb{Z}+\frac{I-1}{2}$ are strictly increasing in $i$, and one of $p^{(N-1)}_i$ is zero for even $N$.
The bound (\ref{qmod_weight}) is $N(N-1)-n_0$.
The quiver consists of one tail $U(1)-\cdots-U(N-2)$ attached to the node $U(N-1)$.
By (\ref{tail_identity}) with $n=N-1$, the tail and the vector multiplet of $U(N-1)$ give
$\pm\prod_{i<j}(p_j-p_i)(q^{p_j}-q^{p_i})\prod_iq^{-\frac{N-2}{2}p_i}/\prod_{k=1}^{N-2}k!$ with $p=p^{(N-1)}$,
and the Cauchy-Binet formula gives
\begin{align}
\label{TSUN_det}
\mathcal{Q}^{T[SU(N)]}&=\pm\frac{1}{\prod_{k=1}^{N-2}k!}\det_{\substack{0\le k\le N-2\\ 1\le b\le N-1}}\Bigl(\sum_{p\in\mathbb{Z}+\frac{N}{2}}\frac{p^kq^{bp}}{(1-q^p)^N}\Bigr).
\end{align}
For even $N$ the sums omit $p=0$, and the terms with $p_i=0$ replace the row $k=0$ by $(1,\cdots,1)$.
Since $q^{bp}/(1-q^p)^N=\sum_{d>0}\binom{d-b+N-1}{N-1}q^{pd}$ for $p>0$ and $1\le b\le N-1$,
the entries are linear combinations of the double sums (\ref{double_sum}).
The polynomials $\binom{d-b+N-1}{N-1}$ vanish at $d=0$, and $d\rightarrow-d$ maps them to $(-1)^{N-1}\binom{d-(N-b)+N-1}{N-1}$.
Hence in the basis $d,d^2,\cdots,d^{N-1}$ the sums over $\pm p$ give $2\sum_{p,d>0}p^kd^lq^{pd}$ for odd $k+l$ and $0$ for even $k+l$.
The change of basis has the determinant $\pm1/\prod_{k=1}^{N-1}k!$, since the values at $d=1,\cdots,N-1$ form a triangular matrix with unit diagonal.
For even $N$ the row $(1,\cdots,1)$ becomes $(N-1)!$ times the row of $d^{N-1}$.
The matrix splits into two blocks, and we find
\begin{align}
\label{TSUN_blocks}
\mathcal{Q}^{T[SU(2n+1)]}&=\pm C_{2n+1}\det_{0\le i,j\le n-1}\Bigl(\sum_{p,d>0}p^{2i}d^{2j+1}q^{pd}\Bigr)\det_{0\le i,j\le n-1}\Bigl(\sum_{p,d>0}p^{2i+1}d^{2j+2}q^{pd}\Bigr),
\\*
\label{TSUN_blocks_even}
\mathcal{Q}^{T[SU(2n+2)]}&=\pm C_{2n+2}\det_{0\le i,j\le n-1}\Bigl(\sum_{p,d>0}p^{2i+1}d^{2j+2}q^{pd}\Bigr)^2,
\end{align}
where $p\in\mathbb{Z}_{\ge0}+\frac12$ for $T[SU(2n+1)]$ and $p\in\mathbb{Z}_{>0}$ for $T[SU(2n+2)]$,
and the constants are
\begin{align}
C_{2n+1}&=\frac{2^{2n}}{\prod_{k=1}^{2n}k!\prod_{k=1}^{2n-1}k!},&
C_{2n+2}&=\frac{2^{2n}}{\bigl(\prod_{k=1}^{2n}k!\bigr)^2}.
\end{align}
The second determinant for $T[SU(2n+1)]$ is obtained by applying $D$ to the entries of the first one,
and for $T[SU(2n+2)]$ the two determinants coincide by the symmetry under $p\leftrightarrow d$.
By (\ref{double_sum}) the double sums are quasimodular forms of pure weight $k+l+1$ for $p^kd^l$.
Hence $\mathcal{Q}^{T[SU(N)]}$ is a quasimodular form of pure weight $N(N-1)-n_0$, in agreement with (\ref{qmod_weight}).
It is a form for $SL(2,\mathbb{Z})$ for even $N$ and for $\Gamma^0(2)$ for odd $N$.
For example, we find up to signs
\begin{align}
\label{TSUN_qmod3}
\mathcal{Q}^{T[SU(3)]}&=\frac{1}{6912}\bigl(E_2-E_2(q^{\frac12})\bigr)\bigl(2E_2^2-2E_4-E_2(q^{\frac12})^2+E_4(q^{\frac12})\bigr),
\\*
\label{TSUN_qmod4}
\mathcal{Q}^{T[SU(4)]}&=\frac{1}{82944}(E_2^2-E_4)^2,
\\*
\label{TSUN_qmod6}
\mathcal{Q}^{T[SU(6)]}&=\frac{1}{5184}\Bigl(\frac{5E_2^6-27E_2^4E_4+32E_2^3E_6-9E_2^2E_4^2+15E_4^3-16E_6^2}{9953280}\Bigr)^2.
\end{align}
In contrast to $\mathcal{Q}^{U(N)+N_f}$, the sums for the mirror quivers and $T[SU(N)]$ have pure weights.

\paragraph{ADHM theory and necklace quiver}
For the Nahm pole boundary condition $\mathrm{Nahm}'$ of the $U(N)$ ADHM theory with $N\ge2$ and $l$ flavors at $x=1$ we have $\mathrm{rk}\,G=N$.
The adjoint hypermultiplet has the integer spins $0,1,\cdots,N-1$, and the $l$ fundamental hypermultiplets have the spin $\frac{N-1}{2}$.
Hence $n_0=N+l$ for odd $N$, and $n_0=N$ and $n_{\frac12}=l$ for even $N$.
The non-vanishing weights $p_i-p_j$ of the adjoint hypermultiplet cancel the vector multiplet,
since $(1-q^x)(1-q^{-x})\,q^{\frac{x}{2}}/(1-q^x)\cdot q^{-\frac{x}{2}}/(1-q^{-x})=1$, and (\ref{chi_unflavored}) for (\ref{ADHM_rv_closed}) reads
\begin{align}
\label{ADHM_C_S}
\chi[\mathsf{V}^{U(N)\,\textrm{ADHM-}[l]}_{\mathrm{Nahm}'}]&=\pm q^{\frac{c}{24}}
\begin{cases}
\eta(q)^{2l}\,\mathcal{Q}^{U(N)\,\textrm{ADHM-}[l]}&(N\ \textrm{odd})\\
\frac{\eta(q^{1/2})^{2l}}{\eta(q)^{2l}}\,\mathcal{Q}^{U(N)\,\textrm{ADHM-}[l]}&(N\ \textrm{even})\\
\end{cases},
\nonumber\\*
\mathcal{Q}^{U(N)\,\textrm{ADHM-}[l]}&=\sum_{p_1<\cdots<p_N}\prod_{i:\,p_i\neq0}\frac{q^{\frac{l}{2}p_i}}{(1-q^{p_i})^l},
\end{align}
where $p_i\in\mathbb{Z}+\frac{N-1}{2}$, and one of $p_i$ is zero for odd $N$.
The bound (\ref{qmod_weight}) is $l(N-1)$ for odd $N$ and $lN$ for even $N$, since $\dim R=N^2+lN$ and $|\Delta|=N(N-1)$.
The non-zero $p_i$ split into the negative and the positive ones, and $q^{-\frac{lm}{2}}/(1-q^{-m})^l=(-1)^lq^{\frac{lm}{2}}/(1-q^m)^l$.
Hence $\mathcal{Q}^{U(N)\,\textrm{ADHM-}[l]}$ is the coefficient of $w^{N-1}$ for odd $N$ and of $w^N$ for even $N$ in
\begin{align}
\label{Zcal}
&\prod_m\Bigl(1+w\frac{q^{\frac{lm}{2}}}{(1-q^m)^l}\Bigr)\Bigl(1+(-1)^lw\frac{q^{\frac{lm}{2}}}{(1-q^m)^l}\Bigr)
\nonumber\\*
&=\exp\Bigl(\sum_{r>0}\frac{(-1)^{r-1}}{r}\bigl(1+(-1)^{lr}\bigr)w^r\sum_m\frac{q^{\frac{lr}{2}m}}{(1-q^m)^{lr}}\Bigr)
\end{align}
with $m\in\mathbb{Z}_{>0}$ for odd $N$ and $m\in\mathbb{Z}_{\ge0}+\frac12$ for even $N$.
Only the terms with even $lr$ survive in the exponent, and the sums over $m$ are $Z_{\frac{lr}{2}}$ in (\ref{Zk_def}) for odd $N$
and $Z_{\frac{lr}{2}}(q^{\frac12})-Z_{\frac{lr}{2}}(q)$ for even $N$.
Hence $\mathcal{Q}^{U(N)\,\textrm{ADHM-}[l]}$ is a quasimodular form of mixed weight at most $l(N-1)$ for odd $N$ and $lN$ for even $N$, in agreement with (\ref{qmod_weight}).
For $m\in\mathbb{Z}_{>0}$ the coefficients of (\ref{Zcal}) are, up to sign, the multiple $q$-zeta values in \cite{okounkov2014hilbert} with all the arguments equal to $l$ or $2l$.
For $l=1$ they are MacMahon's $A_k(q)$, and those for $m\in\mathbb{Z}_{\ge0}+\frac12$ are $C_k(q^{\frac12})$ \cite{MR1576612},
whose quasimodularity is shown in \cite{MR3028756}.
For odd $l$, $\mathcal{Q}^{U(2)\,\textrm{ADHM-}[l]}$ and $\mathcal{Q}^{U(3)\,\textrm{ADHM-}[l]}$ are $-Z_l$ for half-integer and integer $p$.
We find for $N=2$
\begin{align}
\label{Zcal_12}
\mathcal{Q}^{U(2)\,\textrm{ADHM-}[1]}&=\frac{\delta_2}{24},
\\*
\label{Zcal_22}
\mathcal{Q}^{U(2)\,\textrm{ADHM-}[2]}&=\frac{-10\delta_2-\delta_4+5\delta_2^2}{1440},
\\*
\label{Zcal_32}
\mathcal{Q}^{U(2)\,\textrm{ADHM-}[3]}&=\frac{168\delta_2+21\delta_4+2\delta_6}{120960},
\end{align}
for $N=3$
\begin{align}
\label{Zcal_13}
\mathcal{Q}^{U(3)\,\textrm{ADHM-}[1]}&=\frac{E_2-1}{24},
\\*
\label{Zcal_23}
\mathcal{Q}^{U(3)\,\textrm{ADHM-}[2]}&=\frac{16-20E_2-E_4+5E_2^2}{1440},
\\*
\label{Zcal_33}
\mathcal{Q}^{U(3)\,\textrm{ADHM-}[3]}&=\frac{168E_2+21E_4+2E_6-191}{120960},
\end{align}
and for $N=4$
\begin{align}
\label{Zcal_14}
\mathcal{Q}^{U(4)\,\textrm{ADHM-}[1]}&=\frac{-20\delta_2-2\delta_4+5\delta_2^2}{5760},
\\*
\label{Zcal_24}
\mathcal{Q}^{U(4)\,\textrm{ADHM-}[2]}&=\frac{1}{87091200}\bigl(-12960\delta_2-1764\delta_4+8820\delta_2^2-240\delta_6+1260\delta_2\delta_4
\nonumber\\
&\qquad-2100\delta_2^3+80\delta_2\delta_6+3\delta_4^2-36E_4\delta_4-210\delta_2^2\delta_4+175\delta_2^4\bigr),
\\
\label{Zcal_34}
\mathcal{Q}^{U(4)\,\textrm{ADHM-}[3]}&=\frac{1}{20922789888000}\bigl(-157248000\delta_2-23014992\delta_4+20180160\delta_2^2
\nonumber\\
&\qquad-3975400\delta_6+5045040\delta_2\delta_4+480480\delta_2\delta_6-243243\delta_4^2-1117116E_4\delta_4
\nonumber\\
&\qquad+5460\delta_4\delta_6-54600E_4\delta_6-54600E_6\delta_4
\nonumber\\*
&\qquad+1860\delta_6^2-1764\delta_4^3-5292E_4\delta_4^2-5292E_4^2\delta_4-2000E_6\delta_6\bigr).
\end{align}
Here $Z_5$ and $Z_6$ follow from (\ref{double_sum}) with $E_{10}$ and $E_{12}$ in section \ref{sec_qmod_eisenstein}.
Normalizing the characters by $q^{-\frac{c}{24}}$ with the central charge $c=l(1-3N^2)$ in (\ref{c_ADHM_rv}), we find
\begin{align}
\label{necklace_qMZV}
q^{-\frac{c}{24}}\chi[\mathsf{V}^{U(N)\,\textrm{ADHM-}[l]}_{\mathrm{Nahm}'}]
&=(-1)^{l\lfloor\frac{N}{2}\rfloor}\begin{cases}
\eta(q)^{2l}\,\mathcal{Q}^{U(N)\,\textrm{ADHM-}[l]} & (N\ \textrm{odd})\\
\left(\frac{\eta(q^{1/2})}{\eta(q)}\right)^{2l}\mathcal{Q}^{U(N)\,\textrm{ADHM-}[l]} & (N\ \textrm{even})\\
\end{cases}.
\end{align}
By (\ref{ADHM_rv_dual}) this is also the character for the Neumann boundary condition of the necklace quiver.
For $l=1$ it reproduces the supercharacters of the vertex operator superalgebras $\mathsf{V}_{S_N}$ in \cite{Arakawa:2023cki}
times that of the complex fermions from the Fermi multiplet $\tilde{\eta}_{1,1}$.

For the Nahm pole boundary condition $\mathrm{Nahm}^l$ of the necklace quiver $U(N)^l$ in (\ref{ADHM_closed}) we have $\mathrm{rk}\,G=lN$,
and the twisted bifundamental hypermultiplets have the integer spins $0,1,\cdots,N-1$.
Hence $n_0=lN+1$ for odd $N$, and $n_0=lN$ and $n_{\frac12}=1$ for even $N$.
The set $P_0$ consists of the $p$ with $p^{(1)}=\cdots=p^{(l)}$, as in the localization of (\ref{ADHM_closed_Fermi}).
At these $p$ the vector multiplets cancel the non-vanishing weights of the twisted bifundamental hypermultiplets as for the ADHM theory,
and the remaining factors are those of the twisted fundamental hypermultiplet.
Hence (\ref{chi_unflavored}) reads
\begin{align}
\label{ADHM_H_S}
\chi[\mathsf{V}^{\wtilde{U(N)^l}}_{\mathrm{Nahm}^l}]&=\pm q^{\frac{c}{24}}
\begin{cases}
\eta(q)^{2}\,\mathcal{Q}^{U(N)^l}&(N\ \textrm{odd})\\
\frac{\eta(q^{1/2})^{2}}{\eta(q)^{2}}\,\mathcal{Q}^{U(N)^l}&(N\ \textrm{even})\\
\end{cases},
\nonumber\\*
\mathcal{Q}^{U(N)^l}&=\mathcal{Q}^{U(N)\,\textrm{ADHM-}[1]}
\end{align}
for any $l$.
Hence $\mathcal{Q}^{U(N)^l}$ has mixed weight at most $N-1$ for odd $N$ and $N$ for even $N$,
in agreement with (\ref{qmod_weight}), since $\dim R=lN^2+N$ and $|\Delta_{\rho}|=lN(N-1)$.
By (\ref{ADHM_dual}) and (\ref{Schur_Fermi}), the unflavored Schur indices of 4d $\mathcal{N}=4$ $U(N)$ SYM theory are thus eta-quotients times $\mathcal{Q}^{U(N)\,\textrm{ADHM-}[1]}$,
in agreement with \cite{Beem:2021zvt,Hatsuda:2022xdv}.

\paragraph{$USp(2N)$ ADHM theory}
For the Nahm pole boundary condition of the $USp(2N)$ ADHM theory with $l$ flavors at $x=1$ we have $\mathrm{rk}\,G=N$.
The antisymmetric hypermultiplet has the integer spins $0,2,\cdots,2N-2$, and the $l$ fundamental hypermultiplets have the spin $N-\frac12$.
Hence $n_0=N$ and $n_{\frac12}=l$.
Since $\dim R=N(2N-1)+2lN$ and $|\Delta|=2N^2$, the bound (\ref{qmod_weight}) is $2N(l-1)$,
the quaternionic dimension of the Higgs branch, i.e. the moduli space of $N$ $SO(2l)$ instantons.
By (\ref{USp_rv_closed}) with $x=y_a=1$, (\ref{chi_unflavored}) reads
\begin{align}
\label{USp_C_S}
\chi[\mathsf{V}^{USp(2N)\,\textrm{ADHM-}[2l]}_{\mathrm{Nahm}'}]&=q^{\frac{c}{24}}\frac{\eta(q^{\frac12})^{2l}}{\eta(q)^{2l}}\,\mathcal{Q}^{USp(2N)\,\textrm{ADHM-}[2l]},
\\*
\label{USp_C_S_Q}
\mathcal{Q}^{USp(2N)\,\textrm{ADHM-}[2l]}&=\sum_{0<p_1<\cdots<p_N}\prod_{i=1}^N\frac{q^{(l-2)p_i}(1+q^{p_i})^2}{(1-q^{p_i})^{2l-2}},
\end{align}
where $p_i\in\mathbb{Z}+\frac12$ and $c=l-12(l-2)N^2$ is the central charge in (\ref{c_USp}).
Since $(1+q^p)^2=(1-q^p)^2+4q^p$, the factor for each $p$ is $q^{(l-2)p}/(1-q^p)^{2l-4}+4q^{(l-1)p}/(1-q^p)^{2l-2}$.
Hence $\mathcal{Q}^{USp(2N)\,\textrm{ADHM-}[2l]}$ is the coefficient of $w^N$ in
\begin{align}
\label{Zcal_USp}
\prod_{p}\Bigl(1+w\frac{q^{(l-2)p}(1+q^p)^2}{(1-q^p)^{2l-2}}\Bigr)
&=\exp\Bigl(\sum_{r>0}\frac{(-1)^{r-1}}{r}w^r\sum_{j=0}^{r}\binom{r}{j}4^jZ_{r(l-2)+j}\Bigr),
\end{align}
where $Z_k$ in (\ref{Zk_def}) is taken for half-integer $p$, i.e. it is $Z_k(q^{\frac12})-Z_k(q)$.
Thus it is a quasimodular form for $\Gamma^0(2)$ of mixed weight at most $2N(l-1)$, in agreement with (\ref{qmod_weight}).
We find for $N=1$
\begin{align}
\label{Zcal_USp1}
\mathcal{Q}^{USp(2)\,\textrm{ADHM-}[6]}&=\frac{\delta_4-5\delta_2}{360},
\\*
\label{Zcal_USp1_8}
\mathcal{Q}^{USp(2)\,\textrm{ADHM-}[8]}&=\frac{21\delta_2-\delta_6}{15120},
\\*
\label{Zcal_USp1_10}
\mathcal{Q}^{USp(2)\,\textrm{ADHM-}[10]}&=\frac{-360\delta_2-21\delta_4+10\delta_6+3\delta_4^2+6E_4\delta_4}{1814400},
\end{align}
and for $N=2$
\begin{align}
\label{Zcal_USp2}
\mathcal{Q}^{USp(4)\,\textrm{ADHM-}[6]}&=\frac{-540\delta_2+42\delta_4+40\delta_6+175\delta_2^2-70\delta_2\delta_4+\delta_4^2-12E_4\delta_4}{1814400}.
\end{align}
The components of weight $2N(l-1)$ are non-zero.
By (\ref{USp_rv_dual}) these are also the characters for the Neumann boundary condition of the affine $D_l$ quiver.

For the Nahm pole boundary condition $\mathrm{Nahm}^{l+1}$ of the affine $D_l$ quiver in (\ref{USp_fw_closed}),
the unflavored limit sets $u^{(2)}=\cdots=u^{(l-2)}$, so that by the localization above the unflavored character is the same for all $l\ge4$.
For $l=4$ we have $\mathrm{rk}\,G=6N$, $|\Delta_{\rho}|=8N^2-6N$ and $\dim R=8N^2+N$.
The twisted fundamental hypermultiplet has the spin $\frac{N-1}{2}$, and the twisted bifundamental hypermultiplets have the spins $\frac{N}{2},\cdots,\frac{3N}{2}-1$.
For odd $N$ we have $n_0=1$ and $n_{\frac12}=4N$, and the bound (\ref{qmod_weight}) is $7N-1$.
The set $P_0$ consists of the $p$ with $p^{(0)}_i=0$ for some $i$.
For $N=1$ the sums over the end nodes $\alpha_1$, $\alpha_3$ and $\alpha_4$ are given by (\ref{tail_identity}) with $n=2$,
and with $p=p^{(2)}$ we find
\begin{align}
\label{USp_H_N1}
\mathcal{Q}^{\hat{D}_4}\big|_{N=1}&=\sum_{p_1<p_2}(p_2-p_1)^3\frac{q^{\frac12(p_2-p_1)}}{1-q^{p_2-p_1}}\prod_{i=1}^2\frac{q^{-\frac12p_i}}{1-q^{-p_i}}
=-\frac{DE_4}{240},
\end{align}
where $p_i\in\mathbb{Z}+\frac12$.
By partial fractions the sum over $p_1$ at fixed $d=p_2-p_1$ telescopes to $-dq^{\frac{d}{2}}/(1-q^d)$,
so that (\ref{USp_H_N1}) equals $-\sum_{d>0}d^4q^d/(1-q^d)^2$, a double sum of the form (\ref{double_sum}).
It has pure weight $6$, the bound (\ref{qmod_weight}).
With the central charge $c=-10$ in (\ref{c_USp_fw}), the character is $q^{\frac{c}{24}}\eta(q^{\frac12})^8DE_4/(240\,\eta(q)^{18})$.
By (\ref{Schur_D4}) with the free hypermultiplet removed, the unflavored Schur index of $SU(2)$ gauge theory with four flavors is
\begin{align}
\label{Schur_SU2_4}
\mathcal{I}^{\textrm{Schur}}_{SU(2)+4}(q)&=\frac{DE_4}{240\,q\,(q)_{\infty}^{10}},
\end{align}
which agrees with \cite{Arakawa:2016hkg,Pan:2021mrw,Beem:2021zvt}.

For $N\ge2$ the unflavored character vanishes.
On the Neumann side the Fermi multiplets $\Gamma_{I,J}$ with $I=1,l-1,l$ for even $N$ and $\Gamma_{0,2}$ for odd $N$ have half-integer $\kappa$ or $\kappa'$,
and give $\theta(q^n;q)=0$ with $n\in\mathbb{Z}$.
In (\ref{VOA_USp_fw_id}) the corresponding complex fermions have integer weights, and their zero modes make the supercharacter vanish.
The remaining factor is the Schur index of the rank-$N$ $D_4$ theory, which is obtained from (\ref{Schur_D4}) before the unflavored limit (see Appendix \ref{app_vac}).
On the Nahm pole side, for odd $N$ the sum over $p^{(0)}$ vanishes for each $p^{(2)}$.
Keeping $u^{(0)}$ only in the twisted fundamental hypermultiplet, we evaluate this sum by partial fractions as for (\ref{SQCD_rv_Higgs}).
It localizes to $N$ of the $2N+1$ weights,
and each of the $2N$ weights of the twisted bifundamental hypermultiplet contributes $\sum_{t\in\mathbb{Z}+\frac12}q^{\frac{t}{2}}/(1-q^t)=0$, which vanishes under $t\rightarrow-t$.
For even $N$ the twisted bifundamental hypermultiplets have integer spins,
and $P_0$ consists of the $p$ for which $p^{(I)}$ is contained in $p^{(2)}$ for each end node.
For $N=2$ we have checked numerically that the sum over $p^{(I)}$ vanishes for each end node without the twisted fundamental hypermultiplet and each $p^{(2)}$.

The rank-$N$ $D_4$ theories are of class $\mathcal{S}$ \cite{Beem:2019tfp}, and as for many theories of class $\mathcal{S}$ \cite{Beem:2021zvt},
their unflavored Schur indices for $N=2$ and $3$ can be written as eta-quotients times quasimodular forms for $\Gamma^0(2)$.
We find
\begin{align}
\label{D4_Schur2}
\mathcal{I}^{\textrm{Schur}}_{D_4,2}(q)&=\frac{W_6+W_8}{86400\,q^2\,(q)_{\infty}^2(q^{\frac12};q^{\frac12})_{\infty}^{10}},
\\*
\label{D4_Schur3}
\mathcal{I}^{\textrm{Schur}}_{D_4,3}(q)&=\frac{W_{12}+W_{14}}{143700480000\,q^4\,(q)_{\infty}^{10}(q^{\frac12};q^{\frac12})_{\infty}^{12}},
\end{align}
where $W_k$ are quasimodular forms of weight $k$ for $\Gamma^0(2)$,
\begin{align}
\label{D4_Schur_W}
W_6&=90\delta_4(E_2-\delta_2),
\\*
\label{D4_Schur_W8}
W_8&=-\bigl(60E_2^3\delta_2-115E_2^2\delta_2^2+96E_2^2\delta_4+50E_2\delta_2^3-60E_2\delta_2\delta_4
\nonumber\\
&\qquad+5\delta_2^4-4\delta_2^2\delta_4+4\delta_4^2\bigr),
\\
\label{D4_Schur_W12}
W_{12}&=-30\bigl(7200E_2^5\delta_2-36000E_2^4\delta_2^2+720E_2^4\delta_4+72000E_2^3\delta_2^3-10440E_2^3\delta_2\delta_4
\nonumber\\
&\qquad-72000E_2^2\delta_2^4+27000E_2^2\delta_2^2\delta_4+2763E_2^2\delta_4^2+36000E_2\delta_2^5-25560E_2\delta_2^3\delta_4
\nonumber\\
&\qquad-2212E_2\delta_2\delta_4^2-7200\delta_2^6+8280\delta_2^4\delta_4-551\delta_2^2\delta_4^2-64\delta_4^3\bigr),
\\
\label{D4_Schur_W14}
W_{14}&=-77\bigl(750E_2^4\delta_2^3-3000E_2^3\delta_2^4+675E_2^3\delta_2^2\delta_4-1350E_2^3\delta_4^2+4500E_2^2\delta_2^5
\nonumber\\
&\qquad-2175E_2^2\delta_2^3\delta_4+1530E_2^2\delta_2\delta_4^2-3000E_2\delta_2^6+2325E_2\delta_2^4\delta_4
\nonumber\\*
&\qquad-930E_2\delta_2^2\delta_4^2+12E_2\delta_4^3+750\delta_2^7-825\delta_2^5\delta_4+350\delta_2^3\delta_4^2-36\delta_2\delta_4^3\bigr).
\end{align}
The depths of $W_6$, $W_8$, $W_{12}$ and $W_{14}$, i.e. their degrees in $E_2$ when they are written in terms of $E_2$ and the modular forms $E_2-\delta_2$ and $\delta_4$, are $0$, $2$, $1$ and $3$.
For each component the weight minus the depth is $6$ and $11$, which cancels the weight of the eta-quotient, as for (\ref{Schur_SU2_4}) with $5$.
Let $f=q^{-\frac{c}{24}}\mathcal{I}^{\textrm{Schur}}_{D_4,N}$ with $c=-41$ and $-80$ for $N=2$ and $3$.
Then by (\ref{qmod_transf}) $f(\gamma\tau)$ with $\gamma\in\Gamma^0(2)$ is a polynomial in $\tau$ whose coefficients are $q$-series, as are the logarithmic solutions of modular differential equations.
Indeed, $f$ satisfies exactly the twisted modular differential equations of fourth and seventh order in \cite{Beem:2019tfp}, i.e. (5.5) and (6.2) there,
whose coefficients are modular forms for $\Gamma^0(2)$ \cite{Beem:2017ooy}.
The indicial roots are $\frac{41}{24},\frac{5}{24},-\frac{7}{24},-\frac{7}{24}$ and $\frac{10}{3},\frac13,\frac13,\frac{19}{120},-\frac16,-\frac16,-\frac23$,
and the solutions with the largest roots $-\frac{c}{24}$ are unique.
We have found (\ref{D4_Schur2}) and (\ref{D4_Schur3}) by matching these solutions with eta-quotients times quasimodular forms.
The equations are not derived from null vectors in \cite{Beem:2019tfp}, where the Schur indices are given as $q$-series.
Hence (\ref{D4_Schur2}) and (\ref{D4_Schur3}) are exact if the Schur indices satisfy these equations,
and they agree with the Schur indices evaluated as contour integrals up to $q^{\frac{47}{2}}$ and $q^{\frac{39}{2}}$ (see Appendix \ref{app_checks}).
For $N=2$ the closed form (\ref{D4_Schur2}) also agrees with the H-twist limit of the Neumann half-index (\ref{USp_fw_Neu}) at $x=y_a=1$ divided by the Fermi multiplets.

\subsection{Boundary VOAs}
\label{sec_VOA_general}
We now discuss the boundary VOAs whose vacuum characters reproduce the twist limits.
For the Neumann boundary conditions they are the BRST reductions of the boundary fields with respect to the gauge group \cite{Costello:2018fnz}.
For the Nahm pole boundary conditions the twist limit of the prescription in section \ref{sec_proposal}
takes the form of the character of a reduction of Drinfeld-Sokolov type, which leads us to conjecture the boundary VOAs.
We use the term VOA also for vertex operator superalgebras.
The vacuum characters are weighted by $(-1)^F$ as in (\ref{DEF_hind}), i.e. they are supercharacters.

\subsubsection{Neumann boundary conditions}
\label{sec_VOA_BRST}
Consider the Neumann boundary condition $(\mathcal{N}',N'+\mathcal{F})$ with boundary Fermi multiplets $\mathcal{F}$
for the gauge group $G=\prod_IU(N_I)$ with Lie algebra $\mathfrak{g}$.
Let the hypermultiplets and the boundary Fermi multiplets transform in the representations $R$ and $R_F$ of $G$.
In the H-twist limit the boundary values of the chiral multiplets $X$ and $Y$ in the hypermultiplets (see section \ref{sec_multiplets})
become the symplectic bosons $\mathrm{Sb}(R)$ of weight $\frac12$ valued in $R\oplus R^*$,
and the boundary Fermi multiplets become the complex fermions $\mathrm{Ff}(R_F)$ of weight $\frac12$ \cite{Costello:2018fnz}.
We also write $\mathrm{Ff}(\eta)$ for the complex fermions from a Fermi multiplet $\eta$.
For each node the half-index (\ref{vec_N}) of the vector multiplet becomes
\begin{align}
\label{vec_N_H}
\lim_{t\rightarrow q^{\frac14}}\mathbb{II}^{U(N)}_{\mathcal{N}'}
&=\oint d\mu(s)\,\chi^{bc(\mathfrak{u}(N))}(s), \\*
\label{chi_bc}
\chi^{bc(\mathfrak{u}(N))}(s)&=(q)_{\infty}^{2N}\prod_{\alpha\in\Delta}(qs^{\alpha};q)_{\infty}^2,
\end{align}
where $d\mu(s)=\frac{1}{N!}\prod_{i=1}^N\frac{ds_i}{2\pi is_i}\prod_{\alpha\in\Delta}(1-s^{\alpha})$ and $s^{\alpha}=s_i/s_j$ for the root $\alpha=e_i-e_j$.
The measure $d\mu(s)$ is the Haar measure of $U(N)$ in the Weyl integration formula,
so that the integral projects onto the $G$-invariant states.
The factor $\chi^{bc(\mathfrak{u}(N))}(s)$ is the supercharacter of the $bc$ ghosts of weights $(1,0)$ valued in $\mathfrak{u}(N)$ without the zero modes of $c$,
whose modes $b_{-n}$ and $c_{-n}$ with $n\ge1$ give $\prod_{n\ge1}(1-q^ns^a)^2$ for each weight $a$ of the adjoint representation.
The $b$ ghosts come from the adjoint chiral multiplet with the scalar field $\varphi$, which obeys the Dirichlet boundary condition by (\ref{vec_bc}).
Hence the H-twist limit of the Neumann half-index is the supercharacter of the relative BRST complex
\begin{align}
\label{BRST_complex}
\mathcal{C}&=\bigl(\mathrm{Sb}(R)\otimes\mathrm{Ff}(R_F)\otimes bc(\mathfrak{g})\bigr)^G_{b_0=0},
\end{align}
which consists of the $G$-invariant states annihilated by the zero modes of $b$.

The BRST charge is
\begin{align}
Q&=\oint c_a\Bigl(J^a+\frac12J^a_{\textrm{gh}}\Bigr),
\end{align}
where $J^a$ and $J^a_{\textrm{gh}}$ are the $\mathfrak{g}$ currents of $\mathrm{Sb}(R)\otimes\mathrm{Ff}(R_F)$ and of the ghosts.
It is nilpotent if and only if the level of $J^a$ is the negative of the level of $J^a_{\textrm{gh}}$,
which is the Killing form $\mathrm{Tr}_{\mathfrak{g}}(\mathrm{ad}_A\mathrm{ad}_B)$ for $A,B\in\mathfrak{g}$.
For $\mathfrak{gl}_N$ the Killing form is $2N\,\mathrm{Tr}(AB)-2\,\mathrm{Tr}A\,\mathrm{Tr}B$.
The symplectic bosons and the complex fermions contribute $-\mathrm{Tr}_R(AB)$ and $\mathrm{Tr}_{R_F}(AB)$ to the level of $J^a$.
For $A=B=\mathbf{s}$ in the Cartan subalgebra, $-\mathrm{Tr}_R(\mathbf{s}^2)$, $\mathrm{Tr}_{R_F}(\mathbf{s}^2)$
and $\mathrm{Tr}_{\mathfrak{g}}(\mathrm{ad}_{\mathbf{s}}^2)=\sum_{\alpha\in\Delta}(\alpha\cdot\mathbf{s})^2$
are the contributions of the hypermultiplets, the boundary Fermi multiplets and the vector multiplet to the boundary gauge anomaly in section \ref{sec_anomaly}.
Therefore the nilpotency of $Q$ is equivalent to the cancellation of the boundary gauge anomaly.

Since $Q$ preserves the conformal weight and the charges under the global symmetries,
the Euler-Poincar\'e principle\footnote{
Let $\mathcal{C}=\bigoplus_n\mathcal{C}^n$ be a complex graded by the ghost number $n$ with a differential $Q$ of degree $1$.
If $\mathcal{C}$ is finite-dimensional, then $\sum_n(-1)^n\dim\mathcal{C}^n=\sum_n(-1)^n\dim H^n(\mathcal{C},Q)$.
The states which are not annihilated by $Q$ cancel in pairs with their images under $Q$,
and the remaining states represent the cohomology.
We apply this to each subspace of fixed conformal weight and fixed charges, which is finite-dimensional.
Since the ghosts are fermions, $(-1)^F$ contains the sign $(-1)^n$.
This is the standard method to compute the characters of the BRST and quantum Hamiltonian reductions \cite{MR1174415,MR2060475}.
When the subspaces are infinite-dimensional because of zero modes, a regularization is required (see section \ref{sec_VOA_DS}).
}
implies that the supercharacter of $\mathcal{C}$ is the alternating sum of the supercharacters of the cohomologies over the ghost number.
We assume that the cohomology is concentrated in degree zero.
Then the boundary VOA is the cohomology of degree zero
\begin{align}
\label{VOA_BRST}
\mathsf{V}^{\mathcal{T}}_{\mathcal{N}',N'+\mathcal{F}}&=H^G_{\textrm{BRST}}\bigl(\mathrm{Sb}(R)\otimes\mathrm{Ff}(R_F)\bigr),
\end{align}
and its vacuum character is the H-twist limit of the Neumann half-index.
This is the VOA $\mathfrak{A}_H$ in \cite{Costello:2018fnz}.
The gauge-neutral Fermi multiplets are not affected by the reduction and factor out as decoupled complex fermions.
A pair of symplectic bosons of weight $\frac12$, a complex fermion of weight $\frac12$ and a pair of $bc$ ghosts of weights $(1,0)$
have the central charges $-1$, $1$ and $-2$.
Hence the central charge of (\ref{VOA_BRST}) is
\begin{align}
\label{c_BRST}
c(\mathsf{V}^{\mathcal{T}}_{\mathcal{N}',N'+\mathcal{F}})&=-\dim R+\dim R_F-2\dim G.
\end{align}
The same holds for $G=USp(2N)$.
For the twisted multiplets with the boundary condition $(\mathcal{N},N+\mathcal{F})$ the same holds in the C-twist limit.

The H-twist limits of the half-indices of $\mathcal{N}'$ for the vector multiplet and $N'$ for the hypermultiplets
coincide with the contributions of the 4d $\mathcal{N}=2$ vector multiplet and hypermultiplets to the Schur index \cite{Gadde:2011ik,Gadde:2011uv,Beem:2013sza}.
Indeed, the vector multiplet and the hypermultiplet of weight $\lambda$ give
\begin{align*}
\frac{1}{N!}(q)_{\infty}^{2N}\prod_{\alpha\in\Delta}(s^{\alpha};q)_{\infty}(qs^{\alpha};q)_{\infty},\qquad
\frac{1}{(q^{\frac12}s^{\pm\lambda};q)_{\infty}}.
\end{align*}
This holds for any gauge group and hypermultiplets,
since both sides are the supercharacters of the same BRST complex of the symplectic bosons and the $bc$ ghosts.
Hence the H-twist limit of the Neumann half-index without boundary Fermi multiplets is formally
the Schur index of the 4d $\mathcal{N}=2$ gauge theory with the same gauge group and hypermultiplets,
although the latter is conformal only when the 4d beta function vanishes.\footnote{The analogy with the 4d construction is noted in \cite{Costello:2018fnz},
where the $(0,4)$ Neumann boundary condition is related to the compactification of a 4d $\mathcal{N}=2$ theory on a cigar.}
This leads to a reduction in stages.
The Killing form vanishes on the center $\mathfrak{u}(1)^n$ of $\mathfrak{g}$ generated by $\mathbf{1}_{N_I}$, where $n$ is the number of the nodes.
Suppose that all the nodes are balanced and that all the gauge-charged Fermi multiplets are cross-determinant Fermi multiplets,
which are neutral under $\prod_ISU(N_I)$.
Then the symplectic bosons give the $\mathfrak{sl}(N_I)$ currents at the level $-2N_I=-2h^{\vee}$,
which is the condition for the 4d $\mathcal{N}=2$ $\prod_ISU(N_I)$ gauge theory to be conformal.
The BRST reduction for $\prod_ISU(N_I)$ is then the one which defines the chiral algebra of this 4d theory \cite{Beem:2013sza}.
Performing the BRST reduction for $\prod_ISU(N_I)$ first, we obtain
\begin{align}
\label{VOA_stages}
\mathsf{V}^{\mathcal{T}}_{\mathcal{N}',N'+\mathcal{F}}&\simeq H^{\mathfrak{u}(1)^n}_{\textrm{BRST}}\bigl(\mathcal{V}^{\textrm{4d}}\otimes\mathrm{Ff}(R_F)\bigr),
\end{align}
where $\mathcal{V}^{\textrm{4d}}$ is the chiral algebra of the 4d $\mathcal{N}=2$ $\prod_ISU(N_I)$ gauge theory with the hypermultiplets in $R$ \cite{Beem:2013sza}.
The level of the $\mathfrak{u}(1)^n$ currents of $\mathcal{V}^{\textrm{4d}}$ is given by the Abelian part of the gauge anomaly in section \ref{sec_dualrules},
and the cross-determinant Fermi multiplets cancel it.
This applies to $U(N)$ SQCD, $T[SU(N)]$ and, in the C-twist limit, the mirror of $U(N)$ SQCD.
For the $USp(2N)$ ADHM theory with $l=4$ there is no Abelian part, and the reduction gives the 4d chiral algebra directly (see section \ref{sec_VOA_examples}).
When the Fermi multiplets $\Gamma$ are charged under $SU(N_I)$ or $USp(2N)$, as for the unbalanced SQCD, the ADHM theory and the $USp(2N)$ ADHM theory with $l\ge5$,
the first step does not lead to the 4d chiral algebra.

The $\mathfrak{u}(1)^n$ reduction combines the cross-determinant Fermi multiplets with the charged operators of $\mathcal{V}^{\textrm{4d}}$,
such as the baryons, into gauge-invariant operators.
The gauge-invariant composites of the gauge-charged Fermi multiplets generate current or lattice algebras,
e.g. the $\mathfrak{su}(2)_1$ currents built from the Fermi multiplets $\eta_{\pm}$ for $U(N)$ SQCD
and the lattice VOA generated by the product $\prod_I\eta_I$ of the Fermi multiplets for $T[SU(N)]$.
In the examples below, the boundary fugacities of the Nahm pole side,
combined with the flavor and topological fugacities, become the Cartan fugacities of these algebras.
We test this by the Fourier coefficients of the vacuum characters (see Appendix \ref{app_checks}).

\subsubsection{Nahm pole boundary conditions}
\label{sec_VOA_DS}
We consider the Nahm pole boundary condition $(\mathrm{Nahm}',D')$ in the C-twist limit.
Without the boundary monopole operators, the boundary VOA $\mathsf{V}^{\mathcal{T}}_{\mathcal{D}',D'}$ in (\ref{Dhalf_C}) is the affine VOA of the Lie superalgebra
$\mathfrak{g}_{\partial}\ltimes(\mathfrak{g}_{\partial}^*\oplus\Pi(R\oplus R^*))$ of the Chern-Simons theory which describes the bulk in the C-twist \cite{Costello:2018fnz,Costello:2018swh}.
Here $\Pi$ denotes the odd part, and $\mathfrak{g}_{\partial}^*$ is the coadjoint representation regarded as an Abelian ideal,
so that $[x+x^*,y+y^*]=[x,y]+\mathrm{ad}^*_xy^*-\mathrm{ad}^*_yx^*$ for $x,y\in\mathfrak{g}_{\partial}$ and $x^*,y^*\in\mathfrak{g}_{\partial}^*$.
Identifying $\mathfrak{g}_{\partial}^*$ with $\mathfrak{g}_{\partial}$ by an invariant form, $\mathfrak{g}_{\partial}\ltimes\mathfrak{g}_{\partial}^*\cong\mathfrak{g}_{\partial}[\epsilon]/(\epsilon^2)$,
which is the Takiff algebra of $\mathfrak{g}_{\partial}$ \cite{Takiff:1971,Babichenko:2012uq}.
Thus the even part of the Lie superalgebra is the Takiff algebra.
The odd part $\Pi(R\oplus R^*)$ comes from the hypermultiplets, and the bracket of two odd elements is given by the moment map $R\otimes R^*\rightarrow\mathfrak{g}_{\partial}^*$.
For SQED$_1$ it is $\mathfrak{u}(1|1)$ \cite{Costello:2018swh}.
The boundary monopole operators of non-zero flux modify it by removing null vectors and adding modules \cite{Costello:2018fnz}.
In (\ref{Dhalf_C}) the boundary values of the gauge field and of $\varphi$, which obeys the Neumann boundary condition (see section \ref{sec_B}),
give the factor $(q)_{\infty}^{-2\,\mathrm{rk}\,G}\prod_{\alpha\in\Delta}(q^{1+\alpha(m)}u^{\alpha};q)_{\infty}^{-2}$ of the currents of $\mathfrak{g}_{\partial}$ and $\mathfrak{g}_{\partial}^*$,
the hypermultiplets give the factor of the odd currents,
and the sum over $m$ runs over the lattice of the boundary monopole operators.
The affine VOA has the Sugawara conformal vector, under which all the generators have weight $1$ as in (\ref{Dhalf_C}).
Since the invariant form pairs $\mathfrak{g}_{\partial}$ with $\mathfrak{g}_{\partial}^*$ and is symplectic on the odd part,
while the Killing form vanishes except on $\mathfrak{g}_{\partial}$,
the central charge does not depend on the levels and is the superdimension
\begin{align}
\label{c_Dirichlet}
c(\mathsf{V}^{\mathcal{T}}_{\mathcal{D}',D'})&=2\dim G-2\dim R.
\end{align}
The boundary monopole operators do not change it.

In the limit the specialization (\ref{specialization_gen}) becomes $u=\bar{u}q^{\rho}$, since $q^{\frac12}t^{-2}\rightarrow q$.
This is the principal specialization familiar from the characters of the quantum Hamiltonian reductions \cite{MR2013802}.
It amounts to the shift of the conformal vector by the derivative of the Cartan current of $G_{\partial}$ along $\rho$.
Then in the sector $m$ the components of the $G_{\partial}$ currents and of $\varphi$ along the root $-\alpha$ with $\alpha\in\Delta^+_{\rho}$
contribute the modes of weights $1-\alpha(m)-\mathrm{ht}\,\alpha+n$ with $n\ge 0$, where $\mathrm{ht}\,\alpha$ is the height (\ref{ht_def}).
In the sector $m=0$ those along the negative simple roots have weight $0$, and the vev $\langle\varphi\rangle=\rho(\sigma^+)$ lies in them.
This is the grading of the principal Drinfeld-Sokolov reduction, in which such components are constrained to constants \cite{Drinfeld:1984qv,MR1071340}.
For an affine VOA the constrained currents take the form $\partial+f+b$ of an oper, with a regular nilpotent element $f$ and $b$ in a Borel subalgebra \cite{Beilinson:2005opers}.
Here the vev $\langle\varphi\rangle=\rho(\sigma^+)$ plays the role of $f$, but it lies in the Abelian ideal $\mathfrak{g}_{\partial}^*$ rather than in $\mathfrak{g}_{\partial}$,
in accordance with the ``oper-like'' boundary conditions expected for the Nahm pole boundary conditions in \cite{Costello:2018fnz}.

We conjecture that the boundary VOA for the Nahm pole boundary condition is obtained from $\mathsf{V}^{\mathcal{T}}_{\mathcal{D}',D'}$
by a quantum Hamiltonian reduction of Drinfeld-Sokolov type at the Nahm pole nodes
which constrains the components of the $G_{\partial}$ currents and of $\varphi$ along the negative roots,
those of $\varphi$ to the vev and those of the currents to zero.
These components span a nilpotent subalgebra $\hat{\mathfrak{n}}$ of $\mathfrak{g}_{\partial}\ltimes\mathfrak{g}_{\partial}^*$,
which is isotropic with respect to the invariant forms, i.e. the pairing between $\mathfrak{g}_{\partial}$ and $\mathfrak{g}_{\partial}^*$ and the level of $\mathfrak{g}_{\partial}$.
The constraints define a character of $\hat{\mathfrak{n}}$,
since the vev lies along the negative simple roots while $[\hat{\mathfrak{n}},\hat{\mathfrak{n}}]$ lies along the negative roots $-\alpha$ with $\mathrm{ht}\,\alpha\ge 2$.
Hence the BRST charge with one pair of ghosts for each constraint, i.e. two pairs for each positive root, is nilpotent.

In the C-twist limit the decoupled multiplets in (\ref{Nahm_limit}) become
\begin{align}
I^{\textrm{2d tHM}}(t,q^{-\frac14}tu^{\alpha};q)&\rightarrow\frac{1}{\theta(u^{\alpha};q)^2}, \nonumber\\*
F(q^{\frac{1+\kappa}{2}}t^{-2\kappa}\bar{u}^{w_{\lambda}}x^{f_{\lambda}};q)&\rightarrow\theta(q^{\frac12+\kappa}\bar{u}^{w_{\lambda}}x^{f_{\lambda}};q).
\end{align}
The factor $\theta(u^{\alpha};q)=(u^{\alpha};q)_{\infty}(qu^{-\alpha};q)_{\infty}$ is the supercharacter of the fermionic ghosts $c_{\alpha}$ and $b^{\alpha}$
of weights $0$ and $1$ with the charges of $\alpha$ and $-\alpha$.
At $u=\bar{u}q^{\rho}$ it becomes $\theta(q^{\mathrm{ht}\,\alpha};q)$, and the ghosts have weights $\mathrm{ht}\,\alpha$ and $1-\mathrm{ht}\,\alpha$.
These are the ghosts of the principal Drinfeld-Sokolov reduction, one pair for each positive root \cite{MR1071340,MR1174415}.
For example, for the affine VOA of $\mathfrak{sl}_2$ the specialized vacuum character $1/[(q)_{\infty}(q^2;q)_{\infty}(1;q)_{\infty}]$
times $\theta(q;q)=(q)_{\infty}(1;q)_{\infty}$ gives the Virasoro vacuum character $1/(q^2;q)_{\infty}$ after the zero modes cancel.
Thus the inverse of the Goldstone factor is the supercharacter of the two pairs of ghosts for each positive root required by the reduction above.
The C-twist limit of (\ref{Nahm_limit}) reads
\begin{align}
\label{Nahm_DS}
\chi[\mathsf{V}^{\mathcal{T}}_{\mathrm{Nahm}'}](\bar{u})\prod_{(J,\lambda)}\prod_{\kappa}\theta(q^{\frac12+\kappa}\bar{u}^{w_{\lambda}}x^{f_{\lambda}};q)
&=\Bigl[\chi[\mathsf{V}^{\mathcal{T}}_{\mathcal{D}',D'}](u)\prod_{\alpha\in\Delta^+_{\rho}}\theta(u^{\alpha};q)^2\Bigr]'_{u=\bar{u}q^{\rho}},
\end{align}
where the prime indicates that the sum is restricted to the dominant sectors and that the vanishing factors $1-q^0$ are removed.

The order of the limits is essential.
The prescription specializes $u=\bar{u}(q^{\frac12}t^{-2})^{\rho}$ at generic $t$ and then takes the limit $t\rightarrow q^{-\frac14}$.
At generic $t$ the only modes of weight $0$ which are neutral under the twisted $U(1)_C$
are those of the components of $\varphi$ along the negative simple roots,
and they appear only in the dominant sectors (see step 3 of the prescription in section \ref{sec_proposal}).
Hence only the dominant sectors survive in (\ref{Nahm_limit}).
In the limit $t\rightarrow q^{-\frac14}$ each dominant sector has two vanishing factors for each $\alpha\in\Delta^+_{\rho}$ in the numerator,
from the ghosts $b^{\alpha}$ of weight $1-\mathrm{ht}\,\alpha\le 0$,
and two in the denominator, from the components of the $G_{\partial}$ currents and of $\varphi$ along $-\alpha$.
They cancel in pairs as the zero modes in the Drinfeld-Sokolov reduction.
One pair for each simple root is removed by steps 3 and 4 of the prescription at generic $t$, and the other pairs cancel in the limit $t\rightarrow q^{-\frac14}$, so that each sector has a finite limit.
If instead we set $t=q^{-\frac14}$ first and then take $u\rightarrow\bar{u}q^{\rho}$,
the components along the positive roots $\alpha$ also have modes of weight $0$ in the sectors with $\alpha(m)\le-1-\mathrm{ht}\,\alpha$.
Then all the sectors survive except those in which two of $p^{(I)}_i=m^{(I)}_i+\mu^{(I)}_i$ coincide.
The surviving sectors are the images of the dominant sectors under the Weyl group of $G$ acting on $p$,
and they give $|\mathrm{Weyl}(G)|$ times (\ref{Nahm_DS}).
We have confirmed this for the Nahm pole boundary conditions of $U(2)$ and $U(3)$ SQCD with $4$ and $6$ flavors.
Thus the fugacity $t$ for the twisted $U(1)_C$, which is preserved by the vev,
regularizes the remaining zero modes compatibly with the Higgsing.

The reduction also accounts for the split Fermi multiplets.
Consider a spin-$J$ multiplet $(J,\lambda)$.
Its fermionic currents $\psi_{\mu}$ and $\bar{\psi}_{\mu}$, $\mu=-J,\cdots,J$, have the fugacities $\bar{u}^{w_{\lambda}}x^{f_{\lambda}}q^{\mu}$ and $(\bar{u}^{w_{\lambda}}x^{f_{\lambda}})^{-1}q^{-\mu}$
and the weights $1+\mu$ and $1-\mu$.
Since the bracket of the odd part is the moment map,
the operator product of $\psi_{\mu}$ and $\bar{\psi}_{\mu+1}$ contains the components of $\varphi$ along the negative simple roots.
In the reduction they are replaced by the vev, so that $\psi_{\mu}$ and $\bar{\psi}_{\mu+1}$ acquire a constant simple pole
given by the nilpotent mass $\rho_R(\sigma^+)$ in section \ref{sec_Nahm_hyper}.
Their weights $1+\mu$ and $-\mu$ add up to $1$, and they form a complex fermion of weights $\frac12\pm\kappa$ with $\kappa=\mu+\frac12$.
For $\mu=-J,\cdots,J-1$ these are precisely the $2J$ split Fermi multiplets $\theta(q^{\frac12+\kappa}\bar{u}^{w_{\lambda}}x^{f_{\lambda}};q)$, which are expected to decouple.
The remaining components $\psi_J$ and $\bar{\psi}_{-J}$ of weight $1+J$ are the top components, which contribute to the Nahm pole half-index.
We therefore conjecture
\begin{align}
\label{VOA_DS}
H_{\textrm{DS}}(\mathsf{V}^{\mathcal{T}}_{\mathcal{D}',D'})&\cong\mathsf{V}^{\mathcal{T}}_{\mathrm{Nahm}'}\otimes\mathrm{Ff}_{\textrm{split}},
\end{align}
where $H_{\textrm{DS}}$ is the reduction above applied to the dominant sectors of $\mathsf{V}^{\mathcal{T}}_{\mathcal{D}',D'}$, which are selected by the Higgsing at generic $t$,
and $\mathrm{Ff}_{\textrm{split}}$ are the complex fermions from the split Fermi multiplets.
The constant simple poles show that they are free fermions, while their decoupling as a tensor factor is part of the conjecture.
For integer $J$ they have integer weights, and their zero modes make the unflavored supercharacter of $\mathrm{Ff}_{\textrm{split}}$ vanish.
At $t=q^{-\frac14}$ the graded components of the complex are infinite-dimensional because of the zero modes,
so that the Euler-Poincar\'e principle does not apply directly.
We regard (\ref{Nahm_DS}), in which the zero modes are regularized by $t$ as above, as the character of (\ref{VOA_DS}),
assuming that the cohomology is concentrated in degree zero.

The central charge of $\mathsf{V}^{\mathcal{T}}_{\mathrm{Nahm}'}$ then follows from (\ref{c_Dirichlet}).
The level of the $G_{\partial}$ currents is $k_{\rm eff}$ in (\ref{keff}), the boundary anomaly of $G_{\partial}$.
Hence the shift of the conformal vector by the Cartan current along $\rho$ adds $-12k_{\rm eff}(\mu,\mu)$,
where $k_{\rm eff}(\mu,\mu)=\sum_{\lambda}w_{\lambda}(\mu)^2-\sum_{\alpha\in\Delta}\alpha(\mu)^2$ with $\mu$ in (\ref{specialization_gen}).
The two pairs of ghosts of weights $\mathrm{ht}\,\alpha$ and $1-\mathrm{ht}\,\alpha$ for each $\alpha\in\Delta^+_{\rho}$ add $-4\bigl(6(\mathrm{ht}\,\alpha)^2-6\,\mathrm{ht}\,\alpha+1\bigr)$,
and we subtract the central charges $1-12\kappa^2$ of the decoupled split fermions of weights $\frac12\pm\kappa$ in (\ref{VOA_DS}).
Collecting these contributions for each node and each spin-$J$ multiplet, we find
\begin{align}
\label{c_Nahm}
c(\mathsf{V}^{\mathcal{T}}_{\mathrm{Nahm}'})&=2\sum_{j}(6j^2+6j+1)-2\sum_{(J,\lambda)}(6J^2+6J+1),
\end{align}
where $j$ runs over the spins in the decomposition of the adjoint representation of $\mathfrak{g}$ under $\rho$,
i.e. $0,1,\cdots,N_I-1$ for each node $U(N_I)$ and $1,3,\cdots,2N-1$ for $USp(2N)$.
The summands are the central charges of a pair of bosons of weights $j+1$ and $-j$ and of a pair of fermions of weights $J+1$ and $-J$.
For the trivial $\rho$, (\ref{c_Nahm}) reduces to (\ref{c_Dirichlet}).
For all the dualities in sections \ref{sec_SQCD}--\ref{sec_USpADHM}, including those for the unbalanced SQCD, (\ref{c_Nahm}) agrees with (\ref{c_BRST}) for the dual Neumann boundary conditions,
where for (\ref{USp_fw_dual}) the Fermi multiplets $\Gamma_{I,J}$ with the shifts $(q^{\frac12}t^2)^{\kappa}$ become complex fermions of weights $\frac12\pm\kappa$ in the H-twist limit
and contribute $1-12\kappa^2$ in place of $1$.
Moreover, (\ref{c_Nahm}) gives the power $\frac{c}{24}$ of $q$ in (\ref{chi_unflavored}).
Indeed, since $\theta(q^{-J};q)$ gives $q^{-\frac{J(J+1)}{2}-\frac{1}{12}}$ up to the eta-quotient for both integer and half-integer $J$,
the power of $q$ in (\ref{Nahm_closed}) leads to $\sum_{\alpha\in\Delta^+_{\rho}}\mathrm{ht}\,\alpha+\frac{\mathrm{rk}\,G}{12}-\sum_{(J,\lambda)}\bigl(\frac{J(J+1)}{2}+\frac{1}{12}\bigr)$ in (\ref{chi_unflavored}),
which is $\frac{1}{24}$ times (\ref{c_Nahm}) since $\sum_{\alpha\in\Delta^+_{\rho}}\mathrm{ht}\,\alpha=\sum_j\frac{j(j+1)}{2}$.

The reduction (\ref{VOA_DS}) gives a proposal for the Nahm pole boundary VOAs, but it does not identify them with known VOAs.
The dualities in sections \ref{sec_SQCD}--\ref{sec_USpADHM} answer this identification problem.
In the twist limits they equate the vacuum characters of the Nahm pole boundary VOAs
with those of the BRST reductions (\ref{VOA_BRST}) for the dual Neumann boundary conditions,
and we propose that the VOAs themselves are isomorphic.
We write these identifications for each example below.
The identities of the vacuum characters follow from the conjectural dualities, which we have tested in Appendix \ref{app_numerics}, while the isomorphisms are further conjectures.

\subsubsection{Examples}
\label{sec_VOA_examples}
\paragraph{$U(N)$ SQCD}
In the H-twist limit the Neumann half-index for $U(N)$ SQCD becomes the vacuum character of
$\mathsf{V}^{U(N)+2N}_{\mathcal{N}',N'+\Gamma+\eta}=\mathsf{V}^{U(N)+2N}_{\mathcal{N}',N'+\eta}\otimes\bigotimes_{\Gamma}\mathrm{Ff}(\Gamma)$,
where $\Gamma$ are the $N(N-1)$ gauge-neutral Fermi multiplets in (\ref{SQCD_fw_Neu}).
The boundary VOA for $(\mathcal{N}',N'+\eta)$ is
\begin{align}
\label{VOA_SQCD}
\mathsf{V}^{U(N)+2N}_{\mathcal{N}',N'+\eta}&=H^{U(N)}_{\textrm{BRST}}\left(\mathrm{Sb}(\mathbb{C}^N\otimes\mathbb{C}^{2N})\otimes\mathrm{Ff}(\eta_+)\otimes\mathrm{Ff}(\eta_-)\right).
\end{align}
For $N=2$, $\mathsf{V}^{U(2)+4}_{\mathcal{N}',N'+\eta}$ is precisely the VOA $\mathfrak{A}_H[U(2),\mathbb{C}^{16},\mathbb{C}^4(2)]$ in \cite{Costello:2018fnz}.
Performing the reduction in stages as in (\ref{VOA_stages}), we can also write
\begin{align}
\mathsf{V}^{U(N)+2N}_{\mathcal{N}',N'+\eta}&\simeq H^{\mathfrak{u}(1)}_{\textrm{BRST}}\left(\mathcal{V}^{\textrm{4d}}_{SU(N),N_f=2N}\otimes\mathrm{Ff}(\eta_+)\otimes\mathrm{Ff}(\eta_-)\right),
\end{align}
where $\mathcal{V}^{\textrm{4d}}_{SU(N),N_f=2N}$ is the chiral algebra of 4d $\mathcal{N}=2$ $SU(N)$ SQCD with $2N$ flavors \cite{Beem:2013sza}
and the $\mathfrak{u}(1)$ gauges the baryon number.
For $N=2$ the former is $L_{-2}(\mathfrak{so}_8)$ \cite{Beem:2013sza,Arakawa:2018egx}.
By (\ref{c_BRST}) the central charge is
\begin{align}
\label{c_SQCD_H}
c(\mathsf{V}^{U(N)+2N}_{\mathcal{N}',N'+\eta})&=2-4N^2,
\end{align}
which is $-12c_{4d}$ of 4d $\mathcal{N}=2$ $SU(N)$ SQCD with $2N$ flavors, since the $\mathfrak{u}(1)$ ghosts cancel $\mathrm{Ff}(\eta_{\pm})$.
For $N=2$ it is $-14$, the central charge of $L_{-2}(\mathfrak{so}_8)$.

On the Nahm pole side the half-index becomes the vacuum character of the boundary VOA for $(\mathcal{D},\mathrm{Nahm}^{2N-3},\mathcal{D})$,
i.e. the reduction (\ref{VOA_DS}) at the $2N-3$ Nahm pole nodes of the mirror quiver.
The H-twist limit of the duality (\ref{SQCD_fw_dual}) then gives
\begin{align}
\label{VOA_SQCD_fw_id}
\mathsf{V}^{U(N)+2N}_{\mathcal{N}',N'+\eta}\otimes\bigotimes_{\Gamma}\mathrm{Ff}(\Gamma)
&\cong\mathsf{V}^{\wtilde{(1)-(2)-\cdots-(N)-\cdots-(2)-(1)}}_{\mathcal{D},\mathrm{Nahm}^{2N-3},\mathcal{D}}.
\end{align}
For $N=2$ the left-hand side is $\mathfrak{A}_H[U(2),\mathbb{C}^{16},\mathbb{C}^4(2)]$ times two complex fermions.
By (\ref{c_SQCD_H}) and (\ref{c_Nahm}) both sides have the central charge
\begin{align}
\label{c_SQCD_fw}
c&=2-4N^2+N(N-1)=-(N+1)(3N-2).
\end{align}
The states of weight $1$ are the $\mathfrak{sl}(2N)_{-N}$ currents of the symplectic bosons and the $\mathfrak{su}(2)_1$ currents
built from the bilinears of the Fermi multiplets $\eta_{\pm}$.
The odd states of weight $\frac{N+1}{2}$ are the baryons dressed by the Fermi multiplets $\eta_{\pm}$ and their conjugates (see Appendix \ref{app_vac}).
For $N=2$ they are the $\mathfrak{so}(6)_{-2}\oplus\mathfrak{su}(2)_1$ currents and the spin-$\frac32$ fields in \cite{Costello:2018fnz}.
The baryons and the anti-baryons form doublets of the enhanced topological symmetry $SU(2)$, which acts on the VOA as an outer automorphism.
The $\mathfrak{su}(2)_1$ current algebra is associated with the symmetry $SU(2)_{2d}$ rotating the Fermi multiplets $\eta_{\pm}$ \cite{Costello:2018fnz}.
Its Cartan fugacity $b$ is given by
\begin{align}
\label{xi_SQCD}
b^2&=(u^{(N)})^2\frac{\prod_{\alpha=N+1}^{2N}x_{\alpha}}{z_1z_2\prod_{\alpha=1}^Nx_{\alpha}},
\end{align}
so that it contains the boundary fugacity $u^{(N)}$ of the Nahm pole node $U(N)$ of the mirror quiver.

On the SQCD side the C-twist limit of the Nahm pole half-index (\ref{SQCD_rv_simple}) gives the vacuum character
of the ``oper-like'' boundary VOA $\mathsf{V}^{U(N)+2N}_{\mathrm{Nahm}'}$ in \cite{Costello:2018fnz},
i.e. the reduction (\ref{VOA_DS}) for the Dirichlet boundary condition of SQCD.
The C-twist limit of the duality (\ref{SQCD_rv_dual}) then gives
\begin{align}
\label{VOA_SQCD_rv_id}
\mathsf{V}^{U(N)+2N}_{\mathrm{Nahm}'}
&\cong\mathsf{V}^{\wtilde{(1)-(2)-\cdots-(N)-\cdots-(2)-(1)}}_{\mathcal{N},N+\tilde{\eta}}
=H^{\prod_{I=1}^{2N-1}U(r_I)}_{\textrm{BRST}}\Bigl(\mathrm{Sb}(\widetilde{R})\otimes\bigotimes_{I=1}^{2N}\mathrm{Ff}(\tilde{\eta}_I)\Bigr),
\end{align}
where $\widetilde{R}=\bigoplus_{I=1}^{2N-2}\mathbb{C}^{r_I}\otimes(\mathbb{C}^{r_{I+1}})^*\oplus\mathbb{C}^N\otimes\mathbb{C}^2$ is the representation of the twisted hypermultiplets of the mirror quiver
and $\tilde{\eta}_I$ are the Fermi multiplets in (\ref{SQCD_rv_Neu}).
Since all the Fermi multiplets $\tilde{\eta}_I$ carry determinant charges, no complex fermion decouples.
By (\ref{c_Nahm}) and (\ref{c_BRST}) both sides have the central charge
\begin{align}
\label{c_SQCD_rv}
c&=-2N^3.
\end{align}
The states of weight $1$ are the $\mathfrak{sl}(2)$ currents of the enhanced topological symmetry of SQCD and a $\mathfrak{u}(1)$ current (see Appendix \ref{app_vac}).
The product $\prod_I\tilde{\eta}_I$ of the Fermi multiplets, of weight $N$, generates a lattice VOA of level $2N$.
Its Cartan current combines the $\mathfrak{u}(1)$ current with the Cartan current of $\mathfrak{sl}(2)$,
and its fugacity is $u/z$ for $z_1=z$ and $z_2=z^{-1}$ with $\prod_{\alpha}x_{\alpha}=1$, as we have confirmed for $N=2$.

\paragraph{Unbalanced SQCD}
For $N_f>2N$ the H-twist limit of the Neumann half-index (\ref{SQCD_nonbal_fw_Neu}) becomes the vacuum character of
$\mathsf{V}^{U(N)+N_f}_{\mathcal{N}',N'+\Gamma+\eta}=\mathsf{V}^{U(N)+N_f}_{\mathcal{N}',N'+\Gamma_I+\eta}\otimes\bigotimes_{\Gamma}\mathrm{Ff}(\Gamma)$,
where $\Gamma$ are the $N(N-1)$ gauge-neutral Fermi multiplets and
\begin{align}
\label{VOA_SQCD_nonbal}
\mathsf{V}^{U(N)+N_f}_{\mathcal{N}',N'+\Gamma_I+\eta}&=H^{U(N)}_{\textrm{BRST}}\left(\mathrm{Sb}(\mathbb{C}^N\otimes\mathbb{C}^{N_f})\otimes\mathrm{Ff}(\mathbb{C}^N\otimes\mathbb{C}^{N_f-2N})\otimes\mathrm{Ff}(\eta_+)\otimes\mathrm{Ff}(\eta_-)\right).
\end{align}
The complex fermions $\mathrm{Ff}(\mathbb{C}^N\otimes\mathbb{C}^{N_f-2N})$ come from the Fermi multiplets $\Gamma_I$ in the fundamental representation.
Together with the symplectic bosons they take values in the super vector space $\mathbb{C}^N\otimes\mathbb{C}^{N_f|N_f-2N}$,
on which $\mathfrak{gl}(N_f|N_f-2N)$ acts and commutes with $U(N)$.
Hence the $\mathfrak{sl}(2N)_{-N}$ currents for $N_f=2N$ are replaced by the $\mathfrak{sl}(N_f|N_f-2N)_{-N}$ currents.
The unflavored supercharacter depends only on the superdimension $2N$ of $\mathbb{C}^{N_f|N_f-2N}$,
since the factor of each Fermi multiplet $\Gamma_I$ cancels that of one fundamental hypermultiplet in the integrand.
Thus the unflavored character is the same as for $N_f=2N$, in agreement with (\ref{SQCD_nonbal_H}),
and by (\ref{c_BRST}) so is the central charge $2-4N^2$ in (\ref{c_SQCD_H}).
The H-twist limit of the duality (\ref{SQCD_nonbal_fw_dual}) gives
\begin{align}
\label{VOA_SQCD_nonbal_fw_id}
\mathsf{V}^{U(N)+N_f}_{\mathcal{N}',N'+\Gamma_I+\eta}\otimes\bigotimes_{\Gamma}\mathrm{Ff}(\Gamma)
&\cong\mathsf{V}^{\wtilde{(1)-\cdots-(N)-\cdots-(N)-\cdots-(1)}}_{\mathcal{D},\mathrm{Nahm}^{N_f-3},\mathcal{D}},
\end{align}
and both sides have the central charge (\ref{c_SQCD_fw}) for any $N_f$.

On the SQCD side the C-twist limit of (\ref{SQCD_nonbal_rv_simple}) gives the vacuum character of $\mathsf{V}^{U(N)+N_f}_{\mathrm{Nahm}'}$,
and the C-twist limit of the duality (\ref{SQCD_nonbal_rv_dual}) gives
\begin{align}
\label{VOA_SQCD_nonbal_rv_id}
\mathsf{V}^{U(N)+N_f}_{\mathrm{Nahm}'}
&\cong\mathsf{V}^{\wtilde{(1)-\cdots-(N)-\cdots-(N)-\cdots-(1)}}_{\mathcal{N},N+\tilde{\eta}}
=H^{\prod_{I=1}^{N_f-1}U(r_I)}_{\textrm{BRST}}\Bigl(\mathrm{Sb}(\widetilde{R})\otimes\bigotimes_{I=1}^{N_f}\mathrm{Ff}(\tilde{\eta}_I)\Bigr),
\end{align}
where $\widetilde{R}=\bigoplus_{I=1}^{N_f-2}\mathbb{C}^{r_I}\otimes(\mathbb{C}^{r_{I+1}})^*\oplus\mathbb{C}^{r_N}\oplus\mathbb{C}^{r_{N_f-N}}$
with the last two terms at the $N$-th and the $(N_f-N)$-th nodes,
and $\tilde{\eta}_I$ are the Fermi multiplets in (\ref{SQCD_nonbal_rv_Neu}).
By (\ref{c_Nahm}) and (\ref{c_BRST}) both sides have the central charge
\begin{align}
\label{c_SQCD_nonbal_rv}
c&=4N^3-2N-N_f(3N^2-1),
\end{align}
which reduces to (\ref{c_SQCD_rv}) for $N_f=2N$.
Unlike for the other duality, it depends on $N_f$.
The bare monopole operators of topological charges $\pm1$ have weight $\frac{N_f}{2}-N+1$.
For $N_f=2N$ they are the currents which enhance the topological symmetry $U(1)$ to $SU(2)$,
while for $N_f>2N$ their weight exceeds $1$.
Accordingly, for $U(2)+5$ and $U(2)+6$ we find from (\ref{Nahm_closed}) that the two states of weight $1$ carry no charges,
as expected for the Cartan current of the topological symmetry and the $\mathfrak{u}(1)$ current.
The bare monopole operators give the two even states of weight $\frac32$ for $U(2)+5$, and $2q^{\frac32}$ for $U(3)+7$ in Appendix \ref{app_vac}.

\paragraph{$T[SU(N)]$}
In the H-twist limit the Neumann half-index (\ref{TSUN_Neu}) becomes the vacuum character of
$\mathsf{V}^{T[SU(N)]}_{\mathcal{N}',N'+\Gamma+\eta}=\mathsf{V}^{T[SU(N)]}_{\mathcal{N}',N'+\eta}\otimes\bigotimes_{\Gamma}\mathrm{Ff}(\Gamma)$,
where $\Gamma$ are the $\frac{(N-1)(N-2)}{2}$ gauge-neutral Fermi multiplets in (\ref{TSUN_Neu}).
The boundary VOA for $(\mathcal{N}',N'+\eta)$ is the one proposed in \cite{Costello:2018fnz},
\begin{align}
\label{VOA_TSUN}
\mathsf{V}^{T[SU(N)]}_{\mathcal{N}',N'+\eta}&=H^{\prod_{I=1}^{N-1}U(I)}_{\textrm{BRST}}\Bigl(\mathrm{Sb}(R)\otimes\bigotimes_{I=1}^{N}\mathrm{Ff}(\eta_I)\Bigr),
\end{align}
where $R=\bigoplus_{I=1}^{N-2}\mathbb{C}^{I}\otimes(\mathbb{C}^{I+1})^*\oplus\mathbb{C}^{N-1}\otimes\mathbb{C}^{N}$ is the representation of the hypermultiplets.
By (\ref{VOA_stages}), for $N\ge3$ the VOA is isomorphic to the $\mathfrak{u}(1)^{N-1}$ reduction of the tensor product of the complex fermions from the Fermi multiplets $\eta_I$ and the chiral algebra of 4d $\mathcal{N}=2$ $SU(2)\times\cdots\times SU(N-1)$ linear quiver gauge theory
with one and $N$ fundamental hypermultiplets at the two ends.
For $N=3$ the 4d theory is $SU(2)$ SQCD with four flavors, whose chiral algebra is $L_{-2}(\mathfrak{so}_8)$ as in the $U(N)$ SQCD paragraph.
The Fermi multiplets $\eta_I$ in (\ref{TSUN_Neu}) are identified with the complex fermions $(\chi_I,\psi_I)$ in \cite{Costello:2018fnz}.
By (\ref{c_BRST}) the central charge is
\begin{align}
\label{c_TSUN_H}
c(\mathsf{V}^{T[SU(N)]}_{\mathcal{N}',N'+\eta})&=-N(N^2-N-1).
\end{align}

On the Nahm pole side, the $U(1)$ node of the mirror $T[SU(N)]$ keeps the Dirichlet boundary condition
and the reduction (\ref{VOA_DS}) applies to the nodes $U(2),\cdots,U(N-1)$.
The H-twist limit of the duality (\ref{TSUN_dual}) then gives
\begin{align}
\label{VOA_TSUN_id}
\mathsf{V}^{T[SU(N)]}_{\mathcal{N}',N'+\eta}\otimes\bigotimes_{\Gamma}\mathrm{Ff}(\Gamma)
&\cong\mathsf{V}^{\wtilde{T[SU(N)]}}_{\mathcal{D},\mathrm{Nahm}^{N-2}}.
\end{align}
By (\ref{c_TSUN_H}) and (\ref{c_Nahm}) both sides have the central charge
\begin{align}
\label{c_TSUN}
c&=-N(N^2-N-1)+\frac{(N-1)(N-2)}{2}=1-\frac{N(N-1)(2N-1)}{2}.
\end{align}
For $N=2$ the VOA is $L_1(\mathfrak{psl}_{2|2})$ of central charge $-2$ \cite{Costello:2018fnz}, and (\ref{TSU2_closed}) gives its flavored vacuum character.
Its $\mathfrak{su}(2)_1$ currents are built from the Fermi multiplets $\eta_1$ and $\eta_2$, and their Cartan fugacity is $u^{(1)}x_2$.
For $N=3$ the states of weight $1$ are the $\mathfrak{su}(3)_{-2}\oplus\mathfrak{u}(1)_3$ currents,
and the $20=9+9+2$ odd states of weight $\frac32$ are the operators $O^A_i$, $O_A^i$ and $\tilde{O}^{\pm}$ in \cite{Costello:2018fnz} (see Appendix \ref{app_vac}).
We find that the Cartan fugacity of the lattice VOA $\mathfrak{u}(1)_3$
generated by the products $\tilde{O}^{+}=\prod_I\eta_I$ and $\tilde{O}^{-}=\prod_I\bar{\eta}_I$ of the Fermi multiplets is $u^{(2)}x_3$ with $\prod_az_a=1$.
We expect that more generally $u^{(N-1)}x_N$ is the Cartan fugacity of $\mathfrak{u}(1)_N$ for $T[SU(N)]$.
The combination $u^{(N-1)}x_N$ is singled out as the one which has no mixed anomaly with the $SU(N)$ flavor symmetry,
so that $\tilde{O}^{\pm}$ are singlets of $\mathfrak{su}(N)_{1-N}$.
The other boundary symmetries $U(1)_{u^{(I)}}$ with $I<N-1$ act only on the decoupled neutral fermions.

\paragraph{ADHM theory}
In the H-twist limit the gauge-neutral Fermi multiplet $\eta$ decouples on the Neumann side, and the boundary VOA for $(\mathcal{N}',N',N'+\Gamma)$ is
\begin{align}
\label{VOA_ADHM_H}
\mathsf{V}^{U(N)\,\textrm{ADHM-}[l]}_{\mathcal{N}',N',N'+\Gamma}&=H^{U(N)}_{\textrm{BRST}}\Bigl(\mathrm{Sb}(\mathfrak{gl}_N\oplus\mathbb{C}^N\otimes\mathbb{C}^l)\otimes\mathrm{Ff}(\mathbb{C}^N\otimes\mathbb{C}^l)\Bigr),
\end{align}
where the symplectic bosons come from the adjoint and fundamental hypermultiplets and the complex fermions from the Fermi multiplets $\Gamma_{I,I+1}$.
By (\ref{c_BRST}) its central charge is
\begin{align}
\label{c_ADHM_H}
c(\mathsf{V}^{U(N)\,\textrm{ADHM-}[l]}_{\mathcal{N}',N',N'+\Gamma})&=-3N^2
\end{align}
for any $l$, since the fundamental hypermultiplets and the Fermi multiplets $\Gamma_{I,I+1}$ cancel.
The H-twist limit of the duality (\ref{ADHM_dual}) then conjecturally identifies the boundary VOA for the Nahm pole boundary condition of the necklace quiver,
\begin{align}
\label{VOA_ADHM_H_id}
\mathsf{V}^{U(N)\,\textrm{ADHM-}[l]}_{\mathcal{N}',N',N'+\Gamma}\otimes\mathrm{Ff}(\eta)&\cong\mathsf{V}^{\wtilde{U(N)^l}}_{\mathrm{Nahm}^l}.
\end{align}
By (\ref{c_ADHM_H}) and (\ref{c_Nahm}) both sides have the central charge
\begin{align}
\label{c_ADHM_fw}
c&=1-3N^2.
\end{align}
For $l=1$, $\mathsf{V}^{U(N)\,\textrm{ADHM-}[1]}_{\mathcal{N}',N',N'+\Gamma}$ has the same supercharacter as the vertex operator superalgebra $\mathsf{V}_{S_N}$ constructed in \cite{Arakawa:2023cki},
and we expect them to coincide.
In the construction of $\mathsf{V}_{S_N}$ the fundamental hypermultiplet and the Fermi multiplet $\Gamma_{1,1}$ appear as the framing $\mathbb{C}^{1|1}$ of the Jordan quiver.
The adjoint hypermultiplet gives the level $-2N\,\mathrm{Tr}(AB)+2\,\mathrm{Tr}A\,\mathrm{Tr}B$ of the $\mathfrak{gl}_N$ currents,
which is the negative of the Killing form,
while the fundamental hypermultiplet and the Fermi multiplet $\Gamma_{1,1}$ cancel each other (see section \ref{sec_VOA_BRST}).
The vertex operator superalgebra $\mathsf{V}_{S_N}$ has central charge $-3N^2$ and is a chiral quantization of the Hilbert scheme $\mathrm{Hilb}^N(\mathbb{C}^2)$,
and its associated variety is the Higgs branch $\mathbb{C}^{2N}/S_N$.
It factorizes as $\mathsf{V}_{S_N}\cong\mathsf{W}_{S_N}\otimes\beta\gamma\otimes\mathsf{SF}$ \cite{Arakawa:2023cki}.
Here $\mathsf{W}_{S_N}$ is the vertex operator superalgebra for 4d $\mathcal{N}=4$ $SU(N)$ SYM theory conjectured in \cite{Bonetti:2018fqz},
and $\beta\gamma$ is generated by the traces of the adjoint hypermultiplet.
The symplectic fermions $\mathsf{SF}$ are generated by the gauge-invariant bilinears of the fundamental hypermultiplet and the Fermi multiplet $\Gamma_{1,1}$.
The supercharacter of $\mathsf{V}_{S_N}$ is the Schur index of 4d $\mathcal{N}=4$ $U(N)$ SYM theory \cite{Arakawa:2023cki}.
For $l\ge 2$ the framing is $\mathbb{C}^{l|l}$, formed by the $l$ fundamental hypermultiplets and the Fermi multiplets $\Gamma_{I,I+1}$,
and only the specialized characters reduce to those of $\mathsf{V}_{S_N}$.
Since the Higgs branch of the ADHM theory is the moduli space $\mathcal{M}_{N,l}$ of $N$ $U(l)$ instantons on $\mathbb{C}^2$,
we expect that the associated variety of $\mathsf{V}^{U(N)\,\textrm{ADHM-}[l]}_{\mathcal{N}',N',N'+\Gamma}$ is $\mathcal{M}_{N,l}$
and that (\ref{VOA_ADHM_H_id}) identifies the Nahm pole boundary VOA of the necklace quiver with a chiral quantization of the instanton moduli space.

In the C-twist limit, the vacuum character on the mirror side is that of the boundary VOA
\begin{align}
\label{VOA_necklace_def}
\mathsf{V}^{\wtilde{U(N)^l}}_{\mathcal{N},N,N+\tilde{\Gamma}+\tilde{\eta}}&=H^{U(N)^l}_{\textrm{BRST}}\Bigl(\mathrm{Sb}(\widetilde{R})\otimes\mathrm{Ff}(\tilde{\Gamma})\otimes\bigotimes_{I=1}^l\mathrm{Ff}(\tilde{\eta}_{I,I+1})\Bigr),
\end{align}
where $\widetilde{R}=\bigoplus_{I=1}^{l}\mathbb{C}^N_I\otimes(\mathbb{C}^N_{I+1})^*\oplus\mathbb{C}^N_1$.
Here $\mathbb{C}^N_I$ is the fundamental representation of the node $U(N)^{(I)}$ with $\mathbb{C}^N_{l+1}\equiv\mathbb{C}^N_1$,
and the symplectic bosons come from the twisted bifundamental and fundamental hypermultiplets.
The C-twist limit of the duality (\ref{ADHM_rv_dual}) then conjecturally identifies the boundary VOA for the Nahm pole boundary condition of the ADHM theory as
\begin{align}
\label{VOA_ADHM_rv_id}
\mathsf{V}^{U(N)\,\textrm{ADHM-}[l]}_{\mathrm{Nahm}'}&\cong\mathsf{V}^{\wtilde{U(N)^l}}_{\mathcal{N},N,N+\tilde{\Gamma}+\tilde{\eta}}.
\end{align}
By (\ref{c_Nahm}) and (\ref{c_BRST}) both sides have the central charge
\begin{align}
\label{c_ADHM_rv}
c&=l(1-3N^2),
\end{align}
which is the one used in (\ref{necklace_qMZV}).
For $x=1$ the Fermi multiplet $\tilde{\Gamma}$ cancels the twisted fundamental hypermultiplet in (\ref{ADHM_rv_Neu}).
Writing $s^{(I)}_i=\beta_I\sigma^{(I)}_i$ with $\prod_i\sigma^{(I)}_i=1$, we find that the vacuum character is
\begin{align}
\label{necklace_4d}
\chi[\mathsf{V}^{\wtilde{U(N)^l}}_{\mathcal{N},N,N+\tilde{\Gamma}+\tilde{\eta}}]
&=(q)_{\infty}^{2l}\oint\prod_{I=1}^l\frac{d\beta_I}{2\pi i\beta_I}\,
\mathcal{I}^{\textrm{Schur}}_{SU(N)^l}\Bigl(a_I=\frac{\beta_I}{\beta_{I+1}z};q\Bigr)
\nonumber\\*
&\times\prod_{I=1}^lF\Bigl(q^{\frac12}\Bigl(\frac{\beta_I}{\beta_{I+1}}\Bigr)^N\frac{uy_I}{z^N};q\Bigr),
\end{align}
where $\beta_{l+1}\equiv\beta_1$ and $\mathcal{I}^{\textrm{Schur}}_{SU(N)^l}(a;q)$ is the Schur index of 4d $\mathcal{N}=2$ $SU(N)^l$ necklace quiver gauge theory
with the fugacities $a_I$ for the $U(1)$ flavor symmetries of the bifundamental hypermultiplets \cite{Bourdier:2015sga}.
Since the integrand depends on $\beta_I$ only through $\beta_I/\beta_{I+1}$, the integral over the diagonal $U(1)$ leaves only the factor $(q)_{\infty}^2$ of its vector multiplet,
which is the vacuum character of the symplectic fermions $\mathsf{SF}$.
Hence we expect that $\mathsf{V}^{\wtilde{U(N)^l}}_{\mathcal{N},N,N+\tilde{\Gamma}+\tilde{\eta}}$ is isomorphic to
\begin{align}
\label{VOA_necklace}
\mathsf{SF}\otimes H^{\mathfrak{u}(1)^{l-1}}_{\textrm{BRST}}\Bigl(\mathcal{V}^{\textrm{4d}}_{SU(N)^l}\otimes\bigotimes_{I=1}^l\mathrm{Ff}(\tilde{\eta}_{I,I+1})\Bigr),
\end{align}
where $\mathcal{V}^{\textrm{4d}}_{SU(N)^l}$ is the chiral algebra of the 4d necklace quiver gauge theory \cite{Beem:2013sza}
and $\mathfrak{u}(1)^{l-1}$ gauges the $U(1)$ flavor symmetries of the bifundamental hypermultiplets with $\prod_Ia_I$ fixed.
Unlike for $U(N)$ SQCD, this does not follow simply from the BRST reduction for $SU(N)^l$ performed first.
The reason is that the twisted fundamental hypermultiplet and the Fermi multiplet $\tilde{\Gamma}$ are charged under $SU(N)$ at the node $I=1$,
and their contributions cancel only in the character at $x=1$.
For $l=1$ the vacuum character of $\mathsf{V}^{U(N)\,\textrm{ADHM-}[1]}_{\mathrm{Nahm}'}$ is that of $\mathsf{V}_{S_N}\otimes\mathrm{Ff}(\tilde{\eta}_{1,1})$, the counterpart of (\ref{VOA_ADHM_H_id}).
In this case (\ref{VOA_necklace}) is the factorization $\mathsf{V}_{S_N}\otimes\mathrm{Ff}(\tilde{\eta}_{1,1})\cong\mathsf{SF}\otimes\mathsf{W}_{S_N}\otimes\beta\gamma\otimes\mathrm{Ff}(\tilde{\eta}_{1,1})$
above,
where $\mathsf{W}_{S_N}\otimes\beta\gamma$ plays the role of $\mathcal{V}^{\textrm{4d}}_{SU(N)}$ for the adjoint and singlet hypermultiplets.

The boundary VOA $\mathsf{V}^{\wtilde{U(N)^l}}_{\mathcal{N},N,N+\tilde{\Gamma}+\tilde{\eta}}$ generalizes $\mathsf{V}_{S_N}$ to the cyclic quiver.
The symplectic bosons in (\ref{VOA_necklace_def}) are associated with the representations of the cyclic quiver $\hat{A}_{l-1}$
with dimension vector $N\delta$ and one-dimensional framing at the node $I=1$.
The symplectic bosons from the twisted bifundamental hypermultiplets give the level
$\sum_I[-2N\,\mathrm{Tr}(A_IB_I)+\mathrm{Tr}A_I\,\mathrm{Tr}B_{I+1}+\mathrm{Tr}A_{I+1}\,\mathrm{Tr}B_I]$ for $A=(A_I),B=(B_I)\in\mathfrak{gl}_N^{\oplus l}$.
Those from the twisted fundamental hypermultiplet cancel against the complex fermions from the Fermi multiplet $\tilde{\Gamma}$,
and the complex fermions from the Fermi multiplets $\tilde{\eta}_{I,I+1}$ give $\sum_I(\mathrm{Tr}A_I-\mathrm{Tr}A_{I+1})(\mathrm{Tr}B_I-\mathrm{Tr}B_{I+1})$.
The total level is the negative of the Killing form of $\mathfrak{gl}_N^{\oplus l}$,
in accordance with the anomaly cancellation in section \ref{sec_ADHM_rv} (see section \ref{sec_VOA_BRST}).
Thus the Fermi multiplet $\tilde{\Gamma}$ provides the odd framing at the node $I=1$ as in \cite{Arakawa:2023cki}.
For $l\ge 2$ the Fermi multiplets $\tilde{\eta}_{I,I+1}$ also provide odd coordinates valued in $\det V_I\otimes\det V_{I+1}^{-1}$ with $V_I=\mathbb{C}^N$ at the node $I$,
which are required for the Abelian part.
The corresponding Nakajima quiver variety is the $\mathbb{Z}_l$-equivariant Hilbert scheme of points on $\mathbb{C}^2$,
diffeomorphic to the Hilbert scheme of $N$ points on the minimal resolution of $\mathbb{C}^2/\mathbb{Z}_l$ \cite{MR2372209}.
It resolves the Higgs branch $\mathrm{Sym}^N(\mathbb{C}^2/\mathbb{Z}_l)$ of the necklace quiver gauge theory.
We therefore expect that the boundary VOA is obtained from a sheaf of $\hbar$-adic vertex superalgebras on this quiver variety
as in \cite{Arakawa:2023cki}.
We also expect that its associated variety is $\mathrm{Sym}^N(\mathbb{C}^2/\mathbb{Z}_l)$, so that it is quasi-lisse,
and that the symplectic fermions factor out as in (\ref{VOA_necklace}).

For $N=1$ the complex fermions from the Fermi multiplets $\tilde{\Gamma}$ and $\tilde{\eta}_{I,I+1}$ carry the same gauge charges as the twisted fundamental and bifundamental hypermultiplets.
The boundary VOA is then $\mathsf{SF}$ times the vertex operator superalgebra for the $A_{l-1}$ singularity $\mathbb{C}^2/\mathbb{Z}_l$ constructed in \cite{Arakawa:2026nut}.
The latter is quasi-lisse and is identified there with a fermionic extension of the subregular $\mathcal{W}$-algebra of $\mathfrak{sl}_l$.
By (\ref{c_ADHM_rv}) and $c(\mathsf{SF})=-2$ its central charge is $2-2l$.
For $l=2$ it is $L_1(\mathfrak{psl}_{2|2})$ \cite{Ferrari:2024dst,Arakawa:2026nut}, the boundary VOA for $T[SU(2)]$.
Although $\mathrm{Sym}^N(\mathbb{C}^2/\mathbb{Z}_l)=\mathbb{C}^{2N}/(\mathbb{Z}_l^N\rtimes S_N)$ is the symplectic singularity attached to the complex reflection group $\mathbb{Z}_l^N\rtimes S_N$ in \cite{Bonetti:2018fqz},
for $l\ge 2$ the boundary VOA differs from the vertex operator superalgebra conjectured there.
For $(N,l)=(1,2)$ the latter is the small $\mathcal{N}=4$ superconformal algebra of central charge $-9$, while $L_1(\mathfrak{psl}_{2|2})$ has $-2$.

\paragraph{$USp(2N)$ ADHM theory}
In the C-twist limit, the vacuum character on the mirror side is that of the boundary VOA
\begin{align}
\label{VOA_Dquiver_def}
\mathsf{V}^{\wtilde{\hat{D}_l}}_{\mathcal{N},N,N+\tilde{\Gamma}+\tilde{\eta}}&=H^{\tilde{G}}_{\textrm{BRST}}\Bigl(\mathrm{Sb}(\widetilde{R})\otimes\mathrm{Ff}(\tilde{\Gamma})\otimes\bigotimes_{a=1}^l\mathrm{Ff}(\tilde{\eta}_a)\Bigr),
\end{align}
where $\tilde{G}=\prod_{I=0}^lU(N_I)$ and $\widetilde{R}=\bigoplus_{\alpha_I\cdot\alpha_J=-1}\mathbb{C}^{N_I}_I\otimes(\mathbb{C}^{N_J}_J)^*\oplus\mathbb{C}^N_0$
with the fundamental representation $\mathbb{C}^{N_I}_I$ of the node $\alpha_I$.
The complex fermions from the Fermi multiplets $\tilde{\eta}_a$ give the level $\sum_{I,J}C_{IJ}\mathrm{Tr}A_I\,\mathrm{Tr}B_J$,
and with the twisted bifundamental hypermultiplets the total level is the negative of the Killing form of $\bigoplus_I\mathfrak{gl}_{N_I}$,
in accordance with the anomaly cancellation in section \ref{sec_USp_rv}.
The C-twist limit of the duality (\ref{USp_rv_dual}) then conjecturally identifies the boundary VOA for the Nahm pole boundary condition of the $USp(2N)$ ADHM theory as
\begin{align}
\label{VOA_USp_id}
\mathsf{V}^{USp(2N)\,\textrm{ADHM-}[2l]}_{\mathrm{Nahm}'}&\cong\mathsf{V}^{\wtilde{\hat{D}_l}}_{\mathcal{N},N,N+\tilde{\Gamma}+\tilde{\eta}}.
\end{align}
By (\ref{c_Nahm}) and (\ref{c_BRST}) both sides have the central charge
\begin{align}
\label{c_USp}
c&=l-12(l-2)N^2,
\end{align}
which is the one used in (\ref{USp_C_S}).
For $l=3$ it is one less than (\ref{c_ADHM_rv}) for the $U(N)$ ADHM theory with four flavors,
since the Fermi multiplets $\tilde{\eta}_a$ are three instead of four (see (\ref{theta_A3D3})).
For $x=1$ the Fermi multiplet $\tilde{\Gamma}$ cancels the twisted fundamental hypermultiplet in (\ref{USp_rv_Neu}).
Writing $s^{(I)}_i=\beta_I\sigma^{(I)}_i$ with $\prod_i\sigma^{(I)}_i=1$ as for (\ref{necklace_4d}), we find that the vacuum character is
\begin{align}
\label{Dquiver_4d}
\chi[\mathsf{V}^{\wtilde{\hat{D}_l}}_{\mathcal{N},N,N+\tilde{\Gamma}+\tilde{\eta}}]
&=(q)_{\infty}^{2(l+1)}\oint\prod_{I=0}^l\frac{d\beta_I}{2\pi i\beta_I}\,
\mathcal{I}^{\textrm{Schur}}_{\hat{D}_l}\Bigl(a_{IJ}=\frac{\beta_I}{\beta_J};q\Bigr)
\prod_{a=1}^lF\Bigl(q^{\frac12}y_a\prod_{I=0}^l\beta_I^{N_I(\alpha_I)_a};q\Bigr),
\end{align}
where $\mathcal{I}^{\textrm{Schur}}_{\hat{D}_l}(a;q)$ is the Schur index of the 4d $\mathcal{N}=2$ $SU(N)^4\times SU(2N)^{l-3}$ quiver gauge theory of the affine $D_l$ Dynkin diagram
with the fugacities $a_{IJ}$ for the $U(1)$ flavor symmetries of the bifundamental hypermultiplets.
Since $\sum_IN_I\alpha_I=0$, the diagonal $U(1)$ with $\beta_I=\beta$ acts trivially on the integrand,
and as for the necklace quiver its vector multiplet gives the factor $(q)_{\infty}^2$ of $\mathsf{SF}$.
Hence we expect that $\mathsf{V}^{\wtilde{\hat{D}_l}}_{\mathcal{N},N,N+\tilde{\Gamma}+\tilde{\eta}}$ is isomorphic to
\begin{align}
\label{VOA_Dquiver}
\mathsf{SF}\otimes H^{\mathfrak{u}(1)^{l}}_{\textrm{BRST}}\Bigl(\mathcal{V}^{\textrm{4d}}_{\hat{D}_l}\otimes\bigotimes_{a=1}^l\mathrm{Ff}(\tilde{\eta}_a)\Bigr),
\end{align}
where $\mathcal{V}^{\textrm{4d}}_{\hat{D}_l}$ is the chiral algebra of this 4d theory \cite{Beem:2013sza}
and $\mathfrak{u}(1)^{l}$ gauges the $U(1)$ parts of the nodes modulo the diagonal $U(1)$.
By (\ref{USp_Fermi}) the duality gives a free Fermi gas formula for its vacuum character.

In the H-twist limit the gauge-neutral Fermi multiplets $\eta$ and $\Gamma_{I,J}$ decouple on the Neumann side,
and the boundary VOA for $(\mathcal{N}',N',N'+\Gamma)$, where $\Gamma$ now stands for the Fermi multiplets $\Gamma_{k,k+1}$, is
\begin{align}
\label{VOA_USp_H}
\mathsf{V}^{USp(2N)\,\textrm{ADHM-}[2l]}_{\mathcal{N}',N',N'+\Gamma}&=H^{USp(2N)}_{\textrm{BRST}}\Bigl(\mathrm{Sb}(\Lambda^2\mathbb{C}^{2N}\oplus\mathbb{C}^{2N}\otimes\mathbb{C}^l)\otimes\mathrm{Ff}(\mathbb{C}^{2N}\otimes\mathbb{C}^{l-4})\Bigr),
\end{align}
where the complex fermions come from the Fermi multiplets $\Gamma_{k,k+1}$.
The H-twist limit of the duality (\ref{USp_fw_dual}) then conjecturally identifies the boundary VOA for the Nahm pole boundary condition of the affine $D_l$ quiver,
\begin{align}
\label{VOA_USp_fw_id}
\mathsf{V}^{USp(2N)\,\textrm{ADHM-}[2l]}_{\mathcal{N}',N',N'+\Gamma}\otimes\mathrm{Ff}(\eta)\otimes\mathrm{Ff}(\Gamma_{I,J})&\cong\mathsf{V}^{\wtilde{\hat{D}_l}}_{\mathrm{Nahm}^{l+1}},
\end{align}
where $\mathrm{Ff}(\Gamma_{I,J})$ are the complex fermions of weights $\frac12\pm\kappa$.
By (\ref{c_BRST}) and (\ref{c_Nahm}) both sides have the central charge
\begin{align}
\label{c_USp_fw}
c&=1-4N-3N^2-4N^3
\end{align}
for any $l\ge4$, since the fundamental hypermultiplets with $y_3,\cdots,y_{l-2}$ and the Fermi multiplets $\Gamma_{k,k+1}$ cancel.
For $l=4$ there are no Fermi multiplets $\Gamma_{k,k+1}$,
and (\ref{VOA_USp_H}) is the chiral algebra of the rank-$N$ $D_4$ theory \cite{Beem:2013sza} times the symplectic bosons $\beta\gamma$ of the free hypermultiplet.
Their central charges are $-(6N^2+9N-1)$ and $-1$,
and by $\sum_{\kappa}(1-12\kappa^2)$ the complex fermions contribute $1+5N+3N^2-4N^3$.
For $N=1$ the chiral algebra is $L_{-2}(\mathfrak{so}_8)$ of central charge $-14$ \cite{Beem:2013sza,Arakawa:2018egx}, and (\ref{c_USp_fw}) is $-14-1+5=-10$.
For $N=2$ a VOA of central charge $-41$ is proposed for it in \cite{Beem:2019tfp}, and (\ref{c_USp_fw}) is $-41-1-9=-51$.
For $l\ge5$ only the characters at $v_2=\cdots=v_{l-2}$ reduce to those for $l=4$.

\acknowledgments
The work of T.O. was supported by the Startup Funding no.\ 4007012317 of Southeast University.

\appendix
\section{Checks of the dualities}
\label{app_numerics}
We have checked the identities of the half-indices in two independent ways,
by expanding both sides in $q^{\frac14}$ and by evaluating them numerically at fixed $q$.
The fugacities are generic, except that $\prod_{\alpha}x_{\alpha}=1$ for $T[SU(N)]$ as in section \ref{sec_TSUN}.
The cost of both checks grows with the rank, since the numbers of the flux sectors on the Nahm pole side and of the integration variables on the Neumann side increase,
and this limits the ranks and orders in Table \ref{Table_checks}.

\subsection{\texorpdfstring{Expansions in $q^{\frac14}$}{Expansions in q^(1/4)}}
\label{app_expansions}
We expand both sides in $q^{\frac14}$.
On the Nahm pole side we sum over the fluxes up to a cutoff,
which we take larger than the fluxes of all the sectors contributing up to the given order.
To avoid rounding errors, we carry out the expansions exactly in modular arithmetic.
We fix a large prime $p$ such as $2^{31}-1$ and work in the finite field $\mathbb{F}_p=\mathbb{Z}/p\mathbb{Z}$ of the integers modulo $p$.
Since $p$ is a prime, every non-zero element has an inverse,
so that the sums, products and quotients which appear in the expansions are computed exactly with numbers smaller than $p$.
We replace all the fugacities except for the gauge fugacities, including $t$,
by random elements of $\mathbb{F}_p$, keeping $q^{\frac14}$ as the expansion variable.
Each coefficient of the expansion is then computed exactly as an element of $\mathbb{F}_p$.
On the Neumann side the gauge fugacities remain formal variables,
and each coefficient is a Laurent polynomial in them, whose constant term the contour integrals pick.
We take it node by node.
Two different rational functions of the fugacities agree at a random point only with a negligible probability,
and we have repeated the checks with another prime and other random points.
The orders up to which the two sides agree are listed in Table \ref{Table_checks}.

\begin{table}[!h]
\centering
\footnotesize
\setlength{\tabcolsep}{4pt}
\begin{tabular}{c|c|l}
dual pair & duality & checked up to \\ \hline
$U(N)+2N$ & (\ref{SQCD_fw_dual}) &
$q^{12},q^{6},q^{7/2},q^{2}$ \ ($N=2,\cdots,5$) \\ \hline
$U(N)+2N$ & (\ref{SQCD_rv_dual}) &
$q^{25},q^{9},q^{6},q^{7/2}$ \ ($N=2,\cdots,5$) \\
 & & $q^{5/2},q^{2},q^{2}$ \ ($N=6,7,8$) \\ \hline
$U(N)+N_f$, $N_f>2N$ & (\ref{SQCD_nonbal_fw_dual}) &
$U(2)+5,6,7$: $q^{12}$, \ $U(3)+7,8$: $q^{6}$ \\ \hline
$U(N)+N_f$, $N_f>2N$ & (\ref{SQCD_nonbal_rv_dual}) &
$U(2)+5,6$: $q^{8}$, \ $U(2)+7$: $q^{7}$, \ $U(3)+7,8$: $q^{5}$ \\ \hline
$T[SU(N)]$ & (\ref{TSUN_dual}) &
$q^{16},q^{8},q^{4},q^{3},q^{2}$ \ ($N=3,\cdots,7$) \\ \hline
ADHM, $l=1$ & (\ref{ADHM1_dual}) &
$q^{24},q^{12},q^{8},q^{5},q^{4}$ \ ($N=2,\cdots,6$) \\ \hline
ADHM, $l\ge 2$ & (\ref{ADHM_dual}) &
$l=2$: $q^{10},q^4,q^2,q$ \ ($N=2,\cdots,5$) \\
 & & $l=3$: $q^{6},q^3,q$ \ ($N=2,3,4$) \\
 & & $l=4$: $q^5,q^2$ \ ($N=2,3$) \\
 & & $l=5$: $q^4$ \ ($N=2$) \\ \hline
ADHM & (\ref{ADHM_rv_dual}) &
$N=1$: $q^{12}$ \ ($l=2,3,4$) \\
 & & $N=2$: $q^{9},q^8,q^{8},q^{8}$ \ ($l\le 4$) \\
 & & $N=3$: $q^{7},q^6,q^4$ \ ($l\le 3$) \\ \hline
$USp(2N)$ ADHM & (\ref{USp_fw_dual}) &
$N=1$: $q^{5},q^{3},q^{5/2},q^{2}$ \ ($l=4,5,6,7$) \\
 & & $N=2$: $q^{4},q^{3/2}$ \ ($l=4,5$) \\
 & & $N=3$: $q^{3}$ \ ($l=4$) \\ \hline
$USp(2N)$ ADHM & (\ref{USp_rv_dual}) &
$N=1$: $q^{6},q^{6},q^{4}$ \ ($l=3,4,5$) \\
 & & $N=2$: $q^{5},q^{5},q^{3}$ \ ($l=3,4,5$) \\
 & & $N=3$: $q^{3}$ \ ($l=4$) \\ \hline
$USp(2N)$ ADHM, $l=3$ & (\ref{USp_ADHM4}) &
$q^{29/4},q^{4},q^{3}$ \ ($N=1,2,3$) \\
\end{tabular}
\caption{
The orders of $q$ up to which the identities of the half-indices are checked.
}
\label{Table_checks}
\end{table}

\subsection{\texorpdfstring{Evaluation at fixed $q$}{Evaluation at fixed q}}
\label{app_fixed_q}
We also evaluate both sides numerically at fixed $q$ with random fugacities of unit modulus,
using an independent implementation of the formulas as written in the text.
On the Nahm pole side we truncate the sum over the fluxes.
On the Neumann side we evaluate the contour integrals over $|s_i|=1$ by the trapezoidal rule,
i.e. we replace each integral $\oint\frac{ds}{2\pi is}f(s)$ by the average $\frac1K\sum_{k=0}^{K-1}f(e^{2\pi ik/K})$ over the $K$-th roots of unity.
This picks the sum of the Fourier coefficients of $f$ whose indices are multiples of $K$,
and hence it is exact for Laurent polynomials of degree less than $K$.
Since the integrands are analytic in an annulus $r<|s|<r^{-1}$ around the contour, their Fourier coefficients decay as $r^{|n|}$,
and the error decreases as $r^K$.
For several gauge fugacities we take the product of such grids.
The two sides agree within the numerical accuracy.
For the duality (\ref{USp_rv_dual}) we have also evaluated both sides for $N=1$ with $l=3,4,5$ and for $N=2$ with $l=3$.
For the duality (\ref{USp_fw_dual}) we have evaluated both sides for $N=1$ with $l=4,5$ and for $N=2,3$ with $l=4$,
where the differences decrease with $q$ as expected from the truncation of the fluxes.

\subsection{Alternatives and other checks}
\label{app_checks}
The alternatives to the formulas mentioned in the text are examined in the same ways.
They include dropping the contribution $-\sum_{\alpha}\alpha\alpha^T$ of the vector multiplet to $k_{\rm eff}$,
the sign $(-1)^{m^Tk_{\rm eff}m}$ or the power of $q^{\frac12}t^{-2}$ in (\ref{Nahm_single}),
keeping the decoupled 2d (twisted) hypermultiplets or the split Fermi multiplets,
changing the topological fugacities on the Nahm pole side or the identification of the fugacities,
and dropping or changing the boundary Fermi multiplets.
For all of them the two sides of the identities differ already at low orders between $q^0$ and $q$.
Hence the identities single out the formulas in the text.
In particular, they confirm the assignments of the boundary Fermi multiplets fixed by the anomalies,
including those of the gauge-neutral Fermi multiplets which are fixed only by the pure flavor terms.

The other statements in the text are checked in the same way.
The ratio of the Dirichlet half-index to the decoupled multiplets in (\ref{Nahm_limit}) approaches its limit linearly in the distance from the locus $u=\bar{u}(q^{\frac12}t^{-2})^{\rho}$,
which we have confirmed for the mirror of the $U(2)$ ADHM theory with $l=1$ as an unbalanced example with an adjoint hypermultiplet,
and for the $USp(2N)$ ADHM theory with $l=3$ and $N\le3$, for which the limit agrees with (\ref{USp_simple}).
The formula (\ref{Nahm_simple}) and its explicit forms for the examples agree with the expanded forms given by the prescription in section \ref{sec_proposal}.
The twist limits of both sides of the dualities agree,
and they are reproduced by the closed form (\ref{Nahm_closed}) and the formulas (\ref{SQCD_rv_Higgs}), (\ref{Schur_Fermi}), (\ref{necklace_Fermi}), (\ref{USp_rv_closed}) and (\ref{USp_Fermi}).
For $T[SU(N)]$ we have compared (\ref{TSUN_closed}) with the H-twist limit of (\ref{TSUN_simple}) for $N=3$ and $4$.
For (\ref{USp_fw_dual}) with $l=4$ the H-twist limits of both sides agree up to $q^4$ for $N=1$ and up to $q^2$ for $N=2$,
and (\ref{USp_fw_closed}) agrees with the H-twist limit of (\ref{USp_fw_Nahm}) up to $q^3$ for $N=1$ and up to $q^{\frac32}$ for $N=2$.
Here the twist limits are extracted from the exact dependence on $t$ at each order in $q^{\frac14}$.
We have also compared (\ref{SQCD_fw_closed}) for $N\le4$ and (\ref{ADHM_closed_Fermi}) for $N\le3$ and $l\le3$
with the H-twist limits of the Neumann half-indices (\ref{SQCD_fw_Neu}) and (\ref{ADHM_Neu}),
which we evaluate numerically at fixed $q$ as contour integrals.
The flavored Schur indices of the necklace quiver gauge theories reconstructed from (\ref{necklace_Fermi})
agree with the Fermi gas formula in \cite{Bourdier:2015sga} and with the direct evaluation of the contour integrals.
The right-hand side of (\ref{Dquiver_4d}), with the Schur indices evaluated as contour integrals, agrees with (\ref{USp_Fermi}) for $N=1$ with $l=4$ and for $N=2$ with $l=3$.

The explicit forms of $\mathcal{Q}^{\mathcal{T}}$ in section \ref{sec_qmod_examples} are compared with the direct evaluation of the sums in (\ref{QT_def}).
We have checked (\ref{SQCD_C_Hankel}) by expanding (\ref{SQCD_CauchyBinet}) for $N\le7$ and against the direct evaluation for $N\le5$,
(\ref{SQCD_H_det}) for $N\le6$,
(\ref{SQCD_nonbal_even}) and (\ref{SQCD_nonbal_odd}) for $(N,N_f)=(2,5),\cdots,(2,8)$, $(3,7)$, $(3,8)$, $(3,9)$, $(4,9)$ and $(4,10)$ up to $q^{17}$,
(\ref{TSUN_det}), (\ref{TSUN_blocks}) and (\ref{TSUN_blocks_even}) for $N\le7$,
the expressions for the ADHM theory and the necklace quiver up to $q^{20}$,
those for the $USp(2N)$ ADHM theory with $N\le3$ up to $q^{20}$,
and (\ref{USp_H_N1}) up to $q^{20}$.
The vanishing of the sums over the end nodes for $N=2$ and over $p^{(0)}$ for $N=3$ is checked numerically at fixed $q$.
For $N=1$ the H-twist limit of the unflavored Neumann half-index (\ref{USp_fw_Neu}) agrees with $(q^{\frac12};q)_{\infty}^8$ times (\ref{Schur_SU2_4}) up to $q^{14}$.
The closed forms (\ref{D4_Schur2}) and (\ref{D4_Schur3}) satisfy the modular differential equations of \cite{Beem:2019tfp} as polynomial identities in $E_2$, $\theta_2^4$ and $\theta_3^4$,
and agree with the unflavored Schur indices of $USp(4)$ and $USp(6)$ gauge theories with the antisymmetric and four fundamental hypermultiplets, evaluated as contour integrals, up to $q^{\frac{47}{2}}$ and $q^{\frac{39}{2}}$.
For $N=2$ the H-twist limit of (\ref{USp_fw_Neu}) divided by the Fermi multiplets agrees with (\ref{D4_Schur2}) up to $q^{\frac{13}{2}}$.

We test the Cartan fugacities of the lattice VOAs in section \ref{sec_VOA_examples} as follows.
Let $b$ be a candidate for the Cartan fugacity, given by a monomial in the fugacities as in (\ref{xi_SQCD}).
We write the vacuum character as a function of $b$ with the other independent combinations of the fugacities fixed,
and expand it as $\chi=\sum_{k\in\mathbb{Z}}c_k(q)\,b^k$, where $k$ is the charge under the Cartan generator.
If the VOA contains the lattice VOA of level $K$ with the Cartan fugacity $b$,
it decomposes into the $K$ modules of the lattice VOA labeled by $k$ mod $K$ times the modules of the commutant,
and the states of charge $k$ have the weights $\frac{k^2}{2K}$ plus integers or half-integers determined by $k$ mod $K$.
Hence $(\pm1)^kq^{-\frac{k^2}{2K}}c_k(q)$ depends only on $k$ mod $K$, where the sign reflects the statistics of the generator.
We evaluate $c_k(q)$ at fixed $q$ as the Fourier coefficients in $b$ and confirm this periodicity.

\section{Series expansions}
\label{app_series}
We show the first several terms of the $q$-series expansions of the half-indices of the dual pairs.
We set all the fugacities except for $t$ to unity, so that the expansions do not depend on the conventions for the fugacities.
Then the coefficient of $q^{\frac{n}{4}}$ is a Laurent polynomial in $t$ with integer coefficients
whose powers range from $-n$ to $n$ and have the same parity as $n$.
Since the half-indices depend on $t$ only through $q^{\frac14}t^{-1}$ or $q^{\frac14}t$,
only the non-positive powers of $t$ appear for the Nahm pole boundary conditions of $U(N)$ SQCD and of the $USp(2N)$ ADHM theory, and only the non-negative powers appear for the others.
The expansions are common to the two sides of each dual pair, and we label them by both sides.
All the coefficients are confirmed exactly as in Appendix \ref{app_numerics}.
In Appendix \ref{app_vac} we also list the unflavored characters in section \ref{sec_VOA_qmod}.
\renewcommand{\arraystretch}{1.15}
\setlength{\arraycolsep}{3pt}

\subsection{\texorpdfstring{$U(N)$ SQCD}{U(N) SQCD}}
For the duality (\ref{SQCD_fw_dual}) of $U(N)$ SQCD with $2N$ flavors we find
{\small\setlength{\tabcolsep}{3pt}
\begin{longtable}{>{$}c<{$}|>{$}l<{$}}
\textrm{half-index}&\textrm{expansion}\\ \hline
\endhead
\mathbb{II}^{U(2)+4}_{\mathcal{N}',N'+\Gamma+\eta}
&1+q^{\frac12}(-4+15t^2)+q(9-84t^2+104t^4)\\*
=\mathbb{II}^{\wtilde{(1)-(2)-(1)}}_{\mathcal{D},\mathrm{Nahm},\mathcal{D}}
&+q^{\frac32}(-20+247t^2-672t^4+475t^6)\\*
&+q^2(39-556t^2+2244t^4-3300t^6+1659t^8)\\*
&+q^{\frac52}(-68+1129t^2-5624t^4+11840t^6-12012t^8+4802t^{10})\\*
&+\mathcal{O}(q^{\frac{11}{4}})\\ \hline
\mathbb{II}^{U(3)+6}_{\mathcal{N}',N'+\Gamma+\eta}
&1+q^{\frac12}(-12+35t^2)+q(69-420t^2+594t^4)-80q^{\frac54}t^3\\*
=\mathbb{II}^{\wtilde{(1)-(2)-(3)-(2)-(1)}}_{\mathcal{D},\mathrm{Nahm}^{3},\mathcal{D}}
&+q^{\frac32}(-268+2451t^2-7128t^4+6545t^6)+q^{\frac74}(960t^3-2160t^5)\\*
&+q^2(841-9812t^2+42210t^4-78540t^6+53129t^8)\\*
&+q^{\frac94}(-5440t^3+25920t^5-29008t^7)\\*
&+q^{\frac52}(-2340+31955t^2-173880t^4+472290t^6-637548t^8\\*
&\qquad+341810t^{10})+\mathcal{O}(q^{\frac{11}{4}})\\ \hline
\mathbb{II}^{U(4)+8}_{\mathcal{N}',N'+\Gamma+\eta}
&1+q^{\frac12}(-24+63t^2)+q(279-1512t^2+1952t^4)\\*
=\mathbb{II}^{\wtilde{(1)-(2)-(3)-(4)-(3)-(2)-(1)}}_{\mathcal{D},\mathrm{Nahm}^{5},\mathcal{D}}
&+q^{\frac32}(-2120+17641t^2-47128t^4+39711t^6)\\*
&+q^2(12034-135096t^2+555296t^4-967400t^6+597617t^8)\\*
&+q^{\frac52}(-55392+776062t^2-4311312t^4+11544393t^6\\*
&\qquad-14707200t^8+7109907t^{10})+\mathcal{O}(q^{\frac{11}{4}})\\
\end{longtable}}
and for the duality (\ref{SQCD_rv_dual}) we find
{\small\setlength{\tabcolsep}{3pt}
\begin{longtable}{>{$}c<{$}|>{$}l<{$}}
\textrm{half-index}&\textrm{expansion}\\ \hline
\endhead
\mathbb{II}^{U(2)+4}_{\mathrm{Nahm}'}
&1+3q^{\frac12}t^{-2}+q(1-16t^{-2}+9t^{-4})\\*
=\mathbb{II}^{\wtilde{(1)-(2)-(1)}}_{\mathcal{N},N+\tilde{\eta}}
&+q^{\frac32}(19t^{-2}-48t^{-4}+18t^{-6})\\*
&+q^2(4-48t^{-2}+157t^{-4}-128t^{-6}+35t^{-8})\\*
&+q^{\frac52}(44t^{-2}-288t^{-4}+395t^{-6}-240t^{-8}+57t^{-10})\\*
&+\mathcal{O}(q^{\frac{11}{4}})\\ \hline
\mathbb{II}^{U(3)+6}_{\mathrm{Nahm}'}
&1+3q^{\frac12}t^{-2}+q(1+9t^{-4})-24q^{\frac54}t^{-3}+q^{\frac32}(7t^{-2}+22t^{-6})\\*
=\mathbb{II}^{\wtilde{(1)-(2)-(3)-(2)-(1)}}_{\mathcal{N},N+\tilde{\eta}}
&-72q^{\frac74}t^{-5}+q^2(2+55t^{-4}+47t^{-8})+q^{\frac94}(-48t^{-3}-216t^{-7})\\*
&+q^{\frac52}(14t^{-2}+394t^{-6}+92t^{-10})+\mathcal{O}(q^{\frac{11}{4}})\\ \hline
\mathbb{II}^{U(4)+8}_{\mathrm{Nahm}'}
&1+3q^{\frac12}t^{-2}+q(1+9t^{-4})+q^{\frac32}(7t^{-2}-32t^{-4}+22t^{-6})\\*
=\mathbb{II}^{\wtilde{(1)-(2)-(3)-(4)-(3)-(2)-(1)}}_{\mathcal{N},N+\tilde{\eta}}
&+q^2(2+25t^{-4}-96t^{-6}+51t^{-8})\\*
&+q^{\frac52}(14t^{-2}-64t^{-4}+130t^{-6}-288t^{-8}+104t^{-10})\\*
&+\mathcal{O}(q^{\frac{11}{4}})\\
\end{longtable}}
For the dualities (\ref{SQCD_nonbal_fw_dual}) and (\ref{SQCD_nonbal_rv_dual}) of $U(N)$ SQCD with $N_f>2N$ flavors we find
{\small\setlength{\tabcolsep}{3pt}
\begin{longtable}{>{$}c<{$}|>{$}l<{$}}
\textrm{half-index}&\textrm{expansion}\\ \hline
\endhead
\mathbb{II}^{U(2)+5}_{\mathcal{N}',N'+\Gamma+\eta}
&1+q^{\frac12}(-4+24t^2)-10q^{\frac34}t+q(10-136t^2+275t^4)\\*
=\mathbb{II}^{\wtilde{(1)-(2)-(2)-(1)}}_{\mathcal{D},\mathrm{Nahm}^{2},\mathcal{D}}
&+q^{\frac54}(60t-230t^3)+q^{\frac32}(-28+469t^2-1800t^4+2024t^6)\\*
&+q^{\frac74}(-190t+1540t^3-2420t^5)\\*
&+q^2(66-1368t^2+7102t^4-14216t^6+10976t^8)\\*
&+q^{\frac94}(460t-5430t^3+17160t^5-16380t^7)\\*
&+q^{\frac52}(-128+3458t^2-22708t^4+60593t^6-80164t^8+47552t^{10})\\*
&+\mathcal{O}(q^{\frac{11}{4}})\\ \hline
\mathbb{II}^{U(3)+7}_{\mathcal{N}',N'+\Gamma+\eta}
&1+q^{\frac12}(-12+48t^2)-14q^{\frac34}t+q(70-576t^2+1127t^4)\\*
=\mathbb{II}^{\wtilde{(1)-(2)-(3)-(3)-(2)-(1)}}_{\mathcal{D},\mathrm{Nahm}^{4},\mathcal{D}}
&+q^{\frac54}(168t-798t^3)+q^{\frac32}(-280+3583t^2-13524t^4+17296t^6)\\*
&+q^{\frac74}(-1022t+9576t^3-20412t^5)\\*
&+q^2(916-16116t^2+89994t^4-207552t^6+195707t^8)\\*
&+q^{\frac94}(4424t-59178t^3+244944t^5-325276t^7)\\*
&+q^{\frac52}(-2680+59850t^2-448808t^4+1470530t^6-2348484t^8\\*
&\qquad+1747584t^{10})+\mathcal{O}(q^{\frac{11}{4}})\\
\end{longtable}}
and
{\small\setlength{\tabcolsep}{3pt}
\begin{longtable}{>{$}c<{$}|>{$}l<{$}}
\textrm{half-index}&\textrm{expansion}\\ \hline
\endhead
\mathbb{II}^{U(2)+5}_{\mathrm{Nahm}'}
&1+q^{\frac12}t^{-2}+2q^{\frac34}t^{-3}+q(1-10t^{-2}+2t^{-4})+q^{\frac54}(-10t^{-3}+4t^{-5})\\*
=\mathbb{II}^{\wtilde{(1)-(2)-(2)-(1)}}_{\mathcal{N},N+\tilde{\eta}}
&+q^{\frac32}(3t^{-2}-10t^{-4}+5t^{-6})+q^{\frac74}(24t^{-3}-30t^{-5}+6t^{-7})\\*
&+q^2(2-20t^{-2}+50t^{-4}-40t^{-6}+9t^{-8})\\*
&+q^{\frac94}(-40t^{-3}+100t^{-5}-50t^{-7}+12t^{-9})\\*
&+q^{\frac52}(-2+16t^{-2}-60t^{-4}+143t^{-6}-80t^{-8}+14t^{-10})+\mathcal{O}(q^{\frac{11}{4}})\\ \hline
\mathbb{II}^{U(3)+7}_{\mathrm{Nahm}'}
&1+q^{\frac12}t^{-2}+2q^{\frac34}t^{-3}+q(1+2t^{-4})+q^{\frac54}(-14t^{-3}+4t^{-5})\\*
=\mathbb{II}^{\wtilde{(1)-(2)-(3)-(3)-(2)-(1)}}_{\mathcal{N},N+\tilde{\eta}}
&+q^{\frac32}(3t^{-2}-14t^{-4}+6t^{-6})+q^{\frac74}(4t^{-3}-14t^{-5}+8t^{-7})\\*
&+q^2(2+6t^{-4}-42t^{-6}+11t^{-8})\\*
&+q^{\frac94}(-28t^{-3}+54t^{-5}-56t^{-7}+16t^{-9})\\*
&+q^{\frac52}(6t^{-2}-28t^{-4}+108t^{-6}-84t^{-8}+21t^{-10})+\mathcal{O}(q^{\frac{11}{4}})\\
\end{longtable}}

\subsection{\texorpdfstring{$T[SU(N)]$}{T[SU(N)]}}
For the duality (\ref{TSUN_dual}) we find
{\small\setlength{\tabcolsep}{3pt}
\begin{longtable}{>{$}c<{$}|>{$}l<{$}}
\textrm{half-index}&\textrm{expansion}\\ \hline
\endhead
\mathbb{II}^{T[SU(3)]}_{\mathcal{N}',N'+\Gamma+\eta}
&1+q^{\frac12}(-2+8t^2)+q(2-34t^2+35t^4)\\*
=\mathbb{II}^{\wtilde{T[SU(3)]}}_{\mathcal{D},\mathrm{Nahm}}
&+q^{\frac32}(-6+79t^2-196t^4+111t^6)\\*
&+q^2(11-156t^2+584t^4-708t^6+286t^8)\\*
&+q^{\frac52}(-14+277t^2-1346t^4+2362t^6-1958t^8+637t^{10})\\*
&+\mathcal{O}(q^{\frac{11}{4}})\\ \hline
\mathbb{II}^{T[SU(4)]}_{\mathcal{N}',N'+\Gamma+\eta}
&1+q^{\frac12}(-6+15t^2)+q(16-90t^2+119t^4)-32q^{\frac54}t^3\\*
=\mathbb{II}^{\wtilde{T[SU(4)]}}_{\mathcal{D},\mathrm{Nahm}^{2}}
&+q^{\frac32}(-32+256t^2-714t^4+664t^6)+q^{\frac74}(192t^3-448t^5)\\*
&+q^2(70-576t^2+2216t^4-3984t^6+2924t^8)\\*
&+q^{\frac94}(-576t^3+2688t^5-3328t^7)\\*
&+q^{\frac52}(-152+1322t^2-5680t^4+13761t^6-17544t^8+10814t^{10})\\*
&+\mathcal{O}(q^{\frac{11}{4}})\\
\end{longtable}}

\subsection{\texorpdfstring{$U(N)$ ADHM theory}{U(N) ADHM theory}}
For the duality (\ref{ADHM1_dual}) we find
{\small\setlength{\tabcolsep}{3pt}
\begin{longtable}{>{$}c<{$}|>{$}l<{$}}
\textrm{half-index}&\textrm{expansion}\\ \hline
\endhead
\mathbb{II}^{U(2)\,\textrm{ADHM-}[1]}_{\mathcal{N}',N',N'+\Gamma+\eta}
&1+2q^{\frac14}t+q^{\frac12}(-2+6t^2)+q^{\frac34}(-6t+10t^3)+q(1-20t^2+19t^4)\\*
=\mathbb{II}^{\wtilde{U(2)\,\textrm{ADHM-}[1]}}_{\mathrm{Nahm}}
&+q^{\frac54}(8t-40t^3+28t^5)+q^{\frac32}(-2+31t^2-78t^4+44t^6)\\*
&+q^{\frac74}(-12t+80t^3-126t^5+60t^7)\\*
&+q^2(4-50t^2+164t^4-200t^6+85t^8)\\*
&+q^{\frac94}(20t-144t^3+294t^5-288t^7+110t^9)\\*
&+q^{\frac52}(-4+88t^2-320t^4+477t^6-410t^8+146t^{10})+\mathcal{O}(q^{\frac{11}{4}})\\ \hline
\mathbb{II}^{U(3)\,\textrm{ADHM-}[1]}_{\mathcal{N}',N',N'+\Gamma+\eta}
&1+2q^{\frac14}t+q^{\frac12}(-2+6t^2)+q^{\frac34}(-6t+14t^3)+q(1-20t^2+28t^4)\\*
=\mathbb{II}^{\wtilde{U(3)\,\textrm{ADHM-}[1]}}_{\mathrm{Nahm}}
&+q^{\frac54}(8t-54t^3+52t^5)+q^{\frac32}(-2+31t^2-120t^4+93t^6)\\*
&+q^{\frac74}(-12t+102t^3-244t^5+152t^7)\\*
&+q^2(4-50t^2+260t^4-458t^6+242t^8)\\*
&+q^{\frac94}(20t-176t^3+588t^5-796t^7+370t^9)\\*
&+q^{\frac52}(-4+88t^2-492t^4+1182t^6-1316t^8+546t^{10})+\mathcal{O}(q^{\frac{11}{4}})\\ \hline
\mathbb{II}^{U(4)\,\textrm{ADHM-}[1]}_{\mathcal{N}',N',N'+\Gamma+\eta}
&1+2q^{\frac14}t+q^{\frac12}(-2+6t^2)+q^{\frac34}(-6t+14t^3)+q(1-20t^2+33t^4)\\*
=\mathbb{II}^{\wtilde{U(4)\,\textrm{ADHM-}[1]}}_{\mathrm{Nahm}}
&+q^{\frac54}(8t-54t^3+64t^5)+q^{\frac32}(-2+31t^2-138t^4+127t^6)\\*
&+q^{\frac74}(-12t+102t^3-302t^5+228t^7)\\*
&+q^2(4-50t^2+289t^4-634t^6+404t^8)\\*
&+q^{\frac94}(20t-176t^3+724t^5-1228t^7+672t^9)\\*
&+q^{\frac52}(-4+88t^2-534t^4+1638t^6-2276t^8+1100t^{10})+\mathcal{O}(q^{\frac{11}{4}})\\
\end{longtable}}
For the duality (\ref{ADHM_dual}) with $l=2,3$ we find
{\small\setlength{\tabcolsep}{3pt}
\begin{longtable}{>{$}c<{$}|>{$}l<{$}}
\textrm{half-index}&\textrm{expansion}\\ \hline
\endhead
\mathbb{II}^{U(2)\,\textrm{ADHM-}[2]}_{\mathcal{N}',N',N'+\Gamma+\eta}
&1+2q^{\frac14}t+q^{\frac12}(-2+9t^2)+q^{\frac34}(-12t+22t^3)+q(4-50t^2+55t^4)\\*
=\mathbb{II}^{\wtilde{U(2)^2}}_{\mathrm{Nahm}^2}
&+q^{\frac54}(32t-148t^3+116t^5)+q^{\frac32}(-8+151t^2-398t^4+242t^6)\\*
&+q^{\frac74}(-72t+530t^3-912t^5+448t^7)\\*
&+q^2(16-366t^2+1485t^4-1940t^6+820t^8)\\*
&+q^{\frac94}(142t-1380t^3+3662t^5-3800t^7+1400t^9)\\*
&+q^{\frac52}(-28+787t^2-4138t^4+8057t^6-7048t^8+2334t^{10})\\*
&+\mathcal{O}(q^{\frac{11}{4}})\\ \hline
\mathbb{II}^{U(3)\,\textrm{ADHM-}[2]}_{\mathcal{N}',N',N'+\Gamma+\eta}
&1+2q^{\frac14}t+q^{\frac12}(-2+9t^2)+q^{\frac34}(-12t+26t^3)+q(4-50t^2+73t^4)\\*
=\mathbb{II}^{\wtilde{U(3)^2}}_{\mathrm{Nahm}^2}
&+q^{\frac54}(32t-180t^3+178t^5)+q^{\frac32}(-8+160t^2-546t^4+430t^6)\\*
&+q^{\frac74}(-72t+660t^3-1516t^5+940t^7)\\*
&+q^2(16-416t^2+2231t^4-3820t^6+1998t^8)\\*
&+q^{\frac94}(152t-1840t^3+6684t^5-8992t^7+4008t^9)\\*
&+q^{\frac52}(-28+949t^2-6782t^4+17991t^6-19852t^8+7788t^{10})\\*
&+\mathcal{O}(q^{\frac{11}{4}})\\ \hline
\mathbb{II}^{U(2)\,\textrm{ADHM-}[3]}_{\mathcal{N}',N',N'+\Gamma+\eta}
&1+2q^{\frac14}t+q^{\frac12}(-2+14t^2)+q^{\frac34}(-22t+42t^3)\\*
=\mathbb{II}^{\wtilde{U(2)^3}}_{\mathrm{Nahm}^3}
&+q(9-100t^2+135t^4)+q^{\frac54}(72t-408t^3+370t^5)\\*
&+q^{\frac32}(-18+471t^2-1386t^4+960t^6)\\*
&+q^{\frac74}(-252t+2040t^3-4070t^5+2250t^7)\\*
&+q^2(56-1518t^2+7336t^4-10920t^6+5082t^8)\\*
&+q^{\frac94}(598t-7288t^3+22906t^5-26856t^7+10702t^9)\\*
&+q^{\frac52}(-108+4260t^2-27532t^4+63454t^6-61572t^8+21672t^{10})\\*
&+\mathcal{O}(q^{\frac{11}{4}})\\
\end{longtable}}

\subsection{\texorpdfstring{$USp(2N)$ ADHM theory}{USp(2N) ADHM theory}}
For the duality (\ref{USp2_fw_dual}) we find
{\small\setlength{\tabcolsep}{3pt}
\begin{longtable}{>{$}c<{$}|>{$}l<{$}}
\textrm{half-index}&\textrm{expansion}\\ \hline
\endhead
\mathbb{II}^{USp(2)\,\textrm{ADHM-}[8]}_{\mathcal{N}',N',N'+\Gamma+\eta}
&1+2q^{\frac14}t+q^{\frac12}(-10+31t^2)+q^{\frac34}(-20t+60t^3)\\*
=\mathbb{II}^{\wtilde{U(1)^4\times U(2)}}_{\mathrm{Nahm}^{5}}
&+q(45-310t^2+389t^4)+q^{\frac54}(92t-600t^3+718t^5)\\*
&+q^{\frac32}(-130+1428t^2-3890t^4+2972t^6)\\*
&+q^{\frac74}(-280t+2820t^3-7180t^5+5226t^7)\\*
&+q^2(310-4360t^2+18390t^4-29720t^6+16398t^8)+\mathcal{O}(q^{\frac94})\\
\end{longtable}}
\noindent For the dualities (\ref{USp2_rv_dual}) and (\ref{USp_rv_dual}) we find
{\small\setlength{\tabcolsep}{3pt}
\begin{longtable}{>{$}c<{$}|>{$}l<{$}}
\textrm{half-index}&\textrm{expansion}\\ \hline
\endhead
\mathbb{II}^{USp(2)\,\textrm{ADHM-}[6]}_{\mathrm{Nahm}'}
&1+q^{\frac12}t^{-2}-2q^{\frac34}t^{-1}+q(-12t^{-2}+3t^{-4})-2q^{\frac54}t^{-3}\\*
=\mathbb{II}^{\wtilde{U(1)^4}}_{\mathcal{N},N,N+\tilde{\Gamma}+\tilde{\eta}}
&+q^{\frac32}(18t^{-2}-12t^{-4}+3t^{-6})+q^{\frac74}(-2t^{-1}+24t^{-3}-6t^{-5})\\*
&+q^2(-32t^{-2}+50t^{-4}-24t^{-6}+5t^{-8})+q^{\frac94}(-36t^{-3}+24t^{-5}-6t^{-7})\\*
&+q^{\frac52}(37t^{-2}-88t^{-4}+54t^{-6}-24t^{-8}+5t^{-10})+\mathcal{O}(q^{\frac{11}{4}})\\ \hline
\mathbb{II}^{USp(2)\,\textrm{ADHM-}[8]}_{\mathrm{Nahm}'}
&1-2q^{\frac34}t^{-1}+q(-8t^{-2}+2t^{-4})+q^{\frac32}(2t^{-2}-8t^{-4}+t^{-6})\\*
=\mathbb{II}^{\wtilde{U(1)^4\times U(2)}}_{\mathcal{N},N,N+\tilde{\Gamma}+\tilde{\eta}}
&+q^{\frac74}(-2t^{-1}+16t^{-3}-4t^{-5})+q^2(-8t^{-2}+58t^{-4}-16t^{-6}+3t^{-8})\\*
&+q^{\frac94}(-2t^{-3}+16t^{-5}-2t^{-7})\\*
&+q^{\frac52}(5t^{-2}-80t^{-4}+33t^{-6}-16t^{-8}+2t^{-10})+\mathcal{O}(q^{\frac{11}{4}})\\ \hline
\mathbb{II}^{USp(2)\,\textrm{ADHM-}[10]}_{\mathrm{Nahm}'}
&1-2q^{\frac34}t^{-1}+q(-10t^{-2}+t^{-4})+q^{\frac32}(2t^{-2}+t^{-6})\\*
=\mathbb{II}^{\wtilde{U(1)^4\times U(2)^2}}_{\mathcal{N},N,N+\tilde{\Gamma}+\tilde{\eta}}
&+q^{\frac74}(-2t^{-1}+20t^{-3}-2t^{-5})+q^2(-10t^{-2}+46t^{-4}-20t^{-6}+2t^{-8})\\*
&+q^{\frac94}(-2t^{-3}-2t^{-7})\\*
&+q^{\frac52}(5t^{-2}-20t^{-4}+48t^{-6}-10t^{-8}+t^{-10})+\mathcal{O}(q^{\frac{11}{4}})\\ \hline
\mathbb{II}^{USp(4)\,\textrm{ADHM-}[6]}_{\mathrm{Nahm}'}
&1+q^{\frac12}t^{-2}-2q^{\frac34}t^{-1}+4qt^{-4}-4q^{\frac54}t^{-3}\\*
=\mathbb{II}^{\wtilde{U(2)^4}}_{\mathcal{N},N,N+\tilde{\Gamma}+\tilde{\eta}}
&+q^{\frac32}(3t^{-2}-18t^{-4}+6t^{-6})+q^{\frac74}(-2t^{-1}-14t^{-5})\\*
&+q^2(26t^{-4}-24t^{-6}+14t^{-8})+q^{\frac94}(-8t^{-3}+48t^{-5}-24t^{-7})\\*
&+q^{\frac52}(6t^{-2}-38t^{-4}+78t^{-6}-66t^{-8}+19t^{-10})+\mathcal{O}(q^{\frac{11}{4}})\\
\end{longtable}}

\subsection{Unflavored characters}
\label{app_vac}
We list the $q$-series of the unflavored characters (\ref{chi_unflavored}) of the non-Abelian theories obtained from the formulas in section \ref{sec_qmod_examples}.
The twist limits $t\rightarrow q^{\pm\frac14}$ of the expansions above for these theories agree with them up to the orders shown in the tables.

\paragraph{$U(N)$ SQCD}
From (\ref{SQCD_C_Hankel}) we find
{\small
\begin{align*}
\begin{array}{c|l}
N&\chi[\mathsf{V}^{U(N)+2N}_{\mathrm{Nahm}'}]\\ \hline
2&1+4q-16q^{\frac{3}{2}}+32q^{2}-96q^{\frac{5}{2}}+224q^{3}-512q^{\frac{7}{2}}+1132q^{4}-2336q^{\frac{9}{2}}\\
&+4776q^{5}-9456q^{\frac{11}{2}}+18304q^{6}+\cdots\\ \hline
3&1+4q-6q^{2}-24q^{3}-27q^{4}+48q^{5}+180q^{6}+\cdots\\ \hline
4&1+4q+18q^{2}-32q^{\frac{5}{2}}+64q^{3}-160q^{\frac{7}{2}}+273q^{4}-736q^{\frac{9}{2}}+1364q^{5}\\
&-3040q^{\frac{11}{2}}+6058q^{6}+\cdots\\ \hline
5&1+4q+18q^{2}+24q^{3}+15q^{4}-168q^{5}-422q^{6}+\cdots\\ \hline
6&1+4q+18q^{2}+64q^{3}-48q^{\frac{7}{2}}+215q^{4}-240q^{\frac{9}{2}}+660q^{5}-1104q^{\frac{11}{2}}\\
&+2072q^{6}+\cdots\\ \hline
7&1+4q+18q^{2}+64q^{3}+159q^{4}+380q^{5}+650q^{6}+\cdots\\ \hline
8&1+4q+18q^{2}+64q^{3}+215q^{4}-64q^{\frac{9}{2}}+660q^{5}-320q^{\frac{11}{2}}+1938q^{6}+\cdots\\
\end{array}
\end{align*}}
The coefficient $4$ of $q$ in all the characters is $\chi_{\mathbf{3}}(z)+1$ at $z=1$, where $\chi_{\mathbf{3}}(z)$ is the character of the adjoint representation of the $\mathfrak{sl}(2)$ symmetry.
We have confirmed the $\mathfrak{sl}(2)$ currents for $N=2,\cdots,5$, although only $z_2/z_1$ appears in (\ref{SQCD_rv_simple}).

\paragraph{Mirror quiver}
From (\ref{SQCD_H_Hankel}) we find
{\small
\begin{align*}
\begin{array}{c|l}
N&\chi[\mathsf{V}^{\wtilde{(1)-(2)-\cdots-(N)-\cdots-(2)-(1)}}_{\mathcal{D},\mathrm{Nahm}^{2N-3},\mathcal{D}}]\\ \hline
2&1-4q^{\frac{1}{2}}+24q-104q^{\frac{3}{2}}+390q^{2}-1296q^{\frac{5}{2}}+3968q^{3}-11344q^{\frac{7}{2}}+30573q^{4}+\cdots\\ \hline
3&1-12q^{\frac{1}{2}}+104q-688q^{\frac{3}{2}}+3806q^{2}-18320q^{\frac{5}{2}}+79072q^{3}-312128q^{\frac{7}{2}}\\
&+1143653q^{4}+\cdots\\ \hline
4&1-24q^{\frac{1}{2}}+342q-3632q^{\frac{3}{2}}+31627q^{2}-237616q^{\frac{5}{2}}+1589810q^{3}-9675840q^{\frac{7}{2}}\\
&+54381453q^{4}+\cdots\\ \hline
5&1-40q^{\frac{1}{2}}+882q-14000q^{\frac{3}{2}}+177801q^{2}-1911128q^{\frac{5}{2}}+17997714q^{3}\\
&-151994576q^{\frac{7}{2}}+1170459648q^{4}+\cdots\\
\end{array}
\end{align*}}
By (\ref{SQCD_fw_dual}) they are $(q^{\frac12};q)_{\infty}^{2N(N-1)}\chi[\mathsf{V}^{U(N)+2N}_{\mathcal{N}',N'+\eta}]$,
where the first factor comes from the $N(N-1)$ Fermi multiplets $\Gamma$.
With the Cartan fugacity $b$ in (\ref{xi_SQCD}) the character $\chi[\mathsf{V}^{U(N)+2N}_{\mathcal{N}',N'+\eta}]$ decomposes as
\begin{align}
\chi[\mathsf{V}^{U(N)+2N}_{\mathcal{N}',N'+\eta}](b)&=\chi^{\mathfrak{su}(2)_1}_0(b)A_N(q)+\chi^{\mathfrak{su}(2)_1}_{\frac12}(b)B_N(q),
\\*
\chi^{\mathfrak{su}(2)_1}_j(b)&=\frac{1}{(q)_{\infty}}\sum_{n\in\mathbb{Z}+j}q^{n^2}b^{2n}.
\end{align}
We find
{\small
\begin{align*}
\begin{array}{c|l}
&\textrm{expansion}\\ \hline
\chi[\mathsf{V}^{U(2)+4}_{\mathcal{N}',N'+\eta}]&1+18q-24q^{\frac32}+169q^2-304q^{\frac52}+1202q^3-2392q^{\frac72}+7048q^4\\
&-14256q^{\frac92}+35708q^5-71528q^{\frac{11}{2}}+160961q^6+\cdots\\ \hline
A_2&1+15q+120q^2+775q^3+4125q^4+19179q^5+79625q^6+\cdots\\ \hline
B_2&-12q^{\frac54}-140q^{\frac94}-1020q^{\frac{13}{4}}-5640q^{\frac{17}{4}}-26420q^{\frac{21}{4}}+\cdots\\ \hline
\chi[\mathsf{V}^{U(3)+6}_{\mathcal{N}',N'+\eta}]&1+38q+659q^2+7522q^3+\cdots\\ \hline
A_3&1+35q+630q^2+7805q^3+75740q^4+615174q^5+\cdots\\ \hline
B_3&-40q^{\frac74}-1120q^{\frac{11}{4}}-17024q^{\frac{15}{4}}-185080q^{\frac{19}{4}}+\cdots\\
\end{array}
\end{align*}}
The coefficients of $q$ in $\chi[\mathsf{V}^{U(N)+2N}_{\mathcal{N}',N'+\eta}]$ are $\dim\mathfrak{sl}(2N)+3$.
For $N=2$ the odd states of weight $\frac32$ give $-24q^{\frac32}=-(6\times 2)\times 2\,q^{\frac32}$,
since they transform as $(\Lambda^2\mathbb{C}^{4},\mathbf{2})$ under $\mathfrak{sl}(4)\oplus\mathfrak{su}(2)_1$ and come with their conjugates.
The number of the states of weight $2$ in $A_N$ is smaller by $\dim\mathfrak{sl}(2N)$ than that of the universal affine VOA $V^{-N}(\mathfrak{sl}_{2N})$,
i.e. $135-15=120$ and $665-35=630$.
This indicates a null vector in the adjoint representation,
which is consistent with the Higgs branch $\overline{\mathcal{O}}_{[2^N]}\subset\mathfrak{sl}(2N)$, on which the square of the moment map vanishes.

\paragraph{Unbalanced SQCD}
From (\ref{SQCD_nonbal_C_S}) and (\ref{SQCD_nonbal_25})--(\ref{SQCD_nonbal_38}) we find
{\small
\begin{align*}
\begin{array}{c|l}
(N,N_f)&\chi[\mathsf{V}^{U(N)+N_f}_{\mathrm{Nahm}'}]\\ \hline
(2,5)&1+2q-8q^{\frac32}-3q^2-4q^{\frac52}+4q^3+20q^{\frac72}-8q^4+44q^{\frac92}+20q^5+\cdots\\ \hline
(2,6)&1+2q-12q^{\frac32}+9q^2-48q^{\frac52}+122q^3-232q^{\frac72}+598q^4-1200q^{\frac92}+\cdots\\ \hline
(2,7)&1+2q-14q^{\frac32}+7q^2-40q^{\frac52}+93q^3-92q^{\frac72}+291q^4-414q^{\frac92}+\cdots\\ \hline
(3,7)&1+2q+2q^{\frac32}-7q^2-6q^{\frac52}-21q^3-2q^{\frac72}-18q^4+10q^{\frac92}+\cdots\\ \hline
(3,8)&1+2q-7q^2-38q^3+14q^4+120q^5+339q^6+\cdots\\
\end{array}
\end{align*}}
By (\ref{SQCD_nonbal_H}) the characters for the mirror quivers are those for $N_f=2N$.

\paragraph{$T[SU(N)]$}
From (\ref{TSUN_blocks}) and (\ref{TSUN_blocks_even}) we find
{\small
\begin{align*}
\begin{array}{c|l}
N&\chi[\mathsf{V}^{\wtilde{T[SU(N)]}}_{\mathcal{D},\mathrm{Nahm}^{N-2}}]\\ \hline
3&1-2q^{\frac{1}{2}}+10q-40q^{\frac{3}{2}}+125q^{2}-366q^{\frac{5}{2}}+994q^{3}-2560q^{\frac{7}{2}}+6240q^{4}\\
&-14510q^{\frac{9}{2}}+32556q^{5}+\cdots\\ \hline
4&1-6q^{\frac{1}{2}}+31q-122q^{\frac{3}{2}}+413q^{2}-1250q^{\frac{5}{2}}+3466q^{3}-8968q^{\frac{7}{2}}+21912q^{4}\\
&-51026q^{\frac{9}{2}}+113931q^{5}+\cdots\\ \hline
5&1-12q^{\frac{1}{2}}+91q-532q^{\frac{3}{2}}+2639q^{2}-11648q^{\frac{5}{2}}+47048q^{3}-177060q^{\frac{7}{2}}\\
&+628722q^{4}-2126048q^{\frac{9}{2}}+6894355q^{5}+\cdots\\ \hline
6&1-20q^{\frac{1}{2}}+226q-1880q^{\frac{3}{2}}+12787q^{2}-75124q^{\frac{5}{2}}+394048q^{3}-1885984q^{\frac{7}{2}}\\
&+8363465q^{4}-34752288q^{\frac{9}{2}}+136478158q^{5}+\cdots\\
\end{array}
\end{align*}}
By (\ref{TSUN_dual}) they are $(q^{\frac12};q)_{\infty}^{(N-1)(N-2)}\chi[\mathsf{V}^{T[SU(N)]}_{\mathcal{N}',N'+\eta}]$,
where the first factor comes from the $\frac{(N-1)(N-2)}{2}$ Fermi multiplets $\Gamma$.
We find
{\small
\begin{align*}
\begin{array}{c|l}
N&\chi[\mathsf{V}^{T[SU(N)]}_{\mathcal{N}',N'+\eta}]\\ \hline
3&1+9q-20q^{\frac32}+72q^2-180q^{\frac52}+481q^3-1152q^{\frac72}+2673q^4+\cdots\\ \hline
4&1+16q+122q^2+648q^3+\cdots\\
\end{array}
\end{align*}}
For $N=3$ the character is expected to decompose as in \cite{Costello:2018fnz},
\begin{align}
\chi[\mathsf{V}^{T[SU(3)]}_{\mathcal{N}',N'+\eta}]&=\sum_{\lambda}\chi^{SU(3)}_{\lambda}(z)\chi^{\mathfrak{su}(3)_{-2}}_{\lambda}\chi^{\mathfrak{u}(1)_3}_{\lambda\ \textrm{mod}\ 3},
\end{align}
where $\chi^{SU(3)}_{\lambda}(z)$ is the character of the topological symmetry $SU(3)$,
and $\chi^{\mathfrak{su}(3)_{-2}}_{\lambda}$ and $\chi^{\mathfrak{u}(1)_3}_{\lambda\ \textrm{mod}\ 3}$ are the characters of the corresponding modules of $\mathfrak{su}(3)_{-2}$ and $\mathfrak{u}(1)_3$.

\paragraph{ADHM theory and necklace quiver}
From (\ref{necklace_qMZV}) we find
{\small
\begin{align*}
\begin{array}{c|l}
(l,N)&\chi[\mathsf{V}^{U(N)\,\textrm{ADHM-}[l]}_{\mathrm{Nahm}'}]\\ \hline
(1,2)&1+q-4q^{\frac{3}{2}}+2q^{2}-4q^{\frac{5}{2}}+3q^{3}-8q^{\frac{7}{2}}+14q^{4}-12q^{\frac{9}{2}}\\
&+16q^{5}-20q^{\frac{11}{2}}+29q^{6}+\cdots\\ \hline
(1,3)&1+q-3q^{2}-2q^{3}-5q^{4}+6q^{5}+\cdots\\ \hline
(1,4)&1+q+2q^{2}-6q^{\frac{5}{2}}+3q^{3}-6q^{\frac{7}{2}}+5q^{4}-12q^{\frac{9}{2}}+7q^{5}\\
&-18q^{\frac{11}{2}}+31q^{6}+\cdots\\ \hline
(1,5)&1+q+2q^{2}-4q^{3}-2q^{4}-7q^{5}-10q^{6}+\cdots\\ \hline
(2,2)&1+4q-12q^{\frac{3}{2}}+9q^{2}-24q^{\frac{5}{2}}+40q^{3}-72q^{\frac{7}{2}}+122q^{4}-144q^{\frac{9}{2}}+260q^{5}\\
&-400q^{\frac{11}{2}}+547q^{6}+\cdots\\ \hline
(2,3)&1+4q-7q^{2}-12q^{3}-19q^{4}+28q^{5}+62q^{6}+\cdots\\ \hline
(2,4)&1+4q+9q^{2}-20q^{\frac{5}{2}}+20q^{3}-40q^{\frac{7}{2}}+42q^{4}-120q^{\frac{9}{2}}+134q^{5}-240q^{\frac{11}{2}}\\
&+363q^{6}+\cdots\\ \hline
(2,5)&1+4q+9q^{2}-4q^{3}-6q^{4}-64q^{5}-64q^{6}+\cdots\\ \hline
(3,2)&1-6q^{\frac{3}{2}}-6q^{\frac{5}{2}}+16q^{3}-12q^{\frac{7}{2}}+51q^{4}-46q^{\frac{9}{2}}+66q^{5}\\
&-108q^{\frac{11}{2}}+125q^{6}+\cdots\\ \hline
(3,3)&1-6q^{2}-5q^{3}+3q^{4}+45q^{5}+\cdots\\ \hline
(3,4)&1-6q^{\frac{5}{2}}+q^{3}-6q^{\frac{7}{2}}-12q^{\frac{9}{2}}+15q^{5}-6q^{\frac{11}{2}}+53q^{6}+\cdots\\
\end{array}
\end{align*}}
By (\ref{ADHM_rv_dual}) they are also $\chi[\mathsf{V}^{\wtilde{U(N)^l}}_{\mathcal{N},N,N+\tilde{\Gamma}+\tilde{\eta}}]$.
By (\ref{ADHM_H_S}) the characters $\chi[\mathsf{V}^{\wtilde{U(N)^l}}_{\mathrm{Nahm}^l}]$ are those with $l=1$ for any $l$.
For $l=1$ they are the products of $(q^{\frac12};q)_{\infty}^2$ from the Fermi multiplet $\eta$
and the unflavored Schur indices of 4d $\mathcal{N}=4$ $U(N)$ SYM theory given by (\ref{Schur_Fermi}).

\paragraph{$USp(2N)$ ADHM theory}
From (\ref{USp_C_S}) we find
{\small
\begin{align*}
\begin{array}{c|l}
(l,N)&\chi[\mathsf{V}^{USp(2N)\,\textrm{ADHM-}[2l]}_{\mathrm{Nahm}'}]\\ \hline
(3,1)&1-q-12q^{\frac{3}{2}}+17q^{2}-20q^{\frac{5}{2}}+46q^{3}-44q^{\frac{7}{2}}+108q^{4}\\
&-232q^{\frac{9}{2}}+323q^{5}-516q^{\frac{11}{2}}+734q^{6}+\cdots\\ \hline
(4,1)&1-2q-8q^{\frac{3}{2}}+2q^{2}+56q^{3}-56q^{\frac{7}{2}}+79q^{4}\\
&-128q^{\frac{9}{2}}+110q^{5}-400q^{\frac{11}{2}}+670q^{6}+\cdots\\ \hline
(5,1)&1-2q-10q^{\frac{3}{2}}+q^{2}+10q^{\frac{5}{2}}+46q^{3}-10q^{\frac{7}{2}}+55q^{4}\\
&-240q^{\frac{9}{2}}+200q^{5}-472q^{\frac{11}{2}}+529q^{6}+\cdots\\ \hline
(3,2)&1-q+q^{2}-18q^{\frac{5}{2}}+14q^{3}-14q^{\frac{7}{2}}+51q^{4}\\
&-50q^{\frac{9}{2}}+44q^{5}-60q^{\frac{11}{2}}+217q^{6}+\cdots\\ \hline
(4,2)&1-2q+2q^{2}-8q^{\frac{5}{2}}-4q^{3}+7q^{4}\\
&+8q^{\frac{9}{2}}+80q^{5}-64q^{\frac{11}{2}}+26q^{6}+\cdots\\ \hline
(5,2)&1-2q+q^{2}-10q^{\frac{5}{2}}-q^{3}+10q^{\frac{7}{2}}+2q^{4}\\
&-10q^{\frac{9}{2}}+44q^{5}+40q^{\frac{11}{2}}+58q^{6}+\cdots\\ \hline
(3,3)&1-q+q^{2}-2q^{3}-24q^{\frac{7}{2}}+19q^{4}\\
&-8q^{\frac{9}{2}}+44q^{5}-56q^{\frac{11}{2}}+53q^{6}+\cdots\\
\end{array}
\end{align*}}
By (\ref{USp_rv_dual}) they are also $\chi[\mathsf{V}^{\wtilde{\hat{D}_l}}_{\mathcal{N},N,N+\tilde{\Gamma}+\tilde{\eta}}]$.

The characters $\chi[\mathsf{V}^{\wtilde{\hat{D}_l}}_{\mathrm{Nahm}^{l+1}}]$ are the same for all $l\ge4$.
For $N=1$ we find from (\ref{USp_H_N1})
\begin{align}
\chi[\mathsf{V}^{\wtilde{\hat{D}_l}}_{\mathrm{Nahm}^{l+1}}]\big|_{N=1}
&=1-8q^{\frac12}+56q-288q^{\frac32}+1247q^2-4712q^{\frac52}+16164q^3-51200q^{\frac72}
\nonumber\\*
&+151875q^4-426064q^{\frac92}+1139932q^5+\cdots,
\end{align}
which is $(q^{\frac12};q)_{\infty}^8$ times (\ref{Schur_SU2_4}).
For $N\ge2$ the characters vanish.
The unflavored Schur indices (\ref{Schur_SU2_4}), (\ref{D4_Schur2}) and (\ref{D4_Schur3}) of the rank-$N$ $D_4$ theories are
{\small
\begin{align*}
\begin{array}{c|l}
N&\mathcal{I}^{\textrm{Schur}}_{D_4,N}(q)\\ \hline
1&1+28q+329q^{2}+2632q^{3}+16380q^{4}+85764q^{5}+393589q^{6}+\cdots\\ \hline
2&1+31q+56q^{\frac{3}{2}}+527q^{2}+1526q^{\frac{5}{2}}+7292q^{3}+23002q^{\frac{7}{2}}+84406q^{4}\\
&+258818q^{\frac{9}{2}}+823883q^{5}+2394216q^{\frac{11}{2}}+6943434q^{6}+\cdots\\ \hline
3&1+31q+60q^{\frac{3}{2}}+612q^{2}+1920q^{\frac{5}{2}}+10568q^{3}+36968q^{\frac{7}{2}}+157850q^{4}\\
&+548848q^{\frac{9}{2}}+2036655q^{5}+6798456q^{\frac{11}{2}}+22993464q^{6}+\cdots\\
\end{array}
\end{align*}}
For $N=2$ the coefficients $31=\dim\mathfrak{so}(8)+\dim\mathfrak{su}(2)$ and $56=2\times28$ count the currents and the generators of weight $\frac32$ in the representation $(\mathbf{2},\mathbf{28})$ of $\mathfrak{su}(2)\oplus\mathfrak{so}(8)$ in \cite{Beem:2019tfp}.


\bibliographystyle{utphys}
\bibliography{ref}

\end{document}